\documentclass[final,5p,times,twocolumn,authoryear]{elsarticle}

\usepackage{amssymb}
\usepackage{lipsum}
\usepackage{hyperref}
\usepackage{pdfpages}
\usepackage{amsmath}
\usepackage{newtxtext,newtxmath}
\usepackage{xspace}
\usepackage{seqsplit}  
\usepackage{makeidx}
\usepackage{url}
\usepackage{listings}
\usepackage{xcolor}
\usepackage{caption}
\usepackage{multirow}
\usepackage{pifont}
\usepackage{cleveref}
\usepackage{pdfrender}
\newcommand{\msun}{M_\odot}
\newcommand{\mymathbf}[1]{{\mathbf #1}}
\newcommand{\mysymmathbf}[1]{{\boldsymbol #1}}
\newcommand{\myvec}[1]{{\mathbf #1}}
\newcommand{\mysymvec}[1]{{\boldsymbol #1}}
\newcommand{\mytens}[1]{{\mathbf #1}}
\newcommand{\mysymtens}[1]{{\boldsymbol #1}}
\newcommand{\og}{\textsc{OpenGadget3}\xspace}
\newcommand{\minsub}[2]{#1_{\rm #2}}
\newcommand{\somedot}[2]{\dot{#1}_{\rm #2}}

\newcommand{\myvector}[1]{\ensuremath{\mathbf{#1}}}
\newcommand{\mymatrix}[1]{\ensuremath{\mathbf{#1}}}
\newcommand{\dif}{\ensuremath{\text{d}}}
\newcommand{\klammer}[1]{\ensuremath{\left(#1\right)}}
\newcommand{\abs}[1]{\ensuremath{\left\vert#1\right\vert}}
\newcommand{\mathcases}[1]{\ensuremath{\begin{cases}#1\end{cases}}}
\newcommand{\outerprod}{\ensuremath{\vert}}

\usepackage{dsfont}

\definecolor{ao(english)}{rgb}{0.0, 0.5, 0.0}

\journal{Astronomy $\&$ Computing}

\newcommand\X            {{\color{green}\ding{52}}}
\newcommand\x            {{\color{orange}\ding{52}}}
\newcommand\F            {{\color{red}\ding{56}}}

\newcommand{\dd}{%
  \textpdfrender{TextRenderingMode=FillStroke,%
                 FillColor=blue,StrokeColor=blue,LineWidth=0.5pt}{\ding{46}}%
}          

\makeindex

\begin{document}

\begin{frontmatter}


\title{The \og Code for Cosmological Simulations}

\author[]{OG3 Developers}
\author[USM,MPA]{Klaus Dolag}
\author[USM]{Geray S. Karademir}
\author[INAFTS,ICSC]{Luca Tornatore}
\author[INAFTS]{Antonio Ragagnin}
\author[UNITS,INAFTS,ICSC,IFPU]{Milena Valentini} 
\author[UCHICAGO]{Ludwig M. B\"oss}
\author[C3A,ZJU,USM]{Frederick Groth}
\author[INAFTS,ICSC]{Giuseppe Murante}
\author[UNITS,INAFTS,ICSC,IFPU]{Stefano Borgani}
\author[MPA]{Volker Springel}
\author[OKC,TUM]{Cenanda Arido}
\author[UBA,CONICET]{Javier Badía}
\author[USP,INAFTS]{Tiago Castro}
\author[UNITS,INAFTS,ICSC]{Alice Damiano}
\author[USM]{Laura Di Federico}
\author[DIPC,USM]{Moritz S.\ Fischer}
\author[INAFTS,ICSC]{Gian Luigi Granato}
\author[Intel]{Nitya Hariharan\texorpdfstring{\footnotemark}{}}
\author[GUF]{David Klemmer}
\author[INAFTS,IFPU]{Umberto Maio}
\author[USM,CFA]{Tirso Marin-Gilabert}
\author[ESO,ORIGINS,USM]{Ilaria Marini}
\author[UMN,BITS]{Yashraj Patil}
\author[LRZ]{Margarita Egelhofer}
\author[IATE,INAFTS]{Cinthia Ragone-Figueroa}
\author[USM,ORIGINS]{Luca Sala}
\author[UBA,CONICET]{Cecilia Scannapieco}
\author[TUM]{Ludwig D. Schmidt}
\author[USM]{Ulrich P. Steinwandel}
\author[USM]{Giovanni Tedeschi-Prades}
\author[KIT]{Marc Wiertel}
\affiliation[USM]{organization={University Observatory, Faculty of Physics, Ludwig-Maximilians-Universität München},
            addressline={Scheinerstr. 1},
            city={Munich},
            postcode={81679},
            country={Germany}}

\affiliation[MPA]{organization={Max Planck Institute for Astrophysics},
            addressline={Karl-Schwarzschild-Straße 1},
            city={Garching},
            postcode={85741},
            country={Germany}}

\affiliation[INAFTS]{organization={INAF -- Osservatorio Astronomico di Trieste}, 
            addressline={via Tiepolo 11},
            city={Trieste},
            postcode={I-34143}, 
            country={Italy}}
            
\affiliation[UNITS]{organization={Dipartimento di Fisica, Università di Trieste}, 
            addressline={via Valerio 2},
            city={Trieste},
            postcode={I-34127}, 
            country={Italy}}
            
\affiliation[ICSC]{organization={ICSC - Italian Research Center on High Performance Computing, Big Data and Quantum Computing},
            addressline={via Magnanelli 2},
            city={Casalecchio di Reno},
            postcode={40033},
            country={Italy}}

\affiliation[USP]{organization={Department of Mathematical Physics, Institute of Physics, University of São Paulo},
            addressline={R. do Matão 1371},
            city={São Paulo},
            postcode={05508-090},
            country={Brazil}}
            
\affiliation[UCHICAGO]{organization={Department of Astronomy and Astrophysics, The University of Chicago},
            addressline={William Eckhart Research Center, 5640 S. Ellis Ave},
            city={Chicago, IL},
            postcode={60637},
            country={USA}}
            
\affiliation[IATE]{organization={IATE - Instituto de Astronomía Teórica y Experimental, Consejo Nacional de Investigaciones Científicas y Técnicas (CONICET), Universidad Nacional de Córdoba (UNC)},
            addressline={Laprida 854},
            city={Córdoba},
            postcode={X5000BGR},
            country={Argentina}}     
            
\affiliation[IFPU]{organization={IFPU, Institute for Fundamental Physics of the Universe},
            addressline={via Beirut 2},
            city={Trieste},
            postcode={34014},
            country={Italy}}      

\affiliation[DIPC]{organization={DIPC, Donostia International Physics Center},
            addressline={Paseo Manuel de Lardizabal 4},
            city={Donostia-San Sebastian},
            postcode={20018},
            country={Spain}}

\affiliation[CFA]{organization={Center for Astrophysics $|$ Harvard \& Smithsonian},
            addressline={60 Garden St.},
            city={Cambridge, MA},
            postcode={02138},
            country={USA}}

\affiliation[ESO]{organization={European Southern Observatory},
            addressline={Karl-Schwarzschild-Straße 2},
            city={Garching},
            postcode={85748},
            country={Germany}}

\affiliation[ORIGINS]{organization={Excellence Cluster ORIGINS},
            addressline={Boltzmannstr. 2},
            city={Garching},
            postcode={85748},
            country={Germany}}

\affiliation[TUM]{organization={Physics Department, Technical University of Munich},
            addressline={James-Franck-Straße 1},
            city={Garching},
            postcode={85748},
            country={Germany}}

\affiliation[LRZ]{organization={Leibniz-Rechenzentrum (LRZ)},
            addressline={Boltzmannstrasse 1},
            city={Garching},
            postcode={85748},
            country={Germany}}           

\affiliation[C3A] {organization={Center for Cosmology and Computational Astrophysics, Institute for Advanced Study in Physics, Zhejiang University},
            city={Hangzhou},
            country={People’s Republic of China}}
\affiliation[ZJU]{organization={Institute of Astronomy, School of Physics, Zhejiang University},
            city={Hangzhou},
            postcode={310037},
            country={People’s Republic of China}}
\affiliation[OKC]{organization={The Oskar Klein Centre, Department of Physics, Stockholm University},
            addressline={Albanova University Center},
            city={Stockholm},
            postcode={SE-106 91},
            country={Sweden}}

\affiliation[GUF]{organization={Institute for Theoretical Physics, Goethe University Frankfurt},
            city={Frankfurt},
            postcode={60438},
            country={Germany}}

\affiliation[UMN]{organization={College of Science and Engineering, University of Minnesota}, 
            addressline={117 Pleasant St},
            city={Minneapolis},
            postcode={MN 55455},
            country={USA}}

\affiliation[BITS]{organization={Birla Institute of Technology and Science-Pilani, K. K. Birla Goa campus},
            addressline={NH-17B, Zuarinagar},
            city={Goa},
            postcode={403726},
            country={India}}

\affiliation[KIT]{organization={Institute for Theoretical Particle Physics, Karlsruhe Institute of Technology},
            addressline={Wolfgang-Gaede-Straße 1},
            city={Karlsruhe},
            postcode={76128},
            country={Germany}}

\affiliation[Intel]{organization={Intel Technology India Private Ltd},
            addressline={23-56P, Devarabeesanahalli, Varthur Hobli, Outer Ring Road},
            city={Bangalore, KA},
            postcode={560103},
            country={India}}

\affiliation[UBA]{organization={Departamento de Física, Universidad de Buenos Aires},
            addressline={Pabellón I, Ciudad Universitaria},
            city={Buenos Aires},
            postcode={C1428},
            country={Argentina}}
\affiliation[CONICET]{organization={Consejo Nacional de Investigaciones Científicas y Tecnológicas},
            addressline={Godoy Cruz 2290},
            city={Buenos Aires},
            postcode={C1425FQD},
            country={Argentina}}

\begin{abstract}
We present the public release of \og, a substantially extended and re-organised version of the widely used \textsc{Gadget}-2/3 family of cosmological simulation codes. Since \textsc{Gadget}-2 was made publicly available, the absence of continued official development led to the creation of numerous, mutually incompatible developer versions of \textsc{Gadget}-3 across the community. This made it effectively impossible to track bug fixes, reproduce published results, or consistently credit the many individual contributors. \og addresses this by consolidating these disparate developments into a single, documented, and continuously maintained code base, released under the GNU GPLv3 through a GitLab repository (with an accompanying wiki), regularly synchronised with the active developer branch, and supported by a dedicated continuous-integration (CI) pipeline.

\og retains the core algorithmic backbone of its predecessors, a Barnes\&Hut oct-tree combined with a Particle-Mesh method for gravity, and a choice of Smoothed Particle Hydrodynamics or Meshless-Finite-Mass solvers for gas dynamics, within a hybrid MPI/OpenMP parallelisation framework, extended with GPU offloading via OpenACC and OpenMP. The code integrates an extensive suite of sub-resolution and physical modules developed over more than a decade, including primordial chemistry, radiative cooling, star formation, stellar feedback and chemical enrichment, dust formation and evolution, black hole growth and AGN feedback, magneto-hydrodynamics, thermal conduction, physical viscosity, self-interacting dark matter, and massive neutrinos.

For this release, \og introduces several new features aimed at improving robustness, reproducibility, and usability: a flexible mixed-precision framework spanning 16- to 128-bit representations; expert-level tagged parameters with automatic default handling and restart-time change log; extensive compile-time and runtime consistency checks against invalid configurations; and FAIR-compliant reporting of all settings, parameters, and specific publications associated with each active physics module. A tiered CI pipeline validates gravity and hydrodynamic solvers against various canonical analytic test problems across CPU and GPU-offloaded backends. Together, these developments establish \og as a well-documented, extensible, and publicly accessible platform for cosmological hydrodynamical simulations, developed with support from the EuroHPC-JU CoE SPACE.
\end{abstract}

\end{frontmatter}

\begin{NoHyper}
\footnotetext{Current affiliation: Advanced Micro Devices (AMD) (for identification only; no work on this paper was performed at AMD and AMD had no involvement in the work described)}
\end{NoHyper}

\section{Introduction}
\label{introduction}
Numerical simulations have become one of the most fundamental tools for studying the non-linear formation and evolution of cosmic structures \citep[e.g.][]{Navarro1997, Jenkins2001}. Due to their capability to capture the complexity of the hierarchical assembly of such structures over wide ranges of scales and epochs, they have become an essential part of astrophysical and cosmological research.

A gigantic leap forward in this field has been enabled by the concomitant development of high-performance computing (HPC) facilities and innovative algorithmic solutions and parallelisation paradigms that led to an ever-increasing efficiency and reliability of codes for cosmological simulations. Computational cosmology has always been a leading scientific domain in HPC, with some of the most challenging applications making it to the news as outstanding technical and scientific achievements, like the Millennium \citep{springel05}, Illustris \citep{Vogelsberger2014}, Magneticum \citep{Dolag.etal.2025} and the Frontier-E \citep{Frontiere2025}
simulations, to name a few.

Accurately simulating the evolution of the Universe is a formidable physical, mathematical and computational challenge. In its standard formulation, the dark matter (DM) particles dominating the mass content of the Universe form a collisionless fluid that is described by the Vlasov equation coupled to self-gravity through Poisson's equation. The resulting system of partial differential equations must be solved using methods that are highly adaptive in space and time to bridge the huge dynamic range involved in non-linear structure formation. Additionally, the calculation must be highly scalable on distributed memory architectures despite the long-range coupling of up to trillions of tracer particles via gravity. Besides the DM collisionless fluid, bridging simulations to observations also requires them to include the collisional fluid associated with cosmic diffuse baryons. In addition to feeling their own gravity and that exerted by the dynamically dominant DM fluid, baryons are treated as an ideal gas obeying the equations of fluid dynamics. For this reason, codes for cosmological simulations also need to include a hydrodynamical solver, which follows the evolution of cosmic baryons under the action of pressure and gravity forces. In addition, a variety of physical processes, besides hydrodynamics, have to be included in the calculations for simulations to provide predictions that can be directly compared to observational properties of galaxies, clusters of galaxies and the intergalactic medium. Three among them are considered particularly important for the development of the visible Universe: the condensation of diffuse baryons into stars; their further evolution when the surrounding gas is heated by stellar winds and supernova (SN) explosions and enriched with chemical elements; and third, the feedback of supermassive black holes (BHs) that eject massive amounts of energy into their surroundings. Moreover, the fundamental properties of DM (e.g., its scattering properties), along with the underlying nature of dark energy, can have an impact on the evolution of cosmic structures, and other constituents like neutrinos, magnetic fields, cosmic rays (CRs) or dust also play a role in shaping the dynamics and assembly of galaxies.

\begin{figure*}[th]
\begin{center}
 \includegraphics[width=0.49\textwidth]{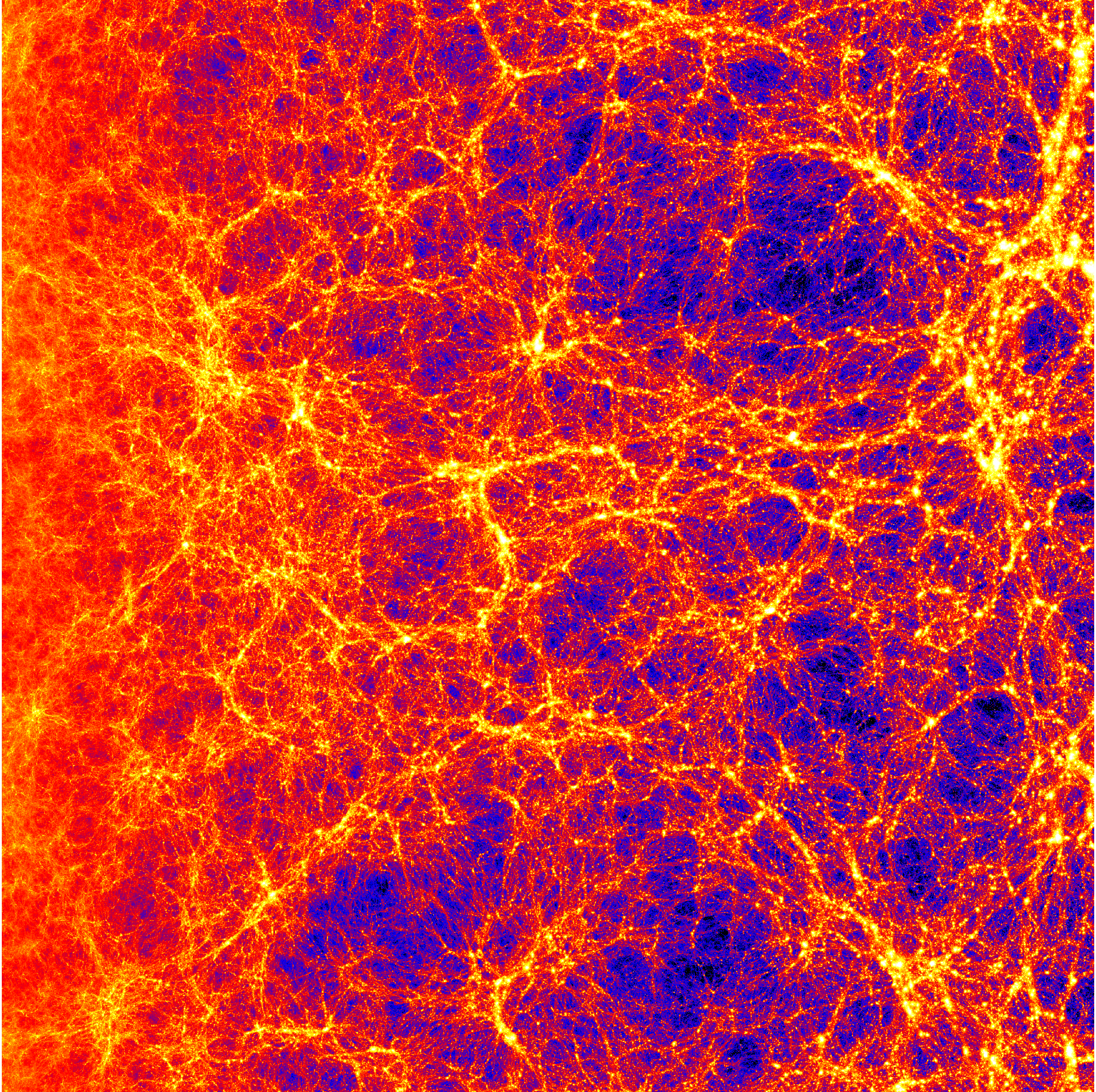}
 \includegraphics[width=0.49\textwidth]{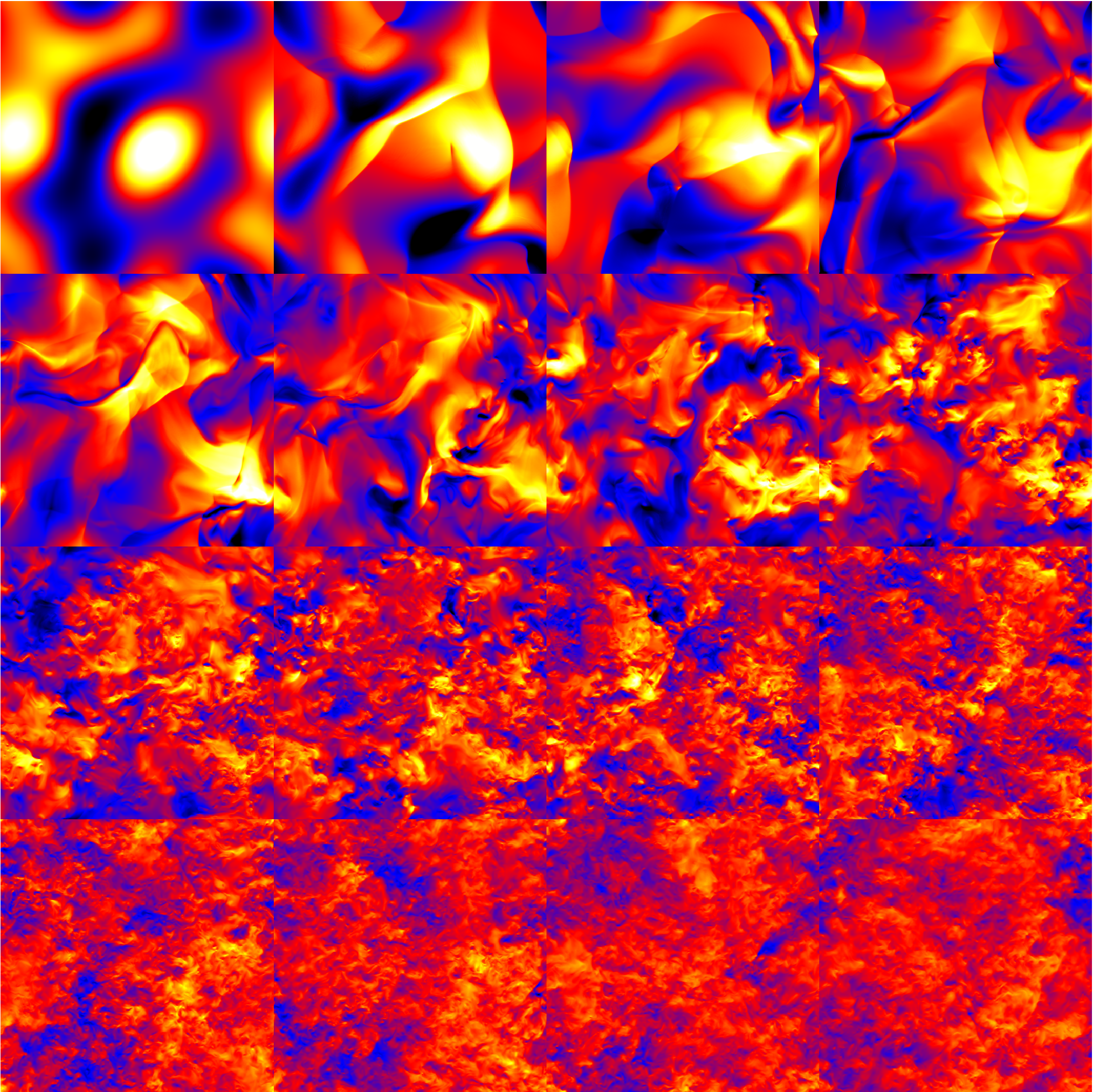} \\
 \includegraphics[width=0.49\textwidth]{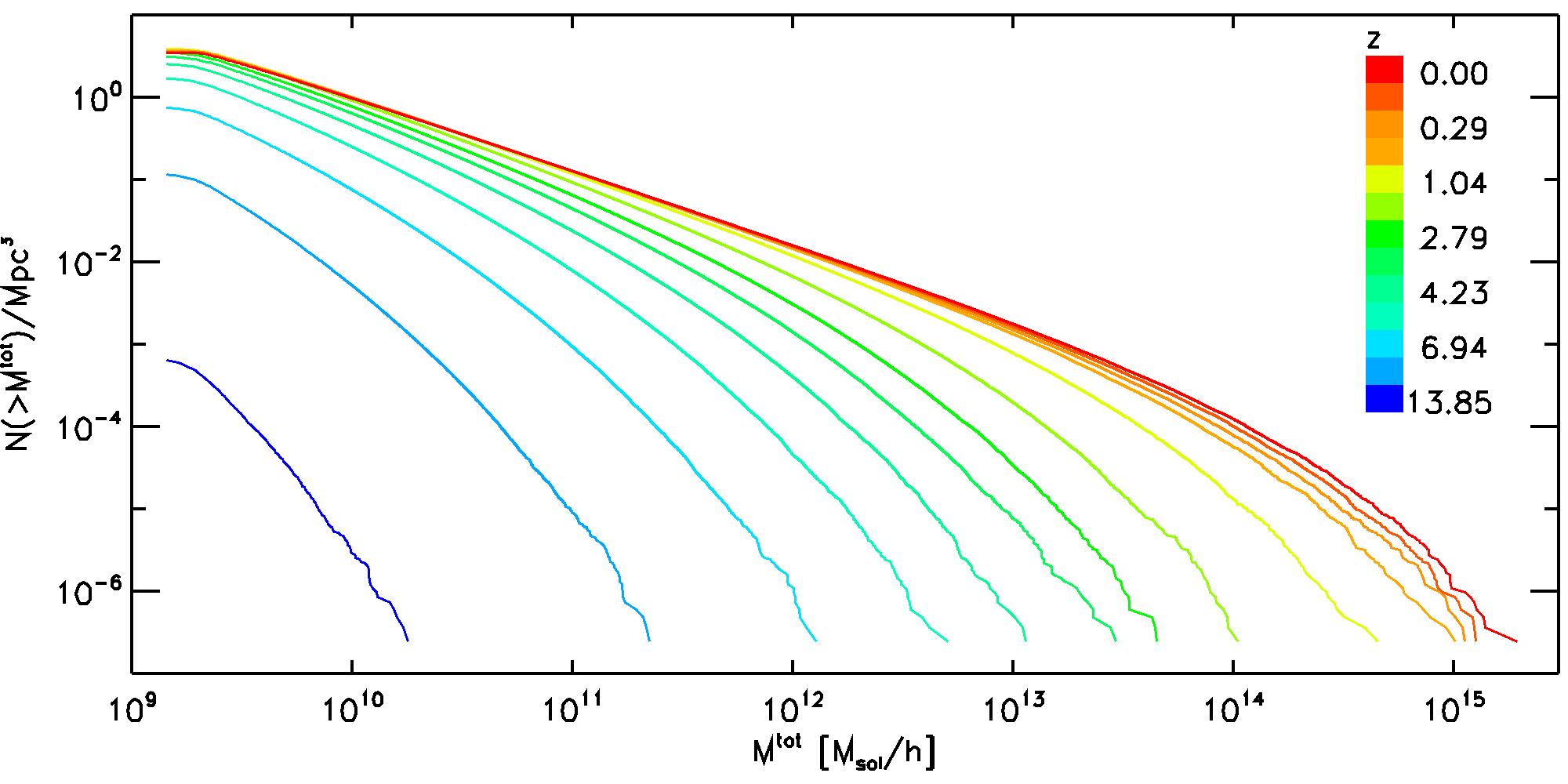}
 \includegraphics[width=0.49\textwidth]{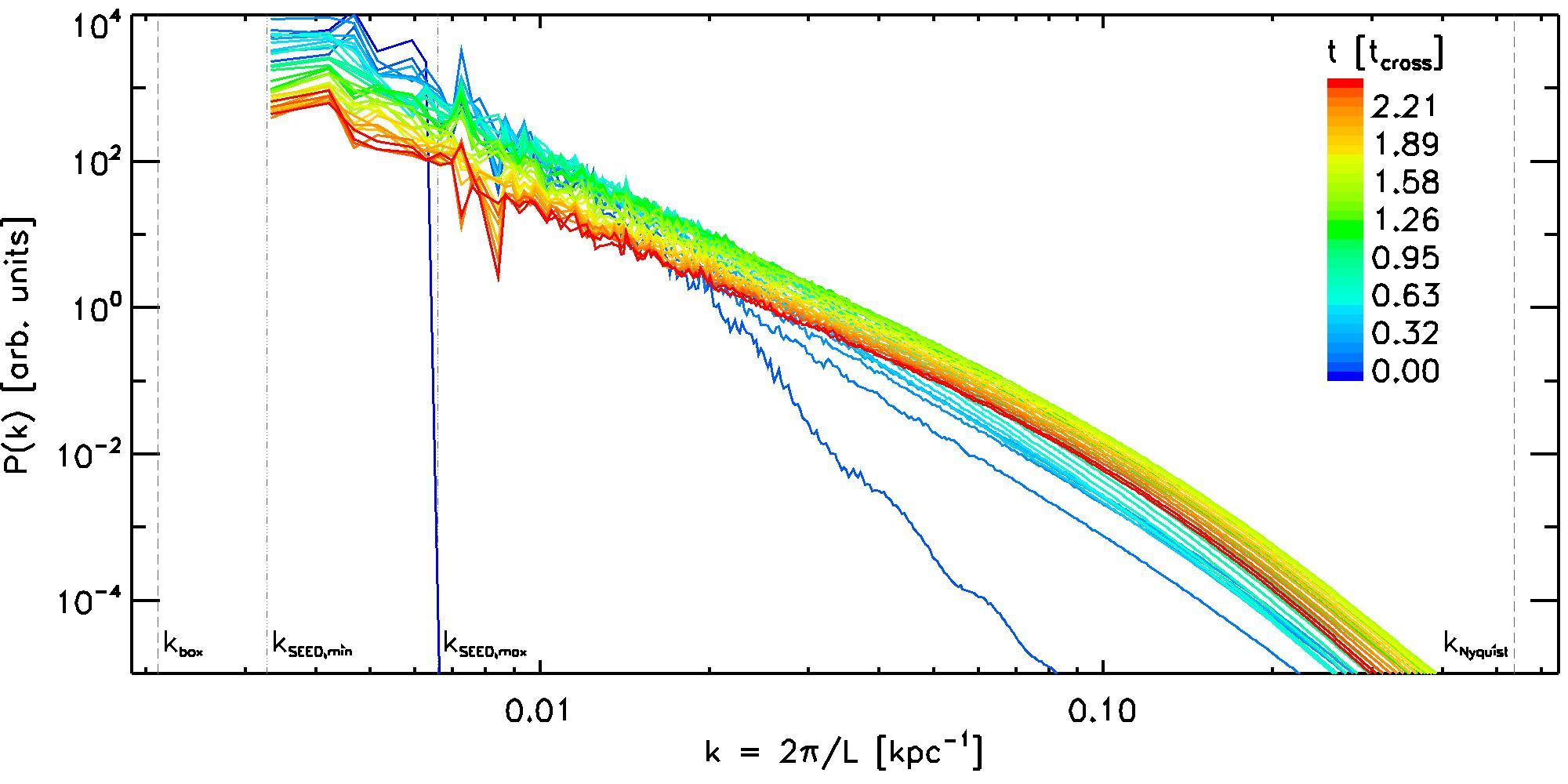}
 \end{center}
 \caption{Left column: The evolution (from left to right) of DM density in a thin slice of a dark-matter-only simulation with a side length of $288 h^{-1} \mathrm{cMpc}$, resolved with $3456^3$ particles and the corresponding evolution of the mass function below. Right column: This slice through an evolving turbulent box resolved with $2048^3$ particles. Here, a subsonic, largest-scale motion was initiated, and the lower part shows the build-up of the spectral energy density of the decaying turbulence, compared to the expected $k^{-11/3}$ slope. Both simulations were performed using GPU offloading via OpenMP on SuperMUC-NG2 at the Leibniz-Rechenzentrum (LRZ).
}
\label{fig:og3sims}
\end{figure*}

Figure \ref{fig:og3sims} illustrates two exemplary simulations. The one that captures the evolution of the universe shows the growth of dark-matter density fluctuations and the corresponding evolution of the mass function in a cosmological dark-matter-only simulation. The other shows the decay of subsonic turbulence and the development of its energy spectrum.

Therefore, state-of-the-art cosmological hydrodynamical simulations must follow these complex, interacting physical processes across scales and volumes spanning extremely large dynamical ranges. Hence, the amount of data required to describe such systems -- whose modelling is implemented through sophisticated, scalable, and highly optimised numerical algorithms -- and the computational power needed to achieve these scientific goals are growing continuously and push current HPC facilities to their limits. 

In the past, platforms like the publicly available \textsc{Gadget}-2 \citep{Springel2005} code allowed groups all over the world to perform outstanding simulations, unveiling the growth and evolution of structures in the Universe. In addition, many other codes of this class were made publicly available over the past decades, like RAMSES \citep{2002A&A...385..337T}, AREPO \citep{Springel2010}, SWIFT \citep{2024MNRAS.530.2378S}, \textsc{Gadget}-4 \citep{Springel2021}, CRK-HACC \citep{2022ApJS..259...15F} and SPH-EXA \citep{2025arXiv250310273C}: however, for many of the aforementioned codes only a reduced version was often released, with a limited number of physics modules.

\begin{figure*}[th]
 \centerline{\includegraphics[trim={3.5cm 0.0cm 2.2cm 0.0cm}, clip, width=1.0\textwidth]{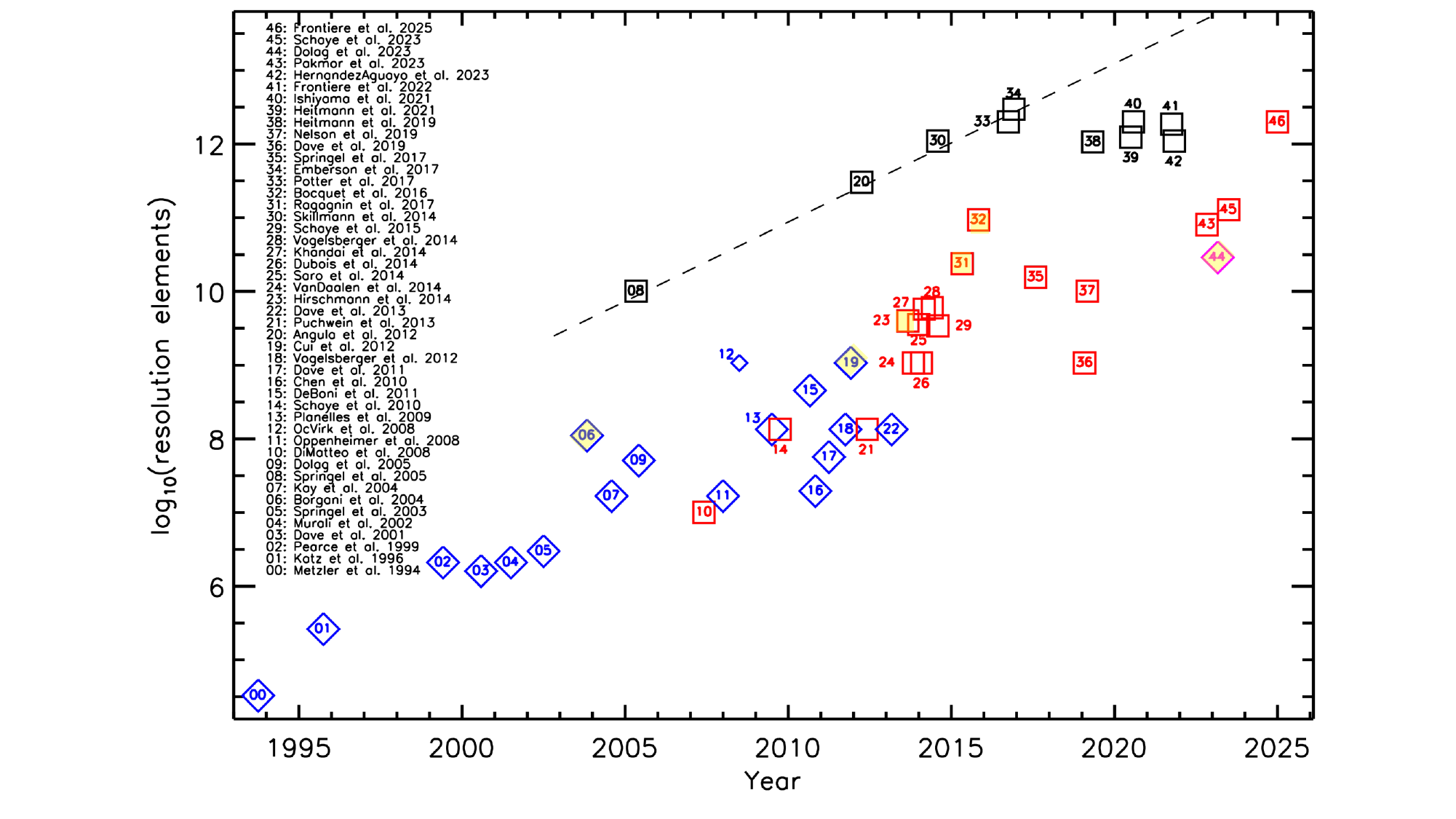}}
 \caption{The evolution of the number of resolution elements in simulations. The black line is a fit to the scaling for pure $N$-body simulations between 2005 and 2015. The black symbols are recent $N$-body simulations: \protect{\cite{2012MNRAS.426.2046A,2014arXiv1407.2600S,2017ComAC...4....2P,2017RAA....17...85E,2019ApJS..245...16H,2021ApJS..252...19H,2021MNRAS.506.4210I,2022ApJS..259...15F}}.
 The coloured symbols pinpoint hydrodynamical, cosmological simulations over the last three decades. Blue data points depict simulations including the effect of cooling and star formation:
\protect{\cite{1994ApJ...437..564M, Katz.etal.1996, 1999ApJ...521L..99P, dave01, 2002ApJ...571....1M, 2003MNRAS.339..312S, borgani2004, 2004MNRAS.355.1091K, 2005MNRAS.363...29D, 2008MNRAS.387..577O, 2009MNRAS.399..410P, 2011MNRAS.415.2758D, 2011MNRAS.415...11D, 2012MNRAS.425.3024V, 2012MNRAS.423.2279C, 2013MNRAS.434.2645D}}, 
while red data points show simulations which also feature the effect of Active Galactic Nuclei (AGN) feedback:
\protect{\cite{DiMatteo2008, 2010MNRAS.402.1536S, 2013MNRAS.428.2966P, Hirschmann.etal.2014, Dubois2014, 2014MNRAS.440.2997V, 2014MNRAS.440.2610S, Khandai2015, Vogelsberger2014, Schaye2015, 2016MNRAS.456.2361B,Ragagnin2017, 2019MNRAS.490.3234N, Dave2019, 2023MNRAS.524.2539P, Schaye2023,Frontiere2025}}. The pink symbol marks a simulation which, in addition, includes the treatment of magnetic fields and spectral cosmic ray electrons and protons \protect{\cite{Dolag2023}}. The yellow background exemplarily shows a simulation performed with the developer versions of \textsc{Gadget}-2/3, including extra physics which is part of the release of \og. A modified version of this figure can be found in \citet{2025arXiv250206954V}.}
\label{fig:sims}
\end{figure*}

\subsection{A brief history}
Since \textsc{Gadget}-2 was made publicly available, many structural changes have been made to the code to efficiently perform simulations by fully exploiting the size and evolving architectures of HPC facilities. In addition, a vast number of various physical modules needed to study galaxy formation, the physics of the intergalactic (IGM) and intracluster (ICM) media, and other astrophysical systems have been developed and included in the code. The lack of continuous updates to the publicly available version of \textsc{Gadget}-2 has led to the creation of an unofficial version of \textsc{Gadget}-3, which has been distributed in various forms and stages across the community over the last decade. 

A sort of Moore's law over more than 30 years of cosmological simulations is shown in figure \ref{fig:sims}, in which the number of resolution elements in each state-of-the-art simulation is plotted against the year of production. Information on the main physics included is reported in this plot, where we also highlight simulations performed with various developer versions of \textsc{Gadget}-2/3. This unsystematic code diffusion and development makes it impossible to ensure that important bug fixes are distributed or tracked. Even more, as there is no way to track which version of the code is used by which group, it is nearly impossible to independently repeat any published simulation and check for reproducibility. A further unpleasant side effect was also that it became quite hard to ensure that the contributions of several young scientists to various parts and modules of the code were acknowledged properly. Although the publication of GIZMO \citep{Hopkins2015} made some of these developments public, it is important to bring together different contributions to the \textsc{Gadget} family from several members of a heterogeneous community of developers to allow potentially interested users to use a well-defined, documented, rich in extra physics and publicly available version of the code. 
Within this framework, the \og project was conceived to have a reference, streamlined version of the code where all the communal effort is funnelled into a single, well-documented code. An additional purpose of this re-organisation work is to make it publicly available. Indeed,  with the support of the HPC-Europe project SPACE\footnote{https://www.eurohpc-ju.europa.eu/research-innovation/our-projects/space\_en} \citep{2025cofr.conf..177S}, we publicly release a largely improved version of \og to the wide community of users interested in carrying out simulations of cosmic structure formation and in becoming part of the community of \og developers.

\subsection{The backbone}

As mentioned above, \og represents a more extended development of the parallel $N$-body codes \textsc{Gadget}-2/3 \citep[][]{Springel2001, Springel2005}. Like its predecessors, it exploits distributing the workload using a domain decomposition based on contiguous chunks of Hilbert-ordered particles. With this parallelisation strategy, an optimal workload balance among different MPI (Message Passing Interface) tasks is achieved by allowing for a memory imbalance. Furthermore, \og inherits the hybrid OpenMP/MPI parallel framework developed within \textsc{Gadget}\footnote{Initialised by the KONWHIR program and in close collaboration with the LRZ and C2PAP of the “Universe” cluster, it has been possible to transform the main parts of \textsc{Gadget}-3 into a hybrid OpenMP/MPI enabled code.}-3. In addition, it supports offloading of the basic parts of the code to GPUs based on an OpenACC and an OpenMP implementation \citep[][]{Ragagnin2020, Ragagnin2026}. 

The gravitational forces are calculated via a Barnes \& Hut tree algorithm \citep[][]{BarnesHut1986} at short ranges and using a Particle Mesh method for long-range interactions \citep{Springel2005}. This is extended by the treatment of self-interacting DM \citep[SIDM;][]{Fischer2026}, of neutrinos as part of the DM fluid \citep{EuclidCollabrotation2024}, as well as non-standard cosmologies \citep{dolag04}.

The code evolves fluid quantities by adopting the Smoothed Particle Hydrodynamics \cite[SPH;][]{Gingold1977} approach to solve the hydrodynamical equations for gas particles: here, the SPH formulation features improvements with respect to standard SPH solvers, as presented by \cite{Beck2016}. In addition, a Meshless Finite Mass \cite[MFM;][]{Groth2023} is also implemented and can be chosen as an alternative to the SPH scheme.

On top of these basic algorithms, \og has implemented various extra physics, such as a primordial chemical network \citep{Maio2007}, thermal conduction \citep{Arth2014}, magneto-hydrodynamics \citep{Dolag2009, Stasyszyn2013}, viscosity \citep{Marin-Gilabert_2022} and cosmic rays \citep{Boess2023}. In addition, a large set of sub-resolution models, e.g. star formation \citep{Springel2003}, stellar evolution/chemical enrichment \citep{Tornatore2004, Tornatore2007}, and the ensuing dust production and evolution \citep{Granato2021, Tedeschi-Prades_2025a}, along with several variants and extensions for BH evolution and AGN feedback \citep{Springel_BHs, Fabjan.etal.2010, Hirschmann.etal.2014, Steinborn2015, Sala2024, Damiano2024}, are also implemented.

\subsection{\og code release}

The \og code will be made available under the GNU General Public License v3 by an open-source repository located on a GitLab instance operated by LRZ \footnote{\url{https://gitlab.lrz.de/AstroCodes/OpenGadget3}}. This repository includes a wiki\footnote{\label{wiki}\url{https://gitlab.lrz.de/AstroCodes/OpenGadget3/-/wikis}} with more information, particularly on how to get started. We intend to keep the public code close to the developer branch by an automatic, monthly update against the developer version. Developers of new features and modules are also encouraged to contact the main authors and join the developer repository directly to release their modules into the public version together with their presentation papers. 

To verify the integrity of the code, a continuous integration (CI) pipeline is run on any merge request as well as additionally on a regular base. The CI pipeline focuses on validating SPH functionality and gravitational forces across canonical benchmarks, including a gravitational free-fall, Taylor–von Neumann–Sedov blast wave, Sod shock tube, and Soundwave test, on CPU-only and OpenACC/OpenMP-offloaded backends. It also tests alternative hydrodynamics solvers (PES and MFM) using the same setups to ensure consistency across numerical methods. A broader weekly CI suite covers long-term and edge-case regressions. It includes MHD benchmarks such as the Brio-Wu shock tube, cosmic ray module tests, and a complete cosmological zoom-in simulation. The current implementation of the CI pipeline reaches a line and function coverage of $\geq75$\%. For more details on the CI tests, see Section \ref{sec:Tests}.

The code performs various cross-checks on the configuration settings and on the choice of parameters both during compilation and at the run start, with the purpose of preventing inconsistent settings and untested combinations of physical modules. 

Following the FAIR (Findable, Accessible, Interoperable, and Reusable) principles, the code outputs all setting configuration options and parameters used at the start, while also explicitly listing the relevant publications associated with the enabled code features which have to be cited, to ensure that proper credit is given to the authors of the different parts of the code used for any specific simulation.

\subsection{Content of the paper}
All these aspects will be described in detail in the next sections, starting with some basics and then describing all physics modules, from gravity and its extension, to the various hydrodynamical aspects, as well as additional processes shaping the different astrophysical aspects of galaxy and structure formation, with some additional details in the appendix. The paper is structured as follows: \tableofcontents


\begin{table*}[ht]   
\centering
    \begin{tabular}{|c|c|c|c|c|c|c|c|c|c|c|}
        \hline
        DOUBLEPRECISION & Label           & -3   & -2  & -1  & 0   & 1   & 2    & 3   & 4    & 5 \\
        \hline
        \hline
        position        & MyLongDouble    & F32  & F32 & F32 & F32 & F64 & F128 & F64 & F128 & F128 \\
        \hline
        gravity force   & MyAtLeastDouble & F64  & F64 & F32 & F64 & F64 & F128 & F64 & F64  & F128 \\
        \hline
        local/exchange  & MyAtLeastDouble & F64  & F64 & F32 & F64 & F64 & F128 & F64 & F64  & F128 \\
        \hline
        PropertiesF32   & MyFloat         & bF16 & F16 & F32 & F32 & F64 & F128 & F32 & F32  & F32 \\
        \hline
        PropertiesF64   & MyDouble        & bF16 & F16 & F32 & F32 & F64 & F128 & F32 & F64  & F64 \\
        \hline
        MHD Test        & [bytes/particle]&  260 & 260 & 388 & 424 & 744 & 1440 & 440 &      &  \\
        \hline
        dm CosmoBox     & [bytes/particle]&  140 & 140 & 140 & 164 & 228 & 412  & 180 &      &  \\
        \hline
    \end{tabular}
    \caption{Precision settings. Variables can be defined in different classes, which then are mapped to different precisions, depending on the settings of the code. Here, the treatment of positions and of the gravitational force is different from all other properties, as they typically need higher precision due to their natural, wide dynamical range (positions) or as they are the result of summations of a large number of varying addenda (gravitational force). In addition, local variables and exchanges can be done in higher precision to minimise the effect of round-off errors during individual computations. The last two rows list the memory footprint of the particles in the two tests presented in \ref{sec:app_back} to better illustrate the effect of the precision setting on the memory consumption.
    Note that bF16 is a recently introduced 16-bit float that reserves $8$ and $7$ bits for the exponent and mantissa, respectively, thereby trading a smaller precision for the same dynamic range as FP32.
    }
    \label{tab:precission}
\end{table*}

\begin{figure}[t]
 \centering
 \includegraphics[width=0.5\textwidth]{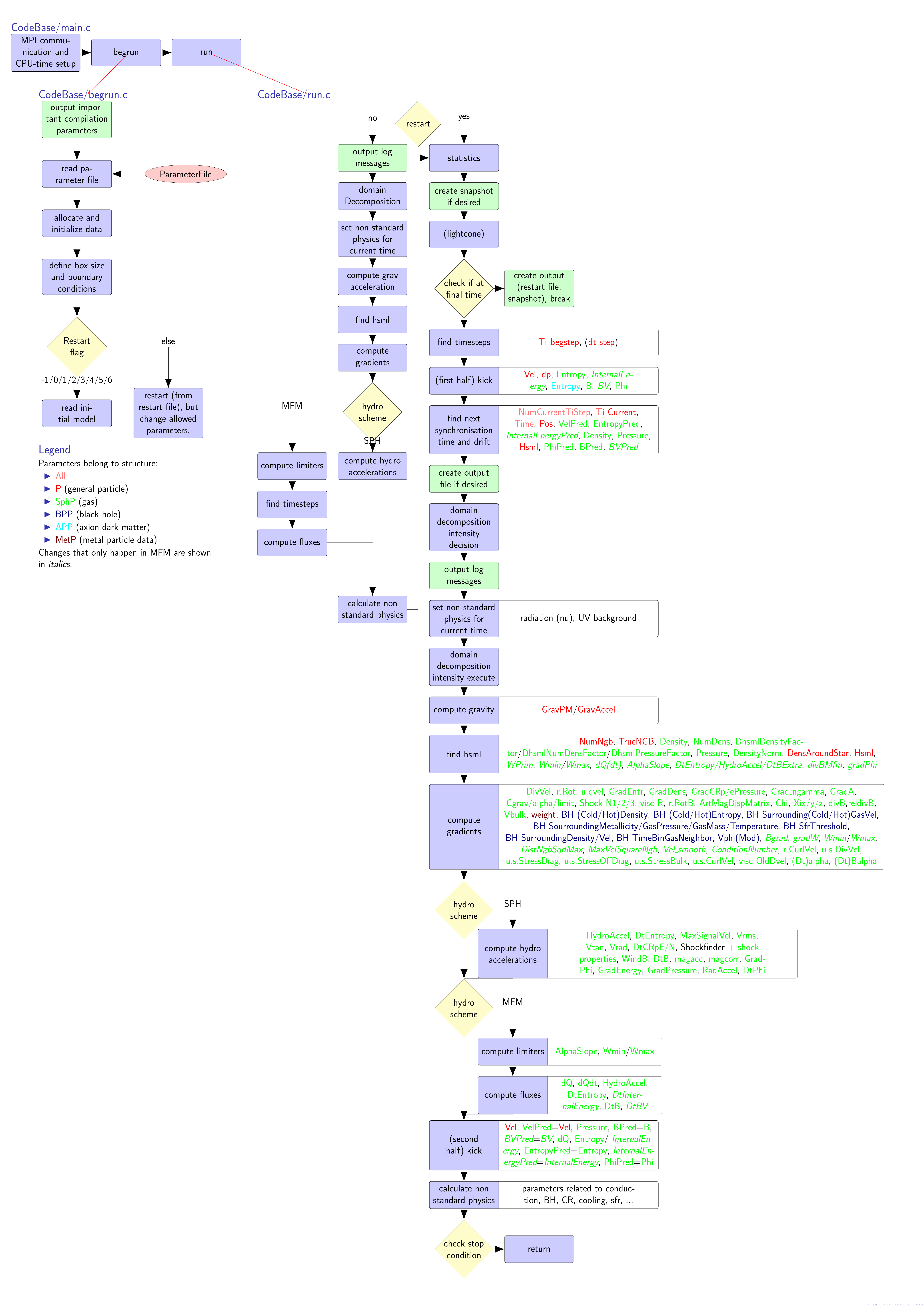}
 \caption{Flowcharts of the operations in \og at the time of release. A maintained version of this flowchart can be found in the code wiki$^{[\ref{wiki}]}$. 
 In the upper left corner, the main part of the code is shown, which splits into the begrun and run functions. As the name suggests, the begrun function handles the simulation setup, such as reading the parameters, ICs, etc. The run function then consists of one main loop iterating over all time-steps, where several operations are executed at each time-step.}
 \label{fig:flowchart}
\end{figure}

\section{Code structure and principles}
\og follows the philosophy of the extended \textsc{Gadget} family, and the order of the different operations is shown in Figure~\ref{fig:flowchart}. A configuration file enables additional physics and code features at compile time and a parameter file controls some crucial input for them at runtime. However, \og comes with some improvements on the original philosophy and handling of the files. First, at compile time it is ensured that flags in the configuration file really exist in the code to prevent typos in the configuration file from changing the code behaviour. Furthermore, the code contains various tests which either at compile time or at execution time detect invalid configuration settings. Following the general strategy of the \textsc{Gadget} family, \og automatically creates restart files (including keeping the previous ones as backup) at given time intervals, and when the code finishes or stops. 

Launching the code also expects a start option (default is zero), which decides whether the code begins from initial conditions (ICs), continues from a restart file, restarts from a snapshot or processes a snapshot in different ways. For the latter ones, in addition, the snapshot number has to be provided.

In case of restarting from restart files, changes of parameters on restart are now recorded, and also restrictions are applied to which parameters can be changed at restart during a simulation. Finally, we changed the philosophy of enabling a physical module at compile time and switching on its effects in the parameter file: in \og it is now possible to enable a physical module at compile time, but have its effect taken into account in the parameter file independently during runtime.

\subsection{Accuracy settings}
\og is based on a flexible scheme to define the accuracy level to be used, which is specified as a setting at compile time. It extends the original scheme adopted in \textsc{Gadget}, which allowed switching between single and double precision floats, to a scheme which also supports 128-bit as well as 16-bit floating-point variables. Furthermore, it allows for mixed-accuracy treatment (as used in the {\it Magneticum} simulations; \citealt{Dolag.etal.2025}), which helps to decrease the memory footprint while maintaining precision in certain variables (like positions) where precise capture of wide dynamical ranges or large summations is needed. An overview of the precision settings available is given in Table \ref{tab:precission}. In principle, additional combinations of accuracy can be easily added by simply defining additional combinations. Although the general recommendation is to use double precision, some of the other modes can be considered in certain cases. The impact of the choice of accuracy on a cosmological, DM only box as well as on an MHD shock tube test is shown in Figure \ref{fig:app_mhd_precission} in \ref{sec:app_back}.

\subsection{Parameter handling}

The handling of parameters is somewhat changed compared to previous \textsc{Gadget} versions. In \og, most parameters are assigned a default value at compile time, so that they do not need to be set in the input parameter file, unless non-standard values are used. In addition, the user has to indicate their expertise level by setting an expert level (default here is 0, 1 for general, 2 for advanced users and 3 for developers) at compile time. Every parameter is tagged by a level, which then defines whether such a parameter can be simply changed by a general user or only by experienced users. This allows us to distinguish between standard parameters, which every user is allowed to change, and more subtle, numerical parameters which users should think carefully about changing from the default values.

The code produces two logs for the parameters when the run starts. One log contains the values which the code used from the parameter file, while a second log contains all values used, including the implicit ones which are set by default. The code also provides a changelog file, which records all parameters which are changed at restart during the run.      

\subsection{Verbosity level and consistency checks}
To ensure that the code can actually be compiled in the configuration chosen by the user, at compilation time a test is performed to ensure that all options set in the configuration file are present in the code; otherwise, the compilation is stopped. This ensures that the user does not expect options which do not exist or are misspelt in the configuration file to be active. In addition, there are various checks at compile time as well as when the code starts, which ensure that configurations and parameters are within the expected or allowed range and that combinations of different physical modules or settings with known conflicts are prevented. A new start option (i.e., $-1$) allows starting the code in a mode where only parameter and configuration checks are performed. In addition, the different physical modules will report the details of their underlying sub-grid models given the parameters and settings. The overall level of progress reporting during runtime can be controlled by setting the verbosity level (default is 0) at compile time. As an option for developers, a debug flag can be set at compile time, which adds various consistency checks at various places in the code. For more details on the options mentioned, see the code wiki$^{[\ref{wiki}]}$.

\subsection{Towards FAIR}

To improve the reproducibility of simulations, the code reports all compile-time as well as runtime settings, including the git hash at the start of the run. In addition, it reports the complete list of references to the publications associated with the currently activated modules and settings. This ensures that proper credit is given for every simulation, something especially crucial for \og, as it comes with a very large list of physical modules and contributing code developers. Although the code is released under the GNU General Public License v3, we encourage PIs starting new projects to consider inviting the authors of the modules they use to participate in those projects. This is particularly important for early-career scientists, whose contributions often represent a substantial part of their scientific work.

\subsection{Continuous Integration}
\label{sec:Tests}
The development of \og is based on a CI pipeline to systematically verify the code functionality. This pipeline consists of multiple tests that the code must pass:

\begin{itemize}
    \item Compilation tests with various configuration files. In these tests, we check the successful build of the code for various common setups.
    
    \item To check the accuracy of the different hydrodynamic solvers, we compare predictions of the code with the analytic solutions (where available) of various test problems. These tests are a Sod shock tube problem \citep[][]{Sod1978}, a Taylor–von Neumann–Sedov blast wave problem \citep[][]{Neumann1963, Taylor1950, Sedov1959} and a sound wave test. For all these tests, the results using the different hydro solvers implemented in \og must (see Sect. \ref{sec:SPH} and \ref{sec:MFM}) reproduce the analytic solutions within a small tolerance due to numerical noise.
    
    \item The accuracy of the coupling of the gravity solver to the underlying hydrodynamical scheme is tested using a free-fall collapse unit test which checks {\em (i)} the correct initial acceleration profile for a uniform sphere, and {\em (ii)} the correct 50\% Lagrangian radius evolution during free-fall collapse. This test is executed for the available SPH, PES (Pressure-Entropy SPH), and MFM implementations.
    
    \item To test the ability of the code to accurately represent the shocks, rarefaction fans, contact discontinuities, and the compound structures of its MHD implementation, a \citet[][]{BrioWu1988} shock tube is used.
    
    \item Tests for isotropic and anisotropic thermal conduction by solving a temperature jump with differently oriented magnetic fields.

    \item A Kepler disk test (see Figure \ref{fig:kepler_disk}), allowing us to study the ability of the code to conserve angular momentum and maintain stable orbits over time \citep[see][]{Hopkins2015, Groth2023}.

    \item Multiple other tests using Cosmological Boxes or Zoom-In simulations including additional physics to verify the ability to run these modules successfully. This also includes tests with GPU offloading.

    \item Several simple tests are performed to verify the SIDM implementation. These tests do not incorporate gravity but consider only the DM self-interactions. They check various aspects, such as the scattering rate, the distribution of the relative velocities at which the particles scatter, energy conservation, comoving integration, and more.
    
\end{itemize}

\subsection{HPC cluster support}
To increase usability, the code comes with presets for various system types. In these system types, the recommended list of modules and libraries, including their paths, is provided, tested and extended continuously. At the time of the code release, it contained more than 12 systems, among them all of which are part of HPC-Europe. The continuously updated list is available in the {\it Build} directory of \og.


\begin{figure*}[ht]
 \centering
 \includegraphics[trim={0.5cm 3.5cm 8.5cm 0.5cm}, clip, width=1.0\textwidth]{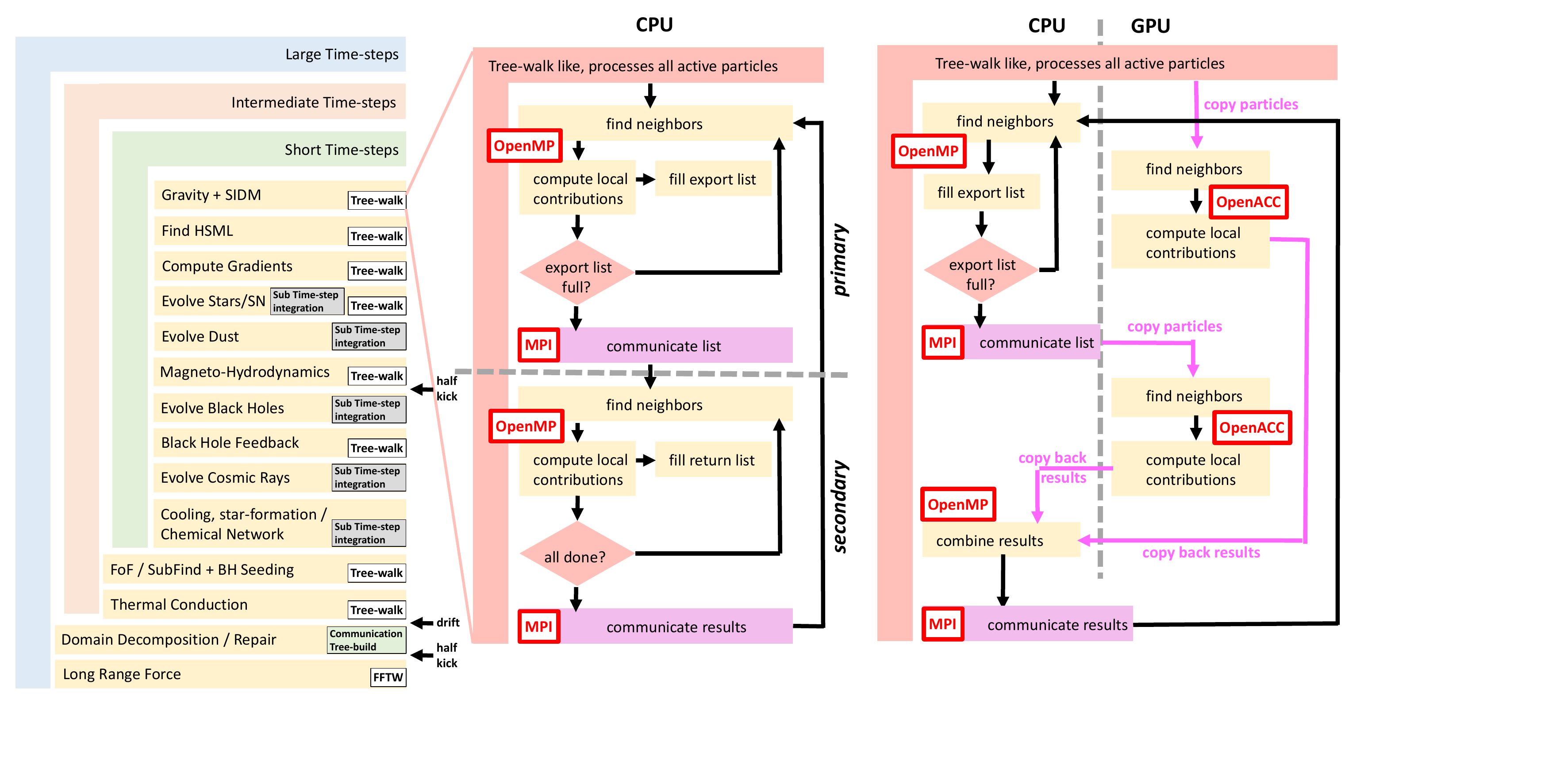}
 \caption{Schematics of \og showing the different physics modules and how they are executed on different time scales (left). Many of them follow an underlying tree walk, which is performed in a primary and secondary phase, as indicated in the centre. The right part shows how this is then split in execution for the case of GPU offloading, and highlights the hybrid CPU/GPU approach used in \og.}
 \label{fig:schematic}
\end{figure*}

\section{Basic Backbone}
\label{sec:Basics}

\og follows a structured execution model consisting of an initialisation phase and a subsequent time-integration loop (see Figure \ref{fig:flowchart}). 

After setting up the parallel communication (MPI) infrastructure, the code starts with the initialisation routine, which performs the following operations in sequence. First, compilation and runtime parameters are reported for reproducibility. Next, the simulation parameter file is parsed, and global data structures are allocated and initialised. The computational domain is subsequently defined, including, e.g., box size and boundary conditions.

The initialisation stage then branches based on the restart flag. In case of a brand-new run, ICs are read from the input model. This typically includes only basic state variables, while additional variables are initialised to their initial values. For runs restarting from a checkpoint, the simulation state is reconstructed from a restart file, which contains all state variables. When restarting from a snapshot file, the code will read the majority of the state variables from the snapshot file, while others will be initialised in the same way as if starting from ICs. In both restart versions, selective parameter updates are permitted. Upon completion of this phase, the system is fully initialised and enters the main evolution loop. The simulation evolves through a sequence of discrete time steps, with each iteration of the loop consisting of several stages, until the final time of the simulation defined in the parameter file is reached.

Following the philosophy of the \textsc{Gadget} family, in \og all 6 different particle types (gas, dark matter, two additional particle types which usually represent boundary or different stellar populations, star particles and black hole particles) evolve on individual time-steps, which are organised in a power-of-2 hierarchy, called time-step levels. While the code progresses at the lowest time-step level, the time at each time step corresponds to a start time within this time-step hierarchy. All particles with time-steps corresponding to this bin and lower are called active particles, for which forces and other variable updates are computed.

The general order of execution proceeds as follows: the Gravitational forces are computed first, followed by the evaluation of the hydrodynamical equations. After the primary force computations are executed, the main loop of the computation proceeds through various physical and structural modules. These may include radiative cooling, star formation, stellar feedback, BH accretion and feedback, or other problem-specific processes. The computation is done through a pipeline of force-like (e.g. including contributions from neighbouring particles) and state-like (i.e. evolving internal properties) evaluations, as indicated in the left part of Figure \ref{fig:schematic}. These different modules are often based on a tree-walk-like structure for finding neighbours to interact with. Such tree-walk-like algorithms are divided into two phases -- {\it primary} and {\it secondary} -- as shown in the middle part of Figure \ref{fig:schematic}. The primary ones process the local particles and fill an export list with the requests for contributions from particles which are hosted on other MPI ranks. After an MPI-based exchange communication phase, the secondary ones then compute the contribution of local particles to requests from other MPI ranks, before the results are sent back in an MPI-based communication. As the export/import buffers are limited, these phases are repeated until all MPI ranks have processed all local, active particles. The local computations in both phases are OpenMP parallelised in many of the modules. In the case of the additional GPU offloading, \og follows a hybrid strategy, where all local computations are offloaded to the GPUs while the communications and the combination of local and remote contributions are handled by the CPUs in parallel. There are several modules which also include a sub-resolution model which is evolved locally for each active particle, often using sub-cycling inside the time-step. A flowchart of the operations of \og is displayed in Figure \ref{fig:flowchart}.

\subsection{Internal units}
\label{sec:units}

The internal variables in \og are usually set in a cosmological framework. Throughout this paper, we will use the symbols commonly adopted for physical quantities (e.g., $\mymathbf{x}$, $\mymathbf{v}$), although they effectively are internal variables, so often co-moving and therefore without a straightforward scaling to their physical counterparts. They are defined as:
\begin{eqnarray}
{\rm Coordinates}:~~    \mymathbf{x} &\equiv& a^{-1}\;\mymathbf{x}_\mathrm{Physical} \\
{\rm Velocities}:~~    \mymathbf{v} &\equiv& a\;\mymathbf{v}_\mathrm{Physical} \\
{\rm Time}:~~    t &\equiv& \ln a \\
                 \delta{}t_\mathrm{Physical} &\equiv& H(a)^{-1} \delta{}(\ln a) \\
{\rm Masses}:~~    m &\equiv& m_\mathrm{Physical} \\  
{\rm Magnetic~field}:~~    \mymathbf{B} &\equiv& a^{-2}\;\mymathbf{B}_\mathrm{Physical} \\
{\rm Density}:~~    \rho &\equiv& a^3\;\rho_\mathrm{Physical} \\
{\rm Pressure}:~~    P &\equiv& a^{3\gamma}\;P_\mathrm{Physical} \\
{\rm Specific internal energy}:~~    u &\equiv& a^{3\gamma-1}\;u_\mathrm{Physical}\\
{\rm Entropy~variable}:~~    A &\equiv& A_\mathrm{Physical}\,,
\end{eqnarray}
where $a$ is the cosmic expansion factor and $\gamma$ is the polytropic index. The units of these variables are then given by:
\begin{eqnarray}
    \left[\mymathbf{x}_\mathrm{Physical}\right] &\equiv& (1+z)^{-1}h^{-1}[\mathrm{LENGTH}] \\
    \left[\mymathbf{v}_\mathrm{Physical}\right] &\equiv& (1+z)[\mathrm{VELOCITY}] \\
    \left[t_\mathrm{Physical}\right] &\equiv&  \frac{[\mathrm{LENGTH}]}{[\mathrm{VELOCITY}]} \equiv \left[\mathrm{TIME}\right]\\
    \left[m_\mathrm{Physical}\right] &\equiv& h^{-1}[\mathrm{MASS}] \\  
    \left[\mymathbf{B}_\mathrm{Physical}\right] &\equiv& (1+z)^{-2}\:h\;[\mathrm{GAUSS}]\,,
\end{eqnarray}
where $z$ is the redshift and $h$ is the Hubble parameter at redshift zero, in units of 100$\,{\rm km\,s^{-1}\, Mpc^{-1}}$.

When the cosmological background is switched off, the aforementioned variables collapse to physical variables.

\subsection{Time integration}
\label{sec:Integration}

To evolve the particle orbits in time, \og uses a Leapfrog kick-drift-kick (KDK) integrator in an implementation described by \cite{Verlet1967} and \cite{Springel2005}. This second-order accurate integrator \cite[e.g.][]{Hernquist1989} has the advantage that for fixed time-steps the phase-space density is preserved and the build-up of long-term secular integration errors in the energy is prevented \citep[see e.g.][]{Saha1992, Hairer2003}.

\subsubsection{Time-step criteria}

Since simulations typically cover regions with highly different dynamics and physical processes, the code allows individual particles to evolve with different time-steps. Therefore, \og employs individual time-steps for all of the particles to increase computational efficiency. These individual time-steps are computed from the local properties defining the dynamics of each particle or to stay within boundaries driven by additional physical processes or numerical approximations. Several time-step criteria are applied. Among them, a criterion based on the acceleration and the gravitational softening of each particle,
\begin{equation}
\Delta{}t_{i}^\mathrm{accel}=\sqrt{2\frac{\mathrm{ErrTolIntAccuracy} \times \epsilon^\mathrm{soft}_i}{|d {\mymathbf v}_i/{d t}|}},
\end{equation}
where the total acceleration of each particle is used and the time step is controlled by an integration error tolerance parameter, which has a default value of $0.05$. A second one is the hydrodynamical time-step for gas particles associated with the Courant condition, defined as
\begin{equation}
\Delta{}t_{i}^\mathrm{courant}=\frac{C\,h_{i}}{v_{i}^\mathrm{sig}},
\end{equation}
where $C$ is the Courant factor (default value is 0.075; specified in the parameterfile) and $v_{i}^\mathrm{sig}$ the maximum signal velocity (see Section \ref{sec:SPH:artvisc} and adaption for MHD in Section \ref{sec:MHD}). Additional time-step criteria apply for physical modules like SIDM, physical viscosity, magnetic field cleaning algorithms, dust-treatment, conduction and more (see the corresponding sections below). From all the different time-step criteria, the minimum is then chosen for each particle.

\subsubsection{Wake-up}
The assignment of individual time steps to particles leads to the distinction between active and inactive particles at any given integration step. Simulations involving spatially overlapping active and inactive regions can become problematic for the gas component, because particles with different time steps may interact with each other. In particular, during rapid changes in velocity or entropy, an active gas particle may enter a region populated by inactive particles. Since these inactive particles do not immediately respond to the sudden appearance of a highly dynamical neighbour, large time-step gradients can develop, potentially resulting in inaccurate force estimates and unphysical outcomes.

To mitigate these issues, \og employs a modified time-step limiting and particle wake-up scheme based on the implementation in the \textsc{Gadget}-3 code by \citet{Pakmor2010, Pakmor2012}. The adopted approach is closely related to the time-step limiter proposed by \citet{Durier2012} and can be viewed as an extension of the mechanism introduced by \citet{Saitoh2009}. In contrast to conventional time-step limiters, our scheme compares signal velocities rather than time steps directly and additionally accounts for errors arising from inaccurate particle state extrapolations.

Each active particle computes the pairwise signal velocity $v_{{\rm sig},ij}$ with all of its gas neighbours and stores the maximum value, which is the same quantity used to determine its Courant time step. If, for an inactive neighbour $j$, the current $v_{{\rm sig},ij}$ exceeds the maximum value previously recorded by $j$ when it was last active by more than a tolerance factor (typically three), the time step assigned to $j$ is no longer considered valid. The particle is therefore woken up, flagged as active, and reassigned the shorter time step required by the updated dynamical conditions. The half-time-step extrapolation previously applied to the particle's hydrodynamical quantities is then reverted and replaced by the contribution from the finer time-step evolution.

In contrast to the original implementation presented by \citet{Beck2016}, \og{} does not assign the new time step to the smallest time step currently present in the simulation. Instead, the woken particle inherits the time step of the highly dynamical particle that triggered the wake-up event. This criterion maintains the required accuracy while avoiding unnecessary reductions in the overall integration efficiency. Furthermore, we modify in \og the original scheme by introducing an auxiliary index list containing particles scheduled for wake-up. This eliminates the need to iterate over the full particle set when identifying particles to be activated and improves computational performance, particularly during time steps with only a small fraction of active particles.

\subsubsection{Treatment of the co-moving background}
\label{sec:Integration:kickfactors}

For the time integration, cosmological pre-factors appearing in the equations are absorbed in the kick factors, which are used as $\Delta t$ in the integration. The adopted formulations are as follows:

\begin{eqnarray}
    \Delta t_\mathrm{grav} &\equiv& \int_{a_i}^{a_{i+1}} \frac{1}{Ha} \frac{1}{a} \mathrm{d}a \\
    \Delta t_\mathrm{hydrokick} &\equiv& \int_{a_i}^{a_{i+1}} \frac{1}{Ha^{3(\gamma-1)}} \frac{1}{a} \mathrm{d}a \\
    \Delta t_\mathrm{mhdkick} &\equiv& \int_{a_i}^{a_{i+1}} \frac{1}{Ha^2} \frac{1}{a} da \\
    \Delta t_\mathrm{SIDM} &\equiv& \int_{a_i}^{a_{i+1}} h\frac{1}{Ha^4} \frac{1}{a} \mathrm{d}a    
\end{eqnarray}

\subsection{Domain decomposition}
\label{sec:domaindecomposition}
The primary parallelisation strategy of \og\ is based on decomposing the full simulation into smaller subdomains that can be distributed efficiently across multiple tasks.

This is achieved by partitioning the computational volume into a set of spatial domains, each assigned to an individual MPI rank. Subsequently, every MPI rank can spawn OpenMP threads to exploit the available shared-memory parallelism and maximise core utilisation. This hybrid approach enables an optimal balance between the number of subdomains, the computational resources (e.g., memory capacity) available to each MPI rank, and the memory locality of individual tasks.

To optimise the spatial compactness of the domains, \og follows the strategy employed by the \textsc{Gadget} family and uses a space-filling Peano-Hilbert curve to map the three-dimensional particle distribution onto a one-dimensional representation. The resulting curve can be partitioned into an arbitrary number of segments, depending on the number of MPI ranks. Multiple segments can be assigned to a single MPI rank, allowing the decomposition to adaptively balance computational workload, memory usage, and communication overhead between ranks.

In contrast to previous \textsc{Gadget} versions, \og\ employs a two-stage domain decomposition scheme with improved control over the frequency of full decompositions. The decision to perform a new domain decomposition is no longer based on the number of force calculations since the previous decomposition. Instead, it is determined automatically by the rank of the current time-bin within the global time-bin hierarchy\footnote{Compile-time options are available to modify this behaviour; see \href{https://gitlab.lrz.de/KlausDolag/OpenGadget3/-/wikis/TimeDomain}.}. If more than two lower time-bin levels exist beneath the time-bin that triggers a full decomposition, an additional repair step runs automatically. During this repair step, only particles that have migrated outside their assigned domain are redistributed to the MPI rank responsible for the corresponding spatial region. Subsequently, all ranks rebuild their tree structures to remove obsolete entries and restore consistency.

\subsection{Neighbour search}
\label{sec:neighbourserach}

\og is using an improved algorithm to search for neighbouring particles compared to the original version implemented in \textsc{Gadget}-3. Instead of searching for neighbouring particles for each particle individually, the code groups nearby particles and performs the neighbour search for a set of particles all at once. This approach reduces the number of searches and therefore results in an improved computational efficiency of the code (see \citet{Ragagnin2016} for details).

\subsection{IO}
Several output formats are supported, including the legacy binary \textsc{Gadget} formats (“format 1” and “format 2”) and HDF5. These formats can be used both for simulation outputs and for initial conditions (ICs), as well as for SubFind outputs (see Section~\ref{sec:subfind}). The I/O scheme allows data from multiple MPI ranks to be consolidated into single files in a flexible manner, while enabling the user to control the number of files written concurrently. This provides an efficient way to optimise parallel I/O by distributing the data across multiple files, while keeping individual file sizes manageable for data transfer and long-term archiving.

The number of concurrent parallel I/O operations is configurable by the user and can be chosen up to the total number of output files. This flexibility allows an optimised balance between maximising I/O parallelism and limiting the number of simultaneous write operations to avoid overloading the underlying filesystem. In \og, the same I/O strategy is applied not only to standard simulation outputs but also to SubFind outputs and other auxiliary data products.

\subsection{SPACE-Timers}
\noindent
{\it Main contributing developer: G. S. Karademir}

The code includes a lightweight profiling tool that measures the time between the start and end of pre-defined regions in the main task and provides insight into the code's and its modules' time consumption. This allows the user and developer to get insights into the actual runtime of the code to identify potential bottlenecks. In the following, we will describe the basic usage of the timers. More details on the tool are described in \cite{Karademir2026}.

\begin{figure*}[ht]
    \centering
    \includegraphics[trim={0.3cm 1cm 0.3cm 0.0cm}, clip,width=\textwidth]{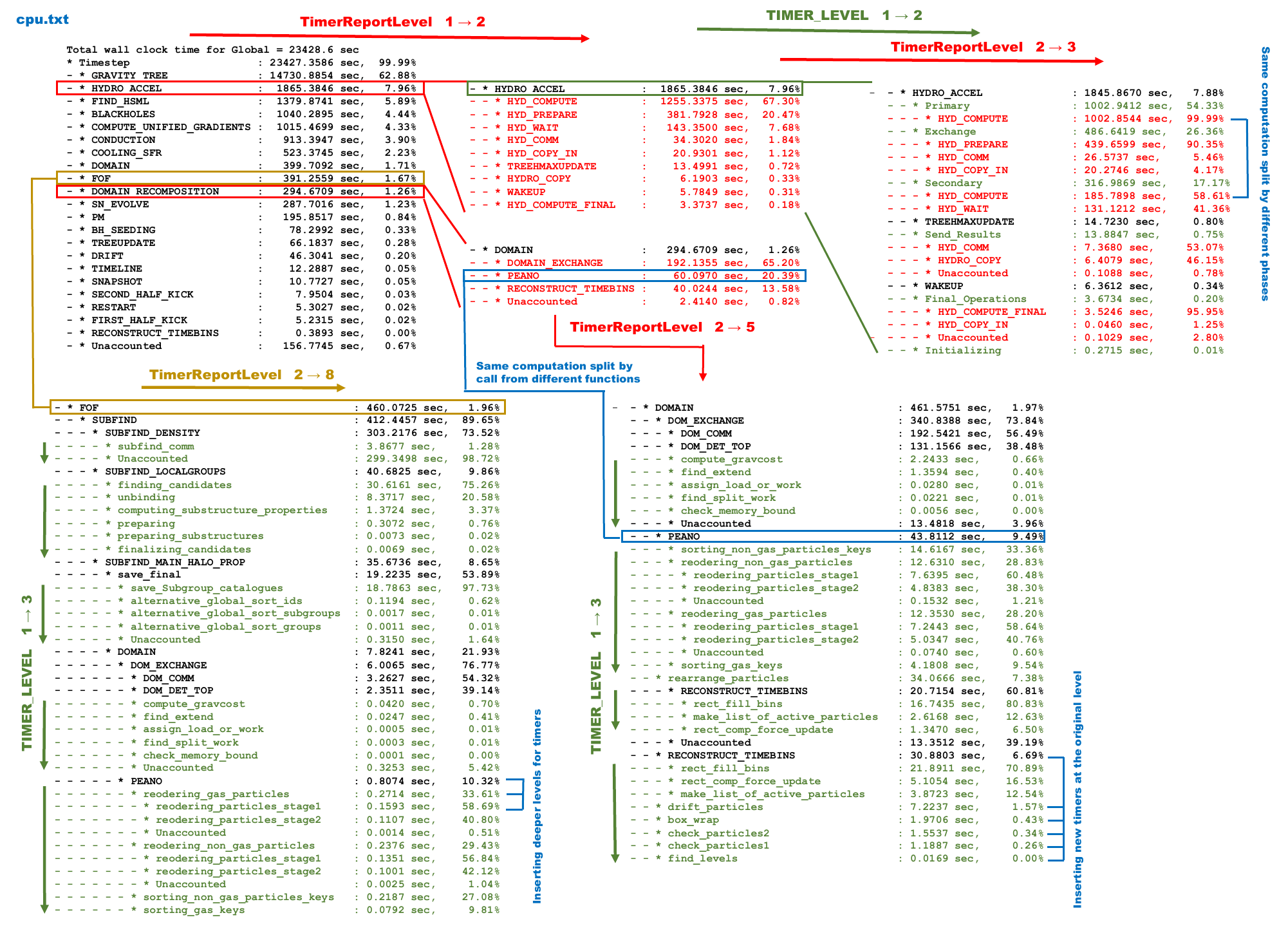}
    \caption{Example of the content of the cpu.txt file, highlighting the interplay between {\tt TIMER\_LEVEL} and {\tt Timer\_report\_level}. Note that increasing the {\tt TIMER\_LEVEL} can be used to split the time measurements of the same part of the code when called from different regions, as the example of {\tt HYD\_COMPUTE} in the upper right part shows, which can be split into the contribution from the call within the primary and secondary loop.
    }
    \label{fig:Timers}
\end{figure*}

The depth of the measurement is set at compilation time by setting a timer level, where larger values activate more and more granular regions. See Figure \ref{fig:Timers} for a schematic view of the timer hierarchy with a level of $2$. While this configuration switch controls the measurement level, parameters at runtime control the depth of reporting as well as additional balance reports. Here, it is important to highlight the difference between the "level" of the measurement and the "depth" of the output. Each region is associated with a level only depending on its importance for the global profiling, while the depth of the reporting is only based on the hierarchy within the timer region tree. This means that even a timer region with $\mathrm{level}=0$ can be omitted from the output if it's embedded in a larger number of other timer regions than the reporting level set.

To inspect the results of the timer regions, the code produces several output files:
\begin{itemize}
    \item cpu.txt: Provides global accumulated timer information based on  MPI rank 0.
    \item balance.txt: This file shows which time regions took which fraction of the total time for each time-step and represents it by an individual symbol. This gives the user a quick overview of which parts of the code are the most time-consuming. As a note, the results are an approximation due to rounding errors.
    \item balance-head.txt: In this file, the legend for the balance.txt shows which symbol corresponds to which timer region can be found.
\end{itemize}

Besides this basic behaviour, it is also possible to measure the regions separately for each MPI rank and in a non-cumulative manner, where each rank will create an individual timing file. This behaviour needs to be set during compilation.

\section{Gravity Solver}
\label{sec:Gravity}
For cosmological simulations, an accurate treatment of gravitational interactions is of fundamental importance. While the gravitational forces can, in principle, be computed through direct summation, the associated computational cost is prohibitive for large particle numbers, as it scales quadratically with the number of particles. In \og, we largely follow the gravity solver implementation introduced in \textsc{Gadget}-2 \citep[][]{Springel2005}, while employing a more effective hybrid approach that combines an Oct-Tree algorithm with a Particle Mesh (PM) solver \citep[see e.g.][]{Xu1995, Bode2000}. The details are described in the following.

\subsection{Oct-Tree}
\label{sec:Tree}
The use of tree algorithms for an accurate and efficient treatment of gravitational interactions was initially proposed by \cite{Appel1985} and \cite{BarnesHut1986}. In this approach, the computational domain is recursively subdivided into eight sub-cubes until each leaf node contains only a single particle. By applying an opening-angle criterion, sufficiently distant groups of particles can be treated collectively rather than individually. Instead of computing pairwise gravitational interactions for every particle, the contribution of a group of particles is approximated through its multipole expansion. This significantly reduces the number of force evaluations compared to direct summation, enabling efficient calculations for large particle distributions.

\subsubsection{Tree construction}
The gravitational tree is organised using a unified index space that distinguishes between leaf particles, internal nodes, and pseudo-particles representing remote MPI domains. Each internal node stores both geometric and dynamical information. The geometric description consists of the cubic cell centre and its side length. During construction, nodes temporarily maintain explicit child pointers. After construction, these are replaced by aggregated physical quantities—total mass and centre of mass—along with traversal metadata (nextnode, sibling, and father). This dual-use memory layout minimises storage overhead while enabling efficient traversal.

A key design feature is the transformation of the tree into a two-pointer linked structure. Each node contains a nextnode pointer, indicating the first child to visit when the node is opened, and a sibling pointer, indicating the next node to visit if the current node is accepted as a single interaction.

Leaf particles are integrated into this structure via a separate nextnode array. After construction, the entire tree can be traversed linearly using a single loop, eliminating the need for recursion or an explicit stack.

Tree construction proceeds in several stages. First, the root node is initialised to cover the entire simulation domain. A top-level skeleton is then created to match the domain decomposition across MPI ranks, ensuring consistency in parallel runs. Particles are inserted sequentially following their Peano–Hilbert ordering, which preserves spatial locality and improves cache efficiency. Once all particles are inserted, a bottom-up pass computes multipole moments (mass and centre of mass) for each node. Subsequently, nodes that are entirely local to a given MPI rank are flagged. For parallel operations, multipole data of top-level nodes are exchanged between ranks, allowing remote regions to be approximated as pseudo-particles.

\subsubsection{Tree walk}
The tree walk evaluates gravitational forces for a single target particle per invocation. The traversal begins either at the root node (for local particles) or at a designated top-level node (for imported particles). The walk proceeds iteratively using the linked-node structure described in the previous section.

At each step, the current index is classified into one of the following three categories:
\begin{itemize}
    \item Leaf particle: The particle is drifted to the current time if necessary, and its contribution to the force is computed directly. The traversal then advances via the nextnode pointer.
    \item Pseudo-particle: These represent remote MPI domains. In the primary (local) walk, encountering such a node triggers export bookkeeping so that the interaction can later be evaluated on the owning rank. No force is computed locally, and traversal continues with the corresponding nextnode.
    \item Internal node: The node’s multipole data (mass and centre of mass) are used to evaluate whether the node can be treated as a single interaction or must be opened. This decision is governed by an opening criterion.
\end{itemize}

For the tree, two opening criteria are supported. The standard Barnes–Hut criterion compares the node size to its distance from the target, opening the node if the angle under which the node is seen is larger than a tolerance parameter. Alternatively, a relative criterion compares the truncation error estimate to the particle’s previous acceleration, opening nodes when higher accuracy is required. Additional checks ensure correctness in regions with mixed gravitational softening or when the target lies close to the node boundary. If a node is opened, traversal proceeds to its first child via nextnode. If it is accepted, the node is treated as a single mass located at its centre of mass, and the traversal skips the entire subtree using the sibling pointer.

\subsubsection{Force evaluation}

Force contributions are computed whenever a leaf particle or an accepted internal node is encountered. Let $r$ denote the distance between the target and the interaction partner, and $h$ the effective gravitational softening length. 

The force transitions smoothly between the following two regimes: for $r\geq h$, the interaction is purely Newtonian, with acceleration proportional to $M/r^3$. Instead, for $r<h$, a spline-softened kernel is used, corresponding to the formulation of \cite{Monaghan1985}. This replaces the point mass with a finite-density distribution, ensuring that the force remains finite and continuous at small separations. The spline kernel is constructed such that both the force and its first derivative are continuous across the transition points, providing stable and accurate force estimates in dense regions. Note that also \og follows the philosophy of the \textsc{Gadget} family, where the softening $\epsilon$ in the parameter files refers to the Plummer equivalent, which means that $h$ from above is $2.8$ times larger. 

\subsection{Particle Mesh}
\label{sec:PM}

An additional option in \og is to use a particle mesh \citep[][]{Eastwood1974} method to calculate the long-range gravitational forces. When used, the global Newtonian potential, $\phi(r)\sim 1/r$, is split into a short-range $\phi_s(r)$ part, evaluated by the Barnes\&Hut oct-tree, and a long-range $\phi_l(r)$ part, evaluated on a mesh. At present, running the particle mesh algorithm on its own is not supported, since by construction it only returns the long-range term. 

The split uses the identity
\begin{equation}
\frac{1}{r} = \frac{\operatorname{erfc}(r/2A_{\rm smth})}{r}
            + \frac{\operatorname{erf}(r/2A_{\rm smth})}{r},
\end{equation}
so that the potential contributed by a particle of mass $m$ decomposes as
\begin{align}
\phi(r) &= \phi_s(r) + \phi_l(r) \\
\phi_s(r) &= -\,\frac{\mathrm{G} m}{r}\,\operatorname{erfc}(u) \\
\phi_l(r) &= -\,\frac{\mathrm{G} m}{r}\,\operatorname{erf}(u) \\
u &\equiv \frac{r}{2 A_{\rm smth}}.
\end{align}
The length scale $A_{\rm smth}$ sets the Gaussian scale of the split between the two regimes.

For the long-range contributions, all particles are assigned to grid cells and the resulting density field is binned onto a mesh, then Fourier transformed and multiplied by the Green's function of the Poisson equation with a Gaussian low-pass filter, and finally transformed back:
\begin{equation}
\tilde{\phi}_l(\mathbf{k}) = -\,\frac{4\uppi \mathrm{G}\,\tilde{\rho}(\mathbf{k})}{k^2}\, \exp\!\left(-k^2 A_{\rm smth}^2\right).
\end{equation}
In the code, the FFTW3 library by \cite{Frigo2005} is used for the implementation of the Fourier transform.

\subsection{Analytic static Potentials} \label{sec:Gravity_analytic_potentials}
For some scientific cases, it is valuable to be able to add an analytically described potential to the simulation. For example, when one wants to study the evolution of satellite galaxies but does not want to resolve the host halo. \og has several functions to describe such analytic potentials. The contribution of these potentials to the gravitational acceleration of each particle is evaluated individually. The following functions are implemented as static potentials centred about the origin of the coordinate system.

The Navarro--Frenk--White (NFW) profile \citep{Navarro1996} can be added to the simulations, for example to represent a DM halo. The corresponding potential as a function of radius $r$ is
\begin{equation} \label{eq:nfw_potential}
    \Phi_\mathrm{NFW}(r) = -4 \uppi \, \mathrm{G} \, \frac{\rho_\mathrm{s} \, r_\mathrm{s}^3}{r} \, \ln\left(1 + \frac{r}{r_\mathrm{s}}\right) \,.
\end{equation}
Here, $\rho_\mathrm{s}$ and $r_\mathrm{s}$ are the density parameter and scale radius describing the NFW profile.

Moreover, the Hernquist profile \citep{Hernquist1990} is available as well. Its potential is
\begin{equation} \label{eq:hernquist_potential}
    \Phi_\mathrm{Hern}(r) = - \frac{\mathrm{G} M_\mathrm{H}}{a_\mathrm{H}} \, \left(1 + \frac{r}{a_\mathrm{H}}\right)^{-1} \, .
\end{equation}
The mass of the halo is given by $M_\mathrm{H}$ and its spatial extend is controlled by $a_\mathrm{H}.$

Finally, \og allows us to add an axisymmetric Miyamoto–Nagai profile \citep{Miyamoto1975}, which can be used to describe a disk-like component.
\begin{equation} \label{eq:disk_potential}
    \Phi_\mathrm{disk}(r) = - \mathrm{G} \, M_\mathrm{d} \left[ R^2 + \left( a_\mathrm{d} + \sqrt{z^2 + b^2_\mathrm{d}} \right)^2 \right]^{-1/2},
\end{equation}
with $R = \sqrt{x^2 + y^2}$. In contrast to the previous potentials, it is not spherically symmetric but depends on the coordinates $x$, $y$, and $z$.
In detail, the potential is controlled by three parameters, $M_\mathrm{d}$, $a_\mathrm{d}$, and $b_\mathrm{d}$.

For a list of all available configuration and parameter options, see the according \href{https://gitlab.lrz.de/AstroCodes/OpenGadget3/-/wikis/Analytic%20Potentials}{section on the code wiki}.
\section{Beyond \texorpdfstring{$\Lambda$}{Lambda}CDM}
\label{sec:Beyond_LCDM}

\subsection{Non standard cosmologies}

\og follows the philosophy of the \textsc{Gadget} family for compatibility reasons and uses by default a Hubble function which neglects the radiation term,
\begin{equation}
H(a) = H_0 \left({\frac{\Omega_0}{a^3} + \frac{1-\Omega_0-\Lambda_0}{a^2}+\Lambda_0}\right)^{1/2}.
\end{equation}
It also allows, at compile time, to switch on the radiation term,
\begin{equation}
H(a) = H_0 \left(\frac{\Omega_0}{a^3} + \frac{1-\Omega_0-\Lambda_0}{a^2}+\Lambda_0+\frac{\Omega_\mathrm{rad}}{a^4}\right)^{1/2}.
\end{equation}
Following \citet{dolag04}, \og allows us to take different dark energy models into account by further modifying the Hubble function. In the simple form, a fixed value of $\Omega_L$ can be replaced with a fixed equation-of-state parameter $\omega$, changing the Hubble function to
\begin{equation}
H(a) = H_0 \left(\frac{\Omega_0}{a^3}+\frac{1-\Omega_0-\Lambda_0}{a^2}+\frac{\Lambda_0}{a^{3(1 + \omega)}}+\frac{\Omega_\mathrm{rad}}{a^4}\right)^{1/2}.
\end{equation}
For more extended dark energy models, \og can load external tables for a time-dependent dark energy equation of state $\omega(a)$ and construct the corresponding Hubble function 
\begin{eqnarray}
H(a) & = & H_0 \bigg[\frac{\Omega_0}{a^3} + \frac{1-\Omega_0-\Lambda_0}{a^2}+ \nonumber \\ 
& + & \Lambda_0\mathrm{exp}\left(-3\int_a^1\frac{1+\omega(a')}{a'}\mathrm{d}a'\right)+\frac{\Omega_\mathrm{rad}}{a^4}\bigg]^{1/2}
\end{eqnarray}
or can read an arbitrary tabulated Hubble function $H(a)$ from an input file. In addition, it allows loading an external auxiliary function $G_\mathrm{cor}(a)$ which modifies the gravitational constant over time.

\subsection{Including Neutrinos}

Massive neutrinos affect structure formation both through their contribution to the homogeneous expansion and through their gravitational clustering. Their large thermal velocities allow them to free-stream out of density perturbations below a mass- and redshift-dependent scale, suppressing the growth of the CDM and baryon density fields. Although neutrinos contribute only a small fraction $f_\nu=\Omega_\nu/\Omega_{\rm m}$ of the order of 5\textperthousand\ of the total matter density~\citep{Planck:2018vyg}, accurately modelling this scale-dependent effect is important for precision predictions of large-scale structure. In the non-relativistic limit, their present-day density is related to the sum of the masses by $\Omega_\nu h^2 \simeq \sum_i m_{\nu, i}/(93.14\,{\rm eV})$, where the sum runs over the number of massive neutrino species.

The neutrino implementation in \og follows the particle-based approach introduced by \citet{Viel2010Neutrinos} and subsequently used for the DEMNUni simulations \citep{Carbone2016DEMNUni}. A separate ensemble of collisionless particles represents neutrinos. The initial-condition generator supplies their positions and bulk velocities using the neutrino density and velocity transfer functions, and adds a thermal component to each bulk velocity by sampling the relic Fermi--Dirac momentum distribution,
\begin{equation}
    \mathcal{P}(q)\,\mathrm{d}q \propto \frac{q^2}{\exp(q)+1}\,\mathrm{d}q,
    \qquad q \equiv \frac{pc}{k_{\rm B}T_\nu(z)} .
\end{equation}
Here, $\mathcal{P}(q)$ is the probability density for drawing the magnitude $q$ of the dimensionless thermal momentum, and $\mathrm{d}q$ denotes the corresponding momentum interval. Furthermore, $p=|\mathbf{p}|$ is the magnitude of the physical neutrino momentum, $c$ is the speed of light, $k_{\rm B}$ is the Boltzmann constant, and $T_\nu(z)=T_{\nu,0}(1+z)$ is the relic-neutrino temperature at redshift $z$, with $T_{\nu,0}\simeq1.95\,\mathrm{K}$. The proportionality constant is fixed by requiring $\int_0^\infty \mathcal{P}(q)\,\mathrm{d}q=1$.
The resulting particle ensemble therefore samples the neutrino phase-space distribution rather than only its first two moments. \og reads these pre-generated type-2 particle velocities and evolves them with the same kick--drift integration as the other collisionless species.

The method is enabled at compile time by switching on neutrino treatment, which designates a specific particle type to represent the neutrino component. Neutrino particles always participate in the long-range particle-mesh calculation. At early times, however, their large thermal velocities make short-range clustering inefficient while greatly increasing the cost of the tree calculation. The parameter \texttt{Time\_tree\_on\_nu} therefore specifies the scale factor after which neutrino particles are included in tree construction and receive short-range tree forces. Before this time, their gravitational coupling is calculated through the PM force alone. The appropriate transition depends on the neutrino mass and on the force and mass resolution of the simulation.

When neutrinos are enabled at compile time, the type-2 candidate time-step is omitted from the minimum that defines the global root-mean-square displacement limit, preventing the high thermal neutrino velocities from shortening this global time-step. Neutrino particles retain their normal individual integration and gravitational evolution. For calculations in which only the response of the cold matter and baryons is required, writing type-2 particles in regular snapshots can be suppressed, while retaining them internally and in special outputs.

The principal limitation of a particle treatment is sampling noise. Since the thermal velocity dispersion is much larger than the coherent peculiar velocity, a finite particle ensemble produces shot noise in the neutrino density field. Although not implemented in \og, this can in principle be mitigated by tracking perturbations to the neutrino phase-space distribution with particles, as described in \citet{2021MNRAS.507.2614E}. Alternatively, increasing the number of neutrino particles reduces this noise but raises the memory and force-calculation costs, potentially making the subdominant neutrino component dominate the simulation expense. The effect is most evident in the neutrino auto-power spectrum, whereas its impact on the total gravitational potential is reduced by the small value of $f_\nu$. Particle number, starting redshift, force resolution, and the sampling of thermal momenta must consequently be tested together for convergence.

The particle implementation was included in the Euclid comparison of massive-neutrino methods \citep{Adamek.etal.2023, EuclidCollabrotation2024}. That study found consistent predictions among independent particle- and mesh-based codes at sub-per cent accuracy for the principal matter statistics when resolution and sampling effects are controlled. It also showed that neutrino auto- and cross-spectra are substantially more sensitive to particle shot noise. These results support the use of the \og implementation for total-matter and halo statistics while motivating explicit convergence tests whenever observables depend directly on the resolved neutrino density field.

For a list of all available configuration and parameter options, see the according \href{https://gitlab.lrz.de/AstroCodes/OpenGadget3/-/wikis/Non-Standard-Cosmology-and-Neutrinos}{section on the code wiki}.


\subsection{Self-interacting DM}
\label{sec:SIDM}
\noindent
{\it Main contributing developers: M.\ S.\ Fischer, C.\ Arido, D.\ Klemmer, Y.\ Patil, L.\ D.\ Schmidt, M.\ Wiertel}

\begin{figure*}
    \centering
    \includegraphics[width=\textwidth]{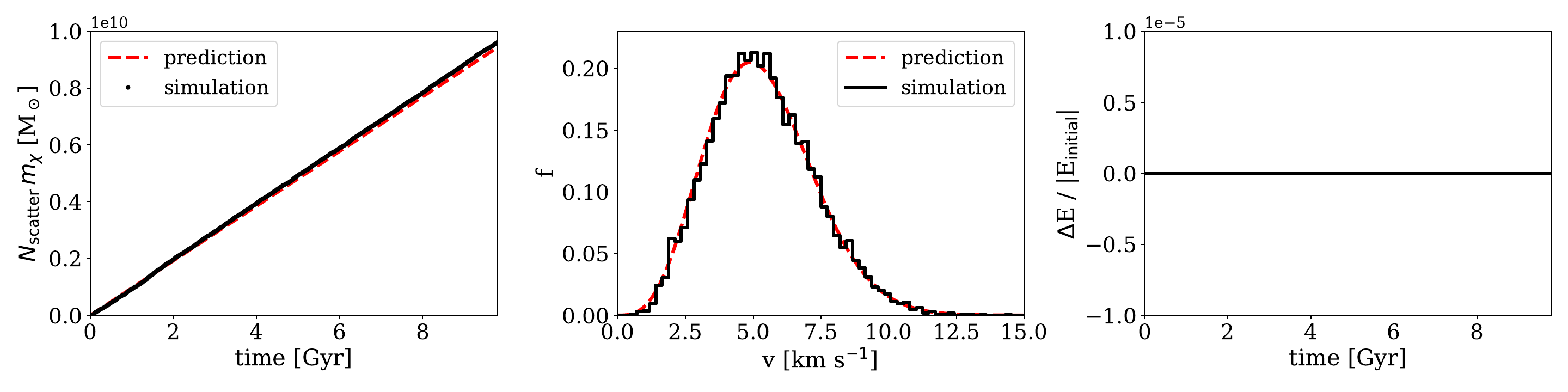}
    \caption{Test problem for the rSIDM implementation simulating a constant DM density with a Maxwell--Boltzmann distribution for the velocities. We compare the simulation results (black) to the analytic solution (red). The left panel gives the number of scatter events times the physical DM particle mass since the start of the simulation as a function of time. In the middle panel, one can see a histogram of the velocities at which the scatterings take place. The right panel gives the error in energy conservation relative to the initial energy of the simulation set-up as a function of time. We note that this figure is a reprint of Fig.~6 in \cite{Fischer2026}. For more information, see that paper.}
    \label{fig:sidm_test}
\end{figure*}

SIDM is a class of particle physics models that allow for non-gravitational interactions of DM particles. Those scatterings between DM particles can alter the matter distribution on small scales and are a promising alternative to the collisionless cold DM (CDM) of the cosmological standard model \citep{Spergel2000}.
They might be more capable of explaining various observations such as those indicating fairly compact dark objects \citep[e.g.][]{Yu2026} or enhancing the growth of BHs \citep[e.g.][]{Sabarish2025}. For more information on SIDM, we refer the reader to the review articles by \cite{Tulin2018} and \cite{Adhikari2025}.

Non-gravitational self-interactions of DM particles with an interaction range below the numerical resolution limit can be simulated with the SIDM module of \og. Its implementation was introduced by \cite{Fischer2021a} and further improved and extended by \cite{Fischer2021b, Fischer2022, Fischer2024a, Arido2025, Patil2025, Klemmer2026, Schmidt2026}. It is fully compatible with the baryonic processes implemented in \og and can be used in full-physics simulations \citep{Ragagnin2024}.
The basic equation that is solved is the Vlasov--Poisson equation with an additional collision term on the right-hand side that describes the DM self-interactions.
\begin{equation} \label{eq:SIDM_vlasov_poisson}
    \frac{\partial f}{\partial t}
    + \mathbf{v} \cdot \nabla_x f
    - \nabla_x \Phi \cdot \nabla_v f
    = \left(\frac{\partial f}{\partial t}\right)_\mathrm{coll} \;.
\end{equation}
Here, $f$ is the phase-space density, $\mathbf{v}$ the velocity of the matter, and $\Phi$ the gravitational potential. The latter is computed as described in Sect.~\ref{sec:Tree} and~\ref{sec:PM}. In the following, we describe how the collision term is modelled in \og{}.

\subsubsection{Basis implementation}

The scheme is based on pairwise interactions of spatially close numerical particles as commonly done in SIDM $N$-body codes \citep[e.g.][]{Kochanek_2000}. Large-angle scattering, also referred to as rarely SIDM (rSIDM), is modelled with a Monte Carlo approach where an interaction probability,
\begin{equation} \label{eq:SIDM_scatter_probability}
    P_{ij} = \frac{\sigma}{m_\chi} \, m \, |\Delta \mathbf{v}_{ij}| \,\Lambda_{ij} \, \Delta t \, ,
\end{equation}
for particles $i$ and $j$ is computed, and particles are scattered based on pseudo-random numbers. Here, $\sigma / m_\chi$ is the total cross-section divided by the physical particle mass; $m$ refers to the mass of the numerical DM particles, and $\Delta \mathbf{v}_{ij}$ is the relative velocity of the two particles. Moreover, a geometric factor $\Lambda_{ij}$ is employed, and $\Delta t$ denotes the time step.

For a differential cross-section $\mathrm{d}\sigma / \mathrm{d}\cos \theta_\mathrm{cms}$ the total cross-section is
\begin{equation} \label{eq:SIDM_total_cross_section}
\sigma=\int_{-1}^{1} \frac{\mathrm{d} \sigma}{\mathrm{d} \cos \theta_{\mathrm{cms}}} \, \mathrm{d} \cos \theta_{\mathrm{cms}} \, .
\end{equation}
Here, $\theta_\mathrm{cms}$ denotes the scattering angle in the centre-of-mass frame.

The geometric factor $\Lambda_{ij}$ is computed with the help of a kernel function $W$.
\begin{equation} \label{eq:SIDM_kernel_overlap}
    \Lambda_{ij} = \int W(|\mathbf{x}-\mathbf{x}_i|, h_i) \, W(|\mathbf{x}-\mathbf{x}_j|, h_j) \, \mathrm{d}^3\mathbf{x} \,.
\end{equation}
The kernel size $h$ of each particle is determined with a nearest-neighbour search, as also done for the hydrodynamic solvers (Sect.~\ref{sec:SPH}). For pairs of close particles, the volume integral over the product of the kernel functions of the two particles is computed to obtain $\Lambda_{ij}$. A detailed description of how $\Lambda_{ij}$ is computed can be found in \citet{Fischer2021a}.

It is of paramount importance to test that the actual implementation reproduces the known solutions for relevant test problems. One of those problems involves a cubic simulation volume with periodic boundary conditions filled with a constant DM density and velocities following a Maxwell--Boltzmann distribution. Only the self-interactions, assuming isotropic, velocity-independent elastic scattering, are computed, and gravity is switched off. For this problem, the expected number of scattering events at the velocities at which they occur is known, as we illustrate in Figure~\ref{fig:sidm_test}. For more details about this test, see Sect.~3.1.1 by \cite{Fischer2026}; for further tests, see their other tests.

A unique feature that makes our SIDM implementation stand out is its ability to simulate forward-dominated cross-sections, i.e.\ DM models where the typical scattering angle is very small.
Small-angle scattering, also referred to as frequently SIDM (fSIDM), in \og is described by an effective drag force and perpendicular momentum diffusion, where the latter is also modelled in a Monte Carlo fashion \citep{Fischer2021a}.
The drag force acting on pairs of numerical DM particles is given by
\begin{equation} \label{eq:SIDM_drag_force}
    F_{\mathrm{drag},ij} = \frac{1}{2} \frac{\sigma_\mathrm{\Tilde{T}}}{m_\chi} \, m^2 \, |\Delta \mathbf{v}_{ij}|^2 \, \Lambda_{ij} \,.
\end{equation}
We note that the total cross-section does not capture the strength of the drag force, but instead the modified momentum transfer cross-section, $\sigma_\mathrm{\Tilde{T}}/m_\chi$, is employed \citep[see also][]{Kahlhoefer2014}.
\begin{equation} \label{eq:SIDM_mod_momentum_transfer_cross_section}
\sigma_\mathrm{\Tilde{T}}= \int_{-1}^{1} \frac{\mathrm{d} \sigma}{\mathrm{d} \cos \theta_{\mathrm{cms}}}\left(1-|\cos \theta_{\mathrm{cms}}| \right) \mathrm{d} \cos \theta_{\mathrm{cms}} \, .
\end{equation}

While Eq.~\eqref{eq:SIDM_scatter_probability} and Eq.~\eqref{eq:SIDM_drag_force} assume that the scattering DM particles have the same mass, the implementation in \og is more general and allows for simulating the scattering between particles with unequal particle masses (\cite{Patil2025}, see also \cite{Fischer2025a}). Importantly, the ratio between the mass of a numerical particle and the mass of the physical DM particles it represents must be the same for all numerical DM particles.

\subsubsection{Full differential cross-section}
Furthermore, \og allows simulating various velocity dependencies \citep{Fischer2024a} and full differential cross-sections \citep{Wiertel2023, Fischer2026}.
This, for example, includes M{\o}ller scattering in the non-relativistic limit using the Born approximation.
\begin{equation} \label{eq:SIDM_moeller_dcs}
    \left.\frac{\mathrm{d}\sigma}{\mathrm{d}\cos\theta_\mathrm{cms}}\right|_\textnormal{M{\o}ller} = \sigma_0 \frac{\left(3 \cos^2\theta_\mathrm{cms} + 1 \right) \frac{v^4}{w^4} + 4 \frac{v^2}{w^2} + 4}{\left( \sin^2\theta_\mathrm{cms} \, \frac{v^4}{w^4} + 4 \frac{v^2}{w^2} + 4 \right)^2} \,.
\end{equation}
Here, the model parameters $\sigma_0$ and $w$ control the scattering rate and the velocity dependence.
To simulate a full differential cross-section, the scattering angle $\theta_\mathrm{cms}$ is drawn in accordance with the differential cross-section (e.g.\ Eq.~\eqref{eq:SIDM_moeller_dcs}).
To do so, the cumulative distribution function of the scattering angles is precomputed and stored in a look-up table used for drawing the scattering angles. Here, we proceed similarly to the pioneering work by \cite{Robertson2017}. 
The details of our implementation are described in \cite{Fischer2026}.
It allows implementing additional differential cross-sections easily.

\subsubsection{Hybrid SIDM scheme}
Furthermore, the hybrid SIDM scheme (hSIDM) \citep{Arido2025} extends the SIDM module to models in which both small- and large-angle scattering contribute significantly, such as light mediator models featuring a Yukawa or Coulomb-like interaction potential. The central idea is to introduce a critical angle $\theta_\mathrm{c}$ that divides the differential cross-section into two complementary parts: scatterings with deflection angles $\theta \leq \theta_\mathrm{c}$ are described by the frequent-SIDM drag force and perpendicular momentum diffusion (Eq.~\eqref{eq:SIDM_drag_force}), while scatterings with $\theta > \theta_\mathrm{c}$ are handled explicitly by the rare-SIDM Monte Carlo scheme (Eq.~\eqref{eq:SIDM_scatter_probability}). For the fSIDM part, the modified momentum transfer cross-section is integrated only over the small-angle regime $[0, \theta_\mathrm{c}]$, and the total cross-section entering the rSIDM scattering probability is integrated over $(\theta_\mathrm{c}, \uppi]$ accordingly. For models with indistinguishable particles, whose differential cross-section is symmetric under $\theta \to \uppi - \theta$ (as in M{\o}ller scattering), an additional backward critical angle $\theta_\mathrm{c,bwd}$ is introduced, so that scatterings near $\theta = \uppi$ are also described by the fSIDM approach, and the rSIDM integration range reduces to $(\theta_\mathrm{c}, \uppi - \theta_\mathrm{c,bwd})$.

The choice of $\theta_\mathrm{c}$ balances accuracy against numerical efficiency: a larger value reduces the number of rSIDM Monte Carlo scatter events and speeds up the simulation, at the cost of approximating a wider range of deflection angles with the effective drag force description. \citet{Arido2025} showed that the scheme is insensitive to the choice of $\theta_\mathrm{c}$ within an expected range and demonstrated speed-ups of several orders of magnitude compared to pure rSIDM for typical light mediator models. The hybrid scheme has furthermore been extended to support fully non-separable differential cross-sections \citep{Fischer2026}, enabling simulations of realistic particle physics models where the velocity and angular dependences are intrinsically coupled, e.g.\ M{\o}ller scattering (Eq.~\eqref{eq:SIDM_moeller_dcs}).

\subsubsection{Dissipative SIDM}
So far, we have assumed the self-interactions to be elastic. DM may, however, also undergo inelastic interactions that radiate energy away and thereby cool overdensities. Such processes are described by a cross-section $\frac{\mathrm{d} \sigma}{\mathrm{d}{\cos \theta_{\mathrm{cms}}}\,\mathrm{d}k^0}$, differential not only in the scattering angle but also in the radiated energy $k^0$. Given this quantity, the dissipative extension of rSIDM is straightforward: the radiated energy is sampled alongside the scattering angle and reduces the speeds of the outgoing DM particles. \og provides a simple model of this type, with an isotropic cross-section and energy loss proportional to the initial CMS kinetic energy \citep[cf.][]{shen21}. The particles emerge back-to-back with relative speed reduced as
\begin{equation}
    |\Delta \boldsymbol{v}_{ij}| \to |\Delta \boldsymbol{v}_{ij}|\sqrt{1-f_\mathrm{diss}},
\end{equation}
where $f_\mathrm{diss}\in[0,1]$ is a constant dissipation factor.

For fSIDM, \citet{Schmidt2026} derived that frequent dissipative small-angle scatterings enhance the elastic drag force (cf. Eq.~\eqref{eq:SIDM_drag_force}) by a factor
\begin{equation}
\begin{split}
    r_\mathrm{diss} = 1 + \frac{4}{\sigma_\mathrm{\Tilde{T}}} &\int\limits_{m_\phi}^{m_\chi|\Delta \boldsymbol{v}_{ij}|^2/4}\mathrm{d}k^0\\ &\int\limits_0^1\mathrm{d}{\cos\theta_\mathrm{cms}}\, \frac{\mathrm{d}{\sigma}}{\mathrm{d}{\cos\theta_\mathrm{cms}}\,\mathrm{d}{k^0}} \frac{k^0}{m_\chi |\Delta \boldsymbol{v}_{ij}|^2}\cos\theta_\mathrm{cms},
\end{split}
\end{equation}
where the lower limit of the $k^0$-integral is set by the mass $m_\phi$ of the particles radiated in a single interaction, and the upper limit by the available CMS kinetic energy. Like the cross-section itself, $r_\mathrm{diss}$ is in general velocity-dependent, although the current implementation is restricted to constant values, corresponding to $k^0\propto m_\chi |\Delta \boldsymbol{v}_{ij}|^2$ as for $f_\mathrm{diss}$ above. The elastic case is recovered for $r_\mathrm{diss}=1$. In contrast to elastic fSIDM, the transverse momentum diffusion no longer restores the full energy removed by drag. Instead, it is adjusted to reproduce the energy loss per interaction of the underlying model, while momentum remains explicitly conserved. For details on this, as well as on the time-step criterion and on combining multiple interaction channels (e.g., elastic and inelastic), we refer the reader to \citet{Schmidt2026}.

\subsubsection{Analytically described host halo}
For simulations where one is interested in the properties of a satellite galaxy orbiting a more massive host, \og allows one to describe the host analytically and only resolve the satellite galaxy explicitly. To model the gravitational influence of the host on the satellite, the host potential is described analytically (see Sect.~\ref{sec:Gravity_analytic_potentials}). Importantly, one needs to take the scattering-induced subhalo–halo interaction (SSHI) into account, i.e.\ the non-gravitational scattering of satellite DM particles with those of the host halo. Doing so was pioneered by \citet{Zeng2022}. In \og, we have a more general description that allows us to model the SSHI for a wider range of SIDM models.
It samples host particles locally on the fly when needed for the DM self-interactions and discards them after the interaction. To sample the particle velocities, \og uses the Eddington Inversion method and assumes an isotropic velocity distribution.
For further details, we refer the reader to \citet{Klemmer2026}.

\subsubsection{Further characteristics}
Additionally, the self-interaction module comes with its own time-step criterion to ensure that the interaction probability (Eq.~\eqref{eq:SIDM_scatter_probability}) and the drag force (Eq.~\eqref{eq:SIDM_drag_force}) are sufficiently small. Moreover, the implementation is MPI- and OpenMP-parallelised, while explicit conservation of energy and linear momentum is ensured. Therefore, the parallelisation follows a different approach than the one employed for gravity or hydrodynamics, ensuring that a particle is sent to one MPI rank at a time only. 
While this had initially been implemented for small-angle scattering, it turned out to be very beneficial for simulating the gravothermal collapse of SIDM halos \citep{Fischer2024b, Fischer2025b}.
A more extensive description of the SIDM implementation is provided in \cite{Fischer2026}. 

For a list of all available configuration and parameter options, see the according \href{https://gitlab.lrz.de/AstroCodes/OpenGadget3/-/wikis/Self-Interacting%20Dark%20Matter}{section on the code wiki}.

\section{Smoothed Particle Hydrodynamics}
\label{sec:SPH}
The SPH approach is a mesh-free Lagrangian method used to solve hydrodynamical equations. As for any Lagrangian method, SPH uses discrete particles and a weighted kernel function to describe the state of the fluid. Based on the values $X_j$ carried by individual particles, the continuous field at the position of an $i$-th target particle is described by a kernel-averaged value 
\begin{equation}
\left<X\right>_i = \sum_j \frac{m_j}{\rho_j} X_j W_{ij}\,,
\end{equation}
with $m_j$ and $\rho_j$ the mass of the $j$th particle and the density at its position, respectively, and $W_{ij}\equiv W(|\mymathbf{x}_i - \mymathbf{x}_j|,h_{i})$ the kernel function with which the $j$th particle contributes to the field variable of the $i$th particle  (see section \ref{sec:SPH:kernels}). We note that density can be obtained by setting $X\equiv \rho$ in the above equation. Adaptive behaviour is obtained by choosing the smoothing length $h_i$ according to the local density of tracer particles, typically by the choice of an effective number of neighbours $N$ and solving for $h_i$ to fulfil 
\begin{equation}
\frac{4\uppi}{3}h_i^3\rho_i = Nm_i .
\end{equation}
Derivatives of quantities can then be formulated as derivatives of the well-known kernel function. Conservative formulations can be obtained by appropriate, symmetric formulation of the pairwise contribution and the computation of the correction factors $\Omega_i$ appearing in the kernel derivatives for non-constant $h_i$ (see section \ref{sec:SPH:formulation}). In some formulations, the symmetrised kernel $W_{ij}\equiv(W_i+W_j)/2$ is also used. For more details on the underlying formalism, see the reviews on SPH by  \citep{Springel2010rev, Price2012rev}.

The advantages of SPH are its automatic spatial adaptivity, robustness, and ability to deal with low densities. In addition, one has the freedom to choose the evolved hydrodynamical variables. Here, different choices have different advantages and disadvantages.

Due to its well-known difficulties in the context of shocks or mixing instabilities \citep[e.g.][]{Morris1996, Agertz2007, Bauer2012}, \og uses an improved implementation of AV, conductivity and magnetic dissipation (in the case of MHD) and makes use of higher-order estimates for gradients. 

\subsection{Entropy-density SPH}
\label{sec:SPH:formulation}
The standard approach in \og is to calculate the hydrodynamic acceleration using the conservative formulation by \cite{Springel2002}, which evolved the entropy as a hydrodynamical variable. In this case, the equation of state is given as
\begin{equation}
P_i = (\gamma-1)\;u_i \;\rho_i = A_i\left(\rho_i\right)^\gamma,
\end{equation}
which relates the pressure $P_i$, the internal energy per unit mass $u_i$ and the entropy $A_i$. This leads to the set of generalised SPH equations
\begin{equation}
\!\!\!\frac{d {\mymathbf v}_i}{d t}\!=\!-\!\sum_j\!m_j\! \left(\frac{P_j}{\Omega_j\rho_j^2}{\mysymmathbf \nabla}_{\!i} W_{ij}(h_i)\!+\!\frac{P_i}{\Omega_i\rho_i^2}{\mysymmathbf \nabla}_{\!i} W_{ij}(h_j)\!+\!\frac{\tilde{\mysymtens{\uppi}}_{ij}}{\rho_{ij}} {\mysymmathbf \nabla}_{\!i} \bar W_{ij}\!\right),
 \label{eqn:moment1}
 \end{equation}
or
 \begin{equation}
\!\!\frac{d {\mymathbf v}_i}{d t}\!=\! -\! \sum_j\! m_j\! \left(\frac{P_j + \rho_j\tilde{\mysymtens{\uppi}}_{ij}}{\Omega_j\rho_j^2}{\mysymmathbf \nabla}_i W_{ij}(h_i)\right.\!+\! \left.\frac{P_i + \rho_i\tilde{\mysymtens{\uppi}}_{ij}}{\Omega_i\rho_i^2}{\mysymmathbf \nabla}_i W_{ij}(h_j)\!\right),
\label{eqn:moment2}
\end{equation}
where $\bar W_{ij}$ is the symmetrized kernel, $\rho_{ij}$ is the average density, and
\begin{equation}
\label{eq:Omega_i}
\Omega_i \equiv \left(1 + \frac{h_i}{3\rho_i} \frac{\partial \rho_i}{\partial h_i}\right) \,.
\end{equation}
Here, $\tilde{\mysymtens{\uppi}}_{ij}$ denotes the AV needed in SPH to properly capture shocks and to guarantee particle order (see section \ref{sec:SPH:artvisc} for details), where Eq. \eqref{eqn:moment1} and \eqref{eqn:moment2} describe two different symmetrisations for the AV term used in the literature. Note that \textsc{Gadget}-2/3 traditionally used Eq. \eqref{eqn:moment1}.
The entropy $A_i$ is then evolved accordingly:
\begin{equation}
\frac{d A_i}{d t} = \frac{1}{2} \frac{\gamma-1}{\rho_i^{\gamma-1}} \sum_j m_j \frac{\tilde{\mysymtens{\uppi}}_{ij}}{\rho_{ij}}
\left({\mymathbf v}_j - {\mymathbf v}_i\right){\mysymmathbf \nabla}_i \bar W_{ij}.
\end{equation}

Following \citet{2013ApJ...768...44S} and in the spirit of \citet{2013MNRAS.428.2840H}, we also maintain a pressure-density formulation in \og. In general, this is only for reference and to allow comparison to other implementations and was never used in any scientific publication. Nevertheless, this might be a valuable test bed, so the basic formulation is still part of the continuous integration testing and maintained accordingly. 

\subsection{Higher order derivatives} \label{sec:SPH:derivatives}

To improve the gradient estimate for vector quantities, \og employs a second-order approximation as proposed in \citet{Cullen2010, Price2012, Hu2014, Beck2016}.
The gradient of a vector quantity $\mathbf{A}$ can generally be written as 
\begin{equation}
    (\widehat{\mysymmathbf \nabla \otimes {\mathbf A}})_{\alpha \beta} = \frac{1}{\rho_i} \sum_j m_j (\mysymtens{A}_j - \mysymtens{A}_i)^\beta \: \mysymmathbf{\nabla}_i^\alpha \bar W_{ij} 
    \label{eq:vector_gradient}
\end{equation}
where $\{i,j\}$ are particle indices and $\{\alpha, \beta, \gamma\}$ are coordinate components.
By expanding $\mathbf{A}_j^\beta$ around $i$ we obtain
\begin{equation}
    \mytens{A}_j^\beta = \mytens{A}_i^\beta + \partial_\gamma \mytens{A}_i^\beta \left( \mymathbf{x}_j - \mymathbf{x}_i \right)^\gamma + \mathcal{O}(h^2)
\end{equation}
which can be inserted into Eq. \eqref{eq:vector_gradient} and gives
\begin{equation}
\sum_j m_j(\mathbf{A}_j - \mathbf{A}_i)^\beta \mathbf{\nabla}_i^\alpha W_{ij} = \partial_\gamma \mathbf{A}_i^\beta \sum_j m_j(\mathbf{x}_j - \mathbf{x}_i)^\gamma \mathbf{\nabla}_i^\alpha W_{ij} \: .
\end{equation}
The improved estimate of $\partial_\gamma \mathbf{A}_i^\beta \equiv\;${\bf X}$_{\gamma\beta}$ can be constructed from a matrix inversion
$\mathbf{X} = \mathbf{M}^{-1} \mathbf{Y}$ where
\begin{align}
    {\bf M}_{\alpha\gamma} &\equiv \sum_j m_j(\mathbf{x}_j - \mathbf{x}_i)^\gamma \mathbf{\nabla}_i^\alpha W_{ij} \\
    &= \sum_j m_j (\mathbf{x}_j - \mathbf{x}_i)^\gamma (\mathbf{x}_i - \mathbf{x}_j)^\alpha \frac{1}{x_{ij}}  \frac{\partial W_{ij}}{\partial x_{ij}}
\end{align}
and
\begin{align}
    {\bf Y}_{\alpha\beta} &\equiv \sum_j m_j(\mathbf{A}_j - \mathbf{A}_i)^\beta \mathbf{\nabla}_i^\alpha W_{ij} \\
    &= \sum_j m_j (\mathbf{A}_j - \mathbf{A}_i)^\beta (\mathbf{x}_i - \mathbf{x}_j)^\alpha \frac{1}{x_{ij}}  \frac{\partial W_{ij}}{\partial x_{ij}} \: .
\end{align}
The divergence and vorticity of any vector field can then be constructed via
\begin{align}
        \mathbf{\nabla}\cdot\mathbf{A} &= \partial_\alpha \mathbf{A}^\alpha \\
        (\mathbf{\nabla}\times\mathbf{A})_\gamma &= \epsilon_{\alpha\beta\gamma}\partial_\alpha \mathbf{A}^\beta
\end{align}
which can be read directly from the matrix $\mathbf{X}$.
\og employs these higher-order derivatives to improve our estimates of the divergence and vorticity of the velocity and magnetic field.


\begin{figure*}[ht]
 \includegraphics[width=0.33\textwidth]{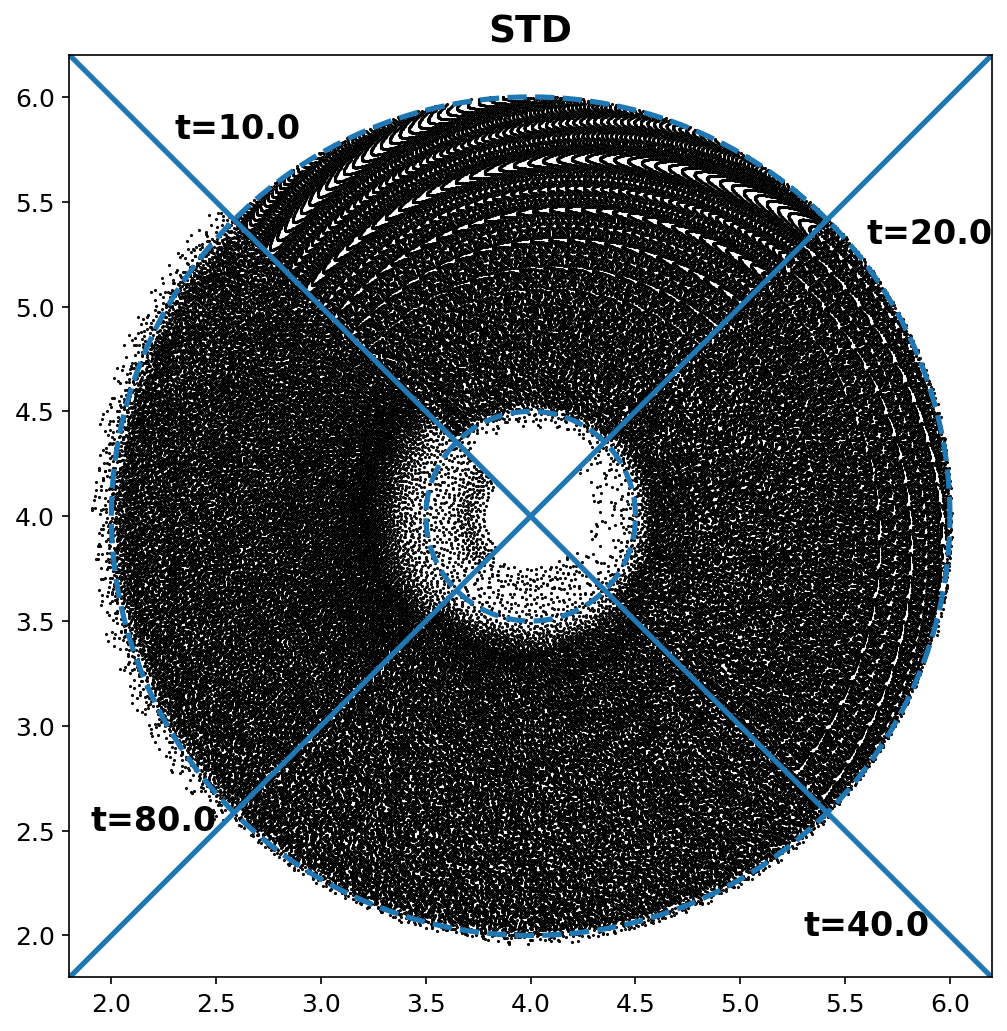}
 \includegraphics[width=0.33\textwidth]{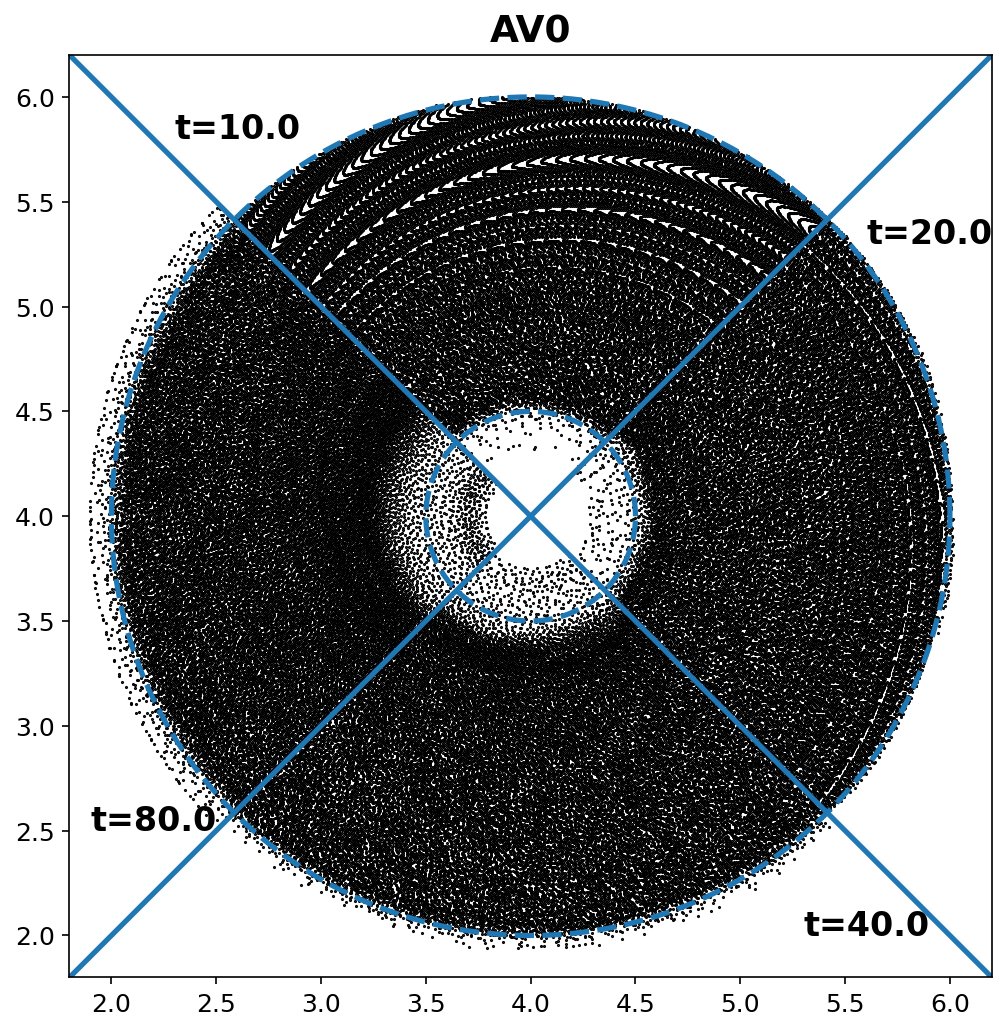}
 \includegraphics[width=0.33\textwidth]{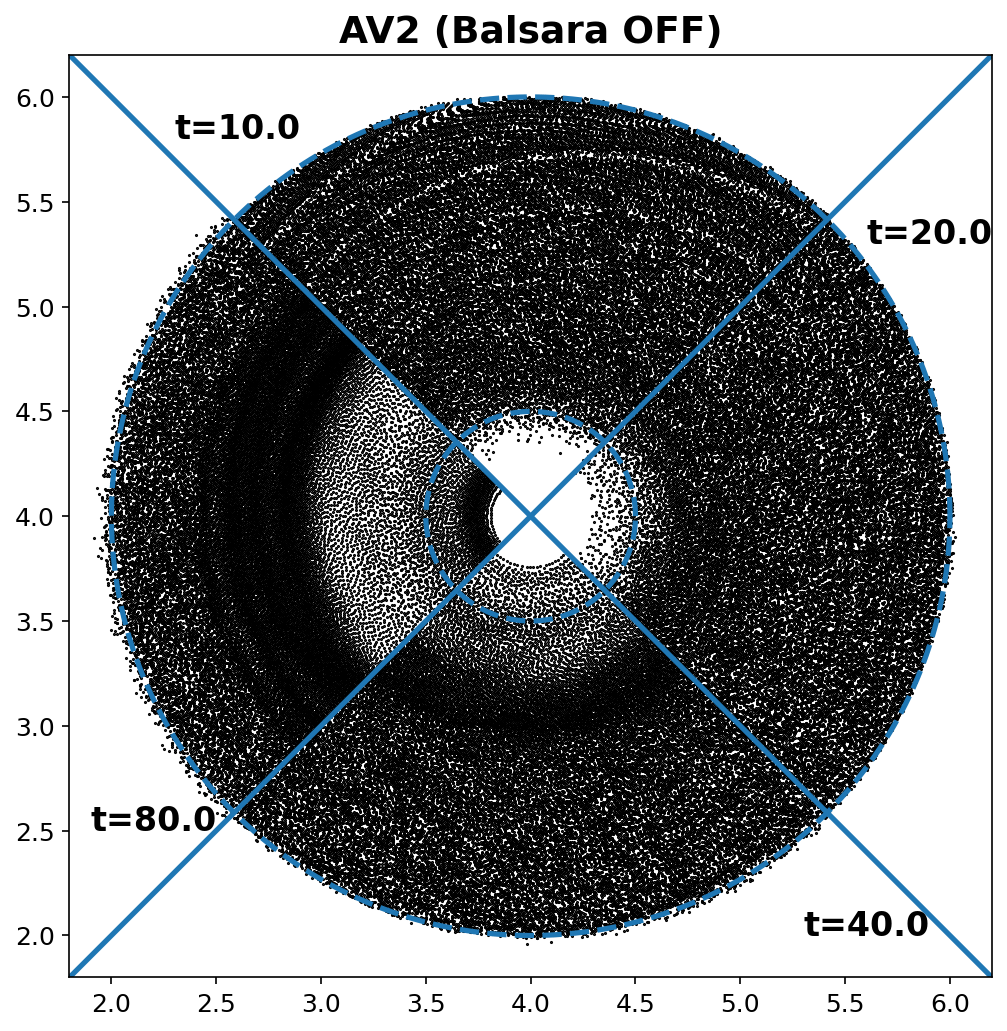}\\
 \includegraphics[width=0.33\textwidth]{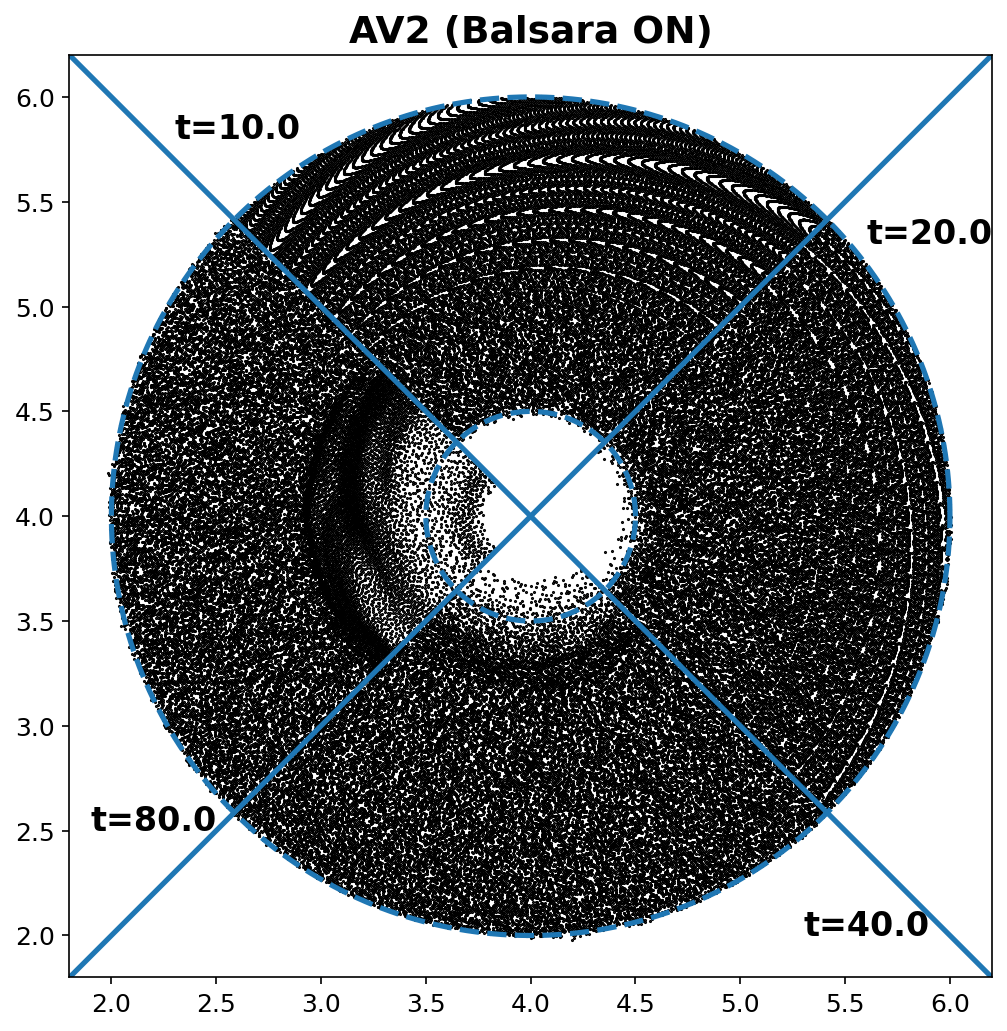}
 \includegraphics[width=0.33\textwidth]{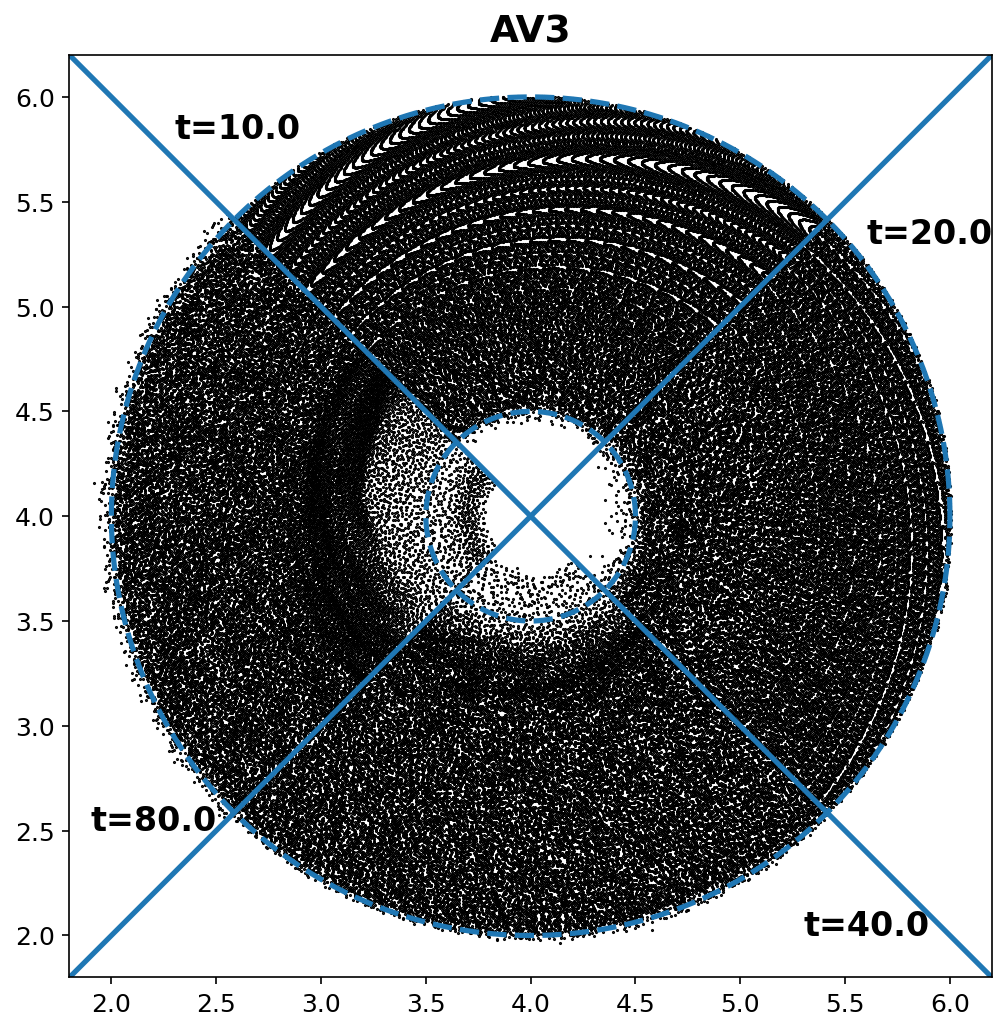}
 \includegraphics[width=0.33\textwidth]{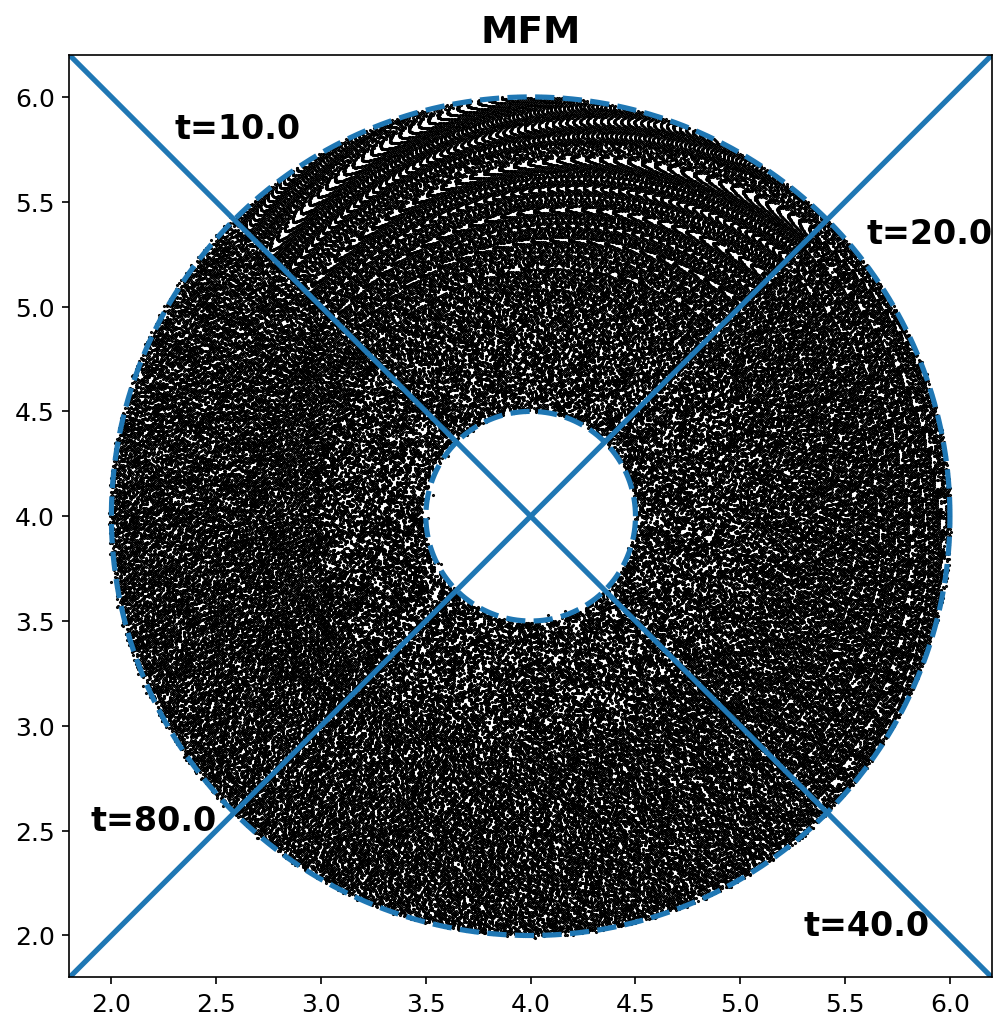}
  \caption{Standard Kepler-disk test with the different SPH viscosity settings (see Table \ref{visctable}) and the MFM version. From top left to bottom right are shown: {\it STD}, {\it AV0}, {\it AV2} without and with the additional Balsara switch, {\it AV3} and the {\it MFM} version.}
 \label{fig:kepler_disk}
\end{figure*}

\subsection{Artificial Viscosity}
\label{sec:SPH:artvisc}

\begin{figure}[htp]
\begin{center}
 \includegraphics[width=0.48\textwidth]{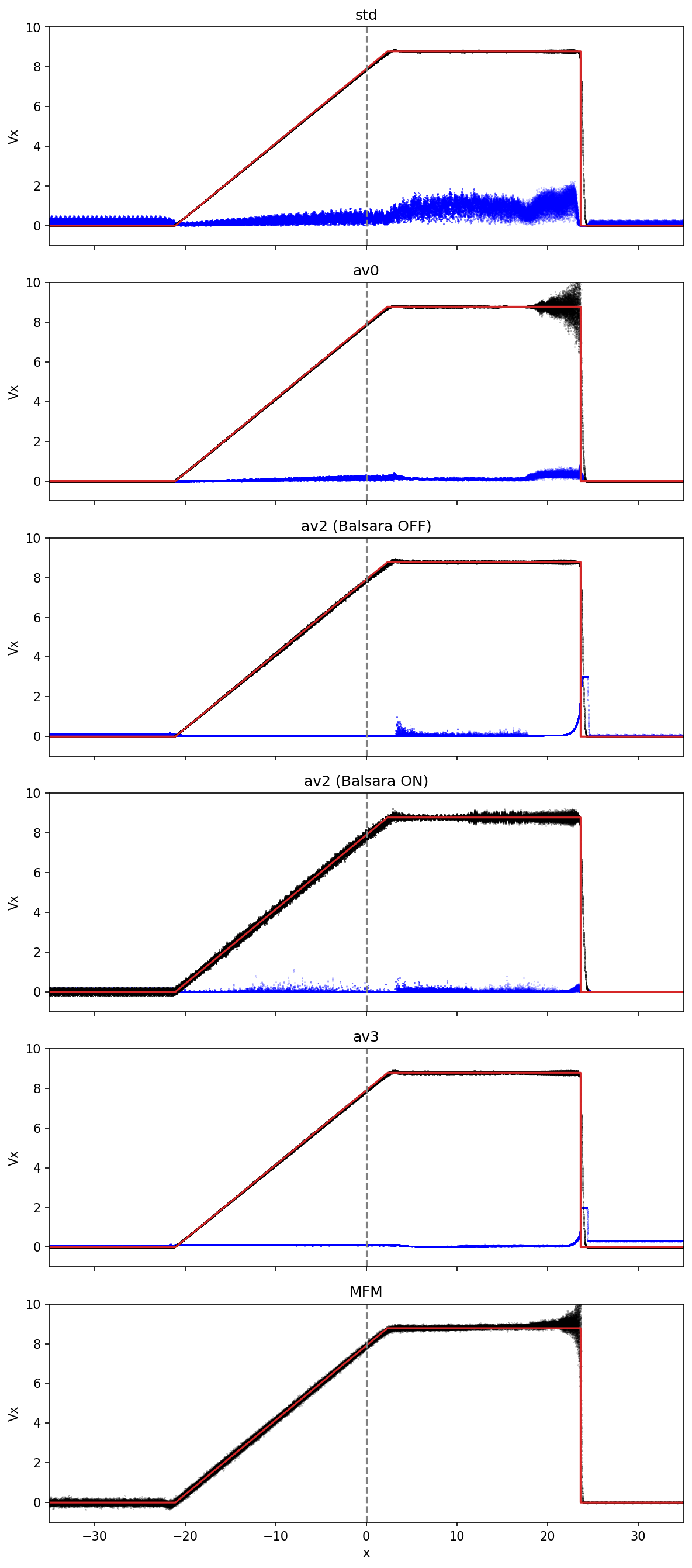}
\end{center}
  \caption{Velocity component of a Mach 10 shock tube in 3D as an example for the performance of the different viscosity schemes (see Table \ref{visctable}) and the MFM version in \og. The blue points show in addition the effective values of $\alpha_i^\mathrm{AV}$ (e.g. including its multiplication with the viscosity-limiter $f_{ij}^\mathrm{shear}$).}
 \label{fig:mach10}
\end{figure}

Here we describe the basic implementations of artificial viscosity (AV) for completeness, but more details about the implementation and various tests of the AV and its time-dependent versions can be found in \cite{Dolag2005, Beck2016, Marin-Gilabert_2022, Groth2023}. For the underlying parametrisation of the AV, we follow a formulation proposed by \cite{1997JCoPh..136....298S} based on an analogy with Riemann solutions of compressible gas dynamics. While the current reference implementation features the formulation and parameters listed in the last row of Table \ref{visctable}, other choices are possible. The latter ones are summarised in rows 1-4 of Table \ref{visctable} and many of them have been used for past simulations as listed in the caption. Their different performance is demonstrated in the evolution of the Kepler disk test case as shown in Figure \ref{fig:kepler_disk} as well as on a Mach 10 shock tube as shown in Figure \ref{fig:mach10}. Note that we deliberately show the strong, Mach 10 test case to highlight the differences. In a more traditional test case like the Sod shock tube, all implementations perform very well, as shown in the Appendix in Figure \ref{fig:mach2}.  

If the particles are approaching each other ($\mathbf{r}_{ij}\cdot\mathbf{v}_{ij}\le 0$), the pairwise viscosity $\tilde{\mysymtens{\uppi}}_{ij}$ is non-zero, and the resulting viscosity term in the {\bf std} formulation can be constructed using the relative velocity projected onto the separation vector $\mu_{ij}=\mathbf{v}_{ij}\cdot\mathbf{r}_{ij}/|\mathbf{r}_{ij}|$; otherwise, $\mu_{ij}$ (and therefore $\tilde{\mysymtens{\uppi}}_{ij}$) is zero. With this, we can define 
\begin{equation}
\tilde{\mysymtens{\uppi}}_{ij}=\left(-\alpha_\mathrm{Bulk} c_{ij} \mu_{ij} + \beta^\mathrm{AV}\mu_{ij}^2\right)f_{ij}^\mathrm{shear} = -0.5\alpha_\mathrm{Bulk}v_{ij}^\mathrm{sig}\mu_{ij}f_{ij}^\mathrm{shear}  \label{eqn:visc2}
\end{equation}
with the signal velocity\footnote{Note that for the sound speed $c_i$ only the thermal pressure is used, so for example pressure contributions from cosmic rays are subtracted, but in case of MHD the fastest of the magnetic waves is added.} 
\begin{equation}
v_{ij}^\mathrm{ sig} = c_i + c_j - \tilde{\beta}^\mathrm{AV}\mu_{ij},
\end{equation}
with $c_i=\sqrt{\gamma P_i/\rho_i}$ denoting the sound velocity and $c_{ij}$ the average sound velocity. Note that \textsc{Gadget}-2/3 traditionally uses the second formulation of the AV term in Eq. \eqref{eqn:visc2}, which implies $\beta^\mathrm{AV}\propto\alpha_\mathrm{Bulk}$. We point the reader to the discussion in Sect. 2.2.7 of \citet{2018PASA...35...31P}, where it is argued that they should be treated separately to ensure particle ordering in certain situations. Here we have also included an averaged viscosity-limiter $f_{ij}^\mathrm{shear}$, which is often used to suppress the viscosity locally in regions of strong shear flows, as measured by
\begin{equation}
f_{i}^\mathrm{ shear} = \frac{|\left<\mathbf{\nabla}\cdot\mathbf{v}\right>_i|}{|\left<\mathbf{\nabla}\cdot\mathbf{v}\right>_i| + |\left<\mathbf{\nabla}\times\mathbf{v}\right >_i|+\sigma c_i/h_i},
\label{eq:balsara}
\end{equation}
where $\left<\mathbf{\nabla}\cdot\mathbf{v}\right>_i$ and $\left<\mathbf{\nabla}\times\mathbf{v}\right >_i$ denotes the SPH estimate of the divergence and of the curl of the velocity around the particle $i$, respectively. The introduction of this limiter helps to avoid spurious angular momentum and vorticity transport in gas disks \citep{1995JCoPh.121..357B,1996IAUS..171..259S}. A common choice for the additional numerical factors are $\alpha_\mathrm{Bulk}=1$, $\beta^\mathrm{AV}=2$ and $\sigma=0.0001$. Note that since the introduction of the higher-order gradients to compute the operators in Eq. \eqref{eq:balsara}, we use $\alpha_\mathrm{Bulk}=3$ and $\tilde{\beta}^\mathrm{AV}=3$, corresponding to $\beta^\mathrm{AV}=4.5$. But note that here $\tilde{\beta}^\mathrm{AV}$ is also multiplied by the averaged viscosity-limiter $f_{ij}^\mathrm{shear}$. 
In \og we use $\sigma_i=0.01$ and, depending on which additional viscosity suppression description is adopted, different settings and combinations of equations as well as values for $\alpha_\mathrm{Bulk}$ and $\beta^\mathrm{AV}$ are used (see Table \ref{visctable} for the exact values and settings).

Following idea originally proposed by \cite{1997JCoPh..136....41S} and following the implementation from \citet{Dolag2005}, in the {\bf AV0} implementation we assign every particle its own viscosity parameter $\alpha_i^\mathrm{AV}$, which is allowed to evolve with time according to
\begin{equation}
\frac{{\rm d}\alpha_i^\mathrm{AV}}{{\rm d}t}=-\frac{\alpha_i^\mathrm{AV}-\alpha_{\rm min}}{\tau}+S_i.
\end{equation}
This causes $\alpha_i^\mathrm{AV}$ to decay to a minimum value $\alpha_{\rm min}$ with an e-folding time $\tau$, while the source term $S_i$ is meant to make $\alpha_i$ rapidly grow when a particle approaches a shock. For the decay timescale, \cite{1997JCoPh..136....41S} proposed to use
\begin{equation}
\tau=h_{i}\,/\,(c_i\,l),
\end{equation}
where $l$ is a free (dimensionless) parameter which determines how many crossing times the viscosity decays. For an ideal gas and a strong shock, this time scale can be related to a length scale $L_\mathrm{Decay}=\sqrt{(\gamma-1)/2\gamma}/l$ (in units of the smoothing length $h_i$) on which the viscosity parameter decays behind the shock front. For the source term $S_i$, we follow
\cite{1997JCoPh..136....41S} and adopt
\begin{equation}
S_i = \alpha_\mathrm{Source}\,\, \mathrm{ln}\left(\frac{\gamma-1}{\gamma+1}\right)\, f_i\,\, \mathrm{max}\left(0,-\left|\left<\vec{\nabla}\cdot\vec{v}\right>_i\right|\right).
\end{equation}
In addition, the viscosity tensor $\tilde{\mysymtens{\uppi}}_{ij}$ is still multiplied by the viscosity-limiter $f_{ij}^\mathrm{shear}$. Note that the numerical values for the parameters as listed in Table \ref{visctable} are those of the original implementation and should be adapted to reflect the changes to different kernels and to the general usage of second-order derivatives. 

\begin{table*}[ht]
    \begin{tabular}{|c|c|c|c|c|c|c|c|c|c|c|c|}
        \hline
        mode   & $\alpha_\mathrm{Bulk}$ & Switch$^+$ & 2nd & Balsara & reduce & Decay & Sym & $\alpha_\mathrm{S}$ & $\alpha_\mathrm{min}$ & $L_\mathrm{D}$ & Used for \\
         & & & order & & $\beta^\mathrm{AV}$ & $\alpha_i^\mathrm{AV}$ & $\dot{v}$ & & & & \\
        \hline
        \hline
        STD    &  1 & Off & Yes & All  & Yes & No        & No  & -- & --   & --  & --\\
        \hline
        AV0    &  1 & 0   & No  & All  & Yes & Exp       & No  & 5  & 0.1  & 2  & Hutt, Magneticum\\
        \hline
        AV1    &  3 & 1   & Yes & Off & Yes & No        & No  & 3  & 0.05 & 3 & --\\
        \hline
        AV2    &  3 & 2   & Yes & All/Off & Yes & Exp       & No  & 3  & 0.05 & 3 & Dianoga, B4/uhr, SLOW\\
        \hline
        AV3    &  2 & 3   & Yes & $\beta^\mathrm{AV}$,$\alpha_\mathrm{min}$ & $\beta_\mathrm{min}=0.3$  & quartic & Yes & 1.5  & 0.2 & 2.0 & \\
        \hline
    \end{tabular}
    \caption{Default parameters for the different viscosity models. $^+$ refers to the presence and the value of the corresponding settings at compile time. The last column gives an overview on which settings was used in very large simulation campaigns, referring to early cluster simulations \citep[HUTT][]{2009MNRAS.399..497D}, recent cluster simulations \citep[DIANOGA][Borgani et al. in prep.]{Bonafede2011, Esposito.etal.2025} and large simulation campaigns like Magneticum and its individual cosmological volumes \citep[][]{Hirschmann.etal.2014,2015ApJ...812...29T} or the SLOW simulation \citep[][]{Dolag2023}.
    }
    \label{visctable}
\end{table*}

An improvement of this scheme was suggested by \citet{Cullen2010}, and the implementation {\bf AV1-3} follows largely what was presented in \citet{Beck2016}. In this implementation, a local viscosity coefficient can be defined as
\begin{equation}
\alpha_{i}^\mathrm{loc}=\alpha_\mathrm{Source}\alpha_\mathrm{Bulk}\frac{h^{2}_{i}A_{i}}{h^{2}_{i}A_{i}+(v_{i}^\mathrm{sig})^{2}}
\label{eqn:alpha_loc}
\end{equation}
with
\begin{equation}
A_i=\xi_{i}\mathrm{max}(0,-(\dot{\mathbf{\nabla}}\cdot\mathbf{v})_{i}),\label{eq:ax}
\end{equation}
which is able to distinguish between pre-shock and post-shock regions. Note that following \citet{2024arXiv240710176P}, we interpret $v_{i}^\mathrm{sig}$ as the sound speed.  We calculate $(\dot{\mathbf{\nabla}}\cdot\mathbf{v})_{i}$ via interpolation between the current and the previous time-step \citep[as suggested by][]{Cullen2010}. Subsequently, we use the shock indicator 
\begin{equation}
\label{eq:Cullen_shock_indicator}
R_i=\frac{1}{\rho_{i}}\sum_{j}{\mathrm{sign}(\mathbf{\nabla}\cdot\mathbf{v})_{j}m_{j}W_{ij}}
\end{equation}
to determine the ratio $\xi_i$ of the shock and shear strengths in quadratic form via
\begin{equation}
\xi_i = \frac{|2(1-R_i)^{4}(\mathbf{\nabla}\cdot\mathbf{v})_{i}|^{2}}{|2(1-R_i)^{4}(\mathbf{\nabla}\cdot\mathbf{v})_{i}|^{2}+|\mathbf{\nabla}\times{}\mathbf{v}|_{i}^2},
\end{equation}
which is proposed by \citet{Cullen2010} as an additional limiting factor for AV in Eq. (\ref{eq:ax}) and was experimentally determined. The coefficient $\alpha_\mathrm{Source}$ allows here for a boost, to ensure that $\alpha_{i}^\mathrm{loc}$ can reach the value of $\alpha_\mathrm{Bulk}$. In addition, $\alpha_i$ is always limited to range between $\alpha_\mathrm{min}$ and $\alpha_\mathrm{Bulk}$. In principle, an accurate calculation of $R_i$ for every particle requires computing first $(\mathbf{\nabla}\cdot\mathbf{v})_{i}$ for every particle. Therefore, an extra SPH summation loop is added between the calculation of density (where velocity divergence can also be computed) and hydro forces. For computational reasons, we use the velocity divergence calculated in the previous time-step. Furthermore, the condition of convergent flow is also indicated by a high velocity divergence, which, however, does not distinguish between pre-shock and post-shock regions. Therefore, we employ the time derivative of velocity divergence to determine a directional shock indicator. This {\bf AV1} implementation, where the current viscosity coefficient $\alpha_{i}^\mathrm{AV}$ is just set to $\alpha_{i}^{\mathrm{loc}}$ immediately, was never used in \textsc{Gadget}-2/3.

Instead, in the {\bf AV2} implementation, current viscosity coefficient $\alpha_{i}^\mathrm{AV}$ will decay if it is larger than the currently estimated $\alpha_{i}^{\mathrm{loc}}$. The decay time is computed according to
\begin{equation}
\frac{{\rm d}\alpha_i^\mathrm{AV}}{{\rm d}t}=(\alpha_{i}^{\mathrm{loc}}-\alpha_{i}^\mathrm{AV})\frac{v_{i}^\mathrm{sig}}{\ell{}_\mathrm{decay}h_{i}},
\label{eqn:dt_alpha}
\end{equation}
which \og integrates in time together with the hydrodynamical quantities. Here $v_{i}^\mathrm{sig}$ denotes the maximum signal velocity of each particle and the additional coefficient $\ell{}_\mathrm{decay}$ controls the length scale over which the decay happens. Note that in contrast to \citet{Cullen2010} and \citet{Price2012rev}, in previous versions of \textsc{Gadget}-2/3 we always used the viscosity limiter $f_{ij}^\mathrm{shear}$, unless MHD was used. Therefore, we present the results for both settings of the {\bf AV2} implementation in the Kepler disk and Mach 10 shock tube test as shown in Figures \ref{fig:kepler_disk} and \ref{fig:mach10}, respectively. Note that to handle certain configurations in a cosmological context better, we add the Hubble flow to the signal velocity in Eq. \eqref{eqn:alpha_loc} when using comoving coordinates. 

The newest {\bf AV3} implementation in \og extends this treatment by switching to the viscosity formulation of Eq. \eqref{eqn:moment2}
and switching to a separate treatment of $\alpha_i^\mathrm{AV}$ and $\beta^\mathrm{AV}$ in Eq. \eqref{eqn:visc2}. Here we do a compromise by modulating $\beta^\mathrm{AV}$ in the same way than $\alpha_{i}^{\mathrm{loc}}$, however with a separate minimum value for $\beta_\mathrm{min}=0.3$ and applying the viscosity-limiter $f_{ij}^\mathrm{shear}$ only to $\beta^\mathrm{AV}$ but not to $\alpha_i^\mathrm{AV}$. Note that to handle low-temperature and early-time configurations in the cosmological context better, we add, in addition, a velocity term of the form $h_i|\mathbf{\nabla}\times\mathbf{v}|_i$ to the signal velocity in Eq. \eqref{eqn:alpha_loc} when using co-moving coordinates. Finally, for the {\bf AV3} implementation, we switch from an exponential decline to a quadratic-like decline for $\alpha_i^\mathrm{AV}$ whenever $\alpha_i^\mathrm{AV}$ is larger than half of $\alpha_\mathrm{Bulk}$. This has the advantage that the shape of the decline is opposite to an exponential one, meaning it first declines slowly, then transitions into the exponential decline. For a given power $\beta$ of the decline,  Eq. \eqref{eqn:dt_alpha} for the evolution of $\alpha_i^\mathrm{AV}$ takes the expression
\begin{equation}
\frac{{\rm d}\alpha_i}{{\rm d}t}=
\alpha_\mathrm{Bulk}\frac{\beta}{2}\left(1-0.999\frac{\alpha_{i}^\mathrm{AV}}{\alpha_\mathrm{Bulk}}\right)^{(\beta-1)/\beta}\frac{v_{i}^\mathrm{sig}}{\ell{}_\mathrm{decay}h_{i}}\,,
\end{equation}
where the numerical factor close to one is included just to prevent having a zero value inside the brackets, and we typically use a power-law exponent $\beta=4$. 

For a list of all available configuration and parameter options, see the according \href{https://gitlab.lrz.de/AstroCodes/OpenGadget3/-/wikis/Artificial%20Viscosity}{section on the code wiki}.
\subsection{Artificial conduction}
\label{sec:SPH:artcond}

In \og, the mixing problem in SPH is addressed by introducing a kernel-scale exchange term for internal energy transport. We include artificial conduction (AC) for purely numerical reasons to treat discontinuities in the internal energy (similar to the capturing of velocity jumps by AV), which arise from our 'density-entropy' formulation of SPH. We note that a 'pressure-entropy' formulation of the EoM is also able to address the mixing problem, but it also requires the presence of AC in order to smooth noise in internal energy behind shocks \citep{2013MNRAS.428.2840H, Hu2014}. Thus, in either flavour of SPH, the inclusion of AC is recommended, and many different formulations of the AC equation have been investigated so far. Although their precise details vary across the literature, they all ensure conservation of internal energy within the kernel. \cite{Price2008}, \cite{Price2012rev} and \cite{2012A&A...546A..45V} propose the diffusion of internal energy, while \cite{2012MNRAS.422.3037R} propose the diffusion of entropy. \cite{2008MNRAS.387..427W} propose a first mixing formulation to resolve the differences in entropy profiles within cosmological comparison simulations \citep{frenk99} between grid and SPH codes. The diffusion coefficient between the $i$-th and the $j$-th particles is approximately proportional to $\alpha^{c}v^\mathrm{sig,c}x_{ij}$ and the numerical coefficient $\alpha^{c}$ is commonly treated as constant through space and time. 

In the AC formulation of \og, we adopt a spatially varying coefficient following \cite{2013arXiv1310.4260T} and additionally calculate a limiter depending on the local hydrodynamical and gravitational states. We compute the gradient of internal energy as
\begin{equation}
(\mathbf{\nabla}u)_{i}=\frac{1}{\rho_{i}}\sum_{j}m_{j}(u_{j}-u_{i})\mathbf{\nabla}_{i}W_{ij}
\end{equation}
and approximate the AC coefficient

\begin{equation}
\alpha_{i}^\mathrm{c}=\frac{h_{i}}{3}\frac{|\mathbf{\nabla}u|_{i}}{|u_{i}|}
\end{equation}
as a measure of noise in internal energy sampling on the kernel scale. The time evolution (i.e. spatially varying SPH discretisation of the second-order diffusion equation) of the internal energy for each particle and its neighbours is then given by
\begin{equation}
\frac{du_{i}}{dt}\bigg|_{\mathrm{cond}} =\sum_{j}\frac{m_{j}}{\rho_{ij}}(u_{j}-u_{i})\alpha_{ij}^{c}v^\mathrm{sig,c}_{ij}\overline{F}_{ij},
\end{equation}
where we employ the choice of \cite{Price2008} for signal velocity depending on the pressure gradient of the form
\begin{equation}
v^\mathrm{sig,c}_{ij}=\sqrt{\frac{|P_{i}-P_{j}|}{\rho_{ij}}}
\end{equation}
and $\alpha^{c}_{ij}=(\alpha^{c}_{i}+\alpha^{c}_{j})/2$ is the symmetrised conduction coefficient, which is individually limited to the interval $[0,1]$. In the literature several other forms of AC \citep[see eg.][]{2008MNRAS.387..427W,2012A&A...546A..45V} or approaches to the mixing problem \citep[see e.g.][]{Hopkins2013} have been proposed.

We note that the amount of AC applied depends on the gradients of internal energy and of pressure. In the case that the thermal pressure gradient is determined by gravitational forces (i.e. hydrostatic equilibrium), this method would incorrectly lead to unwanted conduction. In the following, we determine the contribution of hydrostatic equilibrium to the total thermal pressure gradient and present a method to limit the amount of conduction. Firstly, for every individual active particle, we project the gravitational force $\mathbf{F}^{g}$ onto the hydrodynamical force $\mathbf{F}^{h}_{i}$ and calculate the partial force $\mathbf{F}^{p}_{i}$ of $\mathbf{F}^{h}_{i}$, which is balanced by $\mathbf{F}^{g}_{i}$ to
\begin{equation}
\mathbf{F}^{p}_{i}=\frac{\left(\mathbf{F}^{g}_{i}\cdot\mathbf{F}^{h}_{i}\right)}{|\mathbf{F}^{h}_{i}|^{2}}\mathbf{F}^{h}_{i}.\end{equation}
The sign of $\mathbf{F}^{p}_{i}$ depends on the spatial orientation of the force vectors. Secondly, we subtract/add the partial force $\mathbf{F}^{p}_{i}$ from/to the hydrodynamical force $\mathbf{F}^{h}_{i}$ and obtain $\mathbf{F}^{c}_{i}$, which we call the gravitationally adjusted hydrodynamical force
\begin{equation}
\mathbf{F}^{c}_{i}=\mathbf{F}_{i}^{h}+\mathbf{F}_{i}^{p}\,,
\end{equation}
which we use to determine a limitation factor $\delta^{c}_{i}$ for AC
\begin{equation}
\delta_{i}^{c}=\left(\frac{\left(\mathbf{F}^{c}_{i}\cdot\mathbf{F}^{h}_{i}\right)}{|\mathbf{F}^{h}_{i}|^{2}}\right)^{q}.
\end{equation}
The limiter ensures that AC is only applied to the part of $\mathbf{F}^{h}_{i}$ which is not balanced by $\mathbf{F}^{g}_{i}$. The exponent $q$ represents a scaling for the aggressivity of the gravity correction. We limit our correction factor to the interval $[0,1]$ and directly multiply it onto the individual AC coefficients $\alpha_{i}^{c}$. The limiter performs only as well as the hydrodynamical scheme is able to resolve hydrostatic equilibrium (in the ideal case the angle between force vectors is $180^{\circ}$). However, in SPH simulations, small-scale noise is present at all times within the kernel and thus also in the force vector angles. The exponent $q$ (applied after the boundary verification) then accounts for the noise in the particle distribution and mimics an opening angle of force vectors. After extensive studies and performing a variety of test problems, we settle on $q=5$. The limiter returns zero in the case where no hydrodynamical forces are present and when no gravitational forces are present. We are aware that in the presence of strong pressure gradients and rotational forces, our approach only marginally limits the amount of AC applied. However, we did not encounter major problems in our simulations performed with the `new' scheme so far. Therefore, we assume this issue is not too important at this stage.

For a list of all available configuration and parameter options, see the according \href{https://gitlab.lrz.de/AstroCodes/OpenGadget3/-/wikis/Artificial%20Conduction}{section on the code wiki}.
\subsection{Kernels}
\label{sec:SPH:kernels}

The code allows the user to choose among various kernels, such as a cubic spline \citep{Monaghan1985}, quintic spline \citep{Morris1996} and the Wendland C2, C4, C6 and C8 kernel series \citep[][]{Wendland_1995, Wendland2009, Dehnen2012} as implemented in the \textsc{Gadget} family \citep[][]{Donnert2013,Kummer2019}. 
For SPH, using high-order Wendland kernels allows for using higher neighbour numbers, improving convergence, without triggering the pairing instability \citep{Dehnen2012}. For MFM, in contrast, low-order kernels that can be used with smaller neighbour numbers are sufficient, reducing computational costs.

As pointed out in \citet{Dehnen2012}, particle sampling and especially self-contribution generate a bias between estimated density and true density. To correct for this, \og includes the computation of the bias from the self-contribution for some of the kernels.

The analytic expression of the kernels implemented in \og is described in \ref{app:Kern}. The number of neighbours is set by default depending on the chosen kernel and dimensionality. Default values are provided in Table \ref{tab:kernel}. 

\begin{table}
    \begin{tabular}{|c|c|c|c|c|c|c|c|}
        \hline
        \multirow{2}{*}{Kernel}       & 1D & 2D & 3D  & BC &
      \multicolumn{3}{c|}{$\sigma/h$}\\
      & & & & & \!1D\! & 2D & 3D \\
        \hline
        \hline
        \!Cubic Spline\! &  4     & 12 & 32      & No & 0.29 & 0.28 & 0.27 \\
        \hline
        Quintic          &  7     & 41 & \!195\! & No & 0.24 & 0.23 & 0.23 \\
        \hline
        \!Wendland C2\!  &  5     & 19 &  64     & Yes & 0.31 & 0.26 & 0.26 \\
        \hline
        \!Wendland C4\!  &  7     & 44 & \!216\! & Yes & 0.26 & 0.23 & 0.23 \\
        \hline
        \!Wendland C6\!  &  8     & 54 & \!295\! & Yes & 0.23 & 0.21 & 0.20 \\
        \hline
        \!Wendland C8\!  & \!10\! & 64 & \!384\! & No & $-$ & 0.19 & 0.18 \\
        \hline
    \end{tabular}
    \caption{Recommended values for the number of neighbours for the different kernels and dimensions, which is set by default in \og{}. The fifth column indicates whether a bias correction is implemented. Columns 6-8 indicate the resolution in terms of the footprint over the standard deviation \citep[compare][Tab. 1]{Dehnen2012}.}
    \label{tab:kernel}
\end{table}

For a list of all available configuration and parameter options, see according \href{https://gitlab.lrz.de/AstroCodes/OpenGadget3/-/wikis/SPH}{section on the code wiki}.

\subsection{Multiphase gas model}
\label{sec:SPH:multiphase_gas}
\noindent
{\it Main contributing developers: J. Badía and C. Scannapieco}

\og provides a multiphase treatment designed to improve the description of the complex structure of the interstellar medium, where gas at very different temperatures and densities can coexist within the same spatial regions \citep{Marri2003, Scannapieco2006}. To account for such configurations, particles with sufficiently different thermodynamic properties can be excluded from each other's neighbour lists. This is particularly relevant when hot, diffuse gas lies close to cold, dense clouds, since particles from the dense component can contribute to the SPH density estimate of the hot gas through the smoothing kernel. This can artificially increase the estimated density of the hot component and, consequently, its cooling rate \citep{Pearce2001, Thacker2000}. Specifically, particle $j$ is excluded from the neighbour list of particle $i$ when both of the following conditions are satisfied:
\begin{equation}
A_i > \alpha A_j
\end{equation}
and
\begin{equation}
\mu_{ij} < c_{ij},
\end{equation}
where $A(s)$ is the entropic function 
\begin{equation}
P = A(s)\rho^\gamma,
\end{equation}
and $\mu_{ij}$ and $c_{ij}$ denote the pair-averaged local velocity divergence and sound speed, respectively. The dimensionless parameter $\alpha$ is typically set to 50; the results are largely insensitive to variations over the range $5\lesssim\alpha\lesssim100$, owing to the large entropy contrasts that naturally arise between different thermodynamic components of the ISM. The second condition prevents the neighbour exclusion from operating across shocks, where particles on opposite sides of the shock front need to remain part of each other's neighbour list in order to avoid spurious numerical behaviour.

It is worth noting that the multiphase scheme does not assign particles to a fixed set of phases. Instead, the decision to decouple two particles is made on a pair-by-pair basis according to their instantaneous thermodynamic properties. Thus, the groups of particles that are effectively separated from one another are not fixed and may represent different physical components at different stages of the simulation. Furthermore, the exclusion is applied asymmetrically during the density calculation, as low-entropy particles are excluded from the neighbour list of higher-entropy particles, but not the other way around. The opposite decoupling is not needed since the presence of hot diffuse gas particles does not have a significant effect on the density estimation of dense cold clouds. For the force calculation, however, the decoupling is made symmetric to preserve the conservation properties of the SPH scheme. Therefore, if particle $j$ is excluded from the neighbour list of particle $i$, particle $i$ is also excluded as a neighbour of particle $j$.

As shown in \cite{Scannapieco2006}, this treatment allows particles with substantially different temperatures to coexist more naturally within the same region, producing a spatial distribution that is more consistent with the multiphase structure of the ISM. When combined with thermal feedback schemes, it also facilitates a more efficient deposition of feedback energy into the cold, dense gas associated with star-forming regions.

For a list of all available configuration and parameter options, see the according \href{https://gitlab.lrz.de/AstroCodes/OpenGadget3/-/wikis/SPH}{section on the code wiki}.

\section{Meshless Finite Mass}
\label{sec:MFM}
\noindent
{\it Main contributing developer: F. Groth}

The Meshless Finite Mass (MFM) hydrodynamical solver has been developed by \citet{Lanson&Vila2008, Lanson&Vila2008a}. It aims to combine the advantages of SPH with a moving mesh, calculating hydrodynamical interactions via Riemann solvers based on effective interfaces using an SPH-like neighbour search.
The implementation in \og{}\, is based on the implementation in the \textsc{Gandalf} code \citep[][]{Hubber+2018}. It was originally taken from their code base and then extended and adjusted.
The details and specifics of the implementation in \og have been described by \citet{Groth2023}.

Similar to SPH, the smoothing length is determined adaptively, while the discretisation approach is similar to mesh-based approaches:
\begin{align}
\label{eq:MFM_discretization}
    \frac{\dif}{\dif t}\klammer{V_i\myvector{U}_i} + \sum_{j}\klammer{\mymatrix{F}_{ij}\cdot \myvector{A}_{ij}^\text{eff}} =~& \myvector{S}_iV_i.
\end{align}
The field vector is $\myvector{U}=(\rho,\rho \myvector{v},\rho e)$ and the corresponding flux reads $\mymatrix{F}=(\rho \myvector{v}, \rho \myvector{v}\myvector{v}^T + P\mathds{1}, \klammer{\rho e + P}\myvector{v})$. Source terms in the r.h.s. vanish for pure hydrodynamics: $\myvector{S}=0$.
All interpolations in MFM are calculated to second order, similarly to what is described in Sect.~\ref{sec:SPH:derivatives}:
\begin{align}
    \uppsi_i =~& \frac{1}{\sum_{j} W_j}W_i\\
    \tilde\uppsi_j^\alpha(\myvector{x}_i) =~& \sum_{i}B_i^{\alpha\beta}(x_j-x_i)^\beta\uppsi_j(x_i) \label{eq:tildepsi}
\end{align}
\begin{align}
    \mymatrix{B}_i =~& \mymatrix{E}_i^{-1}\\
    E_i^{\alpha\beta} =~& \sum_{j}(\myvector{x}_j-\myvector{x}_i)^\alpha (\myvector{x}_j-\myvector{x}_i)^\beta \uppsi_j(x_i).
\end{align}
with partition function $\uppsi$\footnote{Note that this is different than the $\psi^{\mathrm{clean}}$ function introduced in Sec.~\ref{sec:dedner_cleaning}}. The symbol for the partition function $\uppsi$ is similar, but was chosen for consistency with \citet{Groth2023} and the variables in the code.
The volume of a particle is
\begin{align}
    V_i = n_i^{-1}
\end{align}
and the effective interface area used for the flux calculation is
\begin{align}
    \myvector{A}_{ij}^\text{eff} =~& V_i\tilde{\uppsi}_j - V_j\tilde{\uppsi}_i.
\end{align}
Inside the Riemann solver, all calculations are based on the primitive fluid vector
\begin{align}
    \myvector{W} =~& \begin{pmatrix} \rho \\ \myvector{v} \\ p \end{pmatrix},
\end{align}
interpolated to the interface using second-order gradients.
Slope-limiters improve the numerical stability of the Riemann problem. By default, the limiters designed for the \textsc{gizmo} code \citep{Hopkins2015} are used, with an additional correction to the pairwise limiter to make it Lagrangian. Several slope-limiters have been implemented suited for specific applications, described in \ref{app:MFM}.

Also, the positions and velocities of the interface are calculated with second-order accuracy as
\begin{align}
    \dif \myvector{r}_{i}^\text{face} =~& \dif \myvector{r}_{ij} s_i\,, \\
    \myvector{v}_{ij}^\text{face} =~& s_j \myvector{v}_j + s_i \myvector{v}_i\,,
    ~~~~~s_{i} = \frac{h_i}{h_i+h_j} \,.
\end{align}

\og{} offers several different Riemann solvers, where by default an iterative exact Riemann solver \citep{Toro2009} is used.

In contrast to SPH, MFM does not rely on additional numerical viscosity or conductivity.
The internal energy is used as the primary variable for the hydrodynamical evolution with MFM instead of entropy for SPH, as it is more closely related to the output of the Riemann solver. It is calculated from the total energy change
: \begin{align}
    \klammer{\frac{\dif u}{\dif t}}_i =~& \klammer{\frac{\dif E_\text{tot}}{\dif t}}_i - \klammer{\myvector{v}_i+\frac{1}{2}\klammer{\frac{\dif \myvector{v}}{\dif t}}_i\Delta t}\cdot \klammer{\frac{\dif \myvector{v}}{\dif t}}_i\Delta t.
\end{align}
The entropy is co-evolved, such that it can be used as input for subgrid models and additional physics.
As the internal energy and entropy changes are re-calculated every time step to ensure correct comoving integration, all contributions to the entropy change by other subgrid models are added to a separate variable, which allows for consistent coupling.

An energy-entropy formulation can be used to better deal with very cold flows, so as to reduce numerical noise. 

For a list of all available configuration and parameter options, see the according \href{https://gitlab.lrz.de/AstroCodes/OpenGadget3/-/wikis/Meshless%20Finite%20Mass%20Solver}{section on the code wiki}.

\section{Magneto-hydrodynamics}
\label{sec:MHD}

The handling of MHD follows mainly \citet{Dolag2009}, except that we treat the magnetic field internally as a comoving magnetic field.  

\subsection{Magnetic extensions to SPH}
A natural generalisation of the signal velocity $v_{ij}^\mathrm{ sig}$ in the framework of MHD is to replace the sound velocity $c_i$ by the fastest magnetic wave as suggested by \citet{2004MNRAS.348..123P}. Therefore the sound velocity $c_i$ is replaced by
\begin{eqnarray}
v_i &=& \frac{1}{\sqrt{2}}\left[\left(c_i^2+\frac{B_i^2}{\mu_0\rho_a}\right) \cdot \right.  \nonumber \\ & &\left.\sqrt{\left(c_i^2+\frac{B_i^2}{\mu_0\rho_i}\right)^2 - 4\frac{c_i^2(\mathbf{B}\cdot\mathbf{r}_{ij}/|\mathbf{r}_{ij}|)^2}{\mu_0\rho_i}} \right]^{0.5}.
\end{eqnarray}
As this new definition of the signal velocity also enters the time-step criteria, no extra time-step criteria due to the magnetic field has to be defined.

In the presence of magnetic fields, the computation of the local viscosity constant $\alpha_{i}^\mathrm{loc}$ (as described in Eq. \eqref{eqn:alpha_loc}) can also be extended at compile time by an additive term: 
\begin{equation}
\alpha_\mathrm{Bfld} \frac{|\mathbf{\nabla}\cdot\mathbf{B}_{i}|}{|\mathrm{B_i}|}\,{h_i}\,,
\end{equation}
 capturing the local value of the divergence of magnetic fields, which triggers AV also at magnetically dominated shocks.

\subsection{Induction equation}

Ignoring Ohmic terms and making use of $\nabla \cdot B = 0$, the induction equation in component form reads
\begin{equation}   
\frac{\mathrm{ d}B_i^k}{\mathrm{ d}t} = \frac{1}{Ha^2} \frac{1}{\Omega_i\rho_i} \left[\sum_j m_j (v^k_{ij}B_i^l - B_i^k v_{ij}^l) \frac{\partial W_i}{\partial u}\frac{\mathbf{r}_{ij}^l}{|\mathbf{r}_{ij}|}\right] \,.
\end{equation}
Note that, as also suggested by \citet{2004MNRAS.348..139P}, we wrote down the equations including the correction factor $\Omega_i$ (see Eq. \eqref{eq:Omega_i}) which reflects the correction terms $\frac{\mathrm{ d}W}{\mathrm{ d}h}$ arising from the variable particle smoothing length.

\subsection{Lorenz force}

The magnetic field acts on the gas via the Lorentz force, which can be written in a symmetric, conservative form involving the magnetic stress tensor \citep{1985MNRAS.216..883P}
\begin{equation}
M_i^{kl} = \left( \mathbf{B}_i^k\mathbf{B}_i^l - \frac{1}{2}|\mathbf{B}_i|^2\delta^{kl}\right).
\label{eq:maxtens}
\end{equation}
The magnetic contribution to the acceleration of the $i$-th particle can therefore be written as
\begin{eqnarray}
\left(\frac{\mathrm{ d}\mathbf{v}_i}{\mathrm{ d}t}\right)^{(\mathrm{mag})} &=& \frac{a^{3\gamma}}{\mu_0}
\sum_{j}m_j \left[f_i^\mathrm{ co}\frac{M_i}{\rho_i^2}                       \cdot\mathbf{\nabla}_i W_i \right.   \nonumber \\
                       & &+\left.f_j^\mathrm{ co}\frac{M_j}{\rho_j^2} \cdot\mathbf{\nabla}_j W_j \right].
\end{eqnarray}
Here $a^{3\gamma} = \frac{\mathrm{ d}t}{\mathrm{ d}\eta}$ is needed to transform the equations to the internal variables for cosmological simulations and is set to one in all other cases. Also, $\mu_0$ has to be formulated in internal units (see Sect. \ref{sec:units}) as 
\begin{equation}
   \mu_0=\frac{[\mathrm{ TIME}]^2[\mathrm{ LENGTH}]}{4\uppi[\mathrm{MASS}]h^2},
\end{equation}
with the Hubble parameter set to $h=1$ for non-cosmological runs.

\subsubsection{The Powell Scheme}

\citet{2001ApJ...561...82B} suggested explicitly subtracting the effect of any numerically non-vanishing divergence of $\mathbf{B}$. Therefore, one can explicitly subtract the term
\begin{eqnarray}
\left(\frac{\mathrm{ d}\mathbf{v}_i^k}{\mathrm{ d}t}\right)^{(\mathrm{corr})} &=& - a^{3\gamma}\frac{1}{\mu_0}\hat\beta\mathbf{B}_i \sum_{j}m_j\left[\frac{\mathbf{B}_i}{\Omega_i\rho_i^2}                        \cdot\mathbf{\nabla}_i W_i \right. \nonumber \\
& & + \left. \frac{\mathbf{B}_j}{\Omega_j\rho_j^2} \cdot\mathbf{\nabla}_j W_j \right]
\end{eqnarray}
from the momentum equation. Here again, $a^{3\gamma} = \frac{\mathrm{ d}t}{\mathrm{ d}\eta}$ and $\mu_0$ are introduced to transform the equation to the internal code units. To be consistent with the other formulations, we included $\Omega_i$, which accounts for the adaptive smoothing length. 
Following the original work \citep{2001ApJ...561...82B}, we use $\hat\beta=1$. However, we apply a limit to the correction so that the magnitude of the correction does not exceed that of the computed magnetic force.

\begin{figure*}[ht]
 \centering
 \includegraphics[width=0.49\textwidth]{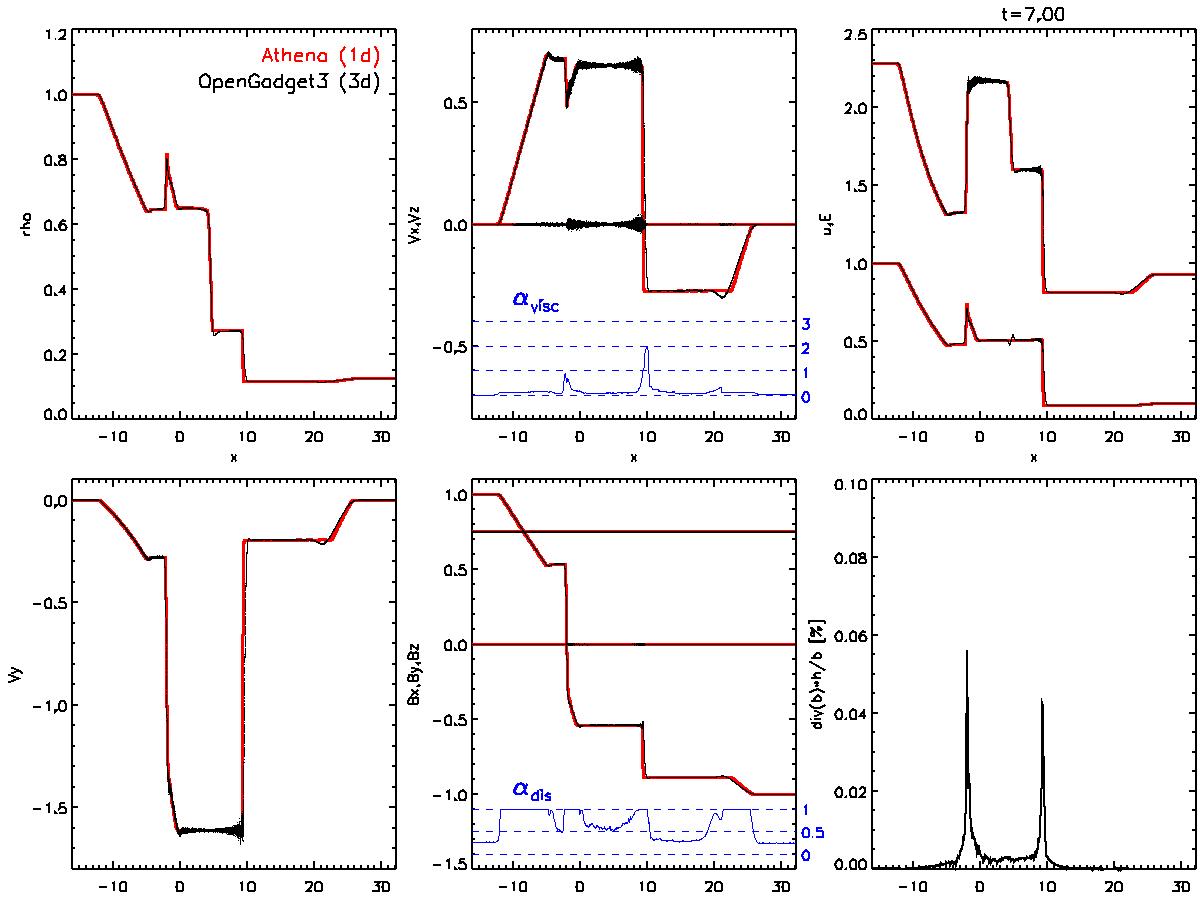}
 \includegraphics[width=0.49\textwidth]{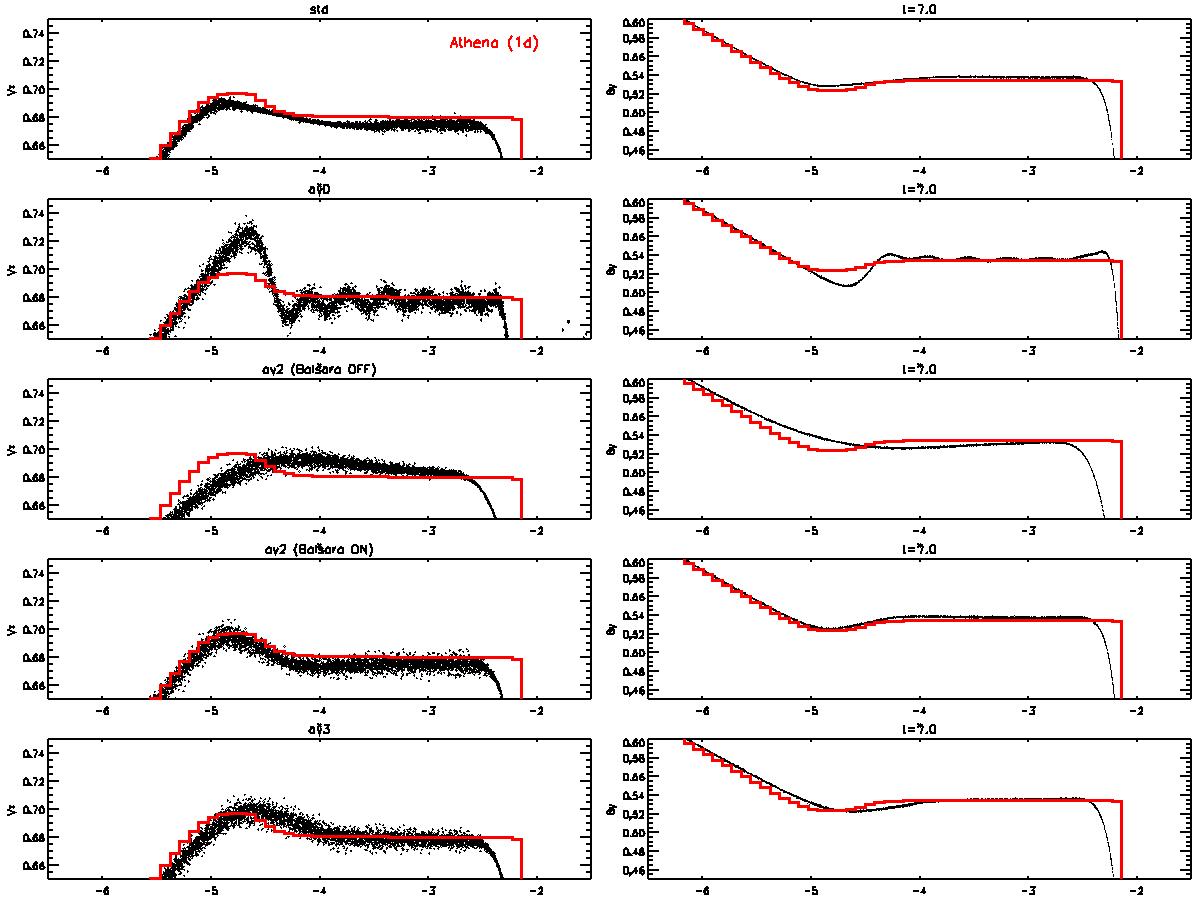}
 \caption{Standard MHD test with the standard code settings, including the Tricco cleaning. The six panels in the left halve show density ($\rho$), velocity components ($\myvec{v}_x$,$\myvec{v}_z$, $\myvec{v}_z$), internal ($u$) and total ($E$) energy, magnetic field components ($\myvec{B}_x$,$\myvec{B}_y$, $\myvec{B}_z$) and divergence of the magnetic field ($\mysymvec{\nabla}\myvec{B}$), where we used the {\bf av3} settings. The blue lines in some of the panels show the corresponding numerical values of the time-dependent coefficients. The red lines are the 1D solution obtained with Athena; black points are all particles from the 3D setup. In the right panel, we zoom onto the region between the fast rarefaction and the slow compound waves. Shown are the x component of the velocity (left part) and the y component of the magnetic field (right panel), where the point clouds in the different columns show the results from the simulations using the different settings for the AV, as labelled in the plots.}
 \label{fig:mhd_shock}
\end{figure*}

\subsection{Magnetic dissipation}
Another possibility to regularise the magnetic field was presented by \citet{2004MNRAS.348..123P}, who suggested including an artificial dissipation for the magnetic field, analogous to the artificial viscosity used in SPH. In \citet{2004MNRAS.348..123P}, it was suggested that the dissipation terms be constructed based on the magnetic field component perpendicular to the line joining the interacting particles. However, to better suppress the small-scale fluctuations within the magnetic field which appear due to numerical effects, especially in multi-dimensional tests, \citet{2004MNRAS.348..139P} suggested basing the artificial dissipation on the change of the total magnetic field rather than on the perpendicular field components only. We also found this to work significantly better in our test cases and therefore adopt this implementation in \og. Such an artificial dissipation term can be included in the induction equation as
\begin{eqnarray}
  \left(\frac{\mathrm{ d}\mathbf{B}_i}{\mathrm{ d}t}\right)^{(\mathrm{ diss})} &=&  \frac{1}{Ha^2}\frac{\rho_i\alpha_B}{2} \nonumber \\ & &
      \sum_{j} \frac{m_jv_{ij}^\mathrm{ sig}}{\hat{\rho}_{ij}^2}\left(\mathbf{B}_i-\mathbf{B}_j\right) \frac{\mathbf{r}_{ij}}{|\mathbf{r}_{ij}|}\cdot\mathbf{\nabla}_i W_i.
\label{eqn:induction_dis}
\end{eqnarray}
The parameter $\alpha_B$ is used to control the strength of the effect, with a typical suggested value around $\alpha_B\sim0.5$. Similar to the AV, this will create entropy at the rate
\begin{eqnarray}
  \left(\frac{\mathrm{ d}A_i}{\mathrm{ d}t}\right)^{(\mathrm{ diss})} &=&  -\frac{\gamma-1}{\rho_i^{\gamma-1}} \frac{\alpha_B}{4\mu_0} \nonumber \\ & &
      \sum_{j} \frac{m_jv_{ij}^\mathrm{ sig}}{\hat{\rho}_{ij}^2}\left(\mathbf{B}_i-\mathbf{B}_j\right)^2 \frac{\mathbf{r}_{ij}}{|\mathbf{r}_{ij}|}\cdot\mathbf{\nabla}_i \bar{W}_{ij}.
\label{eqn:entropy_dis}
\end{eqnarray}
The pre-factor $(\gamma-1)/(\rho_i^{\gamma-1})$ properly converts the dissipation term to a change in entropy.

This method reduces noise significantly. However, depending on the choice of $\alpha_B$, it can also lead to smearing of sharp features. To avoid this outside of strong shocks (e.g. where this is needed), \citet{2005MNRAS.364..384P} proposed evolving $\alpha_B$ for each particle, similar to the handling of the time-dependent viscosity as suggested by \citet{1997JCoPh..136....41S}. Therefore, the evolution of $\alpha_B$ for each particle is followed by integrating
\begin{equation}
\frac{\mathrm{ d} \alpha_B}{\mathrm{ d} t} = - \frac{(\alpha_B-\alpha_B^\mathrm{min})}{\tau} + S,
\end{equation}
where the source term $S$ is
\begin{equation}
   S = S_0 \, \mathrm{max}\left(\frac{|\mathbf{\nabla}\times\mathbf{B}|}{\sqrt{\mu_0\rho}},\frac{|\mathbf{\nabla}\cdot\mathbf{B}|}{\sqrt{\mu_0\rho}}\right)
\end{equation}
\citep[see][]{2005MNRAS.364..384P}. The time-scale $\tau$ defines how fast the dissipation constant decays. The signal velocity can be directly translated into a distance to the shock over which the dissipation constant decays. A useful choice of $\tau$ is:
\begin{equation}
   \tau = \frac{h}{C\;v_\mathrm{sig}},
\end{equation}
where $C\simeq 0.2$ as a typical choice, allowing the dissipation constant to decay within a time-scale that corresponds to the shock travelling 5 kernel lengths \citep[see][]{2004MNRAS.348..123P}.

\subsection{Dedner Cleaning}\label{sec:dedner_cleaning}

One way to remove unwanted divergences of the magnetic field from the simulations is by applying a cleaning scheme as suggested by \citet{Dedner2002}. In \og, we follow the implementation presented in \citet{Stasyszyn2013}:

\begin{equation}
\left.\frac{d\mymathbf{B}}{dt}\right|_{i}^{\mathrm{Ded}}=-(\mysymmathbf{\nabla}\psi^\mathrm{clean})_{i} \,.
\end{equation}

\begin{equation}
\left.\frac{dA}{dt}\right|_{i}^{\mathrm{Ded}}=-\frac{\gamma-1}{\mu_0\rho_{i}^{\gamma-1}}\mymathbf{B}_i \cdot (\mysymmathbf{\nabla}\psi^\mathrm{clean})_{i}\,,
\end{equation}

There are multiple ways to construct the scalar potential $\psi_{i}^\mathrm{clean}$ for each particle. However, we must ensure we have enough information to correct the field towards a divergence-free configuration in the next integration steps.
\cite{Dedner2002} studied this problem deeply, finding that the most practical way should be to construct and evolve $\psi$
in a way that the errors are propagated away from the source (i.e., hyperbolic cleaning) and damped (i.e., parabolic cleaning).

To do so, we must include the following evolution equation
\begin{equation}
\frac{d\psi_{i}^\mathrm{clean}}{dt} = -\left((c_\mathrm{h})_{i}^2(\mysymmathbf{\nabla} \mymathbf{B})_{i} - \frac{\psi_{i}^\mathrm{clean}}{\tau_{i}}\right)
\end{equation}
for $\psi^\mathrm{clean}$, which shows that $\psi$ now satisfies a wave equation propagating the errors outwards from the source with a speed of $c_\mathrm{h}$ (first term of the equation) and decaying them on a timescale of $\tau$ (second term in the equation). It is again natural in \textsc{SPMHD} simulations to relate the propagation speed to the signal velocity, hence using $c_\mathrm{h}=\sigma{}v_{i}^\mathrm{sig}$. Also, the timescale can be related to a typical length scale (smoothing length $h$) and velocity, resulting in $h/\lambda{}v_{i}^\mathrm{sig}$, leaving only dimensionless numerical constants $\lambda$(parabolic) and $\sigma$(hyperbolic) of order unity. We choose values of $\lambda=3$ and $\sigma=1$ to recover the best solution as presented in \cite{2005MNRAS.364..384P}.

\subsection{Tricco Cleaning}

A better cleaning scheme to handle numerical divB was proposed by \citet{Tricco2012}. In \og we follow the implementation presented in \citet{2026MNRAS.550g1121S}.

The symmetrised version of the discretisation in SPH in comoving units for the additional term in the induction equation for cosmological integration then reads  
\begin{eqnarray}
        & & \frac{1}{a^2}\left.\frac{{\rm d}\mathbf{B}_{i}}{{\rm d}t}\right|_i^\mathrm{Tricco} = -\rho_{i} \sum_{j} m_j \\ \nonumber
        &\times & \left[\frac{c_{\mathrm{h},i}\tilde{\psi}_i^\mathrm{clean}}{\Omega_i \rho_i^2} \nabla_i W_{ij}(h_i) + \frac{c_{\mathrm{h},j}\tilde{\psi}_j^\mathrm{clean}}{\Omega_j \rho_j^2} \nabla_i W_{ij}(h_j) \right],
\end{eqnarray}
where $\tilde{\psi}^\mathrm{clean}\equiv\psi^\mathrm{clean}/c_\mathrm{h}$ is the additional field which is evolved following the discretised version in the comoving frame:
\begin{eqnarray}    
    \frac{1}{a^2}\frac{{\rm d}}{{\rm d}t}\tilde{\psi}_i^\mathrm{clean} 
    &=& \frac{c_{\mathrm{h},i}}{\Omega_i \rho_i} \sum_j m_j (\mathbf{B}_i - \mathbf{B}_j) \cdot \nabla_i W_{ij}(h_i) \nonumber \\
    &-& \frac{1}{\tau} \tilde{\psi}_i^\mathrm{clean} \nonumber \\
    &+& \frac{1}{2} \tilde{\psi}_i^\mathrm{clean} \sum_j m_j (\mathbf{v}_i - \mathbf{v}_j) \cdot \nabla_i W_{ij}(h_i),
\end{eqnarray}
where, as before, $c_\mathrm{h} = \sqrt{{\rm v}_A^2 +c_s^2}$ with the Alfven speed $v_A^2$ and the sound speed $c_s$. Finally, Figure \ref{fig:mhd_shock} shows the result of a standard MHD shock tube test, where the Tricco cleaning as well as different treatments of the AV are applied. This demonstrates that, due to the improved treatment of MHD in the SPH implementation, it is possible to combine low-viscosity schemes with MHD shocks and obtain reasonable results.

\subsection{Non-ideal MHD}
\og also allows to follow a resistive term $\eta_m$ \citep[see][for details]{Bonafede2011}, which is included in the induction equation as:
\begin{equation}
\left.\frac{d \mathbf{B_i}}{d t}\right|_{\rm res} = \frac{\eta_m   \rho_i}{Ha^2} \sum_{j}\frac{m_j}{\rho_{ij}}^2\left(\mathbf{B_i}-\mathbf{B_j} \right) \frac{\mathbf{r_{i,j}}}{|\mathbf{r_{i,j}}|} \cdot \mathbf{\nabla_i}W_i\,.
\end{equation}
Here, the factor $(Ha^2)^{-1}= \frac{dt}{da}$ takes into account the internal time variable in \og and is absorbed in the magnetic kick factors as described in Sect. \ref{sec:Integration:kickfactors}. The resistivity term implemented in the induction equation causes a change in the entropy variable $A$ at the rate
\begin{equation}
\left.\frac{d A_i}{dt}\right|_{\rm res}=-\eta_m\frac{\gamma -1}{2 \mu_0 \rho_i^{\gamma-1}} \sum_{j}\frac{m_j}{\rho_{ij}^2}\left(\mathbf{B_i}-\mathbf{B_j} \right)^2\frac{\mathbf{r_{i,j}}}{|\mathbf{r_{i,j}}|} \cdot \mathbf{\nabla_i} \overline{W}_{i,j},
\end{equation}
Typical values for $\eta_m$ within the inter-stellar and intra-cluster media are in the range $\eta_m \approx [10^{25}-10^{29}] \: \mathrm{cm}^2\,\mathrm{s}^{-1}$ (see discussion in \citealt{Bonafede2011}).

\subsection{Astrophysical Sources}
\label{sec:SN_B_seeding}
To seed magnetic fields, \og includes the magnetic SN seeding model by \citet{Beck2013}. In this model, the induction equation is extended with a time-dependent seeding term 
: \begin{equation}
    \left.\frac{\partial \mathbf{B}}{\partial t}\right\vert_\mathrm{seed} = \frac{B_\mathrm{inj}}{\Delta t} \mathbf{e}_B,
\end{equation}
with $B_\mathrm{inj}$ as the magnitude of the injected magnetic field and $\mathbf{e}_B$ its unity vector components. The magnitude of the injected magnetic field is calculated as
\begin{equation}
    B_\mathrm{inj} = \sqrt{N_\mathrm{SN}^\mathrm{eff}} B_\mathrm{SN} \left( \frac{r_\mathrm{SN}}{r_\mathrm{SB}} \right)^2 \left( \frac{r_\mathrm{SB}}{r_\mathrm{inj}} \right)^3
\end{equation}
where $B_\mathrm{SN}$, $r_\mathrm{SN}$ and $r_\mathrm{SB}$ are input parameters specified at the beginning of the simulation: $B_\mathrm{SN} = 10^{-5} - 10^{-4}$ G is the mean magnetic field within a SN remnant, $r_\mathrm{SN} = 5$ pc is the canonical radius of a SN remnant and $r_\mathrm{SB}$ is the radius of the bubble the SN is blown into \citep[these canonical values are based on the work by][]{2012SSRv..166..231R}. Finally, $r_\mathrm{inj}$ is the radius of the resolution element the SN implicitly explodes in, which corresponds to the smoothing length $h_i$ of the SPH particle under consideration. In all this, we assume that the relevant expansion and mixing timescales are shorter than the timescale of the simulation and the size of the SN remnant is smaller than the resolution elements of the simulation. Both these assumptions typically hold in cosmological simulations.

The effective SN rate per time-step $N_\mathrm{SN}^\mathrm{eff}$ is obtained from the predicted star-formation rate of the underlying star-formation model (see Sect. \ref{s:SH03}). 

For the directional components of the seeding $\mathbf{e}_B = \frac{\mathbf{a}}{\vert \mathbf{a} \vert}$, we assume that the magnetic field is frozen into the plasma and its advection is dominated by external acceleration $\mathbf{a}$ of the gas.

Since the injected magnetic field vector needs also to satisfy $\nabla \cdot \mathbf{B} = 0$,
the easiest way to ensure this is to inject it as a dipole with a moment $m$ such that
\begin{equation}
     \left.\frac{\partial \mathbf{B}}{\partial t}\right\vert_\mathrm{seed} = \frac{1}{\vert r \vert^3} \left[ 3 \left( \frac{\partial m}{\partial t} \cdot \mathbf{e}_r \right) \mathbf{e}_r - \frac{\partial m}{\partial t} \right]\,.
\end{equation}
Here $\mathbf{e}_r$ is the unity vector in $r$-direction and
\begin{equation}
    \frac{\partial m}{\partial t} = \sigma \frac{B_\mathrm{inj}}{\Delta t} \mathbf{e}_B \: \,,
\end{equation}
and we introduced the normalisation
\begin{equation}
    \sigma = r_\mathrm{inj}^3 \sqrt{\frac{1}{2} f^3(1 + f^3)}\,,
\end{equation}
with $r_\mathrm{inj}$ the smoothing length of the relevant SPH particle and $f = r_\mathrm{soft}/r_\mathrm{inj}$ the ratio between the dipole smoothing length and the particle smoothing length with a typical value of $f \simeq  0.25$ \citep[see][for details]{Donnert2009}.

For a list of all available configuration and parameter options, see the according \href{https://gitlab.lrz.de/AstroCodes/OpenGadget3/-/wikis/MHD}{section on the code wiki}.


\begin{figure*}[t]
 \centering
 \includegraphics[width=1.0\textwidth]{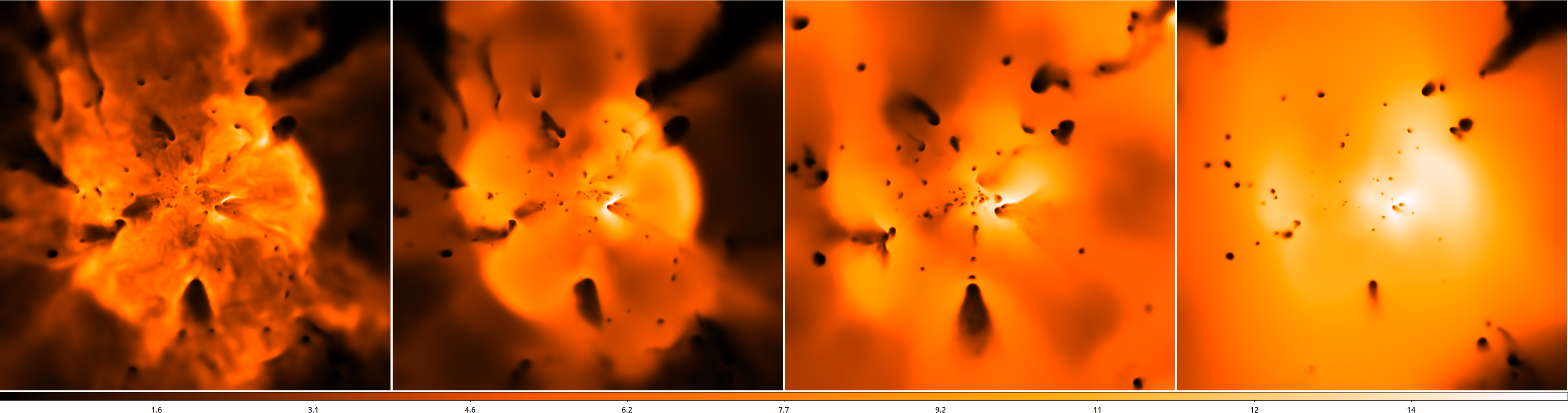}
\caption{Shown is a 5Mpc wide and 1Mpc thick slice through a non-radiative MHD simulation of a massive galaxy cluster, where different descriptions of conduction are used. Visualised is an image of the mass-weighted temperature. The first from the left is a standard simulation, without any thermal conduction. The second from left was performed using an isotropic thermal conduction of $\kappa=0.05$ times the Spitzer value. In the third from the left, anisotropic conduction with the full Spitzer value along the field lines is used. In the right one, full isotropic, Spitzer-conduction is assumed. For more discussion, see \citep{Dolag2004, Jubelgas_2004,Arth2014}.} \label{fig:cluster_slice_conduction}
\end{figure*}

\section{Conduction}
\label{sec:cond}

We can write down a conduction heat flux resulting from a temperature gradient using the common formulation
\begin{equation}
Q = -\kappa\nabla T
\end{equation}
with the conduction coefficient $\kappa$, which can be rephrased in terms of the change in internal energy
\begin{equation}
    \frac{\text{d} u}{\text{d} t} = \frac{1}{\rho}\nabla \cdot \left(\kappa\nabla T\right)
\end{equation}
where we can use $A_i=(\gamma-1)u_i\rho_i^{1-\gamma}$ to link this to change of the entropy $A_i$ within \og.

Following \citet{Jubelgas_2004}, this equation can be cast into an SPH formulation for the variation of the specific internal energy $u_i$ of the $i$-th particle due to the effect of thermal conduction, avoiding the explicit use of second-order derivatives, which results in
\begin{align}
    \frac{\text{d} u_i }{\text{d} t} = 
    \frac{\mu (\gamma-1)}{k_\mathrm{B}} 
    \sum_j \frac{m_j\kappa_{ij}}{\rho_i \rho_j} 
    \frac{\mathbf{x}_{ij} \nabla_i W_{ij}}{|\mathbf{x}_{ij}|^2} 
    \left(
    u_i-u_j
    \right)\,.
    \label{eq:cond_jubelgas1}
\end{align}
This expression can be reformulated for the variation of the entropy variable $A_i$ in the form
\begin{align}
    \frac{\text{d} A_i }{\text{d} t} = 
    \frac{2\mu}{k_\mathrm{B}} 
    \frac{\gamma-1}{\rho_i^{\gamma-1}}
    \sum_j \frac{m_j\kappa_{ij}}{\rho_i \rho_j} 
    \left(
    \frac{A_j}{\rho_j^{\gamma-1}} - \frac{A_i}{\rho_i^{\gamma-1}}
    \right)
    \frac{\mathbf{x}_{ij} \nabla_i W_{ij}}{|\mathbf{x}_{ij}|^2}\,.
    \label{eq:cond_jubelgas2}
\end{align}
In the above equations, $\kappa_{ij}$ is the conduction coefficient between the $i$-th and the $j$-th particle, and $\mu$ is the mean molecular weight. For an idealised Lorentz gas, we can assume Spitzer conductivity \citep{Spitzer1956}, for which the conduction coefficient has the expression
\begin{equation}
    \kappa_\mathrm{Sp} = \left(\frac{2}{\uppi}\right)^{3/2}\frac{(k_\mathrm{B}T_i)^{5/2}k_\mathrm{B}}{m_e^{1/2}e^4\,Z\,\,\ln \Lambda}\,,
\end{equation}
with $Z$ the average proton number of the plasma, and electron mass $m_e$, and elementary charge $e$. We also assume that the temperature $T_i$ of the $i$-th particle corresponds to the electron temperature. The Coulomb logarithm $\ln\Lambda$ is defined as
\begin{equation}
    \ln \Lambda = 37.8 + \ln \left(\left( \frac{T_i}{10^8 \, {\rm K}} \right) \left( \frac{n_e}{10^{-3} \, {\rm cm^{-3}}} \right)^{-1/2} \right) \, ,
\end{equation}
with $n_e$ the electron density.
Applying this expression to a hydrogen-helium plasma with primordial composition gives $Z\approx1.136$.

In the following section, we describe the implementation of this diffusion equation for the cases of isotropic and anisotropic conduction, while figure \ref{fig:cluster_slice_conduction} shows an example of the appearance of the ICM for the different conduction descriptions.

\subsection{Conjugant Gradient Solver} \label{sec:cond_cg_solver}

An explicit discretisation of the diffusion operator in Eq. \eqref{eq:cond_jubelgas2} imposes a stability constraint that in turn requires tight constraints on the time-steps. For this reason, \og solves the diffusion equation by using the conjugate gradient method -- a numerical technique that solves iteratively a system of linear equation. For details, we refer to \cite{Petkova2009} for the conjugate gradient formalism in the \textsc{Gadget} family and \cite{Arth2014} for the application to thermal conduction. In short, we can discretize the time step as such
\begin{equation}
    \frac{\text{d}u_i}{\text{d}t} \rightarrow  \frac{\Delta u_i}{\Delta t} = \frac{u_i^{n+1} - u_i^n}{\Delta t},
\end{equation}
where the final state $u_i^{n+1}$ can be expressed as 
\begin{equation}
    u_i^{n+1} = u_i^n + \sum_{j} \: c_{ij} \left( u_i^{n+1} - u_j^{n+1} \right).
\end{equation}
Here, $c_{ij}$ are the matrix elements of the interaction matrix, with
\begin{equation}
    c_{ij} = 
    \frac{\mu (\gamma-1)}{k_\mathrm{B}} 
    \frac{m_j\kappa_{ij}}{\rho_i \rho_j} 
    \frac{\mathbf{x}_{ij} \nabla_i W_{ij}}{|\mathbf{x}_{ij}|^2} 
\end{equation}
where the subscripts $i$ and $j$ correspond to the quantities of SPH particle $i$ and $j$, respectively and
$\kappa_{ij}$ represents the arithmetic mean of the conduction coefficients of the two particles.

This allows us to write the diffusion equation as a matrix problem
\begin{equation}
    \mathbf{C} \cdot \mathbf{x} = \mathbf{b}
    \label{eq:diff_matrix_eq}
\end{equation}
where
\begin{align}
    C_{ij} &= \delta_{ij} \left( 1 - \sum_k c_{ik} \right) + c_{ij} \\
    x_j &= u_j^{n+1} \\
    b_i &= u_i^n
\end{align}
Knowing the initial quantities $u_i^n$ and explicitly calculating the matrix elements of $C_{ij}$ allows one to effectively solve the diffusion step backwards. 

Formally, this could be done by inverting the matrix $C$ from Eq. \eqref{eq:diff_matrix_eq} using a conjugate gradient solver. This inverted matrix would then be applied to the initial quantities to obtain the updated quantities after a time-step $\Delta t$. However, as the matrix $C$ extremely sparse due to the nature of SPH, the solution is iteratively improved with a matrix-free iterative technique, so that the inverse matrix never has to be explicitly computed in the CG method.

\subsection{Anisotropic extension}
\label{sec:cond_aniso}

The anisotropic extension of the thermal conduction solver follows the implementation by \citet{Arth2014} and \citet{Steinwandel_2022}. In the presence of magnetic fields, the motion of electrons perpendicular to the field lines is suppressed, while heat can still be transported efficiently along the field lines. Following \citet{Braginskii_1965}, the conductive heat flux can be written as
\begin{equation}
    \mathbf{Q} = -\kappa_\parallel \nabla_\parallel T - \kappa_\perp \nabla_\perp T -\kappa_\Lambda \hat{\mathbf{b}} \times \nabla T \, ,
    \label{eq:cond_aniso_heat_flux}
\end{equation}
where $\kappa_\parallel$, $\kappa_\perp$, and $\kappa_\Lambda$ are the parallel, perpendicular, and Hall conduction coefficients, respectively, and $\hat{\mathbf{b}}=\mathbf{B}/|\mathbf{B}|$ is the normalised magnetic field direction. The parallel and perpendicular temperature gradients are defined with respect to $\hat{\mathbf{b}}$. With this, the anisotropic conduction can be written as
\begin{equation}
    \frac{\text{d} u}{\text{d} t} = \frac{1}{\rho} \nabla \cdot \left[ \left( \kappa_\parallel-\kappa_\perp \right) \left( \hat{\mathbf{b}}\cdot\nabla T \right) \hat{\mathbf{b}} + \kappa_\perp \nabla T \right] \, .
    \label{eq:cond_aniso_split}
\end{equation}
The second term has the same form as the isotropic conduction equation and is treated with the solver described above, using $\kappa_\perp$ as the conduction coefficient \citep{Jubelgas_2004}. The first term contains the anisotropic part. Defining the conduction tensor 
\begin{equation}
    \uppi^\mathrm{cond}_{\alpha\beta} = \left( \kappa_\parallel-\kappa_\perp \right) \hat{b}_{\alpha}\hat{b}_{\beta} \, ,
    \label{eq:cond_aniso_A}
\end{equation}
where Greek indices denote spatial components, the anisotropic contribution becomes
\begin{equation}
    \frac{\text{d} u}{\text{d} t}\bigg{|}_{\rm aniso} = \frac{1}{\rho} \sum_{\alpha,\beta} \frac{\partial}{\partial x_\alpha} \left( \uppi^\mathrm{cond}_{\alpha\beta} \frac{\partial T}{\partial x_\beta} \right) \, .
    \label{eq:cond_aniso_component}
\end{equation}

Following \citet{Arth2014} and \citet{Steinwandel_2022}, the anisotropic part is discretised as
\begin{equation}
        \frac{\text{d} u_i}{\text{d} t}\bigg{|}_{\rm aniso} 
        \!\!\!\!\!\!\! 
        =
        \frac{\mu(\gamma-1)}{k_{\rm B}\rho_i}
        \!\sum_{j}\!\frac{m_j}{\rho_j} \frac{ \mathbf{x}_{ij}^{T} \left( \uppi^\mathrm{cond}_i+\uppi^\mathrm{cond}_j \right) \nabla_i W_{ij}}{ |\mathbf{x}_{ij}|^2 } \left( u_j-u_i \right),
    \label{eq:cond_aniso_sph}
\end{equation}
where $\mathbf{x}_{ij}^{T}$ denotes the transposed vector. This form keeps the same dependence on the internal-energy difference between neighbouring particles as in the isotropic conduction scheme.
 
The time integration is performed with the same implicit matrix formalism introduced above. For the anisotropic contribution, the pairwise coefficients entering the interaction matrix are
\begin{equation}
    c_{ij} = - \frac{\mu(\gamma-1)}{k_{\rm B}} \frac{m_j \Delta t}{\rho_i \rho_j} \frac{\mathbf{x}_{ij}^{T}}{|\mathbf{x}_{ij}|^2} \left( \uppi^\mathrm{cond}_i+\uppi^\mathrm{cond}_j \right) \nabla_i W_{ij} \, .
    \label{eq:cond_aniso_cij}
\end{equation}
These coefficients provide the anisotropic contribution to the interaction matrix defined above. The resulting linear system, $\mathbf{C}\cdot\mathbf{x}=\mathbf{b}$, is then solved with the bi-conjugate gradient solver described in Sect.~\ref{sec:cond_cg_solver}.

The Hall term in Eq.~\eqref{eq:cond_aniso_heat_flux} is not explicitly included in the discretised operator. In this SPH formulation, the corresponding tensor is antisymmetric and is contracted with the particle separation vector from both sides, so its contribution vanishes in the pairwise conduction operator.

For strong anisotropies, the tensor $\uppi^\mathrm{cond}_i+\uppi^\mathrm{cond}_j$ is not guaranteed to be positive definite, which can lead to non-physical heat fluxes from cold to hot particles. The code therefore also allows the isotropised discretisation
\begin{equation}
    \uppi^\mathrm{cond} \rightarrow \alpha \uppi^\mathrm{cond} + \frac{1}{3} \left( 1-\alpha \right) {\rm tr} \left( \uppi^\mathrm{cond} \right) \mathbf{1} \, .
    \label{eq:cond_aniso_isotropised}
\end{equation}
where $\alpha$ controls the amount of isotropisation \citep{Arth2014,Steinwandel_2022}. This adds an isotropic component to the transport operator and helps to avoid non-physical heat fluxes from cold to hot particles.


\begin{figure*}[t]
 \centering
 \includegraphics[width=1.0\textwidth]{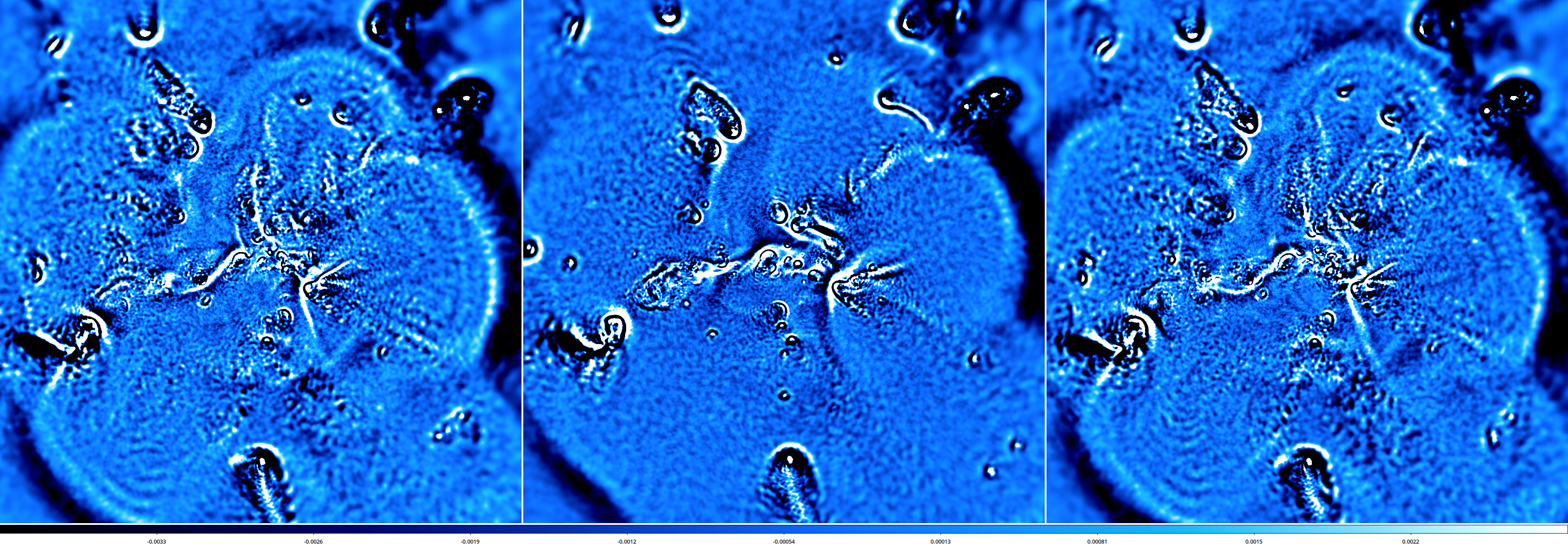}
\caption{Shown is a 5Mpc wide and 1Mpc thick slice through a non-radiative, MHD galaxy cluster simulation where different descriptions of viscosity are used. Visualised is an unsharp-masked image of the thermal SZ effect. The left is a standard simulation without physical viscosity. In the middle one, full isotropic, Spitzer-viscosity is assumed. The right shows a simulation using Braginski-viscosity, including limiters for high $\beta$ plasma. For more discussion, see \citep{Sijacki_2006, Marin-Gilabert_2022, Marin-Gilabert_2024, Marin-Gilabert_2026,2026arXiv260719753G}.} \label{fig:cluster_slice_visc}
\end{figure*}

\section{Navier-Stokes physical viscosity}
\label{sec:NavierStokes}
\noindent
{\it Main contributing developer: T. Marin-Gilabert}

The Navier-Stokes module in \og implements physical shear viscosity in SPH. This is different from the AV described in Sect.~\ref{sec:SPH:artvisc}, which is still required to capture shocks. The isotropic implementation follows \citet{Sijacki_2006} and \citet{Marin-Gilabert_2022}, while the anisotropic implementation follows \citet{Marin-Gilabert_2026}. Throughout this section, repeated Greek indices imply summation over spatial coordinates. An example of the effect of the different viscosity descriptions on the ICM can be seen in figure \ref{fig:cluster_slice_visc}.

The shear viscosity coefficient can either be kept fixed for all gas particles, or computed from a Braginskii-Spitzer parametrisation \citep{Spitzer_1962}. In the latter case, we use
\begin{equation}
    \eta_i = f_{\rm Sp} \eta_{\rm Sp}(T_i) \, ,
    \label{eq:ns_spitzer_fraction}
\end{equation}
where $f_{\rm Sp}$ is a user-defined fraction of the Spitzer value. The Spitzer coefficient is given by
\begin{equation}
    \eta_{\rm Sp} = 0.406 \frac{m_{I}^{1/2} (k_{\rm B}T_{I})^{5/2}} {(Z e)^4 \ln \Lambda} \, ,
    \label{eq:ns_spitzer_coeff}
\end{equation}
where $m_{I}$ is the ion mass, which we approximate by the proton mass $m_p$, $e$ is elementary charge $e$ and we set $Z=1$ accordingly. Here we also set the Coulomb logarithm $\ln \Lambda$ to $37.8$. We also assume that the temperature $T_i$ of the $i$-th particle corresponds to the ion temperature $T_{I}$. \og allows us to perform simulations with a constant viscosity. In this case, the viscosity value is computed from a given temperature according to  Eq.~\eqref{eq:ns_spitzer_coeff} and assigned to all gas particles.

\subsection{Isotropic implementation}

For the isotropic implementation, the rate-of-strain tensor is computed for each gas particle as
\begin{equation}
    \sigma_{\alpha \beta}\Big{|}_i =
    \left. \frac{\partial v_{\alpha}}{\partial x_{\beta}} \right|_i + \left. \frac{\partial v_{\beta}}{\partial x_{\alpha}} \right|_i - \frac{2}{3} \delta_{\alpha \beta}
    \left. \frac{\partial v_{\gamma}}{\partial x_{\gamma}} \right|_i \, .
    \label{eq:ns_shear_tensor}
\end{equation}

The velocity gradients entering Eq.~\eqref{eq:ns_shear_tensor} are computed using the SPH derivative operators described in Sect.~\ref{sec:SPH:derivatives}. The isotropic viscous stress tensor is then
\begin{equation}
    \uppi^{\rm iso}_{\alpha\beta}\Big{|}_i = \eta_i \sigma_{\alpha\beta}\Big{|}_i \, .
    \label{eq:ns_iso_stress}
\end{equation}

The corresponding viscous acceleration is discretised as
\begin{multline}
    \frac{\mathrm{d}v_{\alpha,i}}{\mathrm{d}t}\bigg{|}_{\rm iso} = \sum_j m_j \left[ \frac{\uppi^{\rm iso}_{\alpha \beta}|_i}{\rho_i^2} \left( \nabla_i W_{ij}(h_i) \right)_\beta \right. \\
    \left. + \frac{\uppi^{\rm iso}_{\alpha \beta}|_j}{\rho_j^2} \left( \nabla_i W_{ij}(h_j) \right)_\beta \right] \, .
    \label{eq:ns_iso_acc}
\end{multline}
The associated viscous heating is included as entropy production,
\begin{equation}
    \frac{\mathrm{d} A_i}{\mathrm{d}t}\bigg{|}_{\rm iso} = \frac{1}{2} \frac{\gamma - 1}{\rho_i^{\gamma - 1}} \frac{\eta_i}{\rho_i} \sigma_i^2 \, ,
    \label{eq:ns_iso_entropy}
\end{equation}
where $\sigma_i^2 = \sigma_{\alpha\beta}|_i \sigma_{\alpha\beta}|_i$.

In simulations where very large viscous stresses can arise, the coefficient can be limited using the saturation prescription described in \citet{Sarazin_1986} and implemented following \citet{Marin-Gilabert_2024},
\begin{equation}
    \eta_{\rm sat} = \frac{\eta}{1 + 4.2 \lambda_I/l_v} \, ,
    \label{eq:ns_saturation}
\end{equation}
where $\lambda_{I}$ is the ion mean free path and $l_v = 2c_s/|\sigma|$ is the local velocity length scale, with $|\sigma| = \sqrt{{\rm tr}(\sigma_i^2)}$. This prevents unphysical viscous accelerations and very small time-steps when the velocity varies on scales smaller than the ion mean free path.

\subsection{Anisotropic implementation}
\label{sec:braginski}
The anisotropic viscosity module extends the physical viscosity implementation to magnetised plasmas. It requires the MHD module described in Sect.~\ref{sec:MHD}, since the viscous stress is computed with respect to the local magnetic field direction. This implementation follows the Braginskii description of weakly collisional magnetised plasmas \citep{Braginskii_1965, Marin-Gilabert_2026}.

In the Braginskii limit, particle motion is much less restricted along the magnetic field lines than perpendicular to them. As a result, the viscous stress is determined by the pressure anisotropy
\begin{equation}
    \Delta p_i = \eta_i \left( 3 \, \hat{b}_{\alpha} \hat{b}_{\beta} \left. \frac{\partial v_{\alpha}}{\partial x_{\beta}} \right|_i - \left. \frac{\partial v_{\gamma}}{\partial x_{\gamma}} \right|_i \right) \, ,
    \label{eq:ns_aniso_pressure}
\end{equation}
where $\Delta p = p_\perp - p_\parallel$ and $\hat{\mathbf{b}}=\mathbf{B}/|\mathbf{B}|$ is the magnetic field unit vector. The velocity gradient tensor is computed using the same SPH derivative operators as in the isotropic implementation. The anisotropic viscous stress tensor is given by
\begin{equation}
    \uppi^{\rm aniso}_{\alpha\beta}\Big{|}_i = - \Delta p_i \left( \left. \hat{b}_{\alpha}\hat{b}_{\beta}\right|_i - \frac{1}{3}\delta_{\alpha\beta} \right) \, .
    \label{eq:ns_aniso_stress}
\end{equation}

The corresponding contribution to the acceleration is discretised as
\begin{multline}
    \frac{\mathrm{d}v_{\alpha,i}}{\mathrm{d}t}\bigg{|}_{\rm aniso} = \sum_j m_j \left[ \frac{\uppi^{\rm aniso}_{\alpha\beta}|_i}{\rho_i^2} \left( \nabla_i W_{ij}(h_i) \right)_\beta \right. \\
    \left. + \frac{\uppi^{\rm aniso}_{\alpha\beta}|_j}{\rho_j^2} \left( \nabla_i W_{ij}(h_j) \right)_\beta \right] \, .
    \label{eq:ns_aniso_acc}
\end{multline}
The associated entropy production is
\begin{equation}
    \frac{\mathrm{d} A_i}{\mathrm{d}t}\bigg{|}_{\rm aniso} = \frac{\gamma - 1}{\rho_i^{\gamma - 1}} \frac{\Delta p_i^2}{3 \rho_i \eta_i} \, .
    \label{eq:ns_aniso_entropy}
\end{equation}

For high-$\beta$ plasmas, i.e.\ when thermal pressure dominates over magnetic pressure, pressure anisotropies can excite fire-hose and mirror microinstabilities. When the corresponding limiter is enabled, the pressure anisotropy is restricted to the marginally stable interval
\begin{equation}
    -\frac{B^2}{4\uppi} \leq \Delta p \leq \frac{B^2}{8\uppi} \, .
    \label{eq:ns_aniso_limiter}
\end{equation}
This limiter is applied as an instantaneous hard limit to $\Delta p$, without explicitly evolving a finite scattering rate. The limited value of $\Delta p$ is then used in the anisotropic stress tensor and in the entropy production.

The viscosity module can also impose an additional time-step criterion based on the viscous increase of the entropic function \citep{Sijacki_2006}. This avoids situations in which the Courant time-step is not small enough to accurately integrate large viscous stresses. For each active gas particle, we define
\begin{equation}
    \Delta t_{i,\rm visc} = \alpha_{\rm visc} \frac{A_i} {\left|\dot{A}_{i,\rm visc}\right|} \, ,
    \label{eq:ns_visc_timestep}
\end{equation}
where $\dot{A}_{i,\rm visc}$ is the entropy-production rate due to the physical viscosity and $\alpha_{\rm visc}$ is a dimensionless time-step parameter. The default value is $\alpha_{\rm visc}=0.1$. The particle time-step is then limited by
\begin{equation}
    \Delta t_i \leq \Delta t_{i,\rm visc} \, .
    \label{eq:ns_visc_timestep_limit}
\end{equation}
For the isotropic and anisotropic implementations, $\dot{A}_{i,\rm visc}$ is given by Eqs.~\eqref{eq:ns_iso_entropy} and \eqref{eq:ns_aniso_entropy}, respectively.

For a list of all available configuration and parameter options, see the according \href{https://gitlab.lrz.de/AstroCodes/OpenGadget3/-/wikis/PhysicalViscosity}{section on the code wiki}.


\section{Shock-finder\label{sec:shockfinder}}

\begin{figure*}[t]
    \centering
	\includegraphics[width=\textwidth]{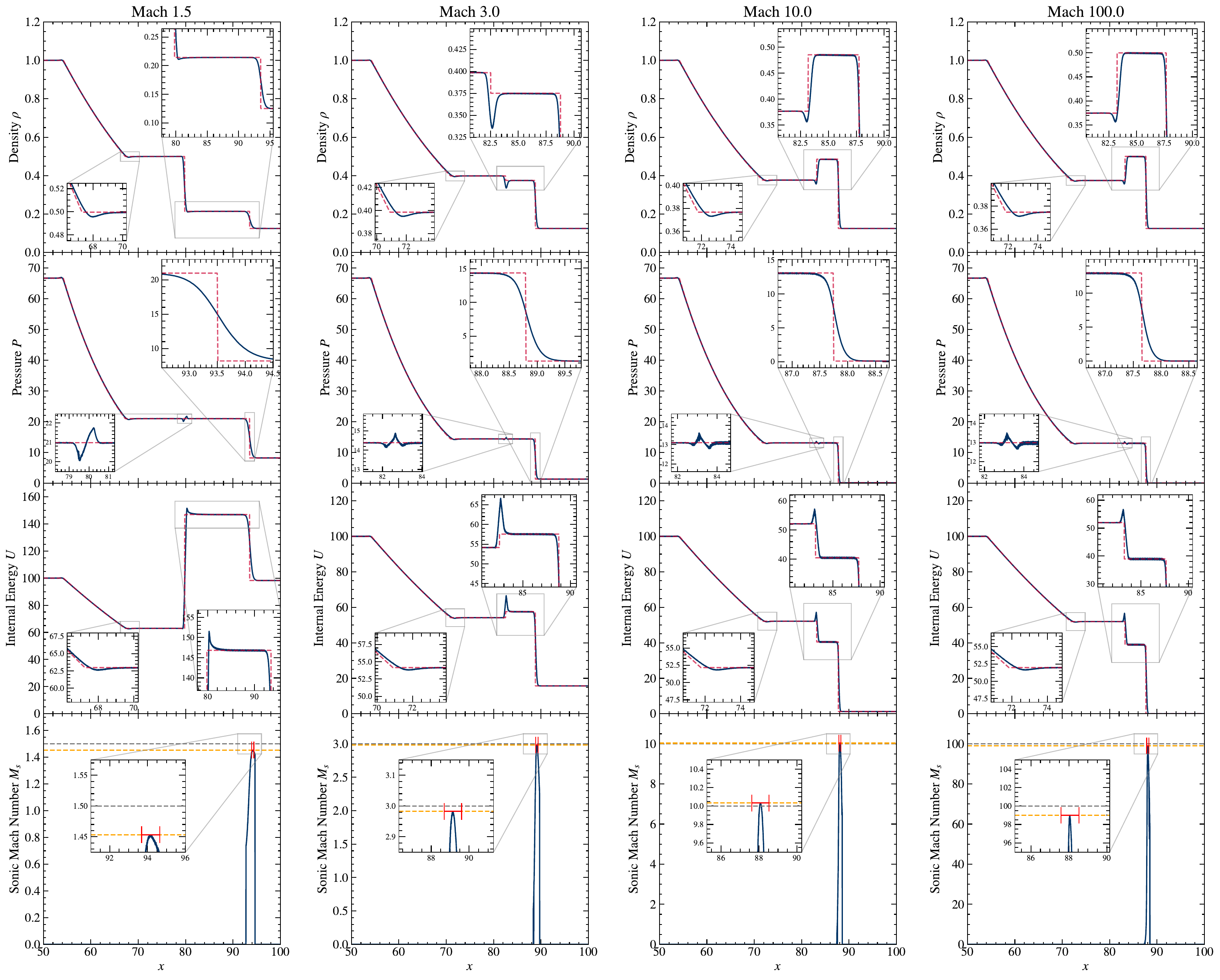}
    \caption{Standard Sod shock tube test with varying Mach numbers (from left to right). Blue lines show the simulation output with all particles plotted, including numerical noise. Red dashed lines show the analytic solution. In the bottom panels, the grey dashed line shows the analytic Mach number, the orange dashed line the maximum of the detected Mach number, and the red error bar indicates the width of the smoothing kernel.}
    \label{fig:sod}
\end{figure*}

{\og} provides an on-the-fly shock finder as introduced in  \citet{Beck2016a}.
We briefly summarise the implementation in this section.
In all these descriptions, the superscript $u$ refers to upstream and $d$ to downstream quantities.
Particles are denoted with the subscripts $\{i,j\}$ and numbered subscripts $\{1,2,3\}$ correspond to the different shock normals.

\subsection{Shock Geometry}

A first fingerprint of a shock in any hydrodynamical system is a jump in pressure.
We therefore use as a first proxy the pressure gradient within the SPH kernel as a potential shock normal
\begin{equation}
    \mathbf{\hat{n}}_{1} = \frac{- \nabla P_i}{\vert \nabla P_i \vert}
\end{equation}
where the pressure gradient is defined as
\begin{equation}
    \nabla P_i = \frac{1}{\rho_i} \sum\limits_{j=1}^{N_\mathrm{ngb}} m_j (P_j - P_i) \: \nabla W_{ij} \: .
\end{equation}
To filter out shear flows and cold fronts, which may show a pressure gradient but not a large density or velocity jump, we construct two other shock normals which are perpendicular to the first one.
These shock normals are constructed from the perpendicular vector to the pressure gradient
$\mathbf{w}_2 = \left( 0, -\hat{n}_{1,x}, \hat{n}_{1,y}\right)$ via
\begin{equation}
    \mathbf{\hat{n}}_2 = \mathbf{w}_2 - \left( \mathbf{\hat{n}}_1 \cdot \mathbf{w}_2 \right) \:  \mathbf{\hat{n}}_1
\end{equation}
and the cross product of the previous two
\begin{equation}
    \mathbf{\hat{n}}_{3} = \mathbf{\hat{n}}_{1} \times \mathbf{\hat{n}}_2
\end{equation}
with a normalisation step between each calculation.

As a subsequent step, we need to account for which neighbouring particles contribute to the up- or downstream properties of the potential shock and how much they should contribute to these quantities, based on their relative position.

\subsection{Filtering}

We employ a number of filtering steps to ensure we are not misidentifying a cold front or a shear flow as a shock, as mentioned above.
The main filtering step is to check the projected velocity jump along the different shock normals
\begin{equation}
    \Delta v = \mathbf{\hat{n}} \cdot \mathbf{v}^d - \mathbf{\hat{n}} \cdot \mathbf{v}^u\,.
    \label{eq:shock_jump}
\end{equation}
For the pressure jump to be classified as a shock, we impose the constraint
\begin{equation}
    \Delta v_1 > 0 \quad \& \quad \Delta v_1 > 2 \Delta v_2 \quad \& \quad \Delta v_1 > 2 \Delta v_3 \,,
\end{equation}
to avoid capturing shear flows.
Next, we check if pressure, velocity and density jumps are sufficiently large by imposing the conditions
\begin{align}
    P_1^d &> 1.05 P_1^u \\
    \Delta v_1 &> 0.025 \left( \mathbf{v}_1^d + \mathbf{v}_1^u \right) \\
    \rho_1^d &> 1.05 \rho_1^u \,.
\end{align}
These conditions limit the minimum sonic Mach number we can capture to $\mathcal{M}_s \geq 1.035$.
Finally, we check if the velocity field is converging with $\mathbf{\nabla} \cdot \mathbf{v} < 0$ and impose an additional shear-flow limiter
\begin{equation}
    \mathcal{F} = \frac{\vert\vert \mathbf{\nabla} \cdot \mathbf{v} \vert\vert}{\vert\vert \mathbf{\nabla} \cdot \mathbf{v} \vert\vert \: \vert\vert \mathbf{\nabla} \times \mathbf{v} \vert\vert} > 0.9\,.
\end{equation}
We note that this last step benefits from a higher-order estimate for $\mathbf{\nabla} \cdot \mathbf{v}$ and $\mathbf{\nabla} \times \mathbf{v}$ as introduced above.

\subsection{Hydrodynamic Shock Properties}

In hydrodynamical simulations, we compute the sonic Mach number from the Rankine-Hugoniot jump condition as follows.
The shock compression ratio $x_r$ can be measured from the density jump within the kernel along the main shock normal
\begin{equation}
    x_s = \frac{\rho_1^d}{\rho_1^u}\,.
\end{equation}
This quantity can in turn be used to compute the shock speed as
\begin{equation}
    v_s = \frac{\Delta v_1}{1 - x_s^{-1}}\,,
\end{equation}
where $\Delta v_1$ is the velocity jump along the main shock normal as given in Eq. \eqref{eq:shock_jump}, and the sonic Mach number $M_s$ is given by
\begin{equation}
    \mathcal{M}_s = \frac{v_s}{c^u}\,,
\end{equation}
with $c^u = \sqrt{\gamma P_1^u / \rho_1^u}$ being the upstream sound speed. Figure \ref{fig:sod} shows an example of several hydrodynamical shocks with strength between $\mathcal{M}=1.5$ and $\mathcal{M}=100$, demonstrating the performance of the on-the-fly shock finder.

\begin{figure*}[ht]
    \centering
	\includegraphics[width=1.0\textwidth]{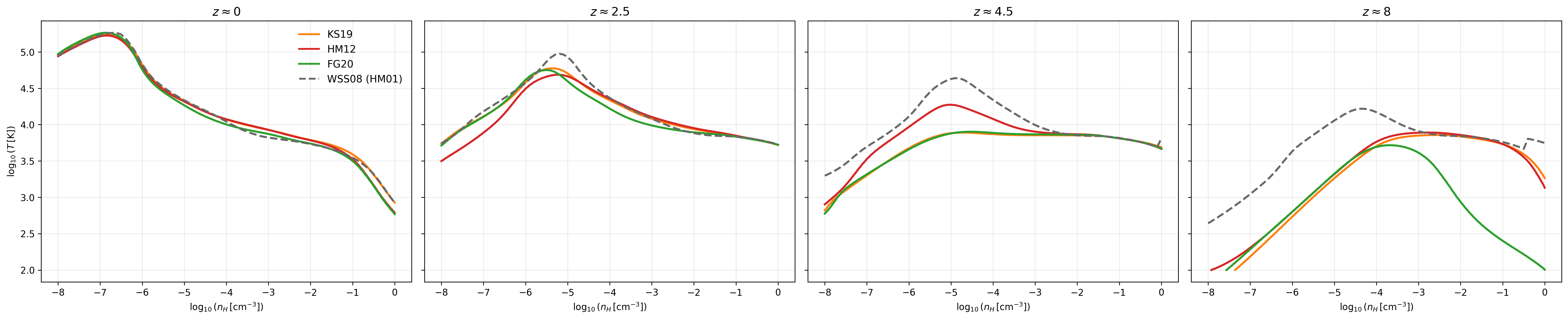}
    \caption{Thermal-equilibrium loci as a function of hydrogen number density for KS19, HM12, FG20, and W09 (HM01), at $z=0.000$, $2.479$, $4.619$, and $8.075$. The curves correspond to the $\Lambda_{\rm net}=0$ contours in the cooling function, here overlaid to facilitate a direct comparison among the models. KS19, HM12, and FG20 use the same adopted Solar reference composition, whereas W09 retains the solar abundance set of the original calculations. For details, see Di Federico et al. 2026.}
    \label{fig:cool}
\end{figure*}

\subsection{Magnetohydrodynamic Shock Properties}
In simulations with MHD, our shock finder captures a number of additional shock properties.
Most notably, the Alfvénic Mach number
\begin{equation}
    \mathcal{M}_A = \frac{v_s}{v_A^u} \,,
\end{equation}
where $v_A^u = \sqrt{\vert\vert \mathbf{B}^u \vert\vert/(\mu_0 \rho^u)}$ is the upstream Alfvén velocity.
We further construct the shock obliquity \citep[which is relevant in the context of CR acceleration, e.g.][]{Caprioli2014, Caprioli2018, Pais2018, Banfi2020, Boess2023} as the angle between shock normal and magnetic field,
\begin{equation}
    \theta_B = \arccos \left( \frac{\vert\mathbf{\hat{n}}_1 \cdot \mathbf{B}^u \vert}{\vert\vert \mathbf{B}^u \vert\vert}\right)\,,
\end{equation}
where we impose a symmetry at $90^\circ$. For a list of all available configuration and parameter options, see the according \href{https://gitlab.lrz.de/AstroCodes/OpenGadget3/-/wikis/Shock-Finder}{section on the code wiki}.


\section{Radiative cooling}
\label{sec:Cool}

Radiative cooling is one of the key processes driving structure formation,
as it determines the ability of gas to condense within DM halos and form stars. During their evolution and especially in the final stages of their lives, stars release metals into the surrounding gas and photoionize it. The local metal abundance affects the rate at which the gas cools, while photoionisation heats it. In \og{}, the quantity that is ultimately used to update the internal energy of a gas particle due to such processes is the normalised net cooling rate:
\begin{equation}
\frac{\mathcal{H}-\Lambda}{n_{\rm H}^{2}}\,,
\label{eq:net_cooling_general}
\end{equation}
where $\mathcal{H}$ the heating rate, $\Lambda$ is the cooling rate and $n_{\rm H}$ is the hydrogen number density. In the current version of \og, the heating rate is contributed by a redshift-dependent, uniform ultraviolet background (UVB). In this section, we summarise the main assumptions and equations of the implementation, referring the reader to other papers for details.

\subsection{Computing Metal--free cooling}
\label{sec:Katz}

To stay compatible with previous simulations, the original radiative cooling implementation of \textsc{Gadget}-1/2 as presented by \cite{Springel2005} is still available, especially in connection with the star formation description from \cite{Springel2003} (see Sect. \ref{s:SH03}). This assumes metal-free gas and follows the procedure described in \cite{Katz.etal.1996}. In short,  the cooling and heating rates are computed by assuming an optically thin gas with primordial composition, in photoionisation equilibrium with the UVB radiation field but not in thermal equilibrium. The UVB in \og is included through an external input file. The heating rate has the expression
\begin{equation}
\mathcal{H} = n_{\rm HI}\,\epsilon_{\rm HI} + n_{\rm HeI}\,\epsilon_{\rm HeI} + n_{\rm HeII}\,\epsilon_{\rm HeII}\,,
\label{eq:heat_katz}
\end{equation}
i.e.\ the sum of the contributions from the photoionisation of neutral hydrogen, neutral helium, and singly ionised helium, and depends on the intensity of the adopted UVB. 

As for the heating part, for completeness we give a very short summary of how the cooling function is computed. We refer the reader for more details to the extensive discussion in \citet{Katz.etal.1996}. Under the above assumptions, it is possible to evaluate the abundances of the relevant ionic species (H, H$^+$, He, He$^+$, He$^{++}$, e$^-$) for every active particle at every time-step. This allows us to calculate the total cooling rate as
\begin{equation}
\Lambda = \Lambda_{\rm exc} + \Lambda_{\rm ion} + \Lambda_{\rm rec} + \Lambda_{\rm ff} + \Lambda_{\rm Compton}\,,
\label{eq:lambda_katz}
\end{equation}
where $\Lambda_{\rm exc}$, $\Lambda_{\rm ion}$, and $\Lambda_{\rm rec}$ are the contributions from collisional excitation, collisional ionisation, and (standard plus dielectronic) recombination of H and He, respectively. Furthermore, $\Lambda_{\rm ff}$ is the free-free emission (bremsstrahlung radiation), and $\Lambda_{\rm Compton}$ is the inverse Compton cooling off the cosmic microwave background (CMB). The analytic expressions used to compute the rates of these processes are taken from \cite{Black1981} (Table 3) and incorporate the modifications introduced by \cite{Cen1992} (Sections 2.1.1--2.1.3).
 
The net cooling rate is then computed as in Eq.~\eqref{eq:net_cooling_general}. For computational efficiency, $\Lambda$ and $\mathcal{H}$ are tabulated as functions of gas density and temperature in lookup tables at different redshifts, which are then interpolated to assign to each particle the corresponding amount of energy gained or lost through radiative processes.
 
\subsection{Metallicity dependent cooling}
\label{sec:CoolCl}
\noindent
{\it Main contributing developers: L. Di Federico, L. Tornatore}

\og follows the original implementation by \cite{WAL09} in which cooling and heating rates are computed from large grids of photoionisation models generated with \textsc{Cloudy 07.02}. The resulting rates are tabulated and read as an input by \og{}, which then interpolates to compute the correct rates according to the properties -- density, temperature, metallicity, redshift -- of each gas particle. The gas is assumed to be optically thin, dust-free, and in ionisation equilibrium, and is exposed to a redshift-dependent UV/X-ray background from quasars and galaxies \citep[][hereafter HM01]{HM01}, as well as to the CMB (see previous section).
 
The input tables span the redshift range $0 \leq z \leq 9$, each corresponding to a redshift at which the UV background is defined. At a fixed redshift, the cooling and heating rates are tabulated in terms of the normalized net cooling rate, $(\mathcal{H}-\Lambda)/n_{\mathrm{H}}^{2}$ [erg s$^{-1}$ cm$^{3}$]\footnote{Note that the tables themselves store the net rate as $\Lambda-\mathcal{H}/n_{\mathrm{H}}^{2}$, positive for net cooling, following the original \citet{WAL09} convention; the sign is inverted internally by the code to match the convention adopted throughout this section.}, as a function of hydrogen number density in the range $n_{\mathrm{H}} = (10^{-8} - 1)\,\mathrm{cm}^{-3}$, temperature in the range $T = (10^{2} - 10^{9})\,\mathrm{K}$, and helium abundance. The latter quantity is expressed in terms of the helium mass fraction $Y \equiv M_{\mathrm{He}} / (M_{\mathrm{H}} + M_{\mathrm{He}})$, which spans the range $Y = 0.238 - 0.298$. At higher redshifts, we adopt cooling rates computed in collisional ionization equilibrium, for which the dependence on gas density does not appear.
 
The cooling rate $\Lambda$ is computed on an element-by-element basis assuming solar relative abundances (indicated by the subscript $\odot$), including the following 11 elements: H, He, C, N, O, Ne, Mg, Si, S, Ca, and Fe, that are all self-consistently followed by the stellar and chemical evolution model implemented in \og (see Sect. \ref{sec:StEv}). Hydrogen and helium are treated together as a metal-free component, for which the cooling rates are also tabulated over the range of helium fractions given above.
 
The total cooling rate $\Lambda$ entering Eq.~\eqref{eq:net_cooling_general} is then reconstructed by combining the H and He contribution with the individual metal contributions, scaled according to the local abundances, leading to
\begin{equation}
\Lambda = \Lambda_{\mathrm{H,He}} + \sum_{i>\mathrm{He}} \Lambda_{i,\odot}
\left( \frac{n_e}{n_{e,\odot}} \right)
\left( \frac{n_i}{n_{i,\odot}} \right),
\label{eq:lambda_wal}
\end{equation}
where $\Lambda_{i,\odot}$, $n_{e,\odot}$ and $n_{i,\odot}$ are, respectively, the tabulated cooling rate and the electron and ion number densities of element $i$, with $i$ corresponding to elements heavier than He, for a gas of solar composition. This formulation allows us to obtain a consistent reconstruction of the total cooling rate for arbitrary chemical compositions.
 
Besides the original cooling tables based on the UVB by HM01, we provide an updated set of cooling and heating tables (see Di Federico et al. 2026) based on more recent UV/X-ray background models, namely \cite{HM12} (hereafter HM12), \cite{KS19} (hereafter KS19), and \cite{FG20} (hereafter FG20). They cover the redshift ranges $0 \leq z \lesssim 14.98$ (KS19, HM12) and $0 \leq z \lesssim 10$ (FG20).
 
The cooling and heating rates in the tables have been computed with \textsc{Cloudy 23.01} \citep{Cloudy23} and adopt solar abundances from \citet{Grevesse2010}. In addition to the original implementation, $\Lambda$ and $\mathcal{H}$ are also tabulated separately, rather than only as their difference, so that the net cooling rate of Eq.~\eqref{eq:net_cooling_general} can be reconstructed while retaining the individual contributions where needed. An example of where the transition of cooling and heating is located at different redshifts and for the different UVBs can be seen in figure \ref{fig:cool}. The chemical network has been extended to ensure consistency with the adopted stellar evolution model (see Sect. \ref{sec:StEv}). In addition to the previous 11 elements, we include Al, Ni, Na, and Ar, for a total of 15 elements. The total cooling rate is reconstructed as described above, by combining the individual elemental contributions according to the local abundances. We also increase the sampling in helium mass fraction to better cover super-solar regimes, extending the helium mass fraction range up to $Y = 0.338$. The tables also provide the fractions of neutral, ionised, and molecular hydrogen (HI, HII, and H$_2$). Further details on the construction of these tables are presented in Di Federico et al. (2026, in preparation).
 

\begin{figure}[t]
    \centering
	\includegraphics[width=0.48\textwidth]{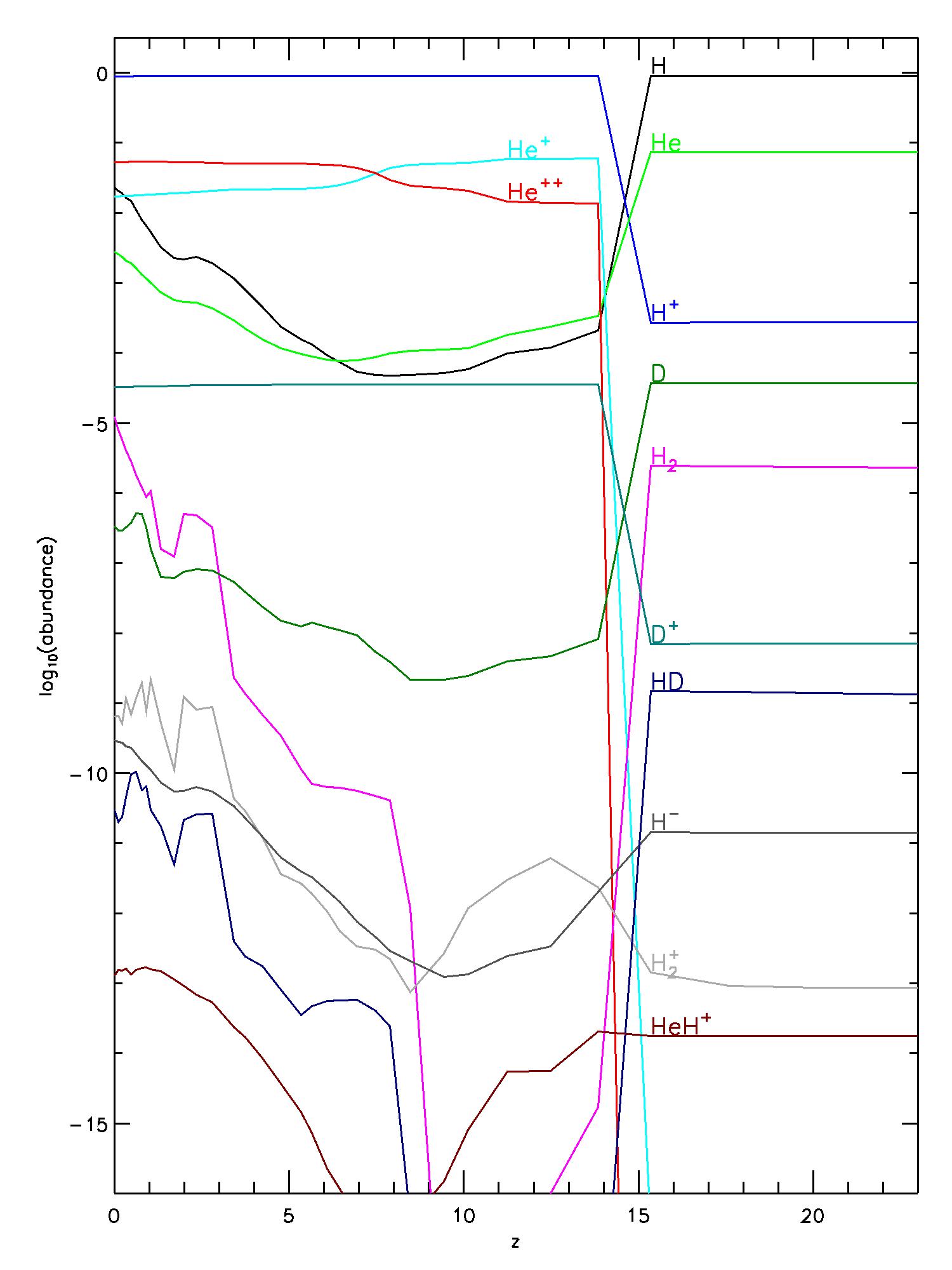}
    \caption{Small cosmological box performed with the chemical network switched on. Shown are the time evolution of the different chemical species. The KS19 background was used, which sets in quite early, as can be seen from reionisation of hydrogen and helium and dissociation of molecules at $z\approx15$.}
    \label{fig:chem}
\end{figure}

\begin{table}[ht]
\begin{center}  
\caption{Set of reactions in the code. Taken from \citet{Maio2007}.}\label{tab:reactions}  
\begin{tabular}{lr}  
\hline  
\hline  
Reactions & References for the coefficients\\  
\hline  
 	H    + e$^-$   $\rightarrow$ H$^{+}$  + 2e$^-$ & A97 / Y06\\  
	H$^+$   + e$^-$  $\rightarrow$ H     + $\gamma$    & A97 / Y06\\  
	He   + e$^-$   $\rightarrow$ He$^+$  + 2e$^-$    & A97 / Y06\\  
	He$^+$  + e$^-$   $\rightarrow$ He   + $\gamma$     & A97 / Y06\\  
	He$^+$  + e$^-$   $\rightarrow$ He$^{++}$ + 2e$^-$    & A97 / Y06\\  
	He$^{++}$ + e$^-$   $\rightarrow$ He$^+$  + $\gamma$ & A97 / Y06\\  
	H    + e$^-$   $\rightarrow$ H$^-$    + $\gamma$     & A97 / Y06\\  
	H$^-$    + H  $\rightarrow$ H$_2$  + e$^-$       & A97 / Y06\\  
        H    + H$^+$ $\rightarrow$ H$_2$$^+$  + $\gamma$ & A97 / Y06\\  
        H$_2$$^+$  + H  $\rightarrow$ H$_2$  + H$^+$     & A97 / Y06\\  
	H$_2$   + H   $\rightarrow$ 3H            & A97\\  
	H$_2$   + H$^+$ $\rightarrow$ H$_2$$^+$  + H     & S04 / Y06\\  
  	H$_2$   + e$^-$   $\rightarrow$ 2H   + e$^-$ & ST99 / GB03 / Y06\\  
      	H$^-$    + e$^-$   $\rightarrow$ H    + 2e$^-$   &A97 / Y06\\  
       	H$^-$    + H   $\rightarrow$ 2H    + e$^-$       & A97 / Y06\\  
       	H$^-$    + H$^+$ $\rightarrow$ 2H                &P71 / GP98 / Y06\\  
       	H$^-$    + H$^+$ $\rightarrow$ H$_2$$^+$  + e$^-$& SK87 / Y06\\  
        H$_2$$^+$  + e$^-$   $\rightarrow$ 2H            &GP98 / Y06\\  
        H$_2$$^+$  + H$^-$  $\rightarrow$ H    + H$_2$   &A97 / Y06\\  
        D    + H$_2$   $\rightarrow$   HD   + H     & WS02\\  
        D$^+$  + H$_2$   $\rightarrow$   HD   + H$^+$  & WS02\\  
        HD   + H   $\rightarrow$   D   + H$_2$         & SLP98\\  
        HD   + H$^+$  $\rightarrow$   D$^+$  + H$_2$   & SLP98\\  
        H$^+$  + D   $\rightarrow$   H    + D$^+$   & S02\\  
        H    + D$^+$  $\rightarrow$   H$^+$  + D    & S02\\  
        He    + H$^+$  $\rightarrow$   HeH$^+$  + $\gamma$    & RD82, GP98\\  
        HeH$^+$    + H $\rightarrow$   He  + H$_2^+$    & KAH79, GP98\\  
        HeH$^+$    + $\gamma$ $\rightarrow$   He  + H$^+$    & RD82, GP98\\  
\hline  
\hline  
\end{tabular}  
\end{center}  
{\footnotesize
Notes -  
P71~=~\cite{Peterson1971};  
KAH79~=~\cite{KAH1979};  
RD82~=~\cite{RD1982};  
SK87~=~\cite{SK1987};  
A97~=~\cite{Abel_et_al1997};  
GP98~=~\cite{GP98};  
SLP98~=~\cite{SLD_1998};  
ST99~=~\cite{ST99};  
WS02~=~\cite{Wang_Stancil_2002};  
S02~=~\cite{Savin_2002};  
GB03~=~\cite{GB03};  
S04~=~\cite{Savin_et_al2004};  
Y06~=~\cite{Yoshida2006_astroph}.}  
\end{table}

\section{Non-equilibrium chemistry}
\label{sec:CoolNEq}
\noindent
{\it Main contributing developer: U. Maio}

\og contains a treatment of non-equilibrium atomic and molecular chemistry, which is modelled as in \cite{Maio2007, maio10, maio11} through a set of chemical reactions including e$^-$, H, H$^+$, He, He$^+$, He$^{++}$, H$_2$, H$_2^+$, H$^-$, D, D$^+$, HD, HeH$^{+}$ for each individual gas particle \cite[following e.g.][]{abel97, GP98, yoshida03} and fully coupled to gas cooling and star formation \cite{Maio2007}.
For each species $i$, the variation in time of its number density $ n_i $ is:
\begin{equation}
    \frac{{\rm d}  n_i}{ {\rm d} t} = \sum_p \sum_q k_{pq,i} n_p n_q - \sum_l k_{li} n_l n_i\,,
\end{equation}
where $k_{pq,i}$ is the creation rate of species $i$ from species $p$ and $q$, and $k_{li}$ is the destruction rate from interactions of species $i$ with species $l$.
The chemistry rates are taken from the literature, as listed in table~\ref{tab:reactions} \citep[see also][]{Maio2007, Petkova2012, maio15}, and coupled with the different UVBs adopted \citep{maio22}.
Chemical abundances are numerically evaluated according to a standard backward difference scheme \cite[][]{anninos97}, while gas cooling includes resonant, fine-structure and molecular line emission in the temperature regime between $\sim 10\,\rm K$ and $10^9\, \rm K$ -- see e.g. Fig.~4 of \cite{Maio2007}.
In this way, different from the common implementations in simulation codes, it is possible to consistently trace the chemical and thermal state of the gas in all its phases (cold, warm, and hot) \citep[][]{maio26} and assess the origin of observational detections of crucial gas line emissions  \citep{casavecchia24, casavecchia25, parente25}.

In more detail, cooling calculations below $ 10^4$ K include line emissions from oxygen, carbon, silicon and iron. They are the most abundant heavy elements released during stellar evolution and, therefore,  play the most important role in chemical enrichment and gas condensation. As in the low-density regime and the heterogeneous conditions of cosmic gas, thermodynamic equilibrium is rarely reached; the Boltzmann distribution for the population of atomic levels can not be adopted. Thus, we use the detailed-balance principle to compute atomic level populations and the resulting emissivities, as in e.g. \citet{Maio2007}, and obtain the correct number fraction of electrons from the ionisation states of the tracked atoms and molecules.

Time stepping is limited by the cooling time and the electron recombination time, while, for convergence reasons, the chemical integration is performed over a sub-cycle of the actual time step, typically 1/10. Wind particles are locally decoupled from hydrodynamics \cite[as in e.g.][]{maio11, maio22} and retain their atomic and molecular content resulting from chemistry cooling calculations and star formation heating (see next Section).

An example of how the averaged abundance (e.g. number fraction of the different species) evolves in a cosmological volume is shown in figure \ref{fig:chem}, which shows a small test based on one of the {\it Mageticum} volumes, namely {\it Box4/hr}.   


\section{Star formation}
\label{s:sf}

\subsection{An effective model of star formation}
\label{s:SH03}

The star formation implemented in \og is largely inspired by that originally introduced by \citet{Springel2003} (SH03 hereafter), to which we refer for a detailed description.
This model has been extended to include the metallicity-dependence of the cooling function, according to the model of chemical enrichment discussed in Sect. \ref{sec:StEv}, and to account for the contribution of magnetic fields, as discussed in Sect. \ref{sec:MHD}.

According to this model, gas particles whose density exceeds a given threshold value, $\rho_{\rm thr}$, are treated as multi-phase particles, in which a cold and a hot phase coexist in pressure equilibrium, with the cold phase being identified with the ISM clouds that provide the reservoir for star formation. Above this density threshold, thermal instability is assumed to operate, so that the specific internal energy of the hot phase, $u_h$, can only change as a result of star formation and feedback. The specific internal energy of the cold phase, $u_c$, is assumed to correspond to a fixed temperature $T_c=10^3$ K. The relative mass fractions of cold and hot phases vary due to the action of star formation and feedback, photo-evaporation and growth of cold clouds due to cooling. At equilibrium, the model predicts that 
\begin{equation}
    u_h\,=\,\frac{u_{SN}}{A+1}+u_c\,,
    \label{eq:uh}
\end{equation}
where $u_{SN}\equiv \beta^{-1}(1-\beta) \epsilon_{SN}$ is a specific internal energy contributed by supernovae (SN). Its value is expressed in terms of the mass fraction $\beta$ of stars that instantaneously die as Type-II SN, and of the average energy returned by SN per unit stellar mass formed. As such, both $\beta$ and $\epsilon_{SN}$ depend on the assumed IMF and on the range of masses of stars that are treated in the Instantaneous Recycling Approximation (IRA). Assuming a Salpeter IMF \citep{Salpeter1955}, typical values are $\beta\sim 0.1$ and $\epsilon_{SN}\sim 5\times 10^{48}\, {\rm erg}\,M_\odot^{-1}$. A corresponding "supernova temperature" is defined as $T_{SN}= 2\mu u_{SN}/(3 k_B)\sim 10^8$ K, where $\mu$ is the mean molecular weight and $k_B$ the Boltzmann constant. As for the factor $A$ appearing in Eq.(\ref{eq:uh}), it represents the efficiency of evaporation of the cold clouds, whose value depends on the local gas density $\rho$ according to 
\begin{equation}
    A(\rho)\,=\,A_0\left(\frac{\rho_{\rm thr}}{\rho}\right)^{4/5}\,.
    \label{eq:A}
\end{equation}
Here the normalisation factor $A_0$ is fixed by requiring that at the onset of thermal instability, i.e. when gas density reaches $\rho_{\rm thr}$, the temperature of the hot phase should reach the value at which the cooling function starts decreasing -- i.e. the condition defining thermal instability. Since this happens at $T\sim 10^5$ K, one obtains $A_0= (10^5K)^{-1}\, T_{SN}=10^3$.

As shown by SH03, any deviation from the equilibrium temperature of the hot phase set by Eq.(\ref{eq:uh}), as induced by star formation and feedback, decays to that value over a time-scale
\begin{equation}
    \tau_h\,=\,\frac{t^*}{\beta (A+1)}\,\frac{\rho_h}{\rho_c}\,
    \label{eq:tauh}
\end{equation}
where $\rho_h$ is the density associated with the hot phase. As for the time scale $t^*$ appearing in the above equation, it corresponds to the characteristic time for the conversion of cold clouds into stars. Under the assumption that it is proportional to the local dynamical time-scale of the gas, it takes the expression
\begin{equation}
    t^*\,=\,t_0^*\left( \frac{\rho}{\rho_{\rm thr}}\right)^{-1/2}\,.
    \label{eq:tstar}
\end{equation}
In this way, $t_0^*$ is a parameter of the model that can be phenomenologically constrained by requiring it to reproduce the normalisation of the relationship between star formation rate per unit area and surface gas mass density -- the so-called Schmidt-Kennicutt law -- observed for normal (i.e. not starbursting) spiral galaxies \citep[e.g.][]{Kennicutt1998}.

\begin{figure}[t]
    \centering
	\includegraphics[width=0.48\textwidth]{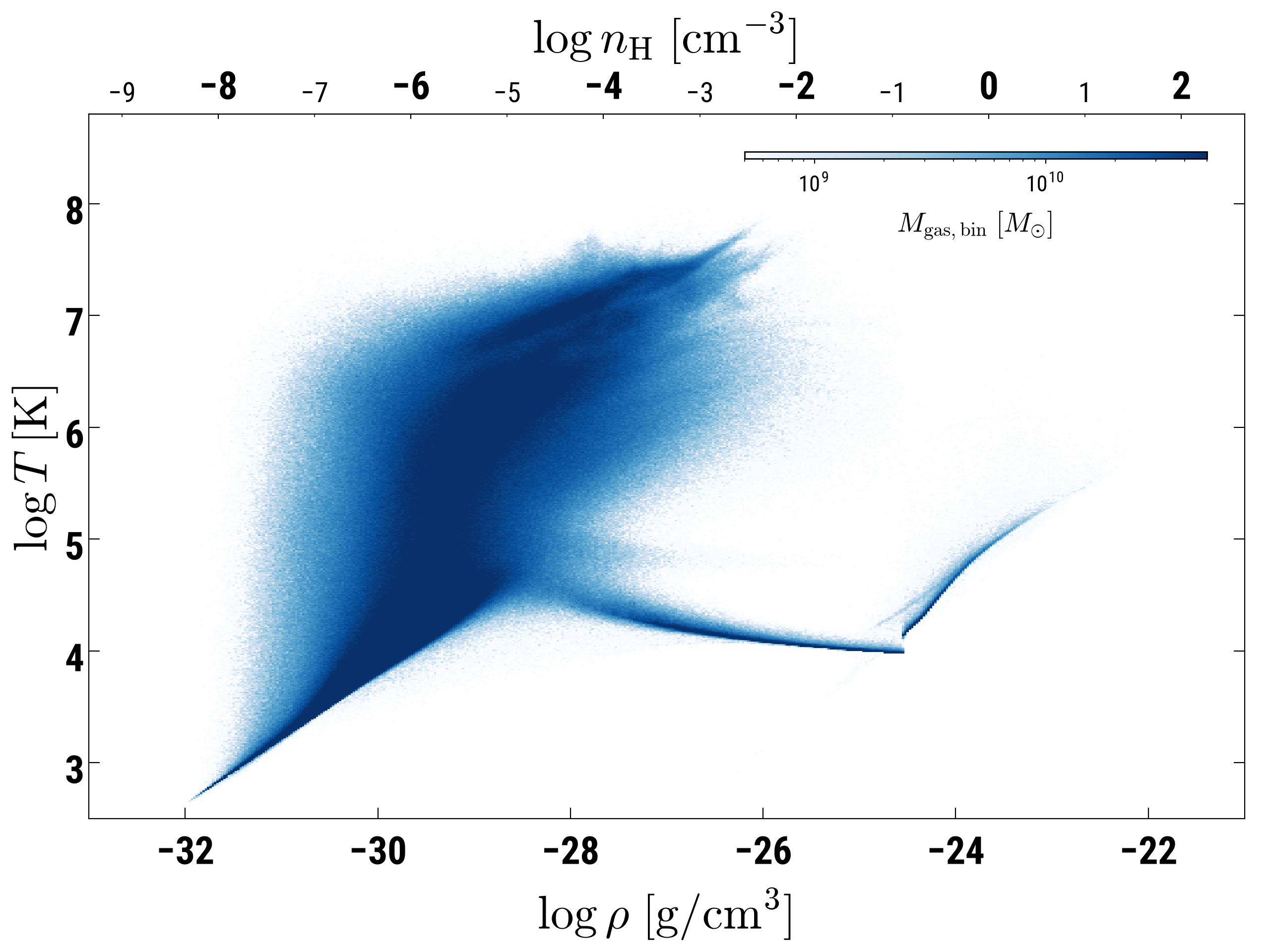}
    \caption{Cosmological box performed with the simple star-formation model. Shown is a phase diagram at redshift zero, colour-coded by mass. For details, see \citet{Springel2003} and \citet{2003MNRAS.339..312S}.}
    \label{fig:sfr_sh}
\end{figure}

In the self-regulated regime, multi-phase gas particles, i.e. those having $\rho > \rho_{\rm thr}$, obey the following effective equation of state:
\begin{equation}
    P_{\rm eff} = (\gamma - 1)\rho [(1-x)u_h + xu_c]=(\gamma - 1)\rho u_{\rm eff}\,,
\label{eq:Peff}
\end{equation}
where $\gamma=5/3$ and $x=\rho_c /\rho$ is the particle mass fraction in the cold phase, and $u_{\rm eff}$ gives the mass-weighted effective temperature of the ISM. The cold fraction $x$ depends on the cooling function and the amount of feedback energy made promptly available by Type-II SN in the instantaneous recycling approximation, according to the expression.
\begin{equation}
    \label{eq:x}
    x\,=\, 1 + \frac{1}{2y} - \left(\frac{1}{y}+\frac{1}{ 4y^2}\right)^{1/2}\,,
\end{equation}
where 
\begin{equation}
    y\,\equiv \,\frac{t^*\,\Lambda_{\rm net}(\rho, u_h, Z)} { \rho [\beta u_{SN} - (1 - \beta)u_c]}\,.
\end{equation}
The metallicity entering as an argument of the cooling function is computed according to the model of stellar evolution described in Sect. \ref{sec:StEv}. 

So far, we have introduced two parameters, $\beta$ and $u_{SN}$, that depend on the IMF, and two parameters, $A_0$ and $t_0^*$, that can be determined phenomenologically. The fifth and last parameter of the model is the density threshold for star formation, $\rho_{\rm thr}$. In the original formulation of SH03, which was developed by assuming no contribution from metals to the cooling function, this threshold was determined by requiring that the normalisation of the effective equation of state matches the limiting temperature, $\simeq 10^4$ K, at which gas can cool at high densities (e.g., Fig. 1 of SH03). This translates into the condition $u_{\rm eff}(\rho_{\rm thr})=u_4$, where $u_4$ is the specific energy corresponding to $10^4$ K. This fixes the threshold to the value
\begin{equation}
    \rho_{\rm thr}\,=\,\frac{x_{\rm thr}}{ (1-x_{\rm thr})^2}\,\frac{\rho^2\left[\beta u_{SN}-(1-\beta)u_c\right]}{ t_0^*\Lambda_{\rm net}(u_{SN}/A_0)}\,,
    \label{eq:rhot}
\end{equation}
where 
\begin{equation}
    x_{\rm thr}=1 + (A_0 + 1)\frac{u_c- u_4}{ u_{SN}}
    \label{eq:xthr}
\end{equation}
is the cold mass fraction at the density threshold, with $\Lambda_{\rm net}(\rho,u)/\rho^2$ losing its dependence on density at temperatures well above $10^4$ K. 

\og extends this model to include the effect of metal enrichment on the cooling function, also self-consistently computing the value of $\beta$ and $u_{SN}$ for each choice of the IMF and of the mass range for the massive stars that instantaneously die as Type-II SNe (see Sect. \ref{sec:StEv}, below).

As for the computation of $\rho_{\rm thr}$, we implement two possibilities depending on whether the code is configured to follow the stellar evolution model of Sect. \ref{sec:StEv}:
\begin{itemize}
    \item If stellar evolution and chemical enrichment are not enabled, the code carries out the original computation of $\rho_{\rm thr}$ by SH03 (see Eq. \eqref{eq:rhot} above), based on assuming a Salpeter IMF and using the metal-free cooling function described in Sect. \ref{sec:Katz};
    \item If stellar evolution and chemical enrichment are enabled, $\rho_{\rm thr}$ becomes a parameter to be specified, which holds for all the gas particles, independently of their metallicity. 
\end{itemize}
At this stage, \og does not include a fully self-consistent computation of a density threshold on a particle-by-particle basis, according to the abundances of the different elements followed by the chemical enrichment model.

In its numerical implementation, a multi-phase star-forming gas particle can spawn collisionless stellar particles according to a stochastic criterion. Let $\dot{M}_*= (1-\beta) x m / t_* $ be the star formation rate of a gas particle having cold gas fraction $x$ and mass $m$. In a given time step $\Delta t$, a stellar particle is spawned by this gas particle if a random number generated in the interval $[0,1]$ falls below the probability
\begin{equation}
    p=\frac{m}{M_*}\,\left\{1-\exp \left[-\frac{(1-\beta)x\Delta t }{ t_*} \right]    \right\}\,.
\end{equation}
Here ${M}_*=m_0/N_g$ is the mass of stellar particles generated by an SPH particle of initial mass $m_0$, while $N_g$ is the number of generations of such stellar particles that can be generated from a single SPH particle. We note that in the original SH03 implementation, all SPH particles have the same $m_0$. This is no longer true when including the model of stellar evolution of Sect. \ref{sec:StEv}. In this case, SPH particles can increase their mass whenever they receive stellar winds from a neighbouring star-forming particle. An example of how the effective equation of state shapes the phase diagram beyond the cooling branch is shown in figure \ref{fig:sfr_sh}, which shows the results from a small test simulation based on one {\it Box4/hr} from the {\it Magneticum} volumes.

\subsubsection{Winds}

In the original SH03 implementation, outflows driven by Type-II SNe are included to prevent an excess of star formation in the presence of only thermal feedback. In that implementation, which is also included in \og, star-forming particles can be uploaded to a wind with a velocity and a probability that depend on its SFR. Defining $\epsilon_{SN}\dot M_*$ as the amount of energy promptly made available in a star-forming particle by Type-II SN, and defining $\chi$ as the fraction of this energy carried by winds,  the wind velocity is
\begin{equation}
    v_w=\left[ \frac{2\beta\chi u_{SN}}{ \eta (1-\beta)} \right]^{1/2}\,.
    \label{eq:vw}
\end{equation}
Here $\eta =\dot M_w/\dot M_*$ is defined as a mass upload parameter. In the original SH implementation, $\chi$ and $\eta$ are treated as two parameters of the model, with typical phenomenological values $\chi \lesssim 1$ and $\eta \sim 2$, which produce typical wind velocities $v_w \simeq [350-500]$ km s$^{-1}$.

In \og we also allow fixing $v_w$ and $\eta$ as parameters of the outflow model, while $\chi$ is computed by inverting Eq. (\ref{eq:vw}), with $\beta$ and $u_{SN}$ self-consistently computed for the assumed stellar IMF and range of stellar masses assumed to be short-lived (see Sect. \ref{sec:StEv} below).

\begin{figure*}[th]
    \centering
	\includegraphics[width=0.4\textwidth]{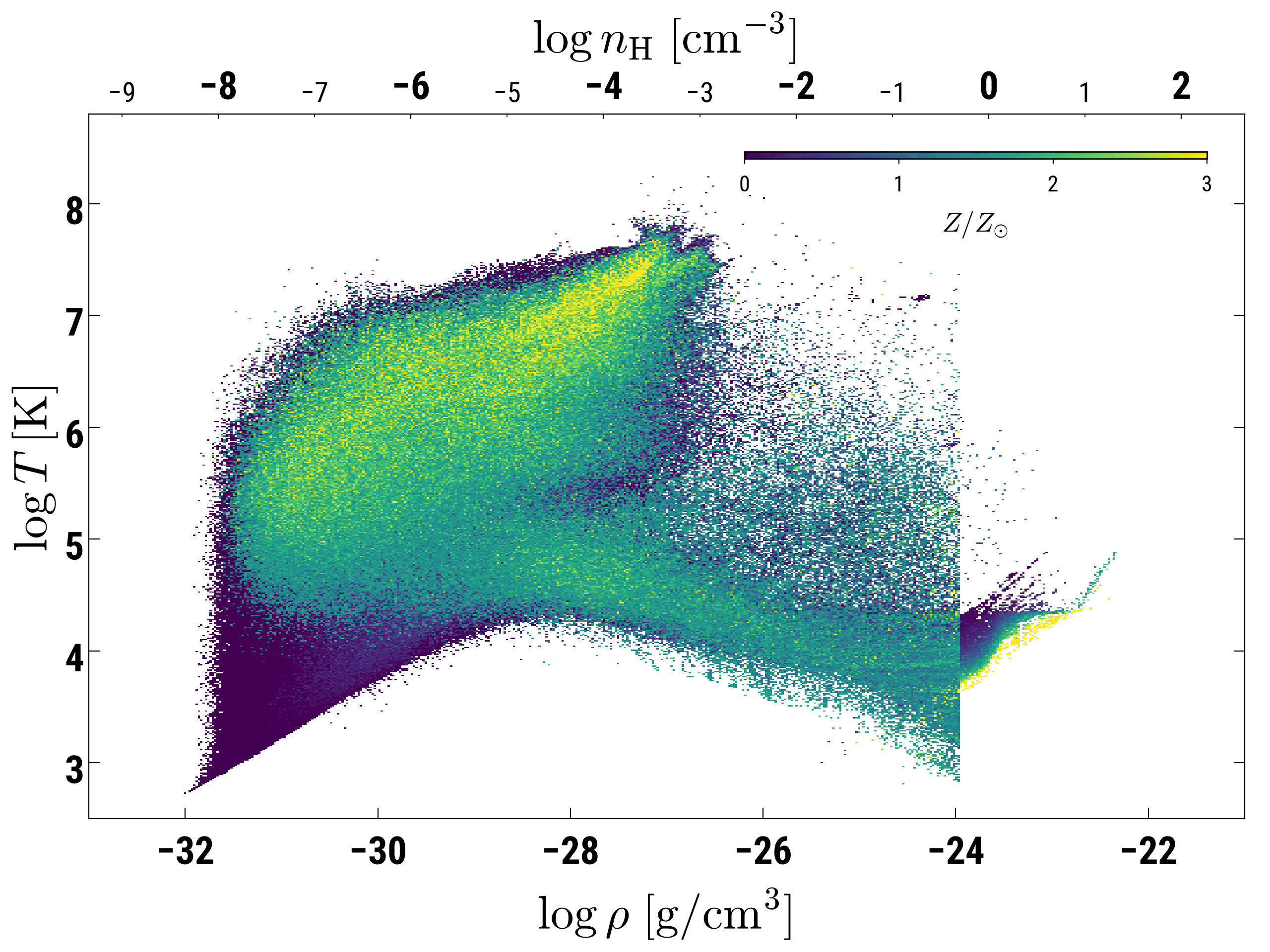}
	\includegraphics[width=0.59\textwidth]{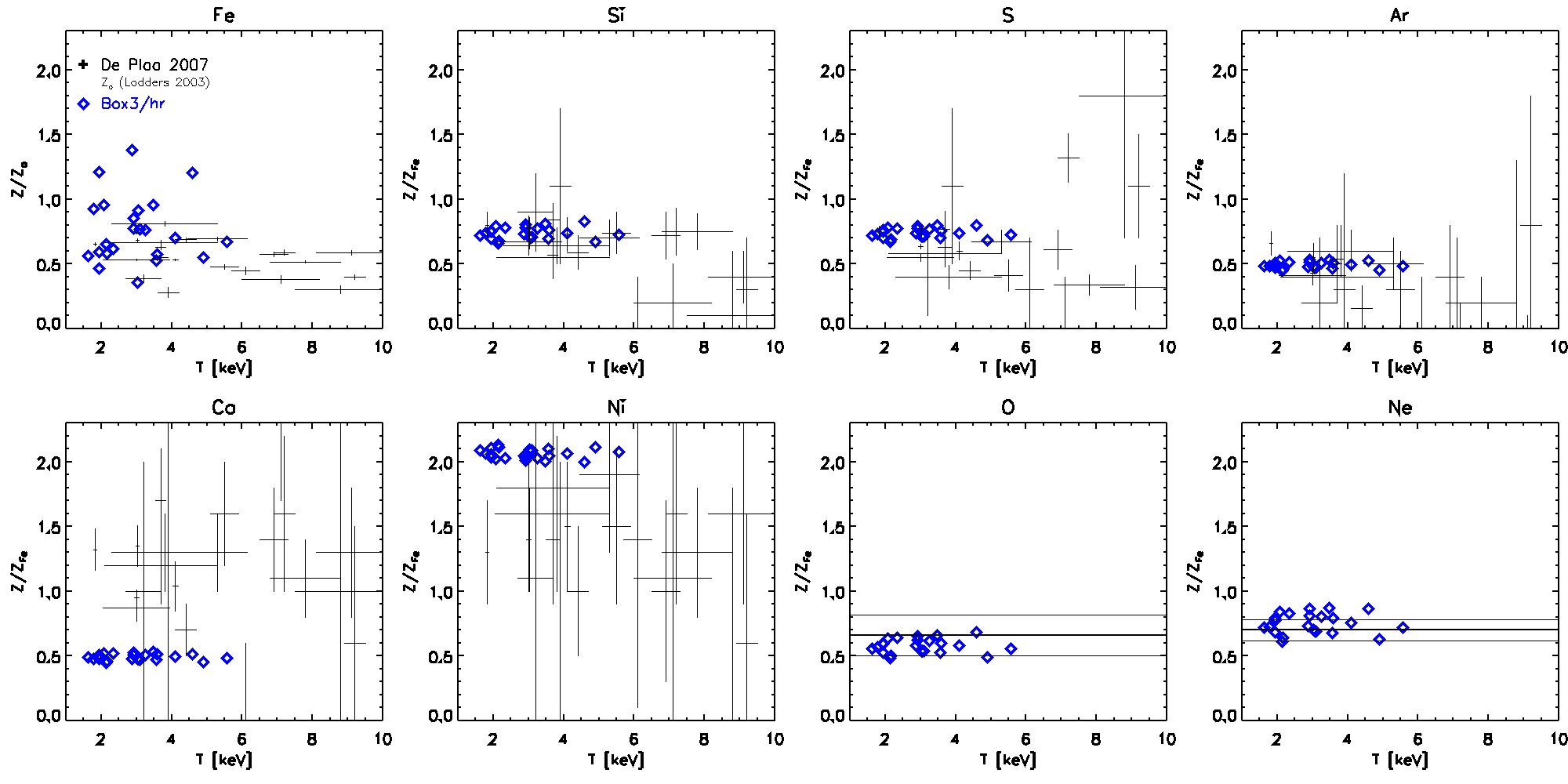}
    \caption{Cosmological box performed with the stellar evolution on. Shown are a phase diagram, colour-coded by the metallicity (left panel) and a comparison of the mean metal content for some of the different metal species tracked compared to observations \citep{2007A&A...465..345D}. For details, see \citet{Tornatore2007}, \citet{2017Galax...5...35D} and \citet{Biffi2018}.}
    \label{fig:stellar_evolution}
\end{figure*}

\section{Stellar evolution and chemical enrichment}
\label{sec:StEv}
\noindent
{\it Main contributing developer: L. Tornatore}

The \og code features a detailed model of stellar evolution and chemical enrichment, based on the original implementation by \cite{Tornatore2004, Tornatore2007} (see also \citealt{Borgani.etal.2008}). In this section we provide a basic description of this model, and we refer to the textbook by \cite{Matteucci2012} for further details.

In cosmological simulations, star particles have a mass far larger than that of a single star, with values of the order of $10^3$--$10^7$~M$_\odot$, depending on the resolution achievable (e.g.,   \citealt{Katz.etal.1996}). As a consequence, each star particle must be treated as a simple stellar population (SSP), i.e. as an ensemble of coeval stars having the same initial metallicity. Every star particle carries all the physical information (e.g. birth time $t_{\mathrm b}$, initial metallicity and mass) that is needed to calculate the evolution of the stellar populations that they represent, once the lifetime function (see Sect.~\ref{lifet}), the IMF (see Sect.~\ref{imf}) and the stellar yields (see Sect.~\ref{yields}) are specified. 

In summary, the main ingredients that define a model of chemical evolution are the following: {\em (a)} the SNe explosion rates, {\em (b)} the adopted lifetime function, {\em (c)} the adopted yields and {\em (d)} the IMF, which fixes the number of stars of a given mass. We describe each of these ingredients in the following.

\subsection{The equations of chemical evolution}

\subsubsection{Type Ia supernovae}

Following  \citet{GreggioRenzini1983}, we assume that SN\, Ia arise from stars belonging to binary systems, having
mass in a given range. Accordingly, in the single--degenerate scenario \citep{Nomoto.etal.2000}, the rate of explosions of SN\,Ia is
\begin{equation}
R_{{\mathrm{SN\,Ia}}}(t) \, = \,
A\,\int\limits_{\displaystyle{M_{\mathrm{B,inf}}}}^{\displaystyle{M_{\mathrm{B,sup}}}}
\phi(m_{\mathrm B})
\int\limits_{\displaystyle{\mu_{\mathrm m}}}^{\displaystyle{\mu_{\mathrm M}}} 
f(\mu)\,
\psi(t-\tau_{m_2})\,{\mathrm d}\mu\,{\mathrm d}m_{\mathrm B}\,. 
\label{eq:snia_rate}
\end{equation}
Here $\phi(m)$ is the IMF, which is defined as the number of stars per unit log-interval of mass, for which we use the following expression.
\begin{equation}
\label{eq:imf}
\phi(m)\,=\,dN/d\log m \propto m^{-x(m)}\,. 
\end{equation}
where the power-law slope can in general be dependent on the mass $m$. Furthermore, $m_{\mathrm B}=m_1+m_2$ is the total mass of the binary system (with $m_1$ and $m_2$ the mass of the primary and secondary companion, respectively). If we define $M_{\mathrm{Bm}}$ and $M_{\mathrm{BM}}$ to be the smallest and the
largest value allowed for the progenitor binary mass $m_{\mathrm B}$, respectively, then the
integral over $m_{\mathrm B}$ in Eq.(\ref{eq:snia_rate}) runs in the range between $M_{\mathrm{B,inf}}$ and
$M_{\mathrm{B,sup}}$, which represent the minimum and the maximum value of the
total mass of the binary system that is allowed to explode at the time
$t$. These values in general are functions of $M_{\mathrm{Bm}}$, $M_{\mathrm{BM}}$, and
$m_2(t)$, where $m_2(t)$ is the mass of the companion which dies at the time $t$. The exact dependence is defined by the SN\, Ia progenitor model. For instance, in the model by \citet{GreggioRenzini1983}, that is implemented in \og, it is $M_{\mathrm{B,inf}} = \max[2m_2(t),
M_{\mathrm{Bm}}]$ and $M_{\mathrm{B,sup}} = 0.5 M_{\mathrm{BM}} + m_2(t)$. 

As for the variable $\mu=m_2/m_{\mathrm B}$, it is distributed according to the function $f(\mu)$, which is assumed to have the expression
\begin{equation}
    \label{eq:fmu}
    f(\mu)=2^{1+\gamma}(1+\gamma)\mu^\gamma\,,
\end{equation}
with $\gamma=2$, as derived from statistical studies of the stellar population in the solar neighbourhood \citep{Tutukov1980, MatteucciRecchi2001}. The quantity $A$ appearing in Eq. \eqref{eq:snia_rate} is the fraction of stars in binary systems of that particular type that are progenitors of SN\, Ia (see \citealt{MatteucciRecchi2001} for more details). For instance, in the model by \citet{GreggioRenzini1983} $\mu$ varies in the range between $\mu_{\mathrm m}$ and $\mu_{\mathrm M}=0.5$, with $\mu_{\mathrm m}=\max\left[m_2(t)/m_{\mathrm B},(m_{\mathrm B}-0.5M_{\mathrm{BM}})/m_{\mathrm B}\right]$.

Finally, the function $\psi(t)$ in Eq.(\ref{eq:snia_rate}) is the star formation rate, with $\tau_m$ the mass--dependent life--time function. Since each star particle describes an SSP, with an instantaneous episode of star formation taking place at the time the star particle is spawned by its parent gas particle, the function $\psi(t)$ becomes a Dirac $\delta$--function. Therefore, using the functional form of $f(\mu)$ of Eq.(\ref{eq:fmu}), the expression for the SN-Ia rate of Eq.(\ref{eq:snia_rate}) can be written as 
\begin{equation}
R_{\mathrm{SN\,Ia}}(t) = -{\frac{\mathrm d}m_2(t)}{ {\mathrm d}t}\bigg|_{m_2\equiv
  \tau^{-1}(t)}
\!\!24\,m_2^2\,A\int_{M_{\mathrm{Bm}}}^{M_{\mathrm{BM}}}\phi(m_{\mathrm B})\frac{1}{m_{\mathrm B}^3}{\mathrm d}m_{\mathrm B}\,.
\label{eq:GR83snia_rate}
\end{equation}
The choice of the parameter $A$ can be phenomenologically derived from the requirement of reproducing a specific observation, once the form of the IMF is fixed. For instance,  \citet{MatteucciGibson1995} found that using $A=0.1$ one can predict realistic values for SNIa rates in galaxy clusters using a \cite{Salpeter1955} IMF. On the other hand, a lower value, $A=0.03$, has been suggested by \cite{Valentini2019} on the scale of galaxies. 

\subsubsection{Supernova Type II and intermediate/low mass stars} 

Computing the rates of SN\,II and ILMS is simpler, since they are determined only by the lifetime function $\tau(m)$, by the star formation history $\psi(t)$ and multiplied by the IMF $\phi(m=\tau^{-1}(t))$. Since $\psi(t)$ is a delta-function for the SSPs, the SN\,II and ILMS rates read 
\begin{equation}
R_{{\mathrm SN\,II}|{\mathrm ILMS}}(t)=\phi(m(t)) \times \left( -\frac{{\mathrm d}\,m(t)}{{\mathrm d}\, t}\right)\,.
\label{eq:my_snii_rate}
\end{equation}
Here $m(t)$ is the mass of the star that dies at time $t$.  The above expression must be multiplied by $(1-A)$
for AGB rates for values of $m(t)$ in the same range of masses, which is relevant for the secondary stars in SN\, Ia binary systems.

\subsection{The lifetime function}
\label{lifet}

Different choices for the mass--dependence of the lifetime function
have been proposed in the literature. \og implements by definition the expression originally proposed by \citet{PadovaniMatteucci1993} PM93; see their Eqs. 3 and 4), while other choices can easily be implemented. Different lifetimes produce different evolution of both absolute and relative abundances (we refer to \citealt{Romano.etal.2005} for a detailed discussion). For instance, we point out that the PM93  functions are independent of metallicity, whereas in principle this dependence can be included in a model of chemical evolution. 

\subsection{Production of metals and stellar yields}
\label{yields}
To compute the metal release by stars (binary systems in case of SN\, Ia) of a given mass, we
need to take into account the yields $p_{Z_i}(m, Z)$, which provide the mass of the element $i$ produced by a star of mass $m$ and {\em initial}
metallicity $Z$. A number of
different sets of yields have been proposed in the literature \citep[e.g.][and references therein]{Valentini2019}.

Once the yields are specified as input tables, \og computes the evolution of the mass $\rho_i(t)$ of element $i$ following Eq. 8 by \cite{Tornatore2007}. In this computation, \og explicitly accounts for the metallicity sink due to the locking of metals in the newborn stars, the metal ejection contributed by Type Ia and II SN, and the enrichment by mass loss from ILMS.

We point out that differences between different sets of yields represent one of the main uncertainties in any modelling of chemical evolution of cosmic structures. An example of how the resulting chemical pattern in cosmological volumes and galaxy clusters is shown in figure \ref{fig:stellar_evolution}, which shows the results based on a simulation where the stellar evolution is switched on\footnote{To obtain realistic cluster and ICM properties, also the black hole treatment was switched on, see next section for details.}, using {\it Box3/hr} from the {\it Magneticum} volumes.

\subsection{The initial mass function}
\label{imf}
The initial mass function (IMF) directly determines the relative ratio between the metals produced by AGB, SN\, II and SN\, Ia and, therefore, the relative
abundance of $\alpha$--elements and Fe--peak elements. The shape of the IMF also determines how many long-lived stars will form with respect to massive short-lived stars. In turn, this ratio affects the amount of energy released by SNe and the luminosity of galaxies, which is dominated by low-mass stars, and the (metal) mass-locking in the stellar phase. 

The IMF $\phi(m)$ is defined as the number of stars of a given mass per unit logarithmic mass interval. As of today, no general consensus has been reached on whether the IMF at a given time is universal or dependent on the environment, e.g. on the local values of temperature, pressure and metallicity in star--forming regions. The most
famous and widely used single power--law IMF is the 
\citet{Salpeter1955} one that has $x=1.35$ in Eq. \eqref{eq:imf}.  
\citet{Arimoto1987} proposed an
IMF with $x=0.95$, which predicts a relatively larger number of massive stars. Different expressions of the IMF have been proposed in order to model a flattening in the low-mass regime that is currently favoured by several observations.
\citet{Kroupa2001} and \citet{Chabrier2003} proposed expressions of the IMF that have different shapes in different stellar mass ranges. Theoretical arguments (e.g. \citealt{Larson1998}) suggest that the present--day characteristic mass scale, where the IMF changes its slope, $\sim 1$~M$_\odot$, should have been larger in the past, so that the IMF at higher redshift was top--heavier than at present. For instance, a variable IMF has been also advocated to reconcile observational evidences for high-redshift massive galaxies with prediction of cosmological models of galaxy formation \citep{Fontanot.etal.2026}.

In the \og code, the IMF is assumed to be universal and described in an input file, where one specifies the number of slopes that describe the IMF and the corresponding mass range of validity of each such slope. 

For a list of all available configuration and parameter options, see the according \href{https://gitlab.lrz.de/AstroCodes/OpenGadget3/-/wikis/Star-Formation-and-evolution}{section on the code wiki}.


\section{Formation and evolution of dust}
In numerical astrophysics, dust modelling generally branches into two paradigms: passive tracer formulations and fully dynamical treatments. Passive tracers treat the dust phase either as a scalar fluid field or as Lagrangian tracer particles that are advected strictly alongside the gas velocity field ($\vec{v}_\text{d} = \vec{v}_\text{g}$). This approach neglects relative aerodynamic drift and the physical back-reaction of dust onto the gas, assuming an instantaneous and perfect coupling limit where the stopping time approaches zero ($t_\mathrm{s} \to 0$). Having instead the advantage of being computationally inexpensive, the passive tracer approach is widely favoured in large-scale cosmological simulations or models of cosmic dust enrichment during galaxy evolution, where the grain population is dominated by sub-micron particles tightly locked to the diffuse interstellar medium (ISM).

A fully dynamical treatment becomes necessary whenever the aerodynamic stopping time is comparable to or exceeds the local dynamical timescale, causing grains to decouple from the gas streamlines. This regime is reached in supersonic turbulence in molecular clouds, where grains cluster into filamentary structures partially decoupled from the gas \citep{Hopkins_2016}; in protoplanetary disks, where the pressure-less dust phase drifts relative to the pressure-supported gas, driving vertical settling, radial drift, and the streaming instability \citep{Weidenschilling_1977, Youdin_2005, Johansen_2007, Birnstiel_2024}; and in radiation-driven outflows around AGB stars and AGN, where radiatively accelerated grains transfer momentum back to the gas via drag \citep{Hofner_2018, Ishibashi_2015}.

By implementing both a passive tracer formulation and a multi-species, two-way coupled dynamical framework featuring explicit coagulation and fragmentation, \og is uniquely equipped to self-consistently bridge these varied astrophysical regimes, tracking the dust from diffuse interstellar scales down to the high-density limits of star and planet formation.

\subsection{Dust production and evolution}
\label{s:dust1}
\noindent
{\it Main contributing developers: G.L. Granato and C. Ragone-Figueroa}

In the formulation where dust is treated as a passive tracer, the treatment of dust formation and evolution in \og is based on the framework presented in \citet{Granato2021}, an updated version of that previously used in \citet{Gjergo2018}. We refer the reader to those works for more details and justifications of the model assumptions.

The dust grain life cycle begins with dust creation, where only large grains are produced in the stellar ejecta of AGB stars, SNII, and SNIa. 
Once released into the ISM, dust grains undergo various evolutionary processes, including accretion, sputtering, coagulation, and shattering. The model also accounts for the cooling of hot gas resulting from collisions between ions and grains.
We account for the chemical composition of dust grains by distinguishing between carbonaceous (C) and silicate dust. For silicate dust, we generally adopt an olivine composition, specifically $\rm{MgFeSiO_4}$ \citep[e.g.,][]{Draine2003}. However, the code allows one to specify a different proportion of the four silicate elements.
The grain size distribution is taken into account using the two-size approximation introduced by \citet{Hirashita2015}, which captures, with a fair level of accuracy, the dependencies of ISM processes on grain size and also provides hints on the evolution of the actual continuous distribution.
Thus, we follow the evolution of four dust grain classes: each gas particle carries a dust mass budget, divided into large and small C grains, and large and small silicate grains.

\subsubsection{Dust produced by AGB, SNe II and Ia}
AGB winds, SNIa, and SNII inject some of the elements into the ISM as solid dust particles rather than as gas.
The relevant elements are C, Si, O, Mg, and Fe.

{\bf AGB winds.}
As proposed by \citet{dwek98}, \og assumes that the formation of carbon and silicate dust in AGB winds is mutually exclusive and determined by the C/O number ratio in the ejecta. Consequently, the maximum possible amount of CO is formed because of the high chemical affinity between carbon and oxygen. If $\text{C/O} > 1$, all oxygen is consumed to produce CO molecules, leaving only excess carbon to condense into dust. In contrast, if $\text{C/O} < 1$, all carbon binds to form CO molecules. The remaining oxygen, although not entirely consumed, becomes potentially available to condense into silicate grains, along with Mg, Si, and Fe.

We denote the oxygen and carbon masses ejected by the AGB winds of a star particle during a time step as $M_{ej, O}^{AGB}$ and $M_{ej, C}^{AGB}$, respectively. If $M_{ej, C}^{AGB} > 0.75 M_{ej ,O}^{AGB}$ (where 0.75 represents the ratio of the atomic weights of carbon to oxygen), the particle forms carbon grains. Therefore, by subtracting the carbon mass utilised in forming CO molecules, the resulting mass of ejected carbon dust is given by:   
\begin{equation}  
M_{dust,C}^{AGB} =  
\max\left[\delta_{AGB,C}\left( M_{ej,C}^{AGB} - 0.75 \, M_{ej,O}^{AGB} \right), 0 \right]\,,
\end{equation}  
where $\delta_{AG B,C}$ is the condensation efficiency of carbon grains in AGB winds. \og adopts $\delta_{AGB,C}=1$, consistent with \citet{dwek98}. 

When instead $M_{ej,C}^{\mathrm{AGB}} < 0.75\,M_{ej,O}^{\mathrm{AGB}}$, silicate grains are formed.  \citet{dwek98} estimated the mass of metals incorporated into silicates by assuming that for each ejected Si, Mg, and Fe atom, one O atom is also locked into dust. When implemented in \og, this prescription produced “silicate grains” with highly variable mass ratios among the four constituent elements and, in particular, an unreasonable oxygen fraction. We therefore generally adopt a different scheme that preserves a given elemental mass partition. The one we usually adopt is that of MgFeSiO$_4$. The procedure assumes that the number of molecules (or groups) of the chosen compound that can form is limited by the availability of a single element, and precisely the one for which, in the stellar ejecta, the number abundance divided by the number of atoms required by the compound is minimal. Following \citet{zhukovska08}, we refer to this limiting species as the \emph{key element}.

Let $M_{ej,X}^{\mathrm{AGB}}$ denote the mass of element $X$ ejected by AGB winds from a star particle over a time step.
For olivine, $X \in \{\mathrm{Mg},\mathrm{Fe},\mathrm{Si},\mathrm{O}\}$. 
Define $N_{\mathrm{mol},\mathrm{sil}}^{\mathrm{AGB}}$ as the number of MgFeSiO$_4$ “molecules” that can form during the time step. 
This is limited by the element for which the ratio between the number of ejected atoms and the stoichiometric requirement $N_{\mathrm{ato}}^X$ (1 for Mg, Fe, Si; 4 for O for our standard adopted composition) is minimal.
Then
\begin{equation}
\label{eq:nmol}
N_{\mathrm{mol},\mathrm{sil}}^{\mathrm{AGB}}
= \delta_{\mathrm{AGB},\mathrm{sil}}\,
\min_{X \in \{\mathrm{Mg},\mathrm{Fe},\mathrm{Si},\mathrm{O}\}}
\left( \frac{M_{ej,X}^{\mathrm{AGB}}}{\mu_X\,N_{\mathrm{ato}}^X} \right),
\end{equation}
where $\mu_X$ is the atomic weight of element $X$, and $\delta_{\mathrm{AGB},\mathrm{sil}}$ is the silicate condensation efficiency (set to 1 in our reference model).
The mass of element $X$ locked into silicate grains is therefore
\begin{equation}
\label{eq:mdust}
M_{\mathrm{dust},X}^{\mathrm{AGB}} =
\begin{cases}
N_{\mathrm{mol},\mathrm{sil}}^{\mathrm{AGB}}\,\mu_X\,N_{\mathrm{ato}}^X, & \text{if } \dfrac{M_{ej,C}^{\mathrm{AGB}}}{M_{ej,O}^{\mathrm{AGB}}} < 0.75,\\[6pt]
0, & \text{otherwise.}
\end{cases}
\end{equation}

{\bf SNe II and Ia.}
In supernova-driven outflows, carbonaceous and silicate dust can condense \emph{simultaneously}, because the ejecta are mixed only on macroscopic scales \citep[e.g.,][]{dwek98}. Thus,
\begin{align}
M_{\mathrm{dust},C}^{\mathrm{SNx}} \; &=\;
\delta_{\mathrm{SNx}}\, M_{ej,C}^{\mathrm{SNx}},\\
M_{\mathrm{dust},X}^{\mathrm{SNx}} \; &=\;
N_{\mathrm{mol},\mathrm{sil}}^{\mathrm{SNx}}\, \mu_X\, N_{\mathrm{ato}}^X,\\
N_{\mathrm{mol},\mathrm{sil}}^{\mathrm{SNx}} \; &=\;
\delta_{\mathrm{SNx},\mathrm{sil}}\;
\min_{X \in \{\mathrm{Mg},\mathrm{Fe},\mathrm{Si},\mathrm{O}\}}
\left(\frac{M_{ej,X}^{\mathrm{SNx}}}{\mu_X\, N_{\mathrm{ato}}^X}\right),
\end{align}
where $\mathrm{SNx}\in\{\mathrm{SNII},\mathrm{SNIa}\}$ and $M_{ej, X}^{\mathrm{SNx}}$ is the mass of element $X$ ejected by the corresponding supernovae from a star particle during a time step.

\subsubsection{Shattering}
Shattering is one of the two possible opposite results of grain–grain collision, the other being coagulation, described in the next section.

In the diffuse ISM, large grains decouple from small-scale turbulent motions \citep{hirashita09}. As a result, mutual collisions occur at sufficiently high velocities ($v \simeq 10~\mbox{km s}^{-1}$; \citealt{yan04}) to induce shattering into small fragments.
Shattering increases the abundance of small grains without changing the total dust mass. The collision timescale sets the timescale
$\tau_{\rm coll}=(v\,\sigma\,n)^{-1}$,
where $v$, $\sigma$, and $n$ are the characteristic collision velocity, cross section, and number density of colliding particles, respectively. Following \citet{yan04}, \og adopts $v=10~\mbox{km s}^{-1}$ for $n_{\rm gas}<1~\mbox{cm}^{-3}$ and let $v$ decrease as $n_{\rm gas}^{-2/3}$ at higher densities, reaching $v=0.1~\mbox{km s}^{-1}$ at $n_{\rm gas}=10^3~\mbox{cm}^{-3}$. Above the latter density, we switch shattering off entirely, assuming instead that low-velocity collisions drive coagulation of small grains into large ones (Section~\ref{sec:acc_coa}).
Following Appendix~B of \citet{aoyama17}, this yields
\begin{equation}
\label{eq:taush}
\tau_{\rm sh}\!=\!
\begin{cases}
\!\tau_{\rm sh,0}\left(\dfrac{0.01}{D_L}\right)\!\left(\dfrac{1~\mbox{cm}^{-3}}{n_{\rm gas}}\right),
& \!\!\!\!\dfrac{n_{\rm gas}}{1\,\mbox{cm}^{-3}} < 1, \\[8pt]
\!\tau_{\rm sh,0}\left(\dfrac{0.01}{D_L}\right)\!\left(\dfrac{1~\mbox{cm}^{-3}}{n_{\rm gas}}\right)\!\left(\dfrac{n_{\rm gas}}{1~\mbox{cm}^{-3}}\right)^{\!2/3}\!\!,
& \!\!\!\!1\le \dfrac{n_{\rm gas}}{1\,\mbox{cm}^{-3}} \le 10^3,
\end{cases}
\end{equation}
where $\tau_{\rm sh,0}=5.41\times 10^7~\mbox{yr}$ assumes a grain radius of $0.1~\mu\mbox{m}$ and a material density of $3~\mbox{g cm}^{-3}$, and $D_L \equiv M_{\rm dust,L}/M_{\rm gas}$ is the dust-to-gas ratio for large grains.

For MP particles, we assume that shattering operates only in the cold phase and insert its density into Eq.~\ref{eq:taush}. Neglecting shattering in the hot phase is justified because it typically comprises $<1\%$ of the particle mass and its temperature, $\gtrsim 10^6$~K, rapidly destroys small grains (Section~\ref{sec:sputtering}).

\begin{table}
\centering
\begin{tabular}{lccccc}
\hline
&\multicolumn{5}{c}{{ELEMENT}}\\
       & C   & O    & Mg   & Si   & Fe   \\
\hline
$f_X$  & 1.0 & 0.37 & 0.14 & 0.16 & 0.32 \\
A$_X${[}$10^3$ yr{]} & 2.5 & 1.56 & 0.73 & 0.91 & 2.52\\
\hline
\end{tabular}
\caption{Mass fractions in the grains and normalization factors used to compute accretion timescale with Eq. \eqref{eq:tauacnum} for each element $X$ participating to grain composition. See text for assumptions.}
\label{tab:ax}
\end{table}

\subsubsection{Accretion and Coagulation}
\label{sec:acc_coa}
Accretion of gas-phase metals onto grains and grain–grain coagulation are efficient only in the densest portions of the cold ISM, $n_H \gtrsim 10^{2}\!-\!10^{3}\ \mathrm{cm^{-3}}$ \citep[e.g.][]{hirashita14}, where hydrogen is predominantly molecular. Such densities are not resolved in most galaxy formation simulations, so \og adopts a subgrid estimate of the local mass fraction of gas in this regime, $F_{\rm dense}$. 

We assume that accretion and coagulation operate only on the fraction of the cold phase of multi-phase star-forming (MP) particles whose number density exceeds 
$n_{\rm th,ac} = 10^{3}\ \mathrm{cm^{-3}}$. We dub this fraction 
$f_{\rm mol}$. High-resolution ($\lesssim 10$ pc) simulations \citep[e.g.][]{wada07,tasker09} show that the probability distribution function of ISM density is described by a log-normal function, characterised by a dispersion $\sigma \simeq 2- 3$,  and a number density normalisation parameter $n_0 \simeq 1.5 -2.5$:  
\begin{equation}
f_{\rm pd}(n) \, dn = \frac{1}{\sqrt{2\uppi} \sigma}\exp\left[ -\frac{\ln(n/n_0)^2 }{2\sigma^2}\right] d \ln n\,.
\label{eq:pdf}
\end{equation}
Therefore  the mass fraction $F(>n_{th})$ of ISM gas above a given density threshold $n_{\rm th}$ is  
\citep[][equation 19]{wada07}
\begin{equation}
F(>n_{\rm th}) = \frac{1}{2}\left[1 - \mathrm{erf}\left(\left[\ln\left(\frac{n_{\rm th}}{n_0}\right) - \sigma^2\right] \left(\sqrt[]{2} \sigma\right)^{-1}\right)\right]
\label{eq:fdense}
\end{equation}

Accordingly, we set $F_{\rm dense}=f_{\rm cold}\, f_{\rm mol}$,
with $f_{\rm cold}$ the cold-gas mass fraction of the SPH particle and $f_{\rm mol}=F(>n_{\rm th,ac})/F(>n_{\rm cold})$.

Being accretion onto pre-existing grains a surface process, within the two-size approximation it directly acts on the small-grain bin, but it indirectly boosts the large-grain mass by enhancing small-grain coagulation. Following \citet{hirashita11}, the mass-growth timescale for accretion of a generic element $X$ onto grains of radius $a$ is
\begin{equation}
\label{eq:tauac}
\tau_{\rm acc,X}
= \frac{a\, f_X\, s\, \mu_X}{3\, n\, Z_X\, \bar{\mu}\, S}
\left(\frac{2\uppi}{m_X k T}\right)^{1/2}
F_{\rm dense}^{-1},
\end{equation}
where $f_X$ is the mass fraction of $X$ in the grain material, $s$ the material density, $\mu_X$ the atomic weight, and $m_X$ the particle mass of element $X$; $n$, $T$, and $\bar{\mu}$ are the gas number density, temperature, and mean molecular weight, respectively; $Z_X$ is the gas-phase mass fraction of $X$; and $S$ is the sticking coefficient. 
The factor of 3 in the denominator converts the radius-growth timescale in \citet{hirashita11} to a mass-growth timescale (for spherical grains, $m/\dot m = a/(3\dot a)$). 
The $F_{\rm dense}$ factor reflects the assumption that accretion proceeds only within the densest molecular pockets of MP particles, as described above.

The previous expression can be recast into a convenient numerical form:
\begin{equation}
\label{eq:tauacnum}
\tau_{\rm acc,X} \;=\;
\frac{A_X\, a_{0.005}}{Z_X\, n_3\, T_{50}^{1/2}\, S_{0.3}\, F_{\rm dense}},
\end{equation}
where $a_{0.005}\!\equiv\!a/(0.005\,\mu{\rm m})$, $n_3\!\equiv\!n/(10^3\,{\rm cm^{-3}})$, $T_{50}\!\equiv\!T/(50\,{\rm K})$, and $S_{0.3}\!\equiv\!S/0.3$. 
The normalisation $A_X$ depends on the grain composition through $f_X$ and $\mu_X$: for pure C grains we take $f_X=1$, while for silicate-forming elements we assume the usual olivine-like composition MgFeSiO$_4$. 
Adopting material densities $s=3.3~{\rm g\,cm^{-3}}$ (silicates) and $2.2~{\rm g\,cm^{-3}}$ (carbonaceous), the resulting $A_X$ values are listed in Table~\ref{tab:ax}.

For silicates, we additionally enforce that accretion preserves the MgFeSiO$_4$ mass partition used for stellar dust production. 
Operationally, we assign the same accretion timescale to all four elements, equal to the \emph{slowest} among the $\tau_{\rm acc,X}$. 
This limiting species is the \emph{key element} of previous literature \citep[e.g.,][]{zhukovska08,hirashita11,asano13,hou19}. 
In our simulations, the key element is determined, at each time step and for each gas particle, from the current abundance pattern; it is often identified with Si \citep[e.g.][]{hirashita11,hou19}.

In the densest ISM, low-velocity collisions among small grains lead to \emph{coagulation} and the production of large grains. 
We adopt the \citet{aoyama17} timescale, modified to act only on the fraction $F_{\rm dense}$ of each SPH particle:
\begin{equation}
\label{eq:tauco}
\tau_{\rm co} \;=\;
\tau_{\rm co,0}\,
\left(\frac{0.01}{D_S}\right)
\left(\frac{0.1~{\rm km\,s^{-1}}}{v_{\rm co}}\right)
F_{\rm dense}^{-1},
\end{equation}
where $D_S \equiv M_{\rm dust,S}/M_{\rm gas}$ is the dust-to-gas ratio of small grains and $v_{\rm co}$ their velocity dispersion. 
We adopt $v_{\rm co}=0.2~{\rm km\,s^{-1}}$ based on \citet{yan04}. 
The normalization $\tau_{\rm co,0}=2.71\times10^5~{\rm yr}$ corresponds to a characteristic small-grain radius of $0.005~\mu{\rm m}$ and a material density of $3~{\rm g\,cm^{-3}}$.

\subsubsection{SNe destruction}
\label{sec:SNedestruction}
Shocks driven by supernovae affect dust grains through erosion caused by a combination of thermal and non-thermal sputtering. We label this channel \emph{SN destruction} and reserve the term \emph{sputtering} for the thermal process acting in diffuse hot gas (see next section). 
For SN destruction, we adopt the treatment of \citet{aoyama17}. 
We do not distinguish here between SN\,II and SN\,Ia; let $N_{\rm SN}$ be their total number within a time step $\Delta t$.

The associated effective timescale is
\begin{align}
\tau_{\rm SN} &= \frac{\Delta t}{1-\left(1-\eta\right)^{N_{\rm SN}}},\\
\eta &= \epsilon_{\rm SN}\,\min\!\left(\frac{m_{\rm SW}}{m_g},\,1\right),
\end{align}
where $m_g$ is the SPH particle gas mass, $m_{\rm SW}$ is the gas mass swept up per SN (typically $m_{\rm SW}\ll m_g$ in our runs), and $\epsilon_{\rm SN}$ is the grain–destruction efficiency in the shock. 
We set $\epsilon_{\rm SN}=0.1$ following \citet{aoyama17} and \citet{Gjergo2018}. 
The swept mass is estimated as in \citep{mckee89}
\begin{equation}
m_{\rm SW}= 6800~M_\odot
\left(\frac{E_{\rm SN}}{10^{51}\,{\rm erg}}\right)
\left(\frac{v_s}{100~{\rm km\,s^{-1}}}\right)^{-2},
\end{equation}
with $E_{\rm SN}$ the energy of a single event and $v_s$ the shock velocity. Following \cite{mckee87}, we set $v_s=200~{\rm km\,s^{-1}}$, neglecting its weak dependence of gas density. Numerically this gives $m_{\rm SW}\simeq1.7\times10^3~M_\odot$. Therefore,
\[
\eta=\epsilon_{\rm SN}\min\!\left(\frac{1.7\times10^3M_\odot}{m_g},\,1\right)\,.
\quad (m_g \gg m_{\rm SW}),
\]
We note that $m_{\rm SW}\ll m_g$ at the typical resolutions of cosmological galaxy formation simulations. 

\subsubsection{Thermal sputtering}
\label{sec:sputtering}
Grains embedded in hot plasma ($T_g \gtrsim 5\times10^5$~K) are efficiently eroded by ion impacts. We treat this \emph{thermal sputtering} using the grain–radius erosion timescale of \citet{tsai95}. From Eqs.~(14)–(15) of \citet{tsai95} one obtains the corresponding \emph{mass}–loss timescale:
\begin{equation}
\label{eq:tausp}
\tau_{\rm sp} \;=\; \tau_{\rm sp,0}\;
\left(\frac{a}{0.1~\mu{\rm m}}\right)
\left(\frac{\frac{0.01}{{\rm cm}^{-3}}}{n_g}\right)
\left[
\left(\frac{T_{\rm sp,0}}{\min\!\left(T_g,\,3\times10^7~{\rm K}\right)}\right)^{\!\omega} + 1
\right],
\end{equation}
with $T_{\rm sp,0}=2\times10^6$~K and $\omega=2.5$. 
This expression reproduces the calculated sputtering efficiencies for both carbonaceous and silicate grains up to $T_g \sim \mathrm{few}\times10^7$~K, beyond which the temperature dependence saturates. 
Here $n_g \equiv \rho/(\bar{\mu} m_p)$ counts ions \emph{and} electrons; adopting $\bar{\mu}=0.59$ for a fully ionized mixture of 75\% H and 25\% He gives the original normalization $\tau_{\rm sp,0}=5.5\times10^6$~yr. 

However, our cluster simulations in \citet{Gjergo2018}, and independently those of \citet{vogelsberger19}, both using this parametrisation, indicate that reproducing the relatively large dust masses inferred in clusters \citep{43planck16} requires a substantially \emph{longer} sputtering timescale (by factors of $\sim$5 and $\sim$10, respectively). 
Accordingly, we adopt $\tau_{\rm sp,0}=2.7\times10^7$~yr (i.e. five times larger). 
For the two-size model we take effective radii $a=0.05~\mu$m (large grains) and $a=0.005~\mu$m (small grains); the former corresponds to the mean radius of a $n(a)\propto a^{-3.5}$ size distribution from  0.03
 to 0.25 $\mu$m (e.g. \citealt{silva98}).

To give an impression of how the dust evolves in a typical cosmological volume, we also switched on the dust evolution in our showcase, based on {\it Box3/hr} from the {\it Mageticum} volumes. Figure \ref{fig:stellar_evolution} shows the phase diagram of the gas in the cosmological volume, but this time colour-coded by the ratio between dust and metals in the gas phase.

\begin{figure}[t]
    \centering
	\includegraphics[width=0.475\textwidth]{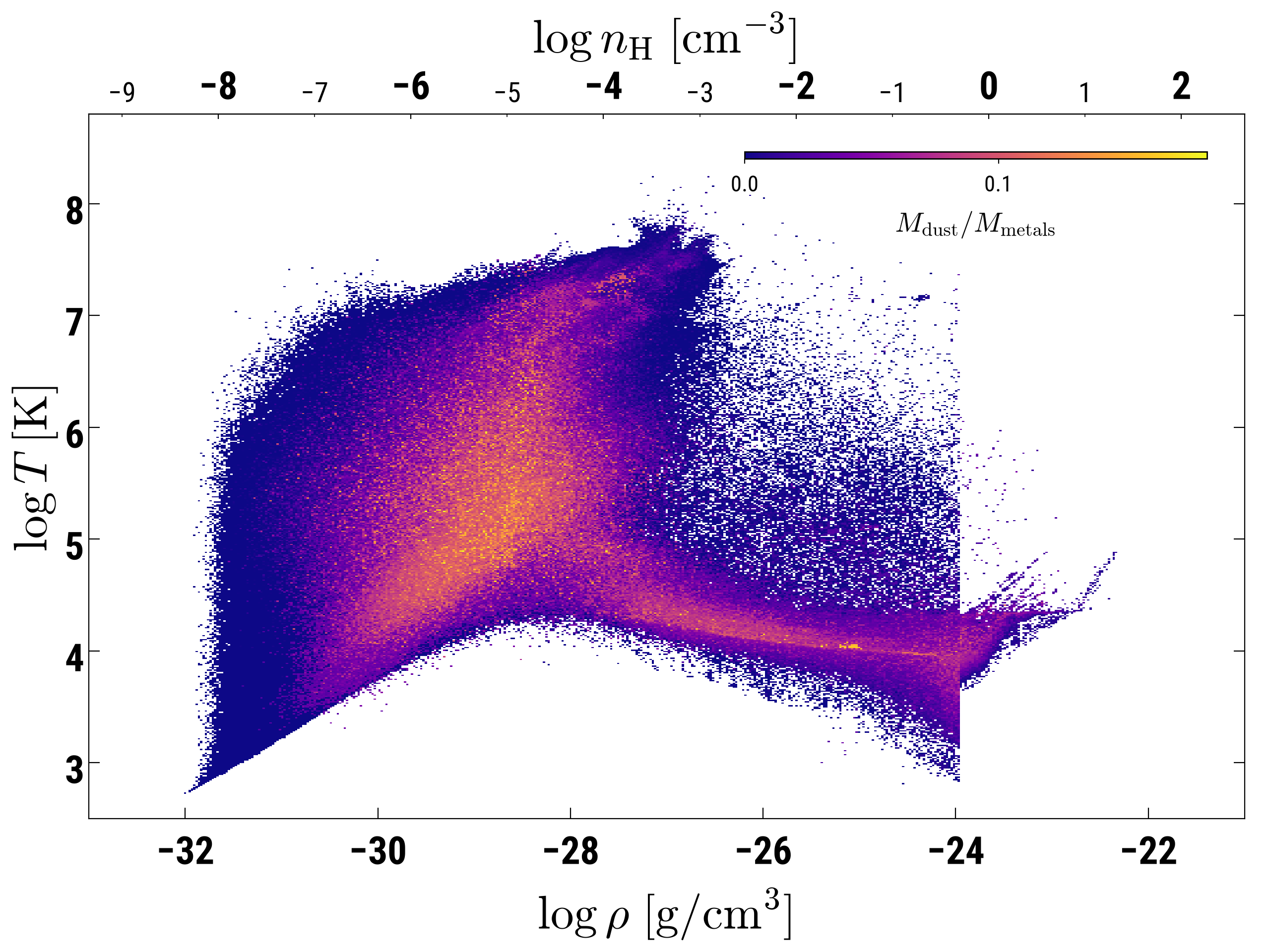}
    \caption{Cosmological box performed with stellar evolution and dust evolution. Shown is a phase diagram, colour-coded by the dust-to-metal ratio. For details, see \citet{Granato2021}.}
    \label{fig:dust_evol}
\end{figure}

\subsubsection{Gas cooling promoted by dust}
Beyond erosion by ion impacts, dust immersed in hot plasma is also \emph{collisionally heated}, predominantly by electrons. Ions at the same kinetic energy move more slowly and collide less frequently, so their contribution to heating is negligible \citep{montier04}. 
The deposited energy is reradiated efficiently in the infrared, yielding a net cooling term for the gas.

The implementation of dust cooling in \og relies on the calculations of \citet{DwekWerner1981}, with the sole difference that we track two grain-size populations. 
For a grain of radius $a$ in a thermal plasma with electron density $n_e$, the heating rate (in erg s$^{-1}$) reads:
\begin{equation}
\label{eq:duscooH}
H(a,T,n_e)=
\begin{cases}
5.38\times10^{-18}\, n_e\, a^2\, T^{1.5}, & x>4.5,\\[4pt]
3.37\times10^{-13}\, n_e\, a^{2.41}\, T^{0.88}, & 1.5 < x \le 4.5,\\[4pt]
6.48\times10^{-6}\, n_e\, a^3, & x \le 1.5,
\end{cases}
\end{equation}
where $x \equiv 2.71\times10^8\, a^{2/3}/T$. 
The associated contribution to the gas cooling function from grains of size $a$ and number density $n_d(a)$ is then
\begin{equation}
\label{eq:duscoo}
\frac{\Lambda_d(a)}{n_H^2}
= \frac{n_d(a)}{n_H^2}\, H(a,T,n_e).
\end{equation}

We evaluate the large- and small-grain components separately, adopting the same effective radii used for sputtering (Section~\ref{sec:sputtering}), namely $a=0.05~\mu$m and $a=0.005~\mu$m. 
For simplicity, when estimating $n_d(a)$ we take a representative grain material density of $3~\mathrm{g\,cm^{-3}}$ for both silicate and carbonaceous species.

\subsection{One fluid approach for dust}
\label{sec:dust_dynamics}
\noindent
{\it Main contributing developer: G. Tedeschi-Prades}

Within this second approach, dust dynamics is implemented in \og using the so-called One-Fluid model introduced in \citet{Laibe_2014a}. In this framework, rather than using additional particle types for dust, the simulation particles that usually represent gas now instead represent the mixture of gas and dust simultaneously. This seemingly counterintuitive choice brings a number of numerical advantages when simulating the dynamics of dust. As demonstrated by \cite{Laibe_2012b}, a multiple fluid method would require both spatial and temporal resolutions that become infinite in the limit of perfectly coupled dust grains. Since, often in astrophysical environments, we deal with dust grains tightly coupled to the gas, this limitation would be too restrictive for hydrodynamical codes. Additionally, this method automatically incorporates the back-reaction of dust onto the motion of gas, without the need for additional interpolated forces acting on the gas particles. 

The mixture particles, as defined in the One-Fluid model, have, as density, the sum of the gas and dust densities
\begin{equation}
\label{eq:OFM_Density}
    \rho = \rho_\text{g} + \sum_{i=1}^{N_\mathrm{d}} \rho_{\text{d},i}
\end{equation}
where the subscripts "g" and "d" indicate gas and dust, respectively, and $N_\text{d}$ is the total number of dust species in the simulation, each determined by its mass. Moreover, the mixture particles move with the velocity of the barycenter of the mixture:
\begin{equation}
    \vec{v} = \frac{1}{\rho}\left(\rho_\text{g}\vec{v}_\text{g} + \sum_{i=1}^{N_\mathrm{d}} \rho_{\text{d},i}\vec{v}_{\text{d},i}\right)
\end{equation}
Additionally, the One-Fluid model evolves two additional quantities, alongside density and velocity. The first is the dust fraction of each dust species
\begin{equation}
    \epsilon_i = \frac{\rho_{\text{d},i}}{\rho}
\end{equation}
and the second is the velocity difference between each dust species and the gas velocity:
\begin{equation}
\label{eq:OFM_DeltaVel}
    \vec{\Delta v}_i = \vec{v}_{\text{d},i} - \vec{v}_{\text{g}}
\end{equation}
These four quantities form a closed system that can be readily inverted to express the individual gas and dust quantities as functions of the One-Fluid model variables. This guarantees a one-to-one correspondence between the simulated system and the properties of the separated gas and dust.

Since the dust fraction is bounded between 0 and 1, a number of parametrisations have been introduced \citep[e.g.]{Ballabio_2018} to prevent it from overflowing. We choose the following parametrisation of the dust fraction
\begin{equation}
    s_i = \sqrt{\frac{\epsilon_i}{1 - \epsilon_i}}
\end{equation}
and evolve the variable $s_i$, so that the corresponding value of $\epsilon_i$ will remain bound between 0 and 1.

\begin{figure*}[ht!]
    \centering
    \includegraphics[width=1.0\linewidth]{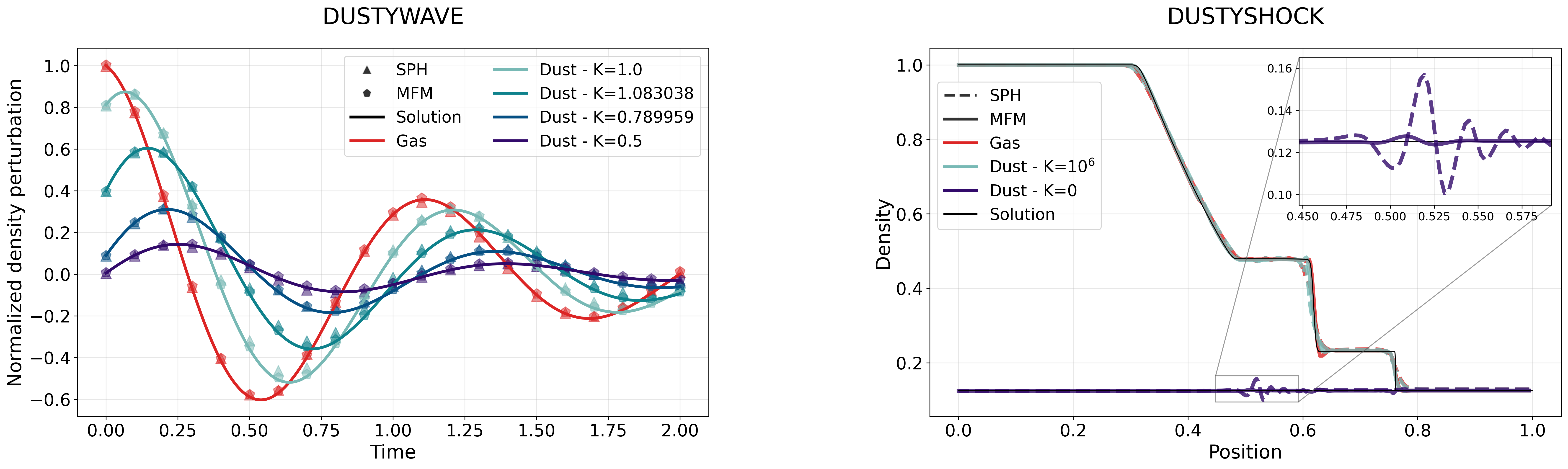}
    \caption{Standard multifluid dust dynamics tests for the SPH and MFM implementations, compared against analytical solutions. \textit{Left}: DUSTYWAVE test, showing the normalised density perturbation of gas (red) and various dust fluids (blue shades). The drag coefficients are selected from \citet{Benitez-Llambay2019}. Both the SPH (triangles) and MFM (pentagons) simulations closely follow the expected analytical solution. \textit{Right}: DUSTYSHOCK test, performed with gas (red) and two dust fluids (blue shades), one tightly coupled to the gas (K=$10^6$), and one fully uncoupled (K=0), in which the dust should remain unperturbed. The inset zooms on the location of the initial discontinuity: the MFM implementation (solid lines) is able to better keep the uncoupled dust fluid at rest, compared to the SPH one (dashed lines).}
    \label{fig:dust_tests}
\end{figure*}

\subsubsection{SPH implementation}
The SPH implementation of the One-Fluid model in \og is described in detail in \citet{Tedeschi-Prades_2025a}. Here we summarise the key elements; readers interested in the full derivation are referred to that paper.

\paragraph{Discretised equations}
Starting from the derivation of \citet{Laibe_2014b}, we derived the first complete set of SPH-discretized equations for the dust fraction, barycenter velocity, velocity difference, and internal energy, for a single dust fluid. The discretised equations conserve total mass, momentum, and energy by construction. 

\paragraph{Artificial viscosity}
In order to treat shocks and prevent particle interpenetration, our implementation includes both a conservative (C-AV) and a non-conservative (NC-AV) formulation of AV, following \citet{Laibe_2014b}. The C-AV formulation has been adapted to be compatible with the Time-Dependent AV described in Sect. \ref{sec:SPH:artvisc}. The NC-AV formulation performs better in shocked regions, while C-AV is preferable away from shocks. Rather than leaving this choice to the user, we introduce a unified framework in which the shock indicator (Eq. \eqref{eq:Cullen_shock_indicator}) is used to define a sigmoid interpolation coefficient $\sigma_{ab}$ that smoothly transitions between the two formulations in and out of shocked regions. In addition, a dissipative term acting on $\vec{\Delta v}$ damps post-shock oscillations in the velocity difference \citep{Laibe_2014b}.
 
\paragraph{Artificial dust diffusion}
Since dust is modelled as a pressureless fluid, it lacks a restoring force and tends to form clumps when the drag coefficient is sufficiently small. Severe clumping can break the fundamental SPH assumption of smooth fields across neighbouring particles, eventually causing code crashes. To stabilise simulations with large dust-to-gas ratios or weak coupling, we introduce a novel artificial diffusion model inspired by the ``pressure-like'' term proposed by \citet{Klahr_2021} and \citet{Binkert_2023}, which enters the momentum equation for dust rather than directly diffusing the dust fraction (which would require mass transfer between SPH particles). The diffusion coefficient is adaptive: it is proportional to the local excess of $\epsilon_a$ over the mean of its neighbours, so that diffusion is active only where large gradients develop while leaving smooth flows unaffected.

\subsubsection{MFM implementation}
The implementation of dust dynamics in the Meshless Finite Mass (MFM) framework of \og is described in \citet{Hutchison_2026}. We summarise the essential ideas here.

\paragraph{Conservative formulation and split Riemann solver}
The One-Fluid equations for a mixture of gas and $N_\mathrm{d}$ dust species can in principle be recast in fully conservative form and discretised as in Eq. \eqref{eq:MFM_discretization}, with suitable conserved variables vector $\mathbf{U}$ and fluxes $\mathbf{F}(\mathbf{U})$. This conservative structure would allow the system to be evolved directly within the MFM framework of \citet{Groth2023} described in Sect. \ref{sec:MFM}. A key complication is that the dust sub-system is only weakly hyperbolic: it admits multi-streaming solutions and $\delta$-shocks that standard Riemann solvers cannot handle. We therefore split the gas and dust components at the level of the Riemann solver. The gas component is evolved with any standard Riemann solver (including straightforward extensions to MHD), while the dust component uses the pressureless Riemann solver of \citet{Huang_2022},
\begin{equation}
    F_{ij} = \begin{cases}
        0,          & u_i < 0 \text{ and } u_j > 0, \\
        F_i,        & u_i > 0 \text{ and } u_j > 0, \\
        F_j,        & u_i < 0 \text{ and } u_j < 0, \\
        F_i + F_j,  & u_i > 0 \text{ and } u_j < 0,
    \end{cases}
\end{equation}
where $u$ is the dust velocity component along the inter-particle direction and $i$ and $j$ indicate the left and right states. This solver correctly captures dust concentration and $\delta$-shock formation without contaminating the gas solution.

\paragraph{Zero-mass-flux reference frame}
In MFM, inter-particle fluxes are computed in a reference frame chosen so that no net mass is exchanged between particles. In the single-fluid case, this frame is identified with the contact discontinuity. In the multi-fluid case, a single frame that simultaneously enforces zero mass flux for each component individually does not exist; moreover, doing so would prevent dust fractions from evolving across particles, undermining the purpose of the One-Fluid formalism. We therefore enforce zero \emph{total} mass flux for the mixture, seeking $v_\mathrm{ZMF}$ such that
\begin{equation}
    F_{\rho_\mathrm{g}} + \sum_k F_{\rho_{\mathrm{d},k}} = 0.
\end{equation}
This condition is found iteratively using a Newton--Raphson-like finite-difference scheme \citep[described in Appendix A of][]{Hutchison_2026}. Individual components are free to exchange mass across the interface, but total particle masses are strictly conserved, preserving the defining property of MFM.

\paragraph{Advantages over the SPH implementation}
A notable advantage of the MFM approach is that no AV is required for the dust component. In the SPH implementation, non-conservative, ad hoc dissipation terms are needed to keep the uncoupled ($K=0$) dust fluid stationary during a gas shock; the pressureless Riemann solver handles this case naturally.

\subsubsection{Aerodynamic drag implicit integration}
Regardless of whether the SPH or MFM framework is used, the aerodynamic drag term coupling gas and dust must be integrated implicitly. Explicit integration requires the time-step to be smaller than the shortest stopping time $t_\mathrm{s}$ in the simulation; for tightly coupled grains this constraint is far more restrictive than the hydrodynamical CFL condition and renders explicit integration impractical.

To integrate implicitly aerodynamic drag for multiple dust species, a matrix
inversion of cost $\mathcal{O}(N_\mathrm{d}^3)$ would normally be
required. \og instead employs the General Implicit Runge-Kutta (GIRK)
integrator derived in \citet{Tedeschi-Prades_2025b}, which builds on the
analytical solution of \citet{Krapp_2020} and the DIRK scheme of
\citet{Krapp_2024} to achieve $\mathcal{O}(N_\mathrm{d})$ complexity
while retaining second-order accuracy. GIRK extends the DIRK framework
by lifting the diagonal constraint and introducing additional free
parameters, which are used to simultaneously enforce:
\begin{itemize}
    \item second-order accuracy for $\Delta t \ll t_\mathrm{s}^{\max}$,
    \item second-order accuracy for $\Delta t \gg t_\mathrm{s}^{\max}$,
    \item second-order convergence to the drag equilibrium in the
          presence of external forces (a property that the DIRK+Strang scheme fails to satisfy).
\end{itemize}
Both the SPH and MFM drag integrators are incorporated into the global
Kick--Drift--Kick time-stepping of \og via Strang operator splitting,
\begin{equation}
    \mathbf{U}^{n+1} = \mathcal{D}_{\Delta t/2}\,\mathcal{H}_{\Delta t}\,\mathcal{D}_{\Delta t/2}\,\mathbf{U}^n,
\end{equation}
where $\mathcal{D}$ denotes the drag update and $\mathcal{H}$ the full
hydrodynamical step. The parameter values for the recommended splitting
scheme are given in Table~1 of \citet{Tedeschi-Prades_2025b}.\\
Figure \ref{fig:dust_tests} shows the performance of both the SPH and MFM implementations of dust dynamics against two standard tests: the DUSTYWAVE \citep{Laibe_2012b, Benitez-Llambay2019} and the DUSTYSHOCK \citep{Laibe_2012b, Hutchison_2026}. While both methods correctly reproduce both tests, we note that, in the DUSTYSHOCK test, MFM is slightly better than SPH in keeping the dust with $K=0$ drag at rest.

\subsubsection{Dust Coagulation and Fragmentation}
Along with the complete dynamics of the coupled dust-gas system, \og also evolves the grain size distribution of dust under grain-grain coagulation and fragmentation. The model implemented in \og follows closely the model described in \citet{Stammler_2022} for the \texttt{Python} package \texttt{DustPy}. This model solves the discretised version of the Smoluchowski equation:
\begin{equation}
    \partial_t n_k = \sum_{i = 1}^{N_\mathrm{d}}\sum_{j=1}^i K_{ijk} R_{ij} n_i n_j  - n_k \sum_{j=1}^{N_\mathrm{d}} n_j R_{ij} (1+\delta_{jk})
\end{equation}
where $K_{ijk}$ is the collision kernel, describing the collision outcomes for each pair of $i$ and $j$ mass bins, $R_{ij}$ are the sticking or fragmentation rates, and the $n_k$ are the discretised number densities integrated over a mass bin
\begin{equation}
    n_k = \int_{m_{k-1/2}}^{m_{k+1/2}} n(m) dm.
\end{equation}
The sticking and fragmentation rates depend on the associated probabilities $p_{ij}^{\mathrm{s/f}}$ and the relative velocities between two mass bins $v_{ij}^{\mathrm{rel}}$:
\begin{equation}
    R_{ij}^{\mathrm{s/f}} = \frac{1}{1+\delta_{ij}} \sigma_{ij}^{\mathrm{geo}} v_{ij}^{\mathrm{rel}} p_{ij}^{\mathrm{s/f}}
\end{equation}
where $\sigma_{ij}^{\mathrm{geo}} = \uppi (a_i^2 + a_j^2)$ is the geometrical cross section, with $a_i$ and $a_j$ the sizes of the dust particles.\\
The relative velocities can have both resolved and sub-grid sources. The fundamental resolved source is simply the drift velocity between two dust fluids:
\begin{equation}
    v_{ij}^{\mathrm{drift}} = \abs{\vec{v}_{\mathrm{d}, i} - \vec{v}_{\mathrm{d}, j}}
\end{equation}
Sub-grid sources are, for instance, the Brownian motion stirring up small grains, or turbulence acting at sub-grid scales, for instance, as modelled by \citet{Ormel_2007}. When building the collision kernel $K_{ijk}$, the model includes the effects of sticking, fragmentation, and erosion. When fragmentation happens, a fragment distribution is formed following a power-law distribution:
\begin{equation}
    n(m) dm = m^{\gamma} dm
\end{equation}
where $\gamma = -11/6$, as determined experimentally by \citet{Dohnanyi_1969}. Unlike full fragmentation, in which the two colliding particles are both destroyed, erosion reduces the size of the larger particle (the target) and removes the projectile. Erosion occurs when the mass ratio of the target to the projectile exceeds 10.\\

For a list of all available configuration and parameter options, see the according \href{https://gitlab.lrz.de/AstroCodes/OpenGadget3/-/wikis/dust}{section on the code wiki}.


\section{Black Hole evolution and AGN feedback}
\label{sec:BH}

The model for the evolution of BHs and the ensuing AGN feedback in its basic formulation is largely inspired to the original implementation by \cite{Springel_BHs} and \cite{Dimatteo.etal.2005}, with the later modifications introduced by \cite{Fabjan.etal.2010,Hirschmann.etal.2014,Steinborn2015,  Sala2024, Damiano2024}. In this section we describe the implementations of the different parts of the BH section in \og, namely BH seeding (Sect. \ref{sec:BHseed}), gas accretion (Sect. \ref{sec:BHaccr}), merging of a BH-BH pair (Sect. \ref{sec:BHmer}), tracking of the BH particles (Sect. \ref{sec:BHtrack}), release of energy feedback (Sect. \ref{sec:BHfeedb}) and evolution of BH spin (Sect. \ref{sec:BHSpin}). The meaning of all the relevant configuration switches and input parameters are described on the according \href{https://gitlab.lrz.de/AstroCodes/OpenGadget3/-/wikis/Black\%20Holes}{section of the code wiki}.

\subsection{BH seeding}
\label{sec:BHseed}

 \og allows for two different possibilities to seed the BH particles. The first option, based upon the original implementation of \citet{DiMatteo2008}, is to seed BHs in halos whenever they first reach a minimum  DM FoF halo mass $M_{\rm DM, seed}$, where the FoF is performed on the DM particles only. 
 In the second option, the FoF algorithm is performed on star particles, grouping them with a linking length of about 0.05 times \footnote{Note that this linking length is thus much smaller than the usually used values of 0.15--0.20 to identify virialised halos.} the mean separation of the DM particles \citep[see][]{Hirschmann.etal.2014}. This method identifies compact stellar bodies in the simulation and guarantees that BHs are seeded only in resolved galaxies. Within this model, the BHs are seeded when the following conditions are fulfilled: $(i)$ the stellar FoF mass $M_{\rm \ast,h}$ reaches a fraction $f_\ast$ of $M_{\rm DM,seed}$, where $f_{\ast}$ is a free parameter; $(ii)$ $M_{\rm \ast,h}$ must exceed a given fraction of the DM mass of the halo; and $(iii)$ the gas-to-stellar mass ratio must exceed a given value. 

Whenever seeded, a BH is assigned a minimum value, $M_0$, in correspondence to the minimum allowed stellar-to-DM fraction, $f_\ast$. For a higher stellar-to-DM mass fraction of the halo, the BH seeding mass scales with:
\begin{equation}
    M_{\rm BH,seed}\,=\,M_0\,\frac{M_{\rm *,h}}{ f_*M_{\rm DM, seed}}\, .
    \label{eq:Mseed}
\end{equation}
 When seeded, a BH is located on the gas particle with the minimum value of the gravitational potential.

\subsection{Gas accretion}
\label{sec:BHaccr}

Once seeded, a BH can increase its mass by gas accretion and by merging with another BH. Gas accretion proceeds according to a Bondi-like criterion, eventually capped at a user-defined factor of the Eddington accretion $b_{\rm Edd}$ (=1 for standard Eddington-limited accretion). As for the Bondi accretion rate, we use the expression:
\begin{equation}
    \dot M_{\rm B}\,=\,\frac{4\alpha \uppi \mathrm{G} M_{\rm BH}^2 \rho_{\rm g} }{ (c_{\rm s}^2+v^2)^{3/2}}\,,
\label{eq:bondi}
\end{equation}
where $M_{\rm BH}$ is the BH mass, while $\rho_{\rm g}$, $c_{\rm s}$ and $v$ are the local gas density, sound speed and relative velocity between gas and BH, all estimated as kernel-averaged quantities over neighbour gas particles, at the BH position. The $\alpha$ factor was originally introduced as a "boost" factor to account for the underestimate of SPH gas density at the Bondi accretion radius \citep[e.g.][]{Springel_BHs}. Based on the results of \cite{Gaspari.etal.2013} and following \cite{Steinborn2015}, \og optionally allows the boost factor to take two distinct values to account for the accretion of cold and hot gas, $\alpha_{\rm c}$ and $\alpha_{\rm h}$, respectively. Typical values of these factors are $\alpha_h, \alpha_c=10, 100$ (\citealt{Steinborn2015}). A temperature $T_{\rm sep}$ is adopted to separate the cold and hot contributions, typically set as $5\times 10^4$ K. Regarding the cold gas accretion, \og includes in the computation both single-phase gas particles with $T<T_{\rm sep}$ and the cold component of the multi-phase gas particles (see Sect. \ref{s:SH03}). Accordingly, the total Bondi accretion rate contributed by hot and cold gas is:
\begin{equation}
   \dot M_{\rm B}\,=\,\dot M_{\rm B,h}+\dot M_{\rm B,c}\,,
    \label{eq:bondi_tot}
\end{equation}
where the two terms in the {\em rhs} of the above expression correspond to the hot and cold accretion rates, which can be obtained from Eq.~\eqref{eq:bondi} using $\alpha_{\rm c}$ and $\alpha_{\rm h}$ for the boost factor, respectively.

The accretion rate computed above is capped at $b_{\rm Edd}$ the Eddington limit:
\begin{equation}
  \dot M_{\rm acc} \,=\,\min{(\dot M_{\rm B}, b_{\rm Edd} \dot M_{\rm Edd})}\,,
  \label{eq:Mdotin}
\end{equation}
where 
\begin{equation}
    \dot M_{\rm Edd}=\frac{M_{ \rm BH}}{ \minsub{\epsilon}{Edd} \minsub{\tau}{S}}\,.
    \label{eq:edd}
\end{equation}
We define $\minsub{\tau}{S}=\sigma_{\rm T}c/(4 \pi G m_{\rm p})\sim 4.5\times10^8$ yr, where $m_p$ is the proton mass, $\sigma_T$ the Thomson cross-section, and $\epsilon_{\rm Edd}$ is the radiative efficiency at the Eddington limit. The mass of the BH is then varied according to:
\begin{equation}
    \dot M_{\rm BH} \,=\,(1-\epsilon_{\rm acc})\,\dot M_{\rm acc}\,,
    \label{eq:MdotBH}
\end{equation}
where $\epsilon_{\rm acc}$ is the accretion efficiency.

The relationship between these efficiency parameters depends on the selected model. In the original implementation \citep{Springel_BHs}, a constant, free parameter was assumed and $\epsilon_{\rm Edd} = \epsilon_{\rm acc} = \epsilon_{\rm r}$, where $\epsilon_{\rm r}$ is the radiative efficiency. Within this model, Eq. \eqref{eq:Mdotin} states that a fraction $\epsilon_r$ of the rest-mass energy is converted into radiation during BH accretion, while the remaining fraction $(1-\epsilon_r)$ contributes to BH mass growth. 

The \og code further includes improved prescriptions where these efficiencies are treated as distinct parameters and either varied according to empirical estimates to better capture the transition between different accretion regimes \citep[e.g.,][]{Hirschmann.etal.2014, Steinborn2015} or varied as a function of the BH spin (see Sect.~\ref{sec:BHSpin} and ~\ref{sec:BHfeedb}).

\og uses two distinct variables to specify the mass of a BH particle, that are varied as a consequence of gas accretion. The actual BH mass, which enters in the above expressions of accretion rates, is varied within a time-step $\Delta t$ by the amount $\Delta M=\dot M_{\rm BH} \Delta t$, thus providing a virtually continuous representation of the gas accretion process. On the other hand, a {\em dynamical} mass is also assigned to each BH particle, that is varied in a discrete way to account for the stochastic description of the process of swallowing onto the BH of surrounding gas particles \citep{Springel_BHs}. In this way, the actual and the dynamical mass of a BH particle can differ by an amount which is related to the coarse-grained representation of the accretion process, due to the finite mass resolution of the simulation. In order to mimic a more continuous process of gas accretion and to decrease the difference between the two BH masses, we assume that in each accretion episode, a gas particle looses a fraction $1/n_s$ of its initial mass, with larger $n_s$ corresponding to a finer "slicing" the gas particle involved in the accretion episode. The number of slices $n_s$ is specified as a parameter.

According to the stochastic model of gas accretion, a swallowing event takes place with a probability given by
\begin{equation}
    \label{eq:stoch}
p = \frac{m_{\rm gas}}{ m_{\rm acc}}\,\left(1-e^{-\dot M_{\rm BH}\Delta t/m_{\rm gas}} \right)\,.
\end{equation}
In the above equation, $m_{\rm gas}$ is the mass of the gas particle to be swallowed, while $m_{\rm acc}$ is the mass to be accreted stochastically, i.e. the initial mass of the gas particles divided by the number of slices $n_s$. 

\subsection{Merging between BH particles}
\label{sec:BHmer}

Setting the criterion for BH-BH merging is a well-known critical issue in cosmological simulations \citep[e.g.][and references therein]{Wurster2013, Tremmel.etal.2015}. In \og a merger between a pair of BHs takes place whenever all the following conditions are satisfied: {\em (i)} the ratio between the BH relative velocity and the local sound speed does not exceed a value $f_{\rm v,merg}$, typically set to 0.5; {\em (ii)} the distance between the two BHs should be lower that $f_{\rm merg}$ times the BH gravitational softening, with $f_{\rm merg}$ typically set to 5; {\em (iii)} the two BHs should satisfy a gravitational binding criteria, such that the quantity
\begin{equation} \label{eq:f_b}
f_b\, = \,\frac{|\Phi_{\rm BH_{\rm cen}}-\Phi_{\rm BH_{\rm sat}}|+v_{\rm rel}^2}{0.5 \ c_s^2}
\end{equation}
does not exceed a value in the range 0.5-1. In the above equation, $\Phi_{\rm BH_{\rm cen}}$ and $\Phi_{ \rm BH_{\rm sat}}$ are the gravitational potentials of the central (more massive) and of the satellite (less massive) BH, respectively.

\subsection{Tracking BH particles}
\label{sec:BHtrack}

In the spirit of $N$-body methods to solve gravity, the orbit of a single particle has little physical meaning, given the chaotic nature of their dynamics and the fact that a particle only represents a local sampling of a continuous fluid. On the other hand, BH particles represent quite an exception: since one single BH particle can determine, through its accretion and feedback action, the general evolution of its host galaxy, its physical position at a given time has a clear and important physical meaning: it determines gas accretion, merging with other BHs, amount and nature of feedback. On the other hand, a BH particle at seeding is typically lighter than surrounding particles and leaves in a relatively poorly sampled halo. As a consequence, two-body scattering with nearby particles and the lack of a proper description of dynamical friction, as a consequence of a poor sampling of the velocity distribution of the surrounding sea of particles \citep{chandrasekhar,binneytremaine}, may cause the BH particle to escape the host halo soon after its seeding \citep[e.g.][]{Tremmel.etal.2015, Damiano2024}. 

For these reasons, different approaches have been introduced to accurately follow the orbits of such BH particles, by accounting, and possibly correcting for spurious numerical effects related to the non-linear gravitational interactions with surrounding particles. Here below we describe three different methods implemented in \og for this purpose. 

\subsubsection{Boosted dynamical mass} 
\label{sec:BHdynm}
To prevent this problem, in \og we followed \citet{CurtisSijacki2015} and included the possibility to assign to a BH a boosted dynamical mass, used only for the computation of gravity force, with the typical value given by the mass of DM particles. This dynamical mass is initially larger than the actual BH mass, which enters in defining the BH accretion rate. As the BH mass increases by gas accretion and merger events, its mass may eventually catch up with the seeding dynamical mass. From that point on, the actual and the boosted dynamical BH masses nearly coincide, the only difference being due to the stochastic nature of gas particle swallowing, which determines the value of the dynamical mass. 

Boosting the dynamical mass of a BH particle with respect to its true mass might produce a local spurious perturbation of the gravitational potential, and it might affect the morphology and the evolutionary history of the BHs, as shown in \cite{Wurster2013}. However, it has the beneficial effect of amplifying the numerically resolved dynamical friction, thereby keeping BHs more efficiently at the centre of host halos. Clearly, this amplified dynamical friction is by no means guaranteed to be a precise proxy of the actual dynamical friction that the same BH particle would experience if the local density field is resolved with a much larger number of lower-mass "sea" particles. 

\subsubsection{Continuous repositioning} 
\label{sec:BHrepos}
This method consists of pinning, at each time step, the BH particle at the position of the most bound particle among its neighbours identified by the SPH kernel. This method has been shown to have major shortcomings during mergers or high-speed close encounters between galaxies \citep[e.g.][]{Tremmel.etal.2015,Damiano2024}. For instance, during a close encounter between two galaxies, one of the two BHs may identify the most bound neighbour particle as a particle belonging to the other galaxy. In this case, at the next time step the BH is suddenly and non-physically relocated to the neighbouring galaxy.

Different implementations have been introduced in the literature to improve the performance of this method, such as searching for neighbours within the BH gravitational softening (instead of within the SPH smoothing) length, or setting limits on the relative velocity between the BH and the particle on which to reposition it \citep[see e.g.][]{DiMatteo2008, booth2009cosmological, vogelsberger2013model, Sijacki.etal.2015, Schaye.etal.2015, Pillepich.etal.2018,Ragone.etal.2018, Bahe.etal.2022}. In its default implementation within \og, the BH neighbour particles are searched with the SPH kernel, and no condition on relative velocity is set. 

\subsubsection{Correcting for unresolved dynamical friction} 
\label{sec:BHDynFr}

\noindent {\it Main contributing developer: A. Damiano}

Besides resorting to the above {\em ad hoc} prescriptions, the \og code also implements a method to explicitly account for the unresolved dynamical friction, which has been proven effective in keeping BH particles at the centre of host galaxies, also when the latter are poorly resolved, and in describing the decay of orbits, at least on macroscopic scales, that anticipates a BH-BH merger event \citep{Damiano2024, Damiano.etal.2025}. According to this approach, for each BH we compute the following dynamical friction correction to its acceleration as provided by the $N$-body solver:
\begin{eqnarray}
\label{df}
\vec{a}_{\rm df} & = &\frac{d \vec{v}_{ \rm BH}}{dt}  
= 
  \sum_j^{N(<h_{\rm BH})}-2 \uppi \mathrm{G} m_j (M_{\rm BH}+ m_j)\,{\tilde n_j}\, \\ \nonumber & \times & \ln\left[1+\Lambda(m_j)^2\right] \frac{\vec{v}_{j}-\vec{v}_{\rm BH}}{|\vec{v}_{j}-\vec{v}_{\rm BH}|^3}\,.
\end{eqnarray} 
In the above equation, the sum is over all the DM and star particles lying within the BH softening length $h_{\rm BH}$; $m_j$ and $\vec{v}_j$ are the mass and the velocity of the $j$-th neighbour particle, while  $\vec{v}_{BH}$ is the BH velocity. Furthermore, $\tilde n_j$ represents the local number density computed at the position of the $j$-th particle, and the term $\Lambda(m_j)$ is calculated as:  
\begin{equation}
\Lambda (m_j)=\frac{h_{\rm BH}(\vec{v}_j-\vec{v}_{\rm BH})^2}{\mathrm{G}({\rm M_{\rm BH}}+m_j)}.
\label{eq:lambda}
\end{equation}
In the derivation of the above expressions, the maximum impact parameter for the particle-BH encounters is identified with the gravitational softening of the BH particle \citep[see also][]{Tremmel.etal.2015}. The validity of this choice is largely discussed in \cite{Damiano.etal.2025}. We note that this method of correcting for unresolved dynamical friction does not rely on the choice of any parameter.

\begin{figure}[t]
\centering
\includegraphics[width=0.45\textwidth]{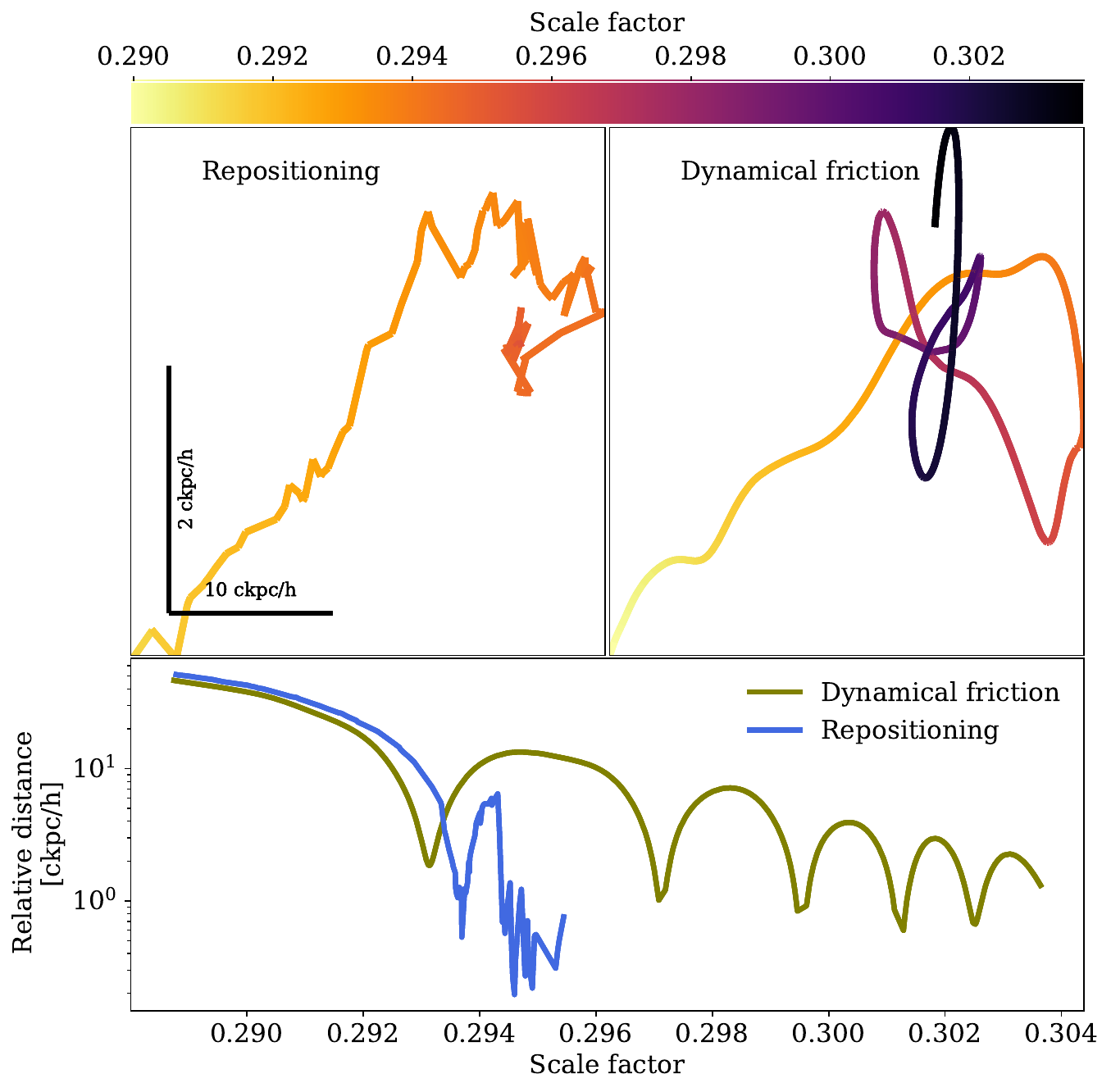}
\caption{Dynamical evolution of BHs during the same BH-BH merger event when using the continuous repositioning of Sect. \protect\ref{sec:BHrepos} and the dynamical friction correction of Sect. \protect\ref{sec:BHDynFr}. 
The orbits of the satellite BH around the central one, which is located at the center of the post-merged galaxy, are displayed in the upper left and right panel and are color-coded according to the expansion factor, as indicated by the colormap. The bottom panel shows the evolution of the relative distance between the two merging BHs when using repositioninig and  dynamical friction in green and blue, respectively.}
\label{fig:BH_merge}
\end{figure}

We show in Figure \ref{fig:BH_merge} the evolution of a "satellite" BH with mass $3\times10^8 \, M_\odot$ sinking toward the core of a galaxy with total mass $M = 1.7 \times 10^{11} \, M_\odot$ hosting a "central" BH with mass $7\times10^8 \, M_\odot$. We compare the merger event between the two BHs as described by the repositioning scheme and by the model to correct for unresolved dynamical friction. The continuous repositioning scheme causes the orbit of the satellite BH to be discontinuous and to rapidly merge. On the other hand, the model based on correcting for the unresolved dynamical friction makes the orbit more regular and the merging to take place after a gradual orbital-decay phase.

\begin{figure*}[t]
    \centering
	\includegraphics[width=0.49\textwidth]{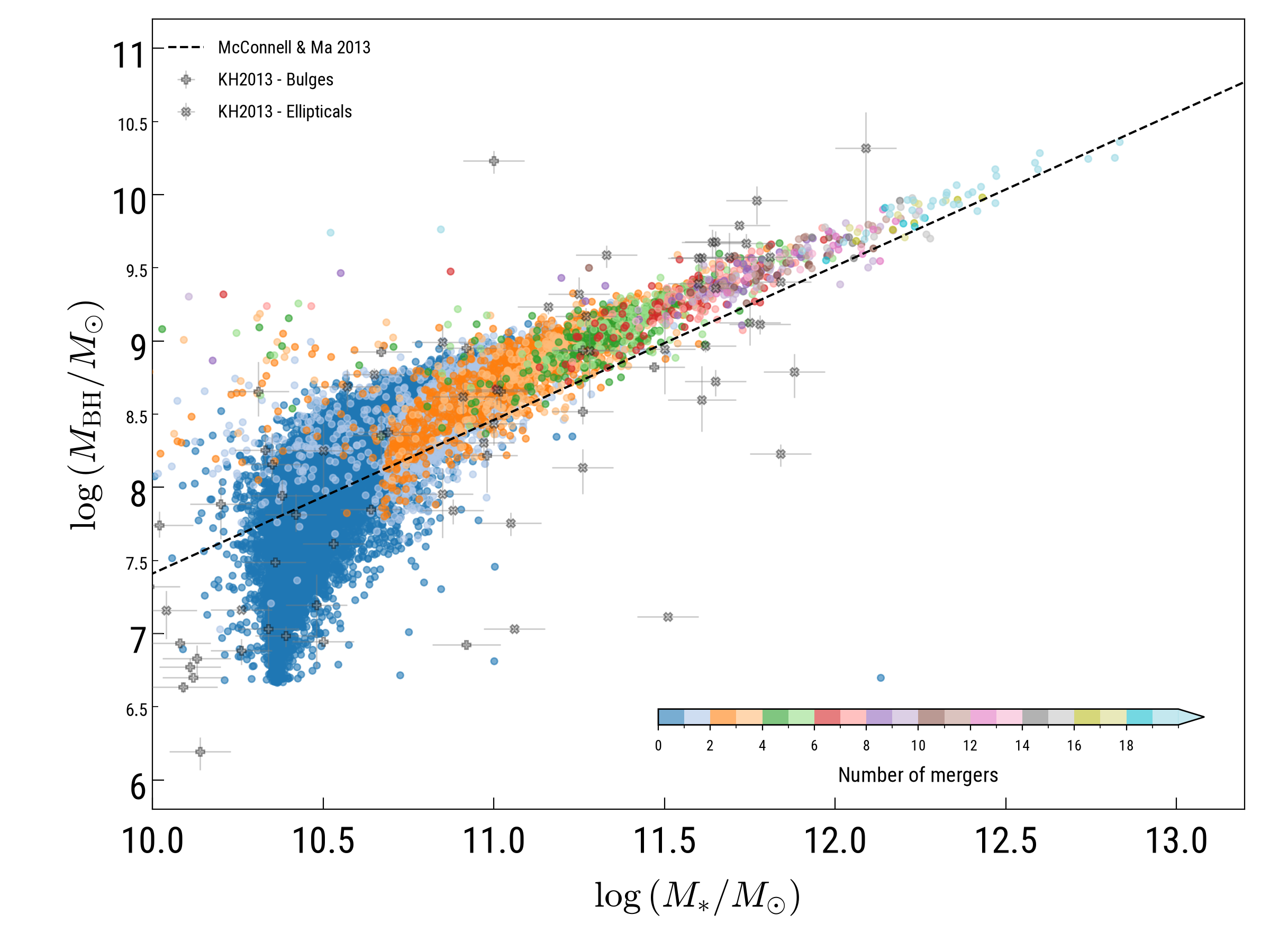}
	\includegraphics[width=0.49\textwidth]{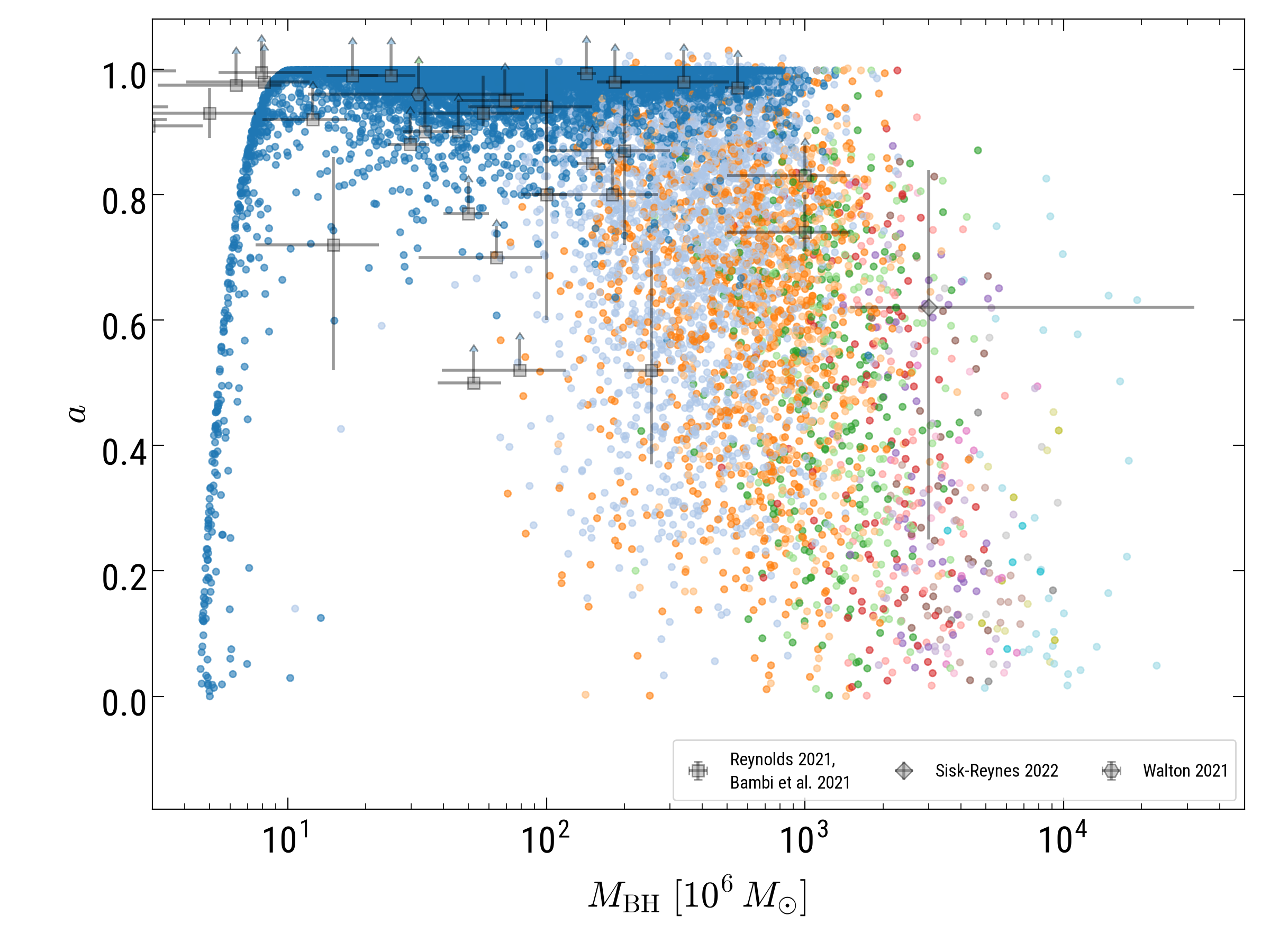}
    \caption{Black hole properties from the cosmological box, where also the dynamical friction and the spin model were switched on. Shown are the stellar mass - black hole mass relation (left) and the spin distribution (right) compared to observations \citep{KormendyHo2013, 2013ApJ...764..184M, Reynolds2021, Bambi+2021, Sisk-Reynes+2022, Walton+2021}. For details, see \citet{Hirschmann.etal.2014}, \citet{Damiano2024} and \citet{Sala+2024}.}
    \label{fig:agn_evol}
\end{figure*}

\subsection{Spin evolution}

\label{sec:BHSpin}
\noindent {\it Main contributing developer: L. Sala}

The \og code features a model for the evolution of the BH spin, which is described in detail in \cite{Sala+2024}. The BH spin is defined as:
\begin{equation}
    \minsub{\myvec{J}}{BH}=\minsub{J}{BH}\cdot \minsub{\myvec{j}}{BH}=a\minsub{J}{max} \minsub{\myvec{j}}{BH}, \label{eq:spindef}
\end{equation}
where $\minsub{J}{max}=\mathrm{G}\minsub{M}{BH}^2/c$, $a$ is the BH dimensionless spin parameter, $\minsub{\myvec{j}}{BH}$ is the unit vector encoding its direction and $\minsub{J}{BH}$ is its magnitude. We define $0 \leq a \leq 1$. Whenever a BH is seeded, its spin is set to the parameter $\minsub{a}{0}$.

The model assumes the presence of a sub-resolution accretion disc which mediates the mass transfer rate from the resolved scales onto the BH, and $\minsub{\myvec{J}}{BH}$ evolves because of its interaction with the distribution of matter in the accretion disc, whose angular momentum is misaligned with respect to $\minsub{\myvec{J}}{BH}$. Lense-Thirring precession induces its innermost region to align with the equatorial plane of the rotating BH, while its outermost region remains unperturbed. We assume that the propagation of vertical perturbation is in the diffusive regime, in which a stable warped configuration is created in the disc \citep[][]{Bardeen&Petterson1975, Martin+2007}. The accretion disc annuli which have the largest misalignment with both the BH spin direction and the disc's outermost region occur at the warp radius $\minsub{R}{w}$, which in units of the Schwarzschild radius $\minsub{R}{BH}=2\mathrm{G}\minsub{M}{BH}/c^2$ reads:
\begin{align}
    \frac{R_{\text {w}}}{R_{\mathrm{BH}}} \simeq 4 \times 10^{2} a^{5 / 8} M_{\mathrm{BH}, 8}^{1 / 8}\left(\frac{\minsub{f}{Edd}}{\epsilon_{\mathrm{r}, 01}}\right)^{-1 / 4}\left(\frac{\nu_{2} / \nu_{1}}{85}\right)^{-5 / 8} \alpha_{\nu_1, 01}^{-1 / 2}, \label{eq:rwarp_numeric}
\end{align}
where $M_{\mathrm{BH}, 8}=\minsub{M}{BH}/(10^8\msun)$, $\epsilon_{\mathrm{r}, 01}=\epsilon_{\mathrm{r}}/0.1$ and we define the Eddington ratio $\minsub{f}{Edd}=\somedot{M}{BH}/\somedot{M}{Edd}$. The radial shear viscosity $\nu_1$ is responsible for the radial inward drift of gas across the disc, while the vertical shear viscosity $\nu_2$ governs the propagation of vertical perturbations. We also assume that the viscosity $\alpha$-parameter \citep{Shakura&Sunyaev1973} is $\alpha_{\nu_1, 01}=\alpha_{\nu_1}/0.1\equiv1$, following \cite{King+2005}, and $\minsub{\nu}{2}/\minsub{\nu}{1}=2(1+7\minsub{\alpha}{\nu_1})/(4+\minsub{\alpha^2}{\nu_1})/\minsub{\alpha^2}{\nu_1}$, following \cite{Ogilvie1999}, leading to a fiducial value for the ratio $\minsub{\nu}{2}/\minsub{\nu}{1}=85$.

\subsubsection{Evolution of the spin magnitude}\label{sec:spin_magnitude_evol}
We define an accretion episode as characterised by a mass $\minsub{M}{d}$, computed as
\begin{equation}
    \minsub{M}{d}\simeq\somedot{M}{BH}\minsub{t}{\nu_1}(\minsub{R}{w})\,,\label{eq:diskmass}
\end{equation}
where $\minsub{t}{\nu_1}(\minsub{R}{w})$ is the accretion timescale at the warp radius:
\begin{align}
    \minsub{t}{\nu_1}(\minsub{R}{w})\!\sim\!3.4 \times 10^{5} a^{7 / 8} M_{\mathrm{BH}, 8}^{11 / 8}\left(\frac{\minsub{f}{Edd}}{\epsilon_{\mathrm{r}, 01}}\right)^{\!-3 / 4}\!\left(\frac{\nu_{2} / \nu_{1}}{85}\right)^{\!-7/8}\!\!\alpha_{\mathrm{\nu_1{}}, 01}^{-3 / 2}\,\mathrm{yr}.\label{eq:tnu1}
\end{align}

When $\minsub{M}{d}$ is accreted onto the BH, $a$ changes due to the accretion of its angular momentum at the innermost stable circular orbit (ISCO), according to the expression from \cite{Bardeen1970}:
\begin{equation}
    \minsub{a}{f}= \frac{1}{3}\frac{\minsub{r^{1/2}}{isco}}{\minsub{M}{ratio}}\left[4-\left(3\frac{\minsub{r}{isco}}{\minsub{M^2}{ratio}}-2\right)^{1/2}\right], \label{eq:spinupdate}
\end{equation} 
where 
\begin{equation}
    \minsub{M}{ratio}=\frac{\minsub{M}{BH}^{i}+\minsub{M}{d}(1-\minsub{\epsilon}{r})}{\minsub{M}{BH}^{i}},\label{eq:mratio}
\end{equation} 
and, normalising to the gravitational radius $\minsub{R}{g}=\minsub{R}{BH}/2=G\minsub{M}{BH}/c^2$, 
\begin{equation}
    \minsub{r}{isco}=\minsub{R}{isco}/\minsub{R}{g}=3+Z_2\pm[(3-Z_1)(3+Z_1+2Z_2)]^{1/2},\label{eq:risco}
\end{equation}
where the positive (negative) sign is for counter-rotating (co) orbits. $Z_1$ and $Z_2$ are functions of the BH dimensionless spin parameter only:
\begin{equation}
    Z_1=1+(1-a^2)^{1/3}[(1+a)^{1/3}+(1-a)^{1/3}],
\end{equation}
\begin{align}
    Z_2=(3a^2+Z_1^2)^{1/2}.
\end{align}
The quantity $\minsub{R}{isco}$ ranges from 1 to 9 $\minsub{R}{g}$, for co- and counter-rotating orbits on a maximally spinning BH, respectively. Physical sizes are therefore between $\sim 5\times10^{-6}M_{\rm BH,8}\text{ pc }$ and $\sim 5\times10^{-5}M_{\rm BH,8}\text{ pc }$. 

We also cap $a$ to 0.998, which is the maximum spin allowed if photon trapping is assumed \citep{Thorne1974}.

The efficiency of the accretion process $\minsub{\epsilon}{acc}$, when the spin evolution model is active, depends on $a$ according to:  
\begin{equation}
    \minsub{\epsilon}{spin} = 1 - \sqrt{1-\frac{2}{3\minsub{r}{isco}}}. \label{eq:accretion_efficiency}
\end{equation}
The BH mass increases as a result of the accretion episode by $\Delta M_{\rm BH}=\minsub{M}{d}(1-\minsub{\epsilon}{spin})$.

\subsubsection{Evolution of the spin direction}\label{sec:spin_direction_evol}
The warped configuration also exerts a torque on the BH spin, which tends to modify the BH spin direction over time. For each accretion episode, we define $\minsub{\myvec{J}}{d}=\minsub{J}{d}\minsub{\myvec{j}}{d}$ as the angular momentum of the disc within $\minsub{R}{w}$, where 
\begin{equation}
    \minsub{J}{d}=\minsub{M}{d}(\mathrm{G}\minsub{M}{BH}\minsub{R}{w})^{1/2}\label{eq:Jdisk}
\end{equation}
and 
\begin{align}
    \minsub{\myvec{j}}{d}=\minsub{\myvec{j}}{g}\equiv\minsub{\myvec{L}}{BH,kernel}/|\minsub{\myvec{L}}{BH,kernel}|\,.\label{eq:disc_dir}
\end{align}
Here, $\minsub{\myvec{L}}{BH, kernel}$ is the angular momentum within the BH smoothing length, kernel-weighted as
\begin{equation}
    \minsub{\myvec{L}}{BH,kernel}=\sum_{j}\minsub{m}{j}(\minsub{\myvec{r}}{j}-\minsub{\myvec{r}}{BH})\times(\minsub{\myvec{v}}{j}-\minsub{\myvec{v}}{BH})w(\minsub{\myvec{r}}{j}-\minsub{\myvec{r}}{BH},\minsub{h}{BH}).
\end{equation}
and is computed taking into account only the cold gas particles -- the component that is able to settle into an accretion disc. During an accretion episode, the acting torque always leads the BH spin to align or counter-align with the total angular momentum $\minsub{\myvec{J}}{tot}$ of the system disc+BH \citep{King+2005}:
\begin{equation}
  \minsub{\myvec{J}}{tot}=\minsub{\myvec{J}}{BH}+\minsub{\myvec{J}}{d}=\minsub{\myvec{J}}{BH}+\minsub{{J}}{d}\minsub{\myvec{j}}{d}  
\end{equation}
We assess whether the innermost part of the disc ends up in a counter-aligned configuration following the condition described in \cite{King+2005}:
\begin{equation}
    \cos{\theta_{\rm BH-d}} < -\frac{\minsub{J}{d}}{2\minsub{J}{BH}}\,,\label{eq:counter-align_condition}
\end{equation} 
where
\begin{align}
        \frac{J_{\mathrm{d}}}{2J_{\mathrm{BH}}} &\sim 6.8 \times 10^{-2} a^{3 / 16} M_{\mathrm{BH}, 8}^{23 / 16} \left(\frac{\minsub{f}{Edd}}{\epsilon_{\mathrm{r}, 01}}\right)^{\!1/8}\!\left(\frac{\nu_{2} / \nu_{1}}{85}\right)^{\!-19/16} \!\alpha_{\minsub{\nu}{1}, 01}^{-7 / 4}. \label{eq:JdJBHratio_numeric}
\end{align}
When the condition is satisfied, the appropriate sign for a counter-rotating episode is chosen in Eqs.~\eqref{eq:mratio} and \eqref{eq:risco}, and the episode has the overall effect of decreasing $a$.

\subsubsection{Sub-cycling and deferred update}
The model is constructed so that BH growth proceeds at the rate given by Eq.~\eqref{eq:MdotBH}. If $M_{d}<\dot{M}_{\rm BH}\Delta t$, the code executes $N=\dot{M}_{\rm BH}\Delta t/M_{d}$ accretion episodes over a time-step, which share the same accretion rate, computed at the beginning of the time-step. At the end of each sub-cycle (i.e. accretion episode), we update BH spin and mass.
Conversely, if $M_{d}>\dot{M}_{\rm BH}\Delta t$, the code splits the accretion episode over $N=M_{d}/(\dot{M}_{\rm BH}\Delta t)$ time-steps using averages for the input quantities of the model equations, deferring the update of the BH mass and spin to the end of the accretion episode.

\subsubsection{Self-gravity regime}
Beyond a radius equal to 
\begin{equation}
    \frac{R_{\mathrm{sg}}}{R_{\mathrm{BH}}} \simeq 5 \times 10^{2}  M_{\mathrm{BH}, 8}^{-52 / 45}\left(\frac{\minsub{f}{Edd}}{\epsilon_{\mathrm{r}, 01}}\right)^{-22 / 45}\alpha_{\minsub{\nu}{1}, 01}^{28 / 45}\,,\label{eq:selfgravradius}
\end{equation}
The outer parts of the accretion disc become unstable due to their own self-gravity. Since only the region within $R_{\mathrm{sg}}$ can be accreted, if $R_{\mathrm{sg}}<R_{\mathrm{w}}$ we set
\begin{equation}
    M_{\mathrm{d}} = M_{\mathrm{sg}} \simeq 6 \times 10^{5} M_{\mathrm{BH}, 8}^{34 / 45}\left(\frac{\minsub{f}{Edd}}{\epsilon_{\mathrm{r}, 01}}\right)^{4 / 45} \alpha_{\mathrm{\nu_1}, 01}^{-1 / 45}\mathrm{M}_{\odot}.\label{eq:selfgravmass}
\end{equation}
and substitute $\minsub{R}{w}$ with $\minsub{R}{sg}$ in Eq.~\eqref{eq:Jdisk}.

\subsubsection{Spin after BH coalescences}
\label{sec:spin_mergers}

To account for spin evolution in the case of BH mergers. We use equations retrieved in a full general-relativistic framework by \citet{Rezzolla+2008} to compute the final spin after a merger event. Once two BHs characterised by masses $\minsub{M}{1}$, $\minsub{M}{2}$ and spins $\minsub{a}{1}$, $\minsub{a}{2}$ have been selected to merge according to the criteria described in Sect.~\ref{sec:BHmer}, the final spin vector of the remnant BH $\myvec{a}^f$ is given by 
\begin{equation}
    \myvec{a}^f=\frac{1}{(1+q)^2}\left(\myvec{a}_1+\myvec{a}_2 q^2+\myvec{\ell} q\right) ,
    \label{eq:merger_a}
\end{equation}
where $\myvec{a}=a\minsub{\myvec{j}}{BH}$, $q=M_2/M_1$, with $M_1\geq M_2$. In this formula, $\minsub{M}{1}$ and $\minsub{M}{2}$ are the BH physical masses. $\myvec{\ell}=\myvec{\ell}'/(M_1 M_2)$, where $\myvec{\ell}'$ is the binary orbital angular momentum that cannot be radiated away in gravitational waves before coalescence. 
The magnitude of $\myvec{\ell}$ reads
\begin{equation}
    \begin{aligned}
        \ell= & \frac{s_4}{\left(1+q^2\right)^2}\left(a_1^2+a_2^2 q^4+2 \myvec{a}_1 \cdot \myvec{a}_2 q^2\right) \\
        & +\left(\frac{s_5 \mu+t_0+2}{1+q^2}\right)\left(a_1 \cos \phi_1+a_2 q^2 \cos \phi_2\right) \\
        & +2 \sqrt{3}+t_2 \mu+t_3 \mu^2.
    \end{aligned}
\end{equation}
Here, $\cos{\phi}=\myvec{a}\cdot\myvec{\ell}/(a\ell)$ identifies the
angle subtended by the spin vector of BH with $\myvec{\ell}$, and $\mu=q/(1+q)^2$. $s_4=-0.129, s_5=-0.384, t_0=-2.686, t_2=-3.454$, $t_3=2.353$ are the parameters of the fit to their numerical results.
We assume that $\myvec{\ell}$ is parallel to the binary angular momentum $\myvec{L}=\myvec{L}_1+\myvec{L}_2$ \citep{Rezzolla+2008}. $\myvec{L}_{i=1,2}$ is the angular momentum vector of BH $i$ with respect to the binary centre of mass (CM), computed as $\myvec{L}_i=M_i(\myvec{r}_i-\myvec{r}_{\rm CM})\times(\myvec{v}_i-\myvec{v}_{\rm CM})$. The values for the radii $\myvec{r}_i$ and velocities $\myvec{v}_i$ are retrieved when the BHs are selected to merge.

\begin{table*}[ht]
\begin{center}
\def\arraystretch{2.0}
\caption{Summary of the efficiency prescriptions in \og for each feedback mode (FM). The $\epsilon$ columns contain the efficiency used for the quantity in \textcolor{blue}{blue}; the last column points to the extended equation(s) in the text where the terms are defined.}
\label{tab:efficiency_modes}

\setlength{\tabcolsep}{3.5pt}
\def\arraystretch{1.5}
\begin{tabular}{l|c|c|c|c|c}
\hline
\textbf{Mode} 
& $\dot M_{\rm Edd}=$ 
& $\dot M_{\rm BH} =$
& $\dot E = $& \textbf{Free parameters} 
& \textbf{Extended}  \\ 
 & $M_{\rm BH} / (\textcolor{blue}{\underbrace{\minsub{\epsilon}{Edd}}_{\parallel}} \tau_{\rm S})$ 
& $(1-\textcolor{blue}{\underbrace{\epsilon_{\rm acc}}_{\parallel}})\,\dot M_{\rm acc}$ 
& $\textcolor{blue}{\underbrace{\epsilon_{\rm tot}}_{\parallel}}\dot M_{\rm acc} c^2$ 
& 
& \textbf{equation(s)} \\ \hline

\textbf{FM0}
  & $\,\,\,\,\,\,\,\,\,\,\,\,\textcolor{blue}{\epsilon_{\rm r}^{\rm std}}$ 
  & $\,\,\textcolor{blue}{\epsilon_{\rm r}^{\rm std}}$ 
  & $\textcolor{blue}{\epsilon_{\rm r}^{\rm std}\,\epsilon_{\rm feed}(...)}$
  & $\epsilon_{\rm r}^{\rm std},\ \epsilon_{\rm f},\ F_\mathrm{boost}$
  & Eq.~\eqref{eq:etot_FM0} \\ \hline
\textbf{FM1}
  & $\,\,\,\,\,\,\,\,\,\,\,\,\textcolor{blue}{\epsilon_{\rm Edd}^{\rm S15}(...)}$
  & $\,\,\,\,\,\,\,\,\,\textcolor{blue}{\epsilon_{\rm r}^{\rm S15}(...)}$
  & $\textcolor{blue}{\epsilon_{\rm r}^{\rm S15}(...)\,\epsilon_{\rm feed}(...)}$
  & $\epsilon_{\rm f},\ F_\mathrm{boost}$
  & Eqs.~\eqref{eq:er_steinborn}, \eqref{eq:etot_FM1} \\ \hline
\textbf{FM2}
  & $\,\,\,\,\,\,\,\,\,\,\,\,\textcolor{blue}{\epsilon_{\rm Edd}^{\rm S15}(...)}$
  & $\,\,\,\,\,\,\textcolor{blue}{\epsilon_{\rm r}^{\rm S15}(...) + \epsilon_{\rm o}^{\rm S15}(...)}$
  & $\textcolor{blue}{\epsilon_{\rm o}^{\rm S15}(...) + \epsilon_{\rm feed}(...)\,\epsilon_{\rm r}^{\rm S15}(...)}$ 
  & $\epsilon_{\rm f}$, $(F_\mathrm{boost}=1)$
  & Eqs.~\eqref{eq:er_steinborn}, \eqref{eq:eoo_steinborn}, \eqref{eq:etot_FM2} \\ \hline
\textbf{FM3}
  & $\,\,\,\,\,\,\,\,\,\,\,\,\textcolor{blue}{0.1}$\,
  & $\,\,\,\,\,\,\,\textcolor{blue}{\epsilon_{\rm r}^{\rm S24}(...)}$
  & $\textcolor{blue}{\epsilon_{\rm r}^{\rm S24}(...)\,\epsilon_{\rm feed}(...)}$
  & $\epsilon_{\rm f},\ F_\mathrm{boost}$
  & Eqs.~\eqref{eq:er_FM3}, \eqref{eq:etot_FM3} \\ \hline
\end{tabular}
\end{center}
{\footnotesize \itshape Note: with spin evolution, $\epsilon_{\rm Edd}=\epsilon_{\rm acc}=\epsilon_{\rm spin}(a)$ (and $\epsilon_{\rm r}=\epsilon_{\rm spin}(a)$ in FM0), while $\epsilon_{\rm tot}$ is unchanged and given by the empirical model of each FM.}
\end{table*}

\subsection{Energy feedback}
\label{sec:BHfeedb}

In \og, BHs exert feedback by injecting thermal energy into the surrounding gas at a rate given by
\begin{equation}
    \dot E =\epsilon_{\rm tot} \dot M_{\rm acc} c^2\,,
\end{equation}
where $\epsilon_{\rm tot}$ depends on the selected feedback model and encapsulates both the efficiency of energy conversion of the accretion power into radiation and/or outflows and the efficiency with which such energy is coupled to the surrounding gas (see below).
The energy produced by a BH gas accretion episode, which is available to heat the surrounding medium during the BH time-step $\Delta t$, is given by
\begin{equation}
  \Delta E\,= \dot E \Delta t\,.
  \label{eq:Eh}
\end{equation}
This energy is used to increase the internal energy of the gas particles surrounding a BH and is distributed using weights given by the SPH interpolating kernel computed at the BH position. 

The code features several alternative prescriptions for $\epsilon_{\rm tot}$, summarised in Table \ref{tab:efficiency_modes}. In the simplest feedback mode \textbf{FM0}, it is set to
\begin{equation}
    \epsilon_{\rm tot} = \epsilon_{\rm r}^{\rm std} \epsilon_{\rm feed}, \label{eq:etot_FM0}
    \end{equation}
where $\epsilon_{\rm r}^{\rm std}$ is the standard, constant radiative efficiency \citep{Springel_BHs}, and $\epsilon_{\rm feed}$ is defined as
\begin{equation} \label{eq:epsfeed_tot}
    \epsilon_{\rm feed} =
    \begin{cases}
            \epsilon_{\rm f} & f_{\rm Edd} \geq f_{\rm Edd,thresh}, \\
            F_\mathrm{boost}\epsilon_{\rm f} & f_{\rm Edd} < f_{\rm Edd,thresh}, \\
    \end{cases}
\end{equation}
$f_{\rm Edd}=\dot M_{\rm BH} / \dot M_{\rm Edd}$ is the Eddington ratio, $\epsilon_{\rm f}$ is a coupling efficiency and $f_{\rm Edd, thresh}$ is a threshold that is introduced to separate between quasar and radio mode. The parameter $F_{\rm boost}$ is introduced to mimic a more efficient thermal coupling of feedback energy expected in the radio-mode phase \citep{Churazov.etal.2005, Sijacki.2006, Fabjan.etal.2010, Hirschmann.etal.2014}. A typical value used for it is $F_{\rm boost}=4$.

Following the sub-resolution model of \cite{Steinborn2015}, we also include a dual-mode AGN feedback scheme where the feedback contribution is partitioned between $\epsilon_{\rm r}$ and a mechanical outflow efficiency $\epsilon_{\rm o}$, both dependent on $M_{\rm BH}$ and $f_{\rm Edd}$:
\begin{equation} \label{eq:er_steinborn}
    \epsilon_{\rm r}^{\rm S15} =
    \begin{cases}
            A\epsilon_{\rm Edd}f_{\text{Edd}}^B & f_{\text{Edd}} < 0.05, \\
            \epsilon_{\text{Edd}} - 10^{-4}\epsilon_{\rm Edd} f_{\text{Edd}}^{-2.8431} & \text{otherwise}.
    \end{cases}
\end{equation}
\begin{equation} \label{eq:eoo_steinborn}
    \epsilon_{\rm o}^{\rm S15} =
    \begin{cases}
            \eta - A\eta f_{\text{Edd}}^B & f_{\text{Edd}} < 0.05, \\
            10^{-4}\eta f_{\text{Edd}}^{-2.8431} & \text{otherwise},
    \end{cases}
\end{equation}
In these equations, the Eddington efficiency $\epsilon_{\text{Edd}}$ is mass-dependent during the radiation-dominated quasar-mode, defined as
\begin{equation}
    \epsilon_{\text{Edd}}^{\rm S15} = 0.089(M_{\rm BH}/10^8 M_\odot)^{0.52}    
\end{equation}
(capped at $0.42$), and assumed to be a constant $\eta = 0.1$ in the radio-mode. Moreover, $B = 0.5$ and $A = 10^{-4} \times 0.05^{-2.8431-B}$.
In feedback mode \textbf{FM1}, the total feedback efficiency is 
\begin{equation}
    \epsilon_{\rm tot} = \epsilon_{\rm feed}\,\epsilon_{r}^{\rm S15}\,,
    \label{eq:etot_FM1}
\end{equation}
while in feedback mode \textbf{FM2} it is
\begin{equation}
    \epsilon_{\rm tot} = \epsilon_{\rm o}^{\rm S15}\,+\,\epsilon_{\rm feed}\,\epsilon_{r}^{\rm S15}\,.
    \label{eq:etot_FM2}
\end{equation}
In both Eqs. \eqref{eq:etot_FM1} and \eqref{eq:etot_FM2},  $\epsilon_{\rm feed}$ follows Eq. \eqref{eq:epsfeed_tot}, and $\epsilon_{\rm r}$ (and $\epsilon_{\rm o}$) follow Eqs. \eqref{eq:er_steinborn} and \eqref{eq:eoo_steinborn}.

Finally, following \citet{salaSupermassiveBlackHole2024}, in feedback mode \textbf{FM3} we prescribe
\begin{align}
     \epsilon_{\rm r}^{\rm S24} &=  C\dot{M}_{\rm acc}^{D}M_{\rm BH}^{E}, \label{eq:er_FM3}\\
     \epsilon_{\rm Edd} &= 0.1, \label{eq:eedd_FM3}\\
     \epsilon_{\rm tot} &= \epsilon_{\rm feed}\,\epsilon_{\rm r}^{\rm S24}\,  \label{eq:etot_FM3},
\end{align}
where $C=4.16\cdot 10^{-6}$, $D=-0.336$, $E=0.535$ are constants that have been fitted using the dataset in \cite{Steinborn2015} and new empirical efficiencies estimates, as described in \citet{salaSupermassiveBlackHole2024}.

When the BH spins are evolved in the simulation, in \textbf{FM0} $\epsilon_{\rm Edd} = \epsilon_{\rm acc} = \epsilon_{\rm r} = \epsilon_{\rm spin}(a)$, while in \textbf{FM1-3} we set $\epsilon_{\rm acc}=\epsilon_{\rm Edd}=\epsilon_{\rm spin}(a)$, while $\epsilon_{\rm tot}$ is independent and set to the values provided by the empirical models (Eq. \eqref{eq:etot_FM1}, \eqref{eq:etot_FM2} and  \eqref{eq:etot_FM3}, for the \textbf{FM1}, \textbf{FM2} and \textbf{FM3}, respectively). This flexibility in the spin-active configuration is physically motivated by the difference in scales between accretion physics and feedback coupling. We assume that $\epsilon_{\rm acc}$ and $\epsilon_{\rm Edd}$ are governed by small-scale processes, specifically the thin-disk approximation where efficiency is tied to the position of the disk and is thus fundamentally dependent on BH spin. Conversely, the total feedback efficiency $\epsilon_{\rm tot}$ operates on much larger scales; we therefore allow it to be decoupled from the spin to enable the use of empirically motivated models that better describe the observed energy injection into the surrounding medium.

To give an impression of the resulting black hole properties in a typical cosmological volume, we also switched on the black hole treatment, including the treatment of dynamical friction and spin in our showcase, based on {\it Box3/hr} from the {\it Magneticum} volumes. Figure \ref{fig:agn_evol} shows the resulting scaling relation of the black holes and stellar mass in galaxies as well as their spin distribution within the whole volume, compared to observations. For a list of all available configuration and parameter options, see the according \href{https://gitlab.lrz.de/AstroCodes/OpenGadget3/-/wikis/Black%20Holes}{section on the code wiki}.

\section{Parallelization}
\og is fully MPI-OpenMP parallelised to make use of hybrid architectures of modern HPC systems. In addition, \og can offload the calculation of gravity, SPH density, hydro-forces and thermal conduction on GPUs using the OpenACC programming model \citep[][]{Ragagnin2020, Ragagnin2026}.
The details of each of these parallelisation options are described in the following.

\subsection{MPI backbone}

\og relies on MPI as its primary parallelisation layer. The computational volume is partitioned among the ranks through the domain decomposition described in Sect. \ref{sec:domaindecomposition}, and the dependencies among the tasks are resolved by an explicit communication layer.

The pure MPI parallelisation reaches the efficiency limit before the algorithmic strong-scaling limit. 
The cost of the decomposition itself grows with the number of subdomains, while their surface-to-volume ratio deteriorates, so that a progressively larger fraction of the runtime is spent importing and exporting boundary data rather than integrating it. In parallel, the memory that does not scale with the local particle load — communication buffers above all — grows with the rank count and competes with the particle data for the memory available per core.  To mitigate these, \og adopts a hybrid model, in which OpenMP exploits the shared memory within a node while MPI is confined to the inter-node decomposition; this is described in the following section.

\subsection{OpenMP}
\label{sec:openmp}

\begin{figure}[t]
 \centering
\includegraphics[width=0.49\textwidth]{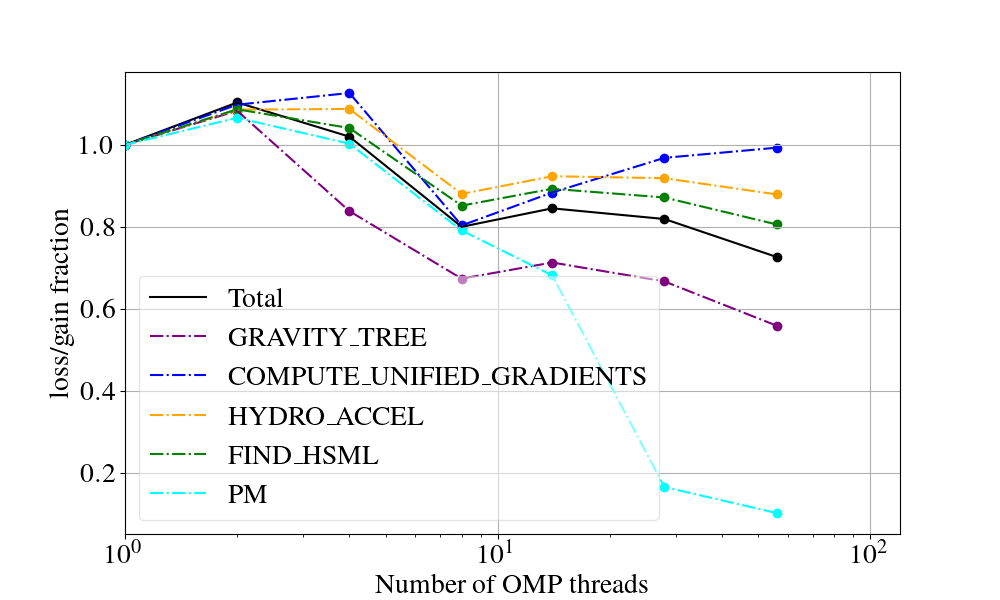}
\caption{\og scaling when replacing MPI tasks, marked by different colours. Curves are normalised to the pure MPI run. Ideal scaling would follow a horizontal line at the unity value. Different curves correspond to different sectors of the code, as indicated in the legend. The curve marking the total time is quite flat as the number of OpenMP threads increases.} \label{fig:scaling_MPI_OpenMP}
\end{figure}

The shared-memory parallelisation framework in \og is OpenMP. Many of the local processes are trivially OpenMP parallelised, while one of the critical uses of OpenMP in \og is the parallelisation of the neighbour search, where each thread processes a chunk of active particles in order to locate neighbours efficiently.
Figure \ref{fig:scaling_MPI_OpenMP} shows the loss/gain in the execution time using different combinations of MPI tasks and OpenMP threads within a single node, keeping fixed to 64 the total number of used cores. We note that for a small number of OpenMP threads ($ \ le 4$), there is a gain contributed by the decrease in communication between a reduced number of MPI tasks. Increasing the weight of the OpenMP part, there is a progressive loss, which however remains relatively modest even at the largest number of OpenMP threads. While the overall efficiency of the OpenMP parallelisation has been observed to be generally similar across various architectures and simulation setups, details can significantly depend on the specific architecture or specific simulation. For instance, we expect such dependency when OpenMP ranks cover memory across sockets or problem sizes exceed (or fit in) different cache levels of the involved processors. 

In this paragraph, we briefly review how MPI and OpenMP work together in the neighbour search. 
In the first phase, the search is limited to the local MPI process, identifying particles within a specified search radius and handling those near the boundaries that may have neighbours in other MPI processes. During this phase, when encountering nodes belonging to a different MPI process, the corresponding particles are added to an export buffer. In \og these buffers are assigned dynamically, depending on the amount of unused memory. Therefore, having more memory available can lead to significantly faster execution. If the buffer exceeds capacity, the simulation is temporarily paused to manage data transfers. The second phase handles particles imported from other MPI processes. For each imported particle, the search is performed again locally to update its physical properties according to the identified neighbours. After the search is complete for a given particle, its updated state is returned to its original process, and the physical quantities are merged.

Both the MPI and OpenMP parallelisation introduce subtle sources of noise that can be minimised.  Even when using summation-conserving settings for the underlying MPI library in the {\tt AllToAll} communication, the additional source of noise is the different internal splitting of the primary and secondary loops, which can lead to a different order of summation. The OpenMP one may be more subtle to understand, since each thread treats a target particle's interaction in a serial way. Almost all OpenMP parallel loops are scheduled in a fixed fashion, so that repeated execution should not change any order. However, some critical regions for additional physical modules could lead to different orders in lists and buffers, and therefore can in principle lead to a different order of summation. One example is that the OpenMP parallelisation may be influenced by the adaptive drifting scheme. In fact, since drifting all particles at each adaptive time-step would be too expensive, non-active particles are drifted only when encountered in a neighbour search tree walk (this operation is not thread-safe and is therefore encapsulated in an OpenMP critical region). This may introduce a subtle source of noise, where a particle that would enter another particle search radius only after drifting may be excluded if it has not yet been drifted. Since this drift may be performed during a neighbour search of another target particle, its inclusion may depend on the thread execution order. These sources of noise are very small and are not expected to lead to sizeable effects. In fact, pure $N$-body simulations and hydrodynamical simulations including only non-radiative physics show almost no chaotic behaviour, which matches theoretical expectations~\citep{Chaitra2026}.

Previously, as demonstrated in~\cite{Ragagnin2016}, this process became a bottleneck, particularly for high-resolution hydrodynamic cosmological simulations, due to its original implementation in \textsc{Gadget}-3 (where a single tree walk was performed for each active particle). In that work, we introduced an optimized neighbour search technique that takes advantage of the already-present space-filling curve particle ordering, specifically the Hilbert curve, to improve memory locality. This allows spatially close particles to also be close in memory, thereby reducing the overhead of the search process. By grouping particles within a certain radius and performing a single search for the group, the algorithm minimizes the number of tree walks, which represents one of the most computationally expensive parts of the code.

\subsection{GPU offloading}
\label{sec:GPU}
\noindent
{\it Main contributing developers: A. Ragagnin, M. Petkova, N. Hariharan}

\subsubsection{OpenACC}

In \og the computation of gravity, SPH density, hydrodynamic forces, and thermal conduction are ported to the GPU using the OpenACC directives \citep[][]{Foley2017}. The OpenACC strategy~\citep{Ragagnin2020, Ragagnin2026} has been chosen to reduce the need for modifications of \og development, while maintaining the code efficiency on CPU-only systems. In the GPU implementation, we overlap computations on the host and the device, such that, for example, the device asynchronously computes physical interactions between particles within the same domain while the CPU performs tree walks (over the domain top tree), fills the export buffer or communicates particles. 
The major issue with this strategy is that each \og OpenMP thread allocates an array of neighbours with a size comparable to the number of particles. This strategy is not feasible for a GPU, where each thread can have only a small private buffer.
To overcome this issue, we process neighbours in chunks of 32 particles. 
This strategy turned out to work well for the short-range gravitational solver (namely, the Barnes-Hut solver), the SPH kernel, and the thermal conduction module. 

\subsubsection{OpenMP6.0}

To enable \og to run on a larger variety of GPUs, including Intel GPUs, \og also supports OpenMP offload. This has been implemented by manually translating the code from OpenACC to OpenMP using the corresponding directives and pragmas.  

\begin{figure}
 \centering
 \includegraphics[width=0.44\textwidth]{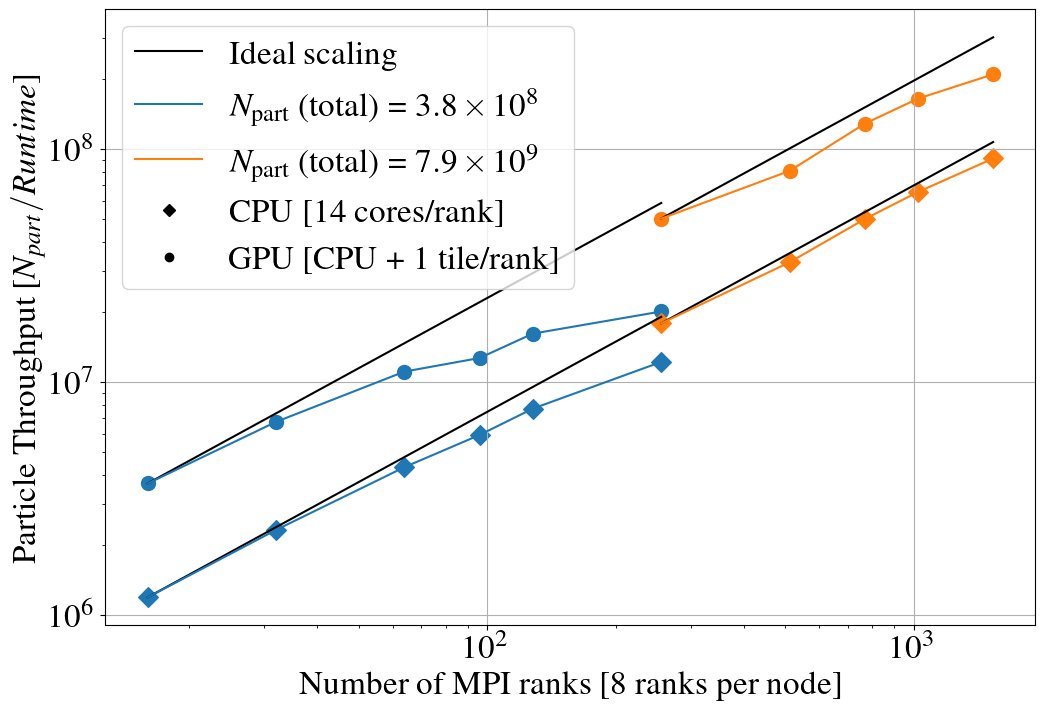}\\
\caption{\og scaling -- Combination of weak and strong scaling tests of \og{}, using the first ten time-steps of three large cosmological simulations including gas and galaxy formation physics, shown as different coloured lines. The upper and lower lines correspond to switching the GPU offloading off and on, respectively.}
\label{fig:scaling_Strong_Weak}
\end{figure}

\subsection{Performance}
\label{sec:Performance}

In Figure~\ref{fig:scaling_Strong_Weak}, we report the results for a combination of strong and weak scaling tests. We used a small number of initial time-steps from a set of hydrodynamical simulations including galaxy formation physics in 3 different cosmological volumes, reaching up to $4.8\times10^{10}$ particles, running up to the full size of {\tt SuperMUC-Phase2} at LRZ. Upper and lower curves correspond to switching off and on the GPU offloading, respectively. A strong scaling test is carried out by varying the number of MPI ranks by about one order of magnitude. We note that efficient weak scaling is always reached. Including the  GPU offloading provides a consistently faster execution, with a speed-up of approximately $2-3$, up to the maximum number of tested GPUs of almost $2000$.    

\begin{figure*}[ht]
 \centering
 \includegraphics[width=1.0\textwidth]{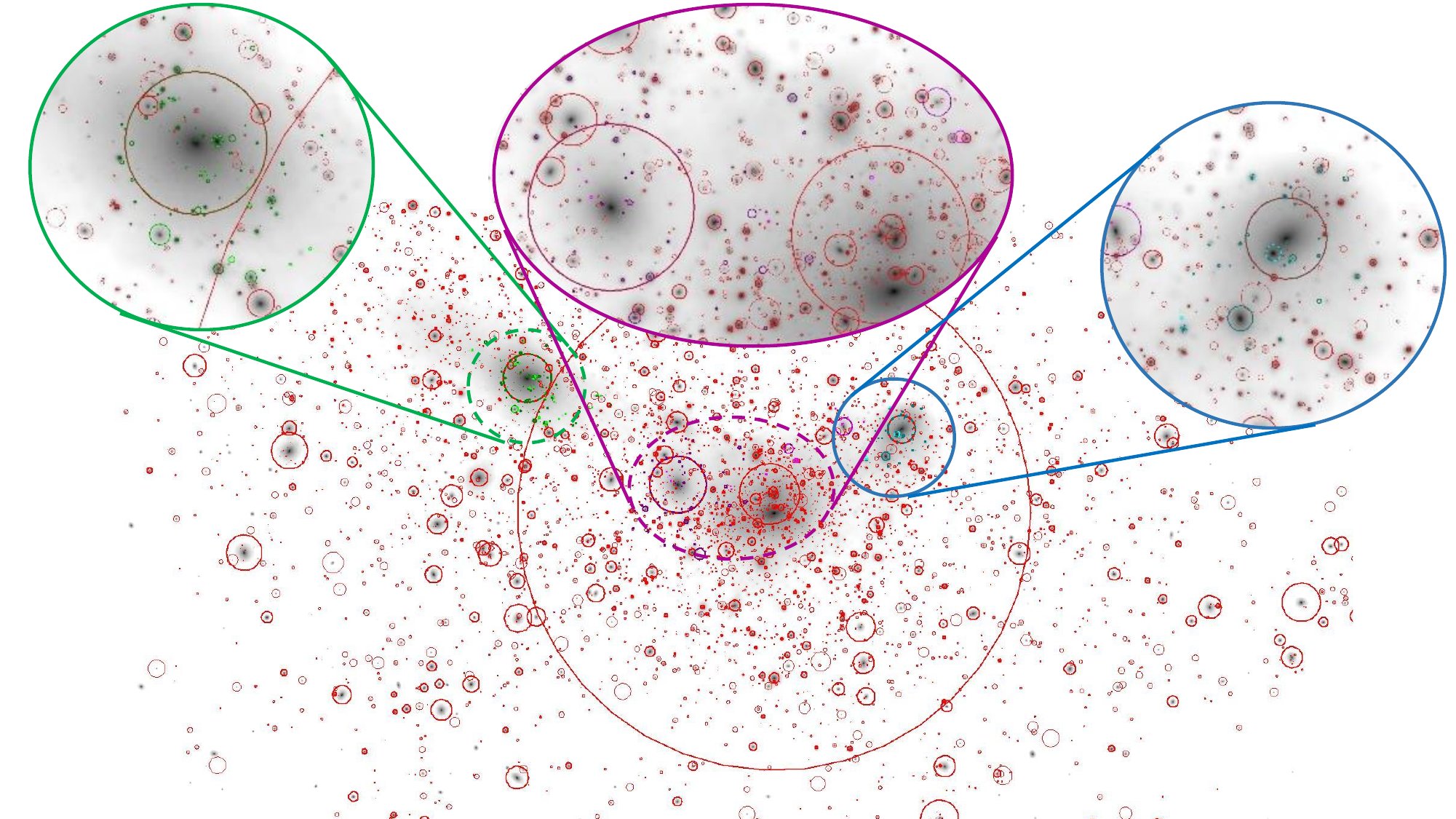}
\caption{Shown in the background is $6\times3$Mpc large stellar mass map, centred on a cluster of a high-resolution zoom-in simulation. The red circles which are over-plotted are corresponding to the half mass radii of the stellar bodies of the sub-structures identified by \textsc{SubFind}. Note that the main body (e.g. BCG) includes also the intra-cluster light. Three large, recently infallen groups are highlighted in green, magenta and blue. In the according zoom-in panels their member galaxies can be clearly seen, as these sub-structures which are associated to the groups and are marked by circles with the same colour. Even some of the these sub-structures have sub-sub-structures identified, marked by the lighter colours, representing tiny satellite galaxies of those group members.} \label{fig:subfind}
\end{figure*}


\section{The {\tt SubFind} halo and substructure finder}
\label{sec:subfind}

\og comes with the on-the-fly \textsc{SubFind} structure finder \citep{2001MNRAS.328..726S,2009MNRAS.399..497D} which identifies gravitationally bound, locally overdense regions within a parent halo. 

\subsection{FoF}
\label{sec:FoF}

Haloes are identified by a friend-of-friend (FoF) group finder, which is typically performed only on DM particles, while all other particle species are attached to their nearest DM particle. In contrast to the original \textsc{SubFind}, the standard FoF linking length is set to $0.16$ times the mean DM particle separation (with the often-used $0.2$). Note that this linking length is obtained when scaling the standard linking length of $0.2$ by $(\Delta_c/\Omega_m)^{1/3}$ according to the adopted cosmology and leads to groups having the overdensity characteristic of virialised objects predicted by the spherical collapse model in a concordance $\Lambda$CDM cosmology \citep[see][]{eke96}. 

\subsection{Sub-Haloes}
\label{sec:SubHaloes}

Density at the position of each particle is estimated from the position of all other particles via adaptive kernel interpolation. Following \cite{2009MNRAS.399..497D}, the density is estimated by summing up the contributions of each particle species independently, allowing for a much sharper contrast if an individual particle species dominates, e.g. the stellar component within a halo hosting a galaxy. 

Like in the original \textsc{SubFind} algorithm, for each particle the nearest neighbours are then considered to identify local overdensities through a topological approach that searches for saddle points in the isodensity contours within the global field of the halo. This is done in a top-down fashion, starting from the particle with the highest associated density and adding particles with progressively lower densities. If a particle has only denser neighbours in a single structure, it is added to this region. If it is isolated, it grows a new density peak, and if it has denser neighbours from two different structures, an isodensity contour that traverses a saddle point is identified. In the latter case, the two involved structures are joined and registered as candidate subhaloes if they contain a minimum of neighbouring particles. These candidates, selected according to the spatial distribution of particles only, are later processed for gravitational self-boundness. Particles with positive total energy are iteratively dismissed until only bound particles remain. In addition, for gas particles, the internal energy is considered when calculating the binding energy. The gravitational potential is computed with a tree algorithm, such that large haloes can be processed efficiently. If the remaining bound number of particles still exceeds the minimum number, the candidate is ultimately recorded as a subhalo. Sub structures which exclusively contain only gas particles are typically discarded. The (non-gas) particle with the smallest binding energy is assumed to be the centre of this sub-structure. Note that the unbinding step here also supports OpenMP parallelisation. In addition, this process can be performed in an MPI parallel fashion, where sub-groups of the MPI ranks compute the candidate lists for different haloes in parallel, while keeping the number of MPI ranks used for each halo to a minimal required number to optimise the performance.  The set of initial substructure candidates forms a nested hierarchy that is processed from inside out, allowing the detection of substructures within substructures. 

In general, a given particle may only become a member of one substructure, i.e. \textsc{SubFind} decomposes the initial group into a set of disjoint self-bound structures. Particles not bound to any genuine substructure are assigned to the “background halo”. This component is also checked for self-boundness, so that some particles that are not bound to any of the structures may remain. For all substructures as well as the main halo, the particle with the minimum gravitational potential is adopted as the (sub)halo centre. This creates a nested hierarchy of substructures, as exemplarily shown in figure \ref{fig:subfind}. 

For each sub-halo, various properties are computed and stored. However, should both gas and a stellar component be present, some of those properties originally defined for DM only simulations do not have a unique definition any longer. We therefore refer the reader to \citet{2009MNRAS.399..497D} for a detailed description of how values, e.g. the maximum circular velocity, are computed in the presence of baryons. These computations also follow a fully OpenMP/MPI hybrid parallelisation. 

For the main halo, \textsc{SubFind} additionally calculates spherical over-density values, taking all particles within such spheres into account. As an extension to the original \textsc{SubFind}, values are computed within 6 different radii, including the virial radius, $R_{500}$ and $R_{200}$, by considering both the critical and mean densities for the calculation. In addition, further observational properties like gas mass, X-ray luminosity, Sunyaev-Zeldovich and mean temperatures are computed for all the 6 different radii. The resulting output can be written in any of the existing output formats, and \og also allows the user to combine the output of different MPI ranks into single files, analogous to the standard snapshot files. 

\subsection{Galaxy properties}
\label{sec:SEDs}
\noindent
{\it Main contributing developer: I. Marini}

If the stellar evolution model is turned on, in addition to stellar half-mass radii, stellar spin, stellar and gas metallicities also luminosities in various bands will be added to the sub-halo properties. 

Assuming that each star particle in a simulation is representative of a single stellar population (SSP), with its mass, metallicity and redshift of formation, we can assign a luminosity measured in given bands. For this purpose, we pre-compiled an interpolation grid of luminosity values from GALAXEV \citep{Bruzual2003} by assuming Chabrier, Kroupa and Salpeter IMF. The grid interpolates for metallicity, initial stellar mass and logarithmic redshift, the latter to account for both rest-frame and observed magnitudes. Consistent with the stellar evolution model implemented in the simulation code, GALAXEV keeps track of stellar mass loss.  Standard values of metallicities in the interpolation are in the range $0.005-2.6\, Z_{\odot}$\footnote{The solar metallicity is set to $Z_{\odot}= 0.02$. }, therefore we decided to set the resulting magnitude to zero as a default answer in cases when the metallicity was above or below this interval. The final set of filters is composed of $97$ filters which include the Large Synoptic Survey Telescope, James Webb Space Telescope, Hubble Space Telescope, Euclid, Herschel, and Spitzer.


\begin{table*}[ht!]   
\centering
\def\arraystretch{1.3}
    \begin{tabular}{|c||c|c|c|c|c|c|c|c|c|c|c|c|c|c|c|c|c|c|}
        \hline
        & {\rotatebox[origin=c]{90}{GPU}} 
        & {\rotatebox[origin=c]{90}{OpenMP}} & 
        {\rotatebox[origin=c]{90}{GRAVITY}} & {\rotatebox[origin=c]{90}{PM (long-range)}} &
        {\rotatebox[origin=c]{90}{SIDM}} & {\rotatebox[origin=c]{90}{SPH}} &
        {\rotatebox[origin=c]{90}{MFM}} & {\rotatebox[origin=c]{90}{MHD}} & 
        {\rotatebox[origin=c]{90}{NS VISCOSITY}} &
        {\rotatebox[origin=c]{90}{CONDUCTION}} &
        {\rotatebox[origin=c]{90}{SHOCKFINDER}} & 
        {\rotatebox[origin=c]{90}{COOLING}} & {\rotatebox[origin=c]{90}{NE CHEMISTRY}} &
        {\rotatebox[origin=c]{90}{SF-SH03}} & {\rotatebox[origin=c]{90}{SF-LT07}} &
        {\rotatebox[origin=c]{90}{DUST (passive)}} & {\rotatebox[origin=c]{90}{~~DUST (dynamics)~~}} & 
        {\rotatebox[origin=c]{90}{AGN}} \\
        \hline
        \hline
        GPU (OpenACC) &
        \multirow{2}{*}{{\tiny
        \begin{tabular}{@{}l@{}}Sect. \\ \ref{sec:GPU} \end{tabular}}}
        & \X & \X & \x & \F & \X & \F & \X & \X & \X & \X & \x & \x & \x & \x & \x & \dd & \x \\
        \cline{1-1} \cline{3-19}
        GPU (OpenMP) 
        & & \X & \X & \x & \F & \X & \F & \X & \X & \X & \X & \x & \x & \F & \F & \F & \dd & \F \\
        \hline
        OpenMP &\multicolumn{2}{l|}{{\tiny\begin{tabular}{@{}l@{}}  
        Sect. \ref{sec:openmp}  \\
        ~
        \end{tabular} }}& \X & \X & \X & \X & \X & \X & \X & \X & \X & \x & \x & \x & \x & \x & \X & \X \\
        \hline
        GRAVITY &\multicolumn{3}{l|}{{\tiny\begin{tabular}{@{}l@{}} 
        Tree-based and/or analytic\\
        potentials, see \ref{sec:Tree} \& \ref{sec:Gravity_analytic_potentials}.
        \end{tabular} }}& \X & \X & \X & \X & \x & \x & \x & \x & \x & \x & \x & \x & \x & \x & \X \\
        \hline
        PM (long-range)&\multicolumn{4}{l|}{{\tiny\begin{tabular}{@{}l@{}} 
        Sect. \ref{sec:PM}. Precision is DOUBLE by\\
        default, can be nested or non-periodic.
        \end{tabular} }}& \X & \x & \x & \x & \x & \x & \x & \x & \x & \x & \x & \x & \x & \x \\
        \hline
        SIDM &\multicolumn{5}{l|}{{\tiny\begin{tabular}{@{}l@{}} 
        Sect. \ref{sec:SIDM}; contains various sub-modules and has \\
        additional time-step constraints. 
        \end{tabular} }} & \x & \x & \x & \x & \x & \x & \x & \x & \x & \x & \x & \x & \x \\
        \hline
        SPH &\multicolumn{6}{l|}{{\tiny\begin{tabular}{@{}l@{}} 
        Sect. \ref{sec:SPH}; various extensions to \\
        improve the hydrodynamical behaviour.
        \end{tabular} }} & \F & \X & \X & \X & \X & \X & \X & \X & \X & \X & \X & \X  \\
        \hline
        MFM &\multicolumn{7}{l|}{{\tiny\begin{tabular}{@{}l@{}} 
        Sect. \ref{sec:MFM}; even with MFM active, the SPH hydro loop\\
        is performed to capture additional fluid processes.
        \end{tabular} }} & \dd & \X & \X & \dd & \X & \x & \X & \X & \X & \X & \X  \\
        \hline       
        MHD &\multicolumn{8}{l|}{{\tiny\begin{tabular}{@{}l@{}} 
        Sect. \ref{sec:MHD}; MHD typically is a quite complex extension of hydrodynamics, so care\\
        is needed, especially as it often requires explicit coupling to additional physics.
        \end{tabular} }} & \X & \X & \X & \x & \x & \X & \X & \x & \dd & \x  \\
        \hline       
        NS VISCOSITY &\multicolumn{9}{l|}{{\tiny\begin{tabular}{@{}l@{}} 
        Sect. \ref{sec:NavierStokes}. Navier-Stokes (NS) viscosity has severe additional time-step constraints, slowing \\
        down the code. Coupling to MHD requires additional settings (see Sect \ref{sec:braginski}).
        \end{tabular} }} & \x & \x & \x & \x & \x & \x & \x & \dd & \x  \\
        \hline       
        CONDUCTION &\multicolumn{10}{l|}{{\tiny\begin{tabular}{@{}l@{}} 
        Sect. \ref{sec:cond}; introduces one more SPH-like loop, thus slowing down the code\\
        Coupling to MHD requires additional settings (see sect \ref{sec:cond_aniso}).
        \end{tabular} }} & \x & \x & \x & \x & \x & \x & \x & \x  \\
        \hline       
        SHOCKFINDER &\multicolumn{11}{l|}{{\tiny\begin{tabular}{@{}l@{}} 
        Sect. \ref{sec:shockfinder}; it reconstructs discontinuities, independent of the hydro solver. A specialized shock-finder for MFM\\
        is under development. Always check the definition of $c_s$ for interpreting $\mathcal{M}_s$ when additional physics is on. 
        \end{tabular} }} & \x & \x & \x & \x & \x & \x & \x  \\
        \hline       
        COOLING &\multicolumn{12}{l|}{{\tiny\begin{tabular}{@{}l@{}} 
        Sect. \ref{sec:Cool}; different star-formation (SF) modules support different treatment of cooling. Modules like CHEMISTRY and\\
        DUST add additional contributions to the standard cooling functions. The code wiki provides cooling tables to download. 
        \end{tabular} }} & \X & \X & \X & \X & \x & \X  \\
        \hline       
        NE CHEMISTRY &\multicolumn{13}{l|}{{\tiny\begin{tabular}{@{}l@{}} 
        Sect. \ref{sec:CoolNEq}; different contributions of the non-equilibrium (NE) chemical network to the cooling must be switched on separately\\
        (see the corresponding section in the code wiki). 
        \end{tabular} }} & \x & \X & \F & \x & \x  \\
        \hline       
        SF-SH03 &\multicolumn{14}{l|}{{\tiny\begin{tabular}{@{}l@{}} 
        Sect. \ref{s:sf}; the SH03 model has only a crude description of metal enrichment and does not support metal-dependent cooling. \\
        Coupling to MHD requires additional settings (see sect \ref{sec:SN_B_seeding}).
        \end{tabular} }} & \F & \F & \x & \X  \\
        \hline       
        SF-LT07 &\multicolumn{15}{l|}{{\tiny\begin{tabular}{@{}l@{}} 
        Sect. \ref{sec:StEv}; the LT07 model needs additional data and input files (see the corresponding section in the code wiki, where also links for downloading all aux-\\
        iliary files are given). In combination with SubFind, this model can generate galaxy luminosities in different photometric bands from pre-computed SEDs.  
        \end{tabular} }} & \X & \x & \X  \\
        \hline       
        DUST (passive) &\multicolumn{16}{l|}{{\tiny \begin{tabular}{@{}l@{}}
        Sect. \ref{s:dust1}; parts and extensions of this module explicitly couple to stellar evolution and chemical enrichment, thus needing LT07 to be active.\\ 
        Additional auxiliary files are provided on the code wiki for download.
        \end{tabular} }} & \F & \x  \\
        \hline       
        DUST (dynamics) &\multicolumn{17}{l|}{{\tiny \begin{tabular}{@{}l@{}}
        Sect. \ref{sec:dust_dynamics}: explicit dust dynamics can be used in both SPH and MFM hydro solvers.\\ 
        With this model on, the gas density block carries the total (dust+gas) density, so this needs to be treated explicitly in other modules.
        \end{tabular} }} & \x  \\
        \hline       
        AGN &\multicolumn{18}{l|}{{\tiny \begin{tabular}{@{}l@{}}
        Sect. \ref{sec:BH}; includes neighbour search for BH particles in SPH-like loops and extends the neighbour list to contain BH particles.\\ 
        The dynamical friction module (Sect. \ref{sec:BHDynFr}) extends the neighbour list to contain star particles in SPH-like loops and modifies the gravitational acceleration of BH particles.
        \end{tabular} }} \\
        \hline
    \end{tabular}
    \caption{Matrix describing the compatibility of different modules within \og. In the upper triangle, each cell reports the level of compatibility between two modules. Different markers correspond to different levels of compatibility. (\X): is working explicitly together or contains some features which work explicitly together; (\x): can be combined but do not talk to each other; (\F): cannot be switched on simultaneously for technical reasons; (\dd): development to enable simultaneous usage is ongoing. 
    Typical examples of (\x) are GPU-Offloading and the hybrid MPI/OpenMP implementations: there are modules that can be switched on when using GPU-Offloading or hybrid MPI/OpenMP, but do not make use of them explicitly. In the lower triangle, we provide specific comments and references to specific sections for each module. 
    }
    \label{tab:module_matrix}
\end{table*}

\section{Conclusion}
\label{sec:Summary}
In this paper, we have presented \og, a code for large-scale hydrodynamical simulations, especially in cosmological contexts. It is designed to follow the coupled evolution of dark matter (either collisionless or self-interacting) and collisional baryons across the wide range of scales involved in non-linear structure formation and allows us to capture a variety of physical processes across a wide range of astrophysical applications. While being based on the original public \textsc{Gadget}-2 \citep{Springel2002} and \textsc{Gadget}-3 code, the \og code extended such codes in many different aspects: besides including a thorough investigation of residual bugs and inconsistencies, it made sure that they are all fixed and propagated in a shared code version; the backbone of the code is based on a hybrid, MPI/OpenMP implementation which makes the code highly optimized to handle very large simulation setups; it allows to offload certain modules to GPUs using the OpenACC or OpenMP6.0 framework. Unlike several existing codes, which have been distributed in reduced form with only a subset of their physics modules, \og is released including a large number of extra modules. In addition, the released version of \og will be regularly updated against the developer version, so that bug fixes, improvements and new modules will constantly be part of the \og release. 

As for the gravity solver, \og combines an Oct-Tree algorithm \citep{BarnesHut1986} as well as a particle mesh (PM) solver for the long-range and short-range gravitational forces. Optional modules allow including the effect of $\Lambda$CDM extensions, such as massive neutrinos, a time-dependent Dark Energy equation of state, and SIDM. 

As for the hydro sector, the code allows for different hydrodynamical solvers, i.e. an advanced SPH scheme and an MFM solver, along with extensions to capture MHD, thermal conduction and viscosity. In addition, \og provides a set of extra modules to describe the physical processes which form the observable Universe: radiative cooling and heating, star formation, stellar evolution with the associated chemical enrichment and supernova feedback, and the growth of supermassive black holes including the effect of unresolved dynamical friction on their orbits and merging, the evolution of their spin, and energy feedback mechanisms. Additional modules allow extending the calculation to chemical networks and describing dust production and evolution. 

Table \ref{tab:module_matrix} provides a summary of the different modules which are part of this initial release, and the current possibilities to combine them, together with some general notes for the individual modules. In addition, several more physical modules are currently under development and will be part of the \og release together with their presentation papers at a later time. The code wiki will always provide an updated version of this table, monitoring improvements of the possible combination of different modules and also reflecting new modules which will be part of the \og release. We also encourage other colleagues who want to develop new physics modules within \og to join the developer branch for their development work, making use of the extended CI pipelines, streamline their code towards the \og philosophy and code etiquette. This will also allow them to make their module part of \og by publicly releasing it at any time, given that the extended developer CI demonstrated compatibility and non-interference with the existing code.

The complexity of simulating astrophysical systems often requires combining various different physical processes. Therefore, we want to remind the reader that although sometimes it is technically possible to combine different physics modules, it is not guaranteed that this always leads to consistent results. Just as a simple example, pre-computed cooling tables provided with the code might be implicitly inconsistent or can lead to double counting of cooling processes when the explicit treatment of additional coolants is used. For this reason, we encourage the users to carefully read the original papers where the physics modules within \og are presented, whenever using them for a specific problem, to ensure that the combination of different modules meaningfully captures the processes of interest. We also remind the code users that some of the individual modules can be highly complex and therefore can contain several sub-levels of details which are controlled at compile- or run-time. The meaning of the corresponding compilation switches and input parameters is described in detail in the original papers which presented such \og modules.
  
The complete source, including all physics modules described here, including several test cases, as well as the documentation, is available at \url{https://gitlab.lrz.de/AstroCodes/OpenGadget3} under the GNU General Public License v3. We consider this of high importance for at least two reasons. First, reproducibility of simulation results requires access to the same code paths that produced them. Second, the highly complex research field of computational astrophysics and cosmology thrives when research groups can use and contribute to a shared, robust code infrastructure, moving past the need to build foundational tools from scratch.

\section*{Acknowledgements}

GSK, AR, LDF and LT acknowledge support from the Scalable Parallel Astrophysical Codes for Exascale (SPACE) Centre of Excellence, which received funding from the European High Performance Computing Joint Undertaking (JU) and Belgium, Czech Republic, France, Germany, Greece, Italy, Norway, and Spain under grant agreement No 101093441.

SB, AD, AR, LT and MV acknowledge support from the Fondazione ICSC National Recovery and Resilience Plan (PNRR) Project ID CN-00000013 "Italian Research Center on High-Performance Computing, Big Data and Quantum Computing" funded by MUR Missione 4 Componente 2 Investimento 1.4: "Potenziamento strutture di ricerca e creazione di campioni nazionali di R\&S (M4C2-19 )" - Next Generation EU (NGEU). 

SB and LT acknowledge support by the National Recovery and Resilience Plan (NRRP), Mission 4, Component 2, Investment 1.1, Call for tender No. 1409 published on 14.9.2022 by the Italian Ministry of University and Research (MUR), funded by the European Union – NextGenerationEU–Project Title "Space-based cosmology with Euclid: the role of High-Performance Computing" – CUP J53D23019100001 - Grant Assignment Decree No. 962 adopted on 30/06/2023 by the Italian Ministry of Ministry of University and Research (MUR). 

KD, LMB, FG, MSF, TMG and GTP acknowledge support from the COMPLEX project from the European Research Council (ERC) under the European Union’s Horizon 2020 research and innovation program grant agreement ERC-2019-AdG 882679.

KD, LMB, FG, MSF, LS, LPS and GTP acknowledge support from the Deutsche Forschungsgemeinschaft (DFG, German Research Foundation) under Germany’s Excellence Strategy - EXC-2094 - 390783311.

SB and MV acknowledge partial financial support from the INFN Indark Grant. 

SB, AD and MV acknowledge the "Astrofisica Fondamentale 2024" INAF Grant "Black Hole Dynamics and Galaxy Formation from Cosmological Simulations". 

MSF gratefully acknowledges the support of the Alexander von Humboldt Foundation through a Feodor Lynen Research Fellowship.

TMG acknowledges the support provided by a Smithsonian Scholar Award.

AR and LT acknowledge the use of the INAF IT framework~\citep{Bertocco_2020, Taffoni_2020}.

AR acknowledges the EuroHPC Joint Undertaking for awarding the project ID EHPC-REG-2024R01-029 access to Leonardo at CINECA(also with EHPC-DEV-2026D07-143), Italy; and for awarding us access to MareNostrum5 at BSC, Spain (ID EHPC-BEN-2025B11-033) and LUMI at CSC, Finland (ID EHPC-DEV-2026D02-240).

AR acknowledges ISCRA for awarding this project access to the LEONARDO supercomputer, owned by the EuroHPC Joint Undertaking, hosted by CINECA, Italy (HP10BUFI59).

LT acknowledges the EuroHPC Joint Undertaking for awarding the project IDs EHPC-DEV-2023D04-035, EHPC-DEV-2024D06-062, EHPC-BEN-2026B06-135, EHPC-DEV-2026D07-227 on LUMI-C/G, MareNostrumV-GPP/ACC, Leonardo DCGP/Booster, Karolina CPU/GPU, Meluxina CPU/GPU.

LMB acknowledges financial support from NASA through grant 80NSSC24K0173 and the NSF through grant AST-2510951.

LDS gratefully acknowledges support from the Deutsche Forschungsgemeinschaft (DFG, German Research Foundation) through the Collaborative Research Centre ``Neutrinos and Dark Matter in Astro- and Particle Physics'' (SFB 1258 -- 283604770).

CS and JB acknowledge the SICyT and its administration team for supercomputing time on CLEMENTINA XXI under project PCI-92.

NH contributions to the work described in this paper were made during their employment at Intel and using Intel resources.

The authors gratefully acknowledge the Gauss Centre for Supercomputing e.V. (www.gauss-centre.eu) for funding this project by providing computing time on the GCS Supercomputer SuperMUC-NG at Leibniz Supercomputing Centre (www.lrz.de).

\onecolumn
\appendix
\begingroup
\renewcommand{\addcontentsline}[3]{}

\section{Additional Information}
\subsection{Accuracy settings}
\label{sec:app_back}
While for cosmological simulation and the gravity computation the usage of Float64 data (especially for position, velocity and acceleration is recommended to stay better than the per cent precision needed for modern simulations (see right part of Figure \ref{fig:app_mhd_precission}, the precision has less impact when doing hydro computation, as the forces are local results with less summation over less dynamical range. So even an MHD shock tube test is not very sensitive to the accuracy of the SPH data, and there is hardly a difference visible between the different choices of base type (beyond the Float16 case, of course), as shown in Figure \ref{fig:app_mhd_precission}.  
\begin{figure*}
 \centering
 \includegraphics[width=0.49\textwidth]{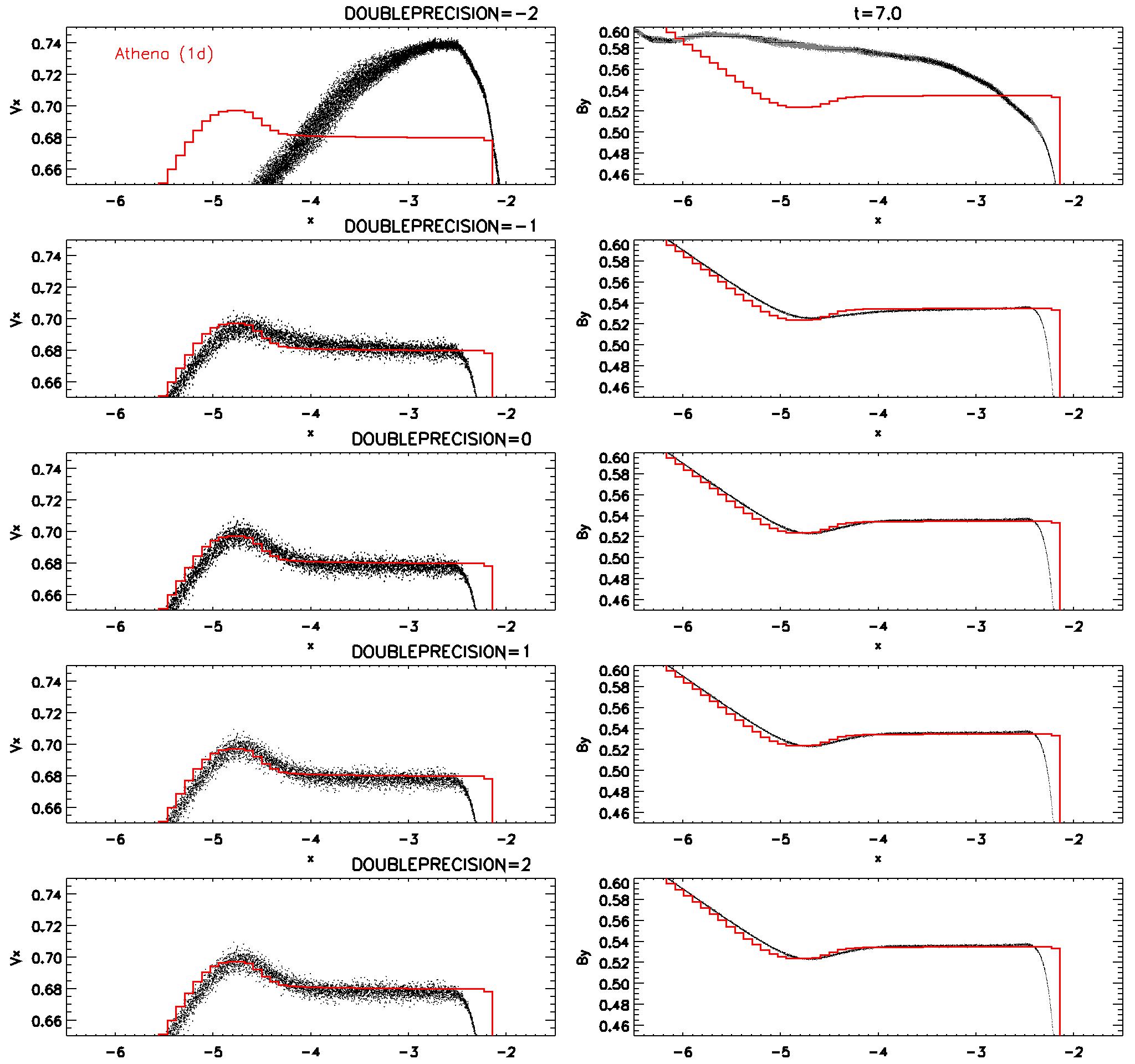}
 \includegraphics[width=0.49\textwidth]{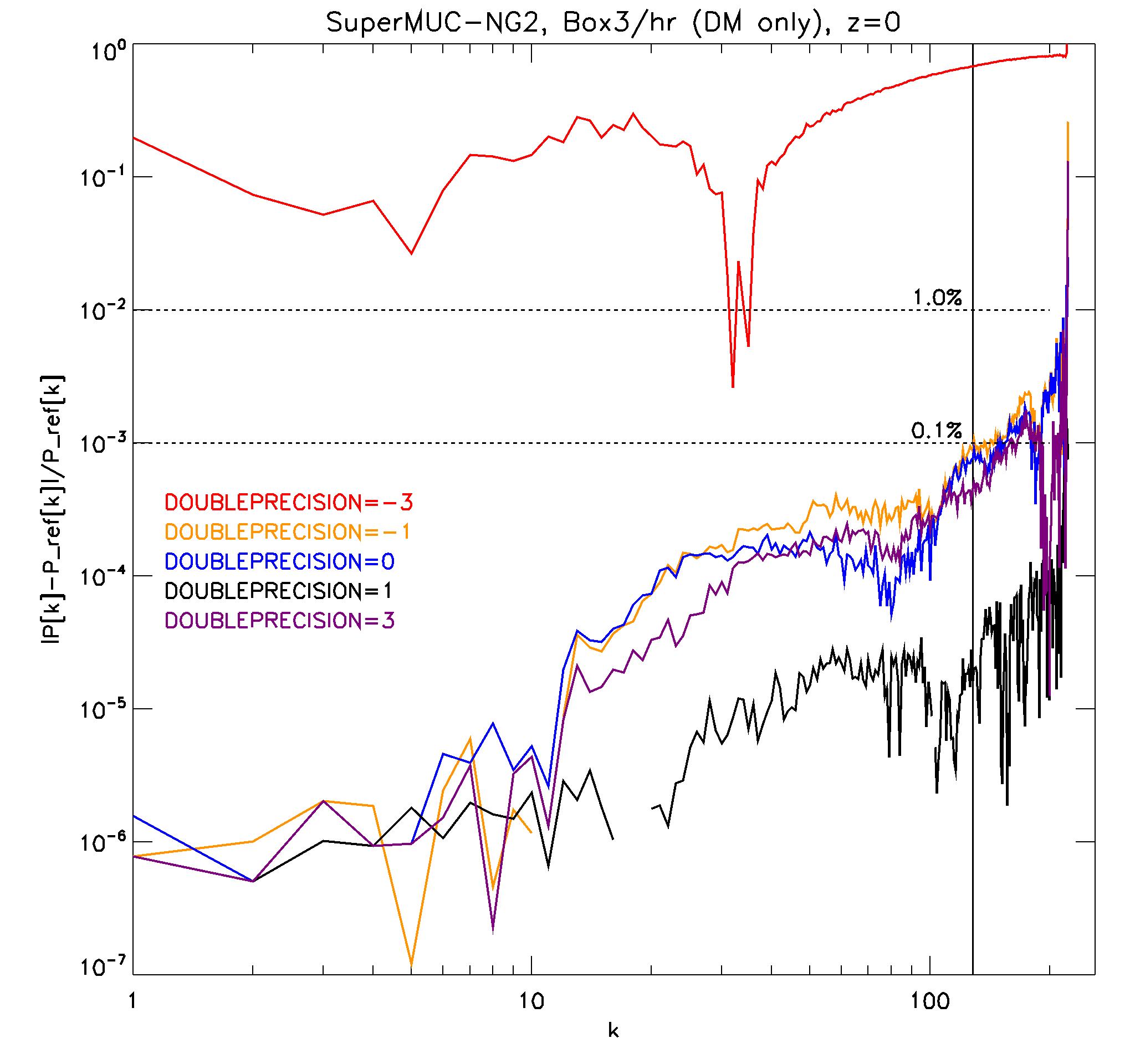}
 \caption{Left panel: Same as Figure \ref{fig:mhd_shock} but highlighting the minor effect (except for Float16) of precision on the local, hydrodynamical computations. Right panel: Comparison of the power spectrum for a cosmological box varying the precision. Note that we used the Float128 setting as a reference here. }
 \label{fig:app_mhd_precission}
\end{figure*}

\subsection{Available kernel functions}
\label{app:Kern}

Below we list the different, available kernel functions of \og. Figure \ref{fig:app_mhd_kernels} shows a zoom onto the MHD test problem to illustrate the effect of the kernel on the reproduction of some of the features. Note that we generally recommend to use at least the \textit{Wendland C4} kernel.

\vspace{0.3truecm}
\textit{Cubic spline:}
\begin{equation}
    W(r,h) = \begin{cases}
        \frac{4}{3h} & 1\mathrm{D} \\
        \frac{40}{7\uppi h^2} & 2\mathrm{D} \\
        \frac{8}{\uppi h^3} & 3\mathrm{D}
    \end{cases}
    \cdot \begin{cases}
        1 + 6 \left(\frac{r}{h}-1\right)\left(\frac{r}{h}\right)^2 & \frac{r}{h}<0.5 \\
        2 \left(1-\frac{r}{h}\right)^3 & \frac{r}{h} \geq 0.5
    \end{cases}
\end{equation}
\begin{equation}
    \nabla W(r,h) = \begin{cases}
        \frac{4}{3h^2} & 1\mathrm{D} \\
        \frac{40}{7\uppi h^3} & 2\mathrm{D} \\
        \frac{8}{\uppi h^4} & 3\mathrm{D}
    \end{cases}
    \cdot \begin{cases}
        \frac{r}{h}\left(18\frac{r}{h}-12\right) & \frac{r}{h} < 0.5 \\
        -6\left(1-\frac{r}{h}\right)^2 & \frac{r}{h} \ge 0.5
    \end{cases}
\end{equation}

\vspace{0.3truecm}
\textit{Quintic spline:}
\begin{equation}
    W(r,h) = \begin{cases}
        \frac{243}{40h} &1\mathrm{D} \\
        \frac{15309}{478\uppi h^2} &2\mathrm{D} \\
        \frac{2187}{40\uppi h^3} &3\mathrm{D}
    \end{cases} \cdot
    \begin{cases}
        \left(1-\frac{r}{h}\right)^5 - 6\left(\frac{2}{3}-\frac{r}{h}\right)^5 +15\left(\frac{1}{3}-\frac{r}{h}\right)^5 & \frac{r}{h} < \frac{1}{3} \\
        \left(1-\frac{r}{h}\right)^5 - 6\left(\frac{2}{3}-\frac{r}{h}\right)^5 & \frac{1}{3} \le \frac{r}{h} < \frac{2}{3} \\
        \left(1-\frac{r}{h}\right)^5 & \frac{2}{3} \le \frac{r}{h}
    \end{cases}
\end{equation}
\begin{equation}
    \nabla W(r,h) = \begin{cases}
        \frac{243}{40h^2} &1\mathrm{D} \\
        \frac{15309}{478\uppi h^3} &2\mathrm{D} \\
        \frac{2187}{40\uppi h^4} &3\mathrm{D}
    \end{cases} \cdot
    \begin{cases}
        -5\left(1-\frac{r}{h}\right)^4 + 30\left(\frac{2}{3}-\frac{r}{h}\right)^4 - 75\left(\frac{1}{3}-\frac{r}{h}\right)^4 & \frac{r}{h} < \frac{1}{3} \\
        -5\left(1-\frac{r}{h}\right)^4 + 30\left(\frac{2}{3}-\frac{r}{h}\right)^4 & \frac{1}{3} \le \frac{r}{h} < \frac{2}{3} \\
        -5\left(1-\frac{r}{h}\right)^4& \frac{2}{3} \le \frac{r}{h}
    \end{cases}
\end{equation}

\vspace{0.3truecm}
\newpage
\textit{Wendland C2:}

\begin{equation}
    W(r,h) = \begin{cases}
        \frac{5}{4 h} \left( 1 - \frac{r}{h} \right)^3 \left(1 + \frac{3r}{h} \right) \quad  \\
        \frac{7}{\uppi h^2} \left( 1 - \frac{r}{h} \right)^4 \left(1 + \frac{4r}{h} \right) \quad  \\
        \frac{21}{2\uppi h^3} \left( 1 - \frac{r}{h} \right)^4 \left(1 + \frac{4r}{h} \right) \quad 
    \end{cases}
    \nabla W(r,h) = \begin{cases}
        \frac{5}{4 h^2} \left(-12 \frac{r}{h} \right) \left( 1 - \frac{r}{h} \right)^2  \quad &1\mathrm{D} \\
        \frac{7}{\uppi h^3} \left(-20 \frac{r}{h} \right) \left( 1 - \frac{r}{h} \right)^3 \quad &2\mathrm{D} \\
        \frac{21}{2\uppi h^4} \left(-20 \frac{r}{h} \right) \left( 1 - \frac{r}{h} \right)^3 \quad &3\mathrm{D}
    \end{cases}    
\end{equation}

\vspace{0.3truecm}
\textit{Wendland C4:}
\begin{equation}
    W(r,h) = \begin{cases}
        \frac{3}{2 h} \left( 1 - \frac{r}{h} \right)^5 \left(1 + 5\frac{r}{h} + 8 \left(\frac{r}{h}\right)^2 \right) \quad  \\
        \frac{9}{\uppi h^2} \left( 1 - \frac{r}{h} \right)^6 \left(1 + 6\frac{r}{h} + \frac{35}{3} \left(\frac{r}{h}\right)^2 \right) \quad  \\
        \frac{495}{32\uppi h^3} \left( 1 - \frac{r}{h} \right)^6 \left(1 + 6\frac{r}{h} + \frac{35}{3} \left(\frac{r}{h}\right)^2 \right) \quad 
    \end{cases}
    \nabla W(r,h) = \begin{cases}
        \frac{3}{2 h^2} \left( 1 - \frac{r}{h} \right)^4 \left(-14 \frac{r}{h} - 56 \left(\frac{r}{h}\right)^2 \right) \quad &1\mathrm{D} \\
        \frac{9}{\uppi h^3} \left( 1 - \frac{r}{h} \right)^5 \left(-\frac{56}{3} \frac{r}{h} - \frac{288}{3} \left(\frac{r}{h}\right)^2 \right) \quad &2\mathrm{D} \\
        \frac{495}{32\uppi h^4} \left( 1 - \frac{r}{h} \right)^5 \left(-\frac{56}{3} \frac{r}{h} - \frac{288}{3} \left(\frac{r}{h}\right)^2 \right) \quad &3\mathrm{D}
    \end{cases}
\end{equation}

\vspace{0.3truecm}
\textit{Wendland C6:}

\begin{equation}
    W(r,h) = \begin{cases}
        \frac{55}{32 h} \left( 1 - \frac{r}{h} \right)^7 \left(1 + 7\frac{r}{h} + 19 \left(\frac{r}{h}\right)^2 + 21 \left(\frac{r}{h}\right)^3 \right) \quad  \\
        \frac{78}{7\uppi h^2} \left( 1 - \frac{r}{h} \right)^8 \left(1 + 8\frac{r}{h} + 25 \left(\frac{r}{h}\right)^2 + 32 \left(\frac{r}{h}\right)^3 \right) \quad  \\
        \frac{1365}{64\uppi h^3} \left( 1 - \frac{r}{h} \right)^8 \left(1 + 8\frac{r}{h} + 25 \left(\frac{r}{h}\right)^2 + 32 \left(\frac{r}{h}\right)^3 \right) \quad 
    \end{cases}
    \nabla W(r,h) = \begin{cases}
        \frac{55}{32 h^2} \left( 1 - \frac{r}{h} \right)^6 \left(-6\frac{r}{h}\right) \left(1 + 18\frac{r}{h} + 35 \left(\frac{r}{h}\right)^2 \right) \quad &1\mathrm{D} \\
        \frac{78}{7\uppi h^3} \left( 1 - \frac{r}{h} \right)^7 \left(-22\frac{r}{h}\right) \left(1 + 7\frac{r}{h} + 16 \left(\frac{r}{h}\right)^2 \right) \quad &2\mathrm{D} \\
        \frac{1365}{64\uppi h^4} \left( 1 - \frac{r}{h} \right)^7 \left(-22\frac{r}{h}\right) \left(1 + 7\frac{r}{h} + 16 \left(\frac{r}{h}\right)^2 \right) \quad &3\mathrm{D}
    \end{cases}
\end{equation}

\begin{figure*}[t]
 \centering
 \includegraphics[width=1.0\textwidth]{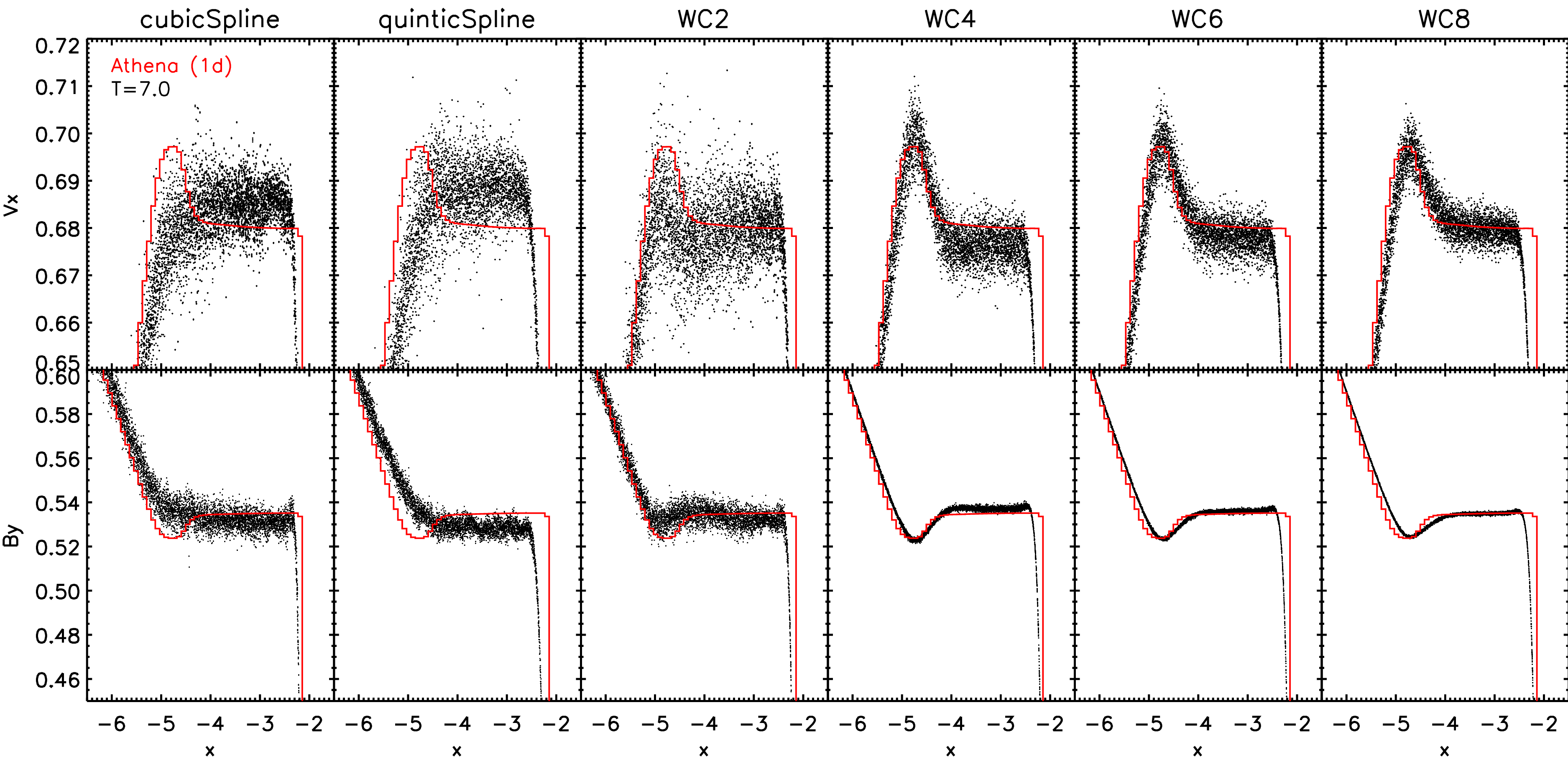}
 \caption{Same as figure \ref{fig:mhd_shock} but highlighting the effect of the choice of the different kernels.}
 \label{fig:app_mhd_kernels}
\end{figure*}

\vspace{0.3truecm}
\textit{Wendland C8:}
\begin{align}
    W(r,h) = \mathcal{N}_\mathrm{dim} \left( 1 - \frac{r}{h} \right)^{10} \left(5 + 50\frac{r}{h} + 210 \left(\frac{r}{h}\right)^2
    + 450 \left(\frac{r}{h}\right)^3 + 429 \left(\frac{r}{h}\right)^4 \right) 
\end{align}

\begin{equation}
    \nabla W(r,h) =  \mathcal{N}_\mathrm{dim} h^{-1} \left( 1 - \frac{r}{h} \right)^{9} \left(-26\frac{r}{h}\right) \left(5 + 45\frac{r}{h} + 159 \left(\frac{r}{h}\right)^2 + 231 \left(\frac{r}{h}\right)^3 \right) \,,
\end{equation}
where the normalisation factor reads
\begin{equation}
    \mathcal{N}_\mathrm{dim} = \begin{cases}
        \frac{8}{3\uppi h^3}\quad &2\mathrm{D} \\
        \frac{357}{64\uppi h^4}  \quad &3\mathrm{D}
    \end{cases}
\end{equation}

\subsection{Available slope limiters for MFM}
\label{app:MFM}
\og contains seven different slope limiters, reducing the slopes according to $\nabla \myvector{W}_{i,k}\to \alpha_{i,k}\nabla \myvector{W}_{i,k},\alpha_{i,k}\in [0,1]$.

At first, zeroth-order interpolation can be used:
\begin{align}
    \alpha_{i,k}^{\text{ZERO SLOPES}} =~& 0
\end{align}
Alternatively, no slope limiter can be employed:
\begin{align}
    \alpha_{i,k}^{\text{NULL}} =~& 1.
\end{align}
The TVD scalar limiter \citep{Duffell&MacFadyen2011} is designed mainly for strong shocks, but is very diffusive:
\begin{align}
    \alpha_{i,k}^{\text{TVD SCALAR}} =~& \min_{j\in \text{Ngb}}\max\mathcases{0 \\ \min\mathcases{1\\ \dif W_{ij,kk}/\dif W_k}}
\end{align}
where $\dif \myvector{W}_{ij}=\myvector{W}_j - \myvector{W}_i$, $\dif \myvector{W}=\dif \myvector{r}_{ij}\cdot \nabla\outerprod \myvector{W}$.

An alternative is the scalar limiter \citep{Balsara2004, Gaburov&Nitadori2011, Hubber+2018}, which loses the TVD behaviour but is less diffusive. It sets
\begin{align}
    \alpha_{i,k}^{\text{SCALAR}} =~& \max\mathcases{0 \\ \min\mathcases{1 \\ \min\mathcases{\frac{\dif W_{k,\max}}{\abs{\dif r}_{\max}\abs{\nabla W_k}}\\ \frac{\dif W_{k,\min}}{\abs{\dif r}_{\max}\abs{\nabla W_k}} }}}
\end{align}
where 
$\dif W_{k,\min/\max}= \abs{W_{i,k}-\min/\max_{j\in \text{Ngb}}W_{j,k}}$, and $\abs{\dif r}_{\max}= \max\left( \max_{j\in\text{Ngb}}\abs{r_{ij}}, h_i \right)$.

In the \textsc{Arepo} code \citep{Springel2010}, the slope is limited using
\begin{align}
    \alpha_{i,k}^{\text{\textsc{Arepo}}} =~& \min_{i\in \text{Ngb}}\mathcases{\dif W_{k,\max}/\dif W_k~&\text{if }\dif W_k>0\\\dif W_{k,\min}/\dif W_k~&\text{if }\dif W_k<0 \\ 1& \dif W_k=0.}
\end{align}

Our default case is based on the gizmo code \citep{Hopkins2015}:
\begin{align}
    \alpha_{i,k}^{\text{\textsc{gizmo}}} =~& \min\mathcases{1\\ \beta_i\min\mathcases{\frac{\dif W_{k,\text{max}}}{0.5h_i\abs{\nabla W_k}}\\ \frac{\dif W_{k,\min}}{0.5h_i\abs{\nabla W_k}}.}} \label{eq:gizmo_limiter}
\end{align}
The parameter $\beta$ has to be $\beta_i>0.5$ to ensure second-order stability. We use the suggested value $\beta=2$ from \citet{Hopkins2015}.

The pairwise limiter described by \citet{Hopkins2015} limits the already interpolated face values. The aim is to directly calculate the face value $W_{ij,k}^{\text{new}}$, starting from the extrapolated value $W_{ij,k}^{\text{face}}$, possibly already with limited gradients. If $W_{i,k}=W_{j,k}$, the face value is just chosen to be the same as the cell values, $W_{ij,k}^{\text{new}}=W_{i,k}$.
Otherwise, the values
\begin{align}
    \delta_1 =~& \psi_1\abs{W_{i,k}-W_{j,k}} \label{eq:gizmo_pairwise_gamma1}\\
    \delta_2 =~& \psi_2\abs{W_{i,k}-W_{j,k}}
\end{align}
are calculated. The free parameters $\psi_{1/2}$ are tuned to $\psi_1=0.5$, $\psi_2=0.25$.
A simple intermediate value used later is given by
\begin{align}
    \bar W_{ij,k} =~& W_{i,k}+\frac{\dif r_{ij}}{\dif r_{i}^{\text{face}}}(W_{j,k}-W_{i,k}).
\end{align}
The maximum/minimum value is $W_{k,\min/\max}=\min/\max(W_{i,k},W_{j,k})$.
Depending on how the two face values compare, the new face value is calculated:
If $W_{i,k}<W_{j,k}$, then
\begin{align}
    W_{ij,k}^{\text{new}} =~& \max\mathcases{\mathcases{W_{k,\min}-\delta_1~&\text{if }\text{SIGN}(W_{k,\min}-\delta_1)=SIGN(W_{k,\min}) \\ \frac{W_{k,\min}}{1+\frac{\delta_1}{\abs{W_{k,\min}}}}~&\text{else}} \\ \min\mathcases{W_{ij,k}^{\text{face}} \\ \bar W_{ij,k}+\delta_2.}}
\end{align}
If $W_{i,k}\ge W_{j,k}$, then
\begin{align}
    W_{ij,k}^{\text{new}} =~& \min\mathcases{\mathcases{W_{k,\max}+\delta_1~&\text{if }\text{SIGN}(W_{k,\max}+\delta_1)=SIGN(W_{k,\max}) \\ \frac{W_{k,\max}}{1+\frac{\delta_1}{\abs{W_{k,\max}}}}~&\text{else}}\\
    \max\mathcases{W_{ij,k}^\text{face} \\ \bar W_{ij,k}-\delta_2.}}
\end{align}
The same limiter is applied for particle $j$. Finally, the \textsc{gizmo} code uses a slightly different pairwise limiter. Depending on the tolerance $t$ chosen, the parameters
\begin{align}
    \psi_1 =~& \mathcases{0 \\ 0.5 \\ 0.75} ~~~~~~
    \psi_2 = \mathcases{0 &~~~t=0 \\ 0.4 &~~~t=1 \\ 0.375~&~~~t=2}
\end{align}
are defined. To calculate $\bar W_{ij,k}$, the factor $\dif r_{ij}/\dif r_i^{\text{face}}$ is approximated by the first order value $0.5$. Except for these differences, the limiter is identical to the already described one.
In our implementation, we apply the limiter in the reference frame of the interface, such that the velocity is a relative velocity. This makes the limiter Lagrangian and increases the symmetry between different directions.

\subsection{Artificial Viscosity for low Mach number shocks}

In section \ref{sec:SPH:artvisc} we used a Mach 10 shock to validate the different artificial viscosity settings. Here we show in figure \ref{fig:mach2} in addition, how the different settings for artificial viscosity perform in a low Mach number shock, where we have chosen a Mach 2 shock setup.

\begin{figure}[h]
 \includegraphics[width=1.0\textwidth]{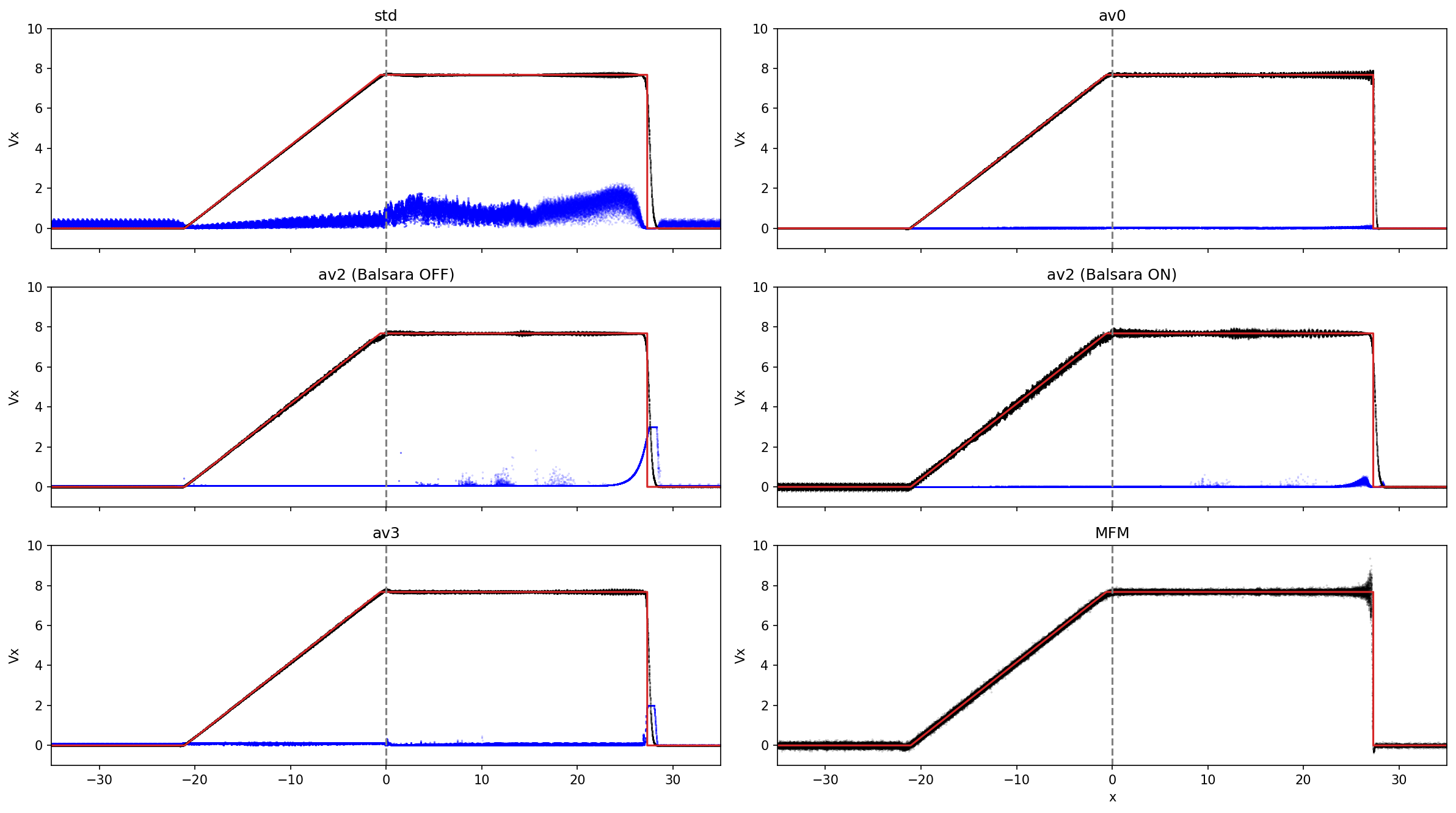}
  \caption{Same as figure \ref{fig:mach10} but for a Mach 2 shock tube. }
 \label{fig:mach2}
\end{figure}

\endgroup
\twocolumn

\bibliographystyle{elsarticle-harv} 
\bibliography{sources}

\end{document}